\documentclass[a4paper]{scrartcl}

\usepackage{amsmath,amsthm,amssymb,dsfont,extarrows,amscd}

\usepackage{tikz}
	\usetikzlibrary{shapes}
	\usetikzlibrary{matrix}
\usepackage{tikz-cd}
\usetikzlibrary{babel}
\usetikzlibrary{decorations.text}
\usepackage{enumitem}
\usepackage[margin=1.2in]{geometry}
\usepackage{scalerel}
\usepackage{microtype} 
\usepackage{array}
\usepackage{makecell}

\usepackage[backend=biber, 
style=alphabetic,
giveninits=true,
isbn=false, 
url=false,
minalphanames=1,
maxalphanames=5,
maxbibnames=99]{biblatex}
\AtEveryBibitem{\clearfield{month}}
\DeclareNameAlias{author}{given-family}
\renewbibmacro{in:}{}
\DeclareFieldFormat{extraalpha}{#1}

\DeclareLabelalphaTemplate{
  \labelelement{
    \field[final]{shorthand}
    \field{label}
    \field[strwidth=3,strside=left,ifnames=1]{labelname}
    \field[strwidth=1,strside=left]{labelname}
  }
}

\usepackage{tabstackengine}
\TABstackMath
\usepackage{multirow}
\usepackage{graphicx}
\definecolor{Myblue}{rgb}{0,0,0.6}  
\definecolor{Mybrown}{RGB}{132, 74, 17}
\usepackage{todonotes}
	\setuptodonotes{size=\footnotesize, color=red!40}

\usepackage{tikz-3dplot}
\usepackage{pgfplots}
\pgfplotsset{width=7cm,compat=1.8}
\usetikzlibrary{decorations.pathreplacing}
\usetikzlibrary{decorations.markings}
\usetikzlibrary{patterns}
\usepgflibrary{shapes.geometric}
\usepackage{subcaption}

\newcommand{\be}{\begin{equation}}
\newcommand{\ee}{\end{equation}}
\newcommand\tikzzbox[1]
{#1}

\tikzset{
	string/.style={draw=#1, postaction={decorate}, decoration={markings,mark=at position .51 with {\arrow[draw=#1]{>}}}},
	costring/.style={draw=#1, postaction={decorate}, decoration={markings,mark=at position .51 with {\arrow[draw=#1]{<}}}},
	ostring/.style={draw=#1, postaction={decorate}, decoration={markings,mark=at position .47 with {\arrow[draw=#1]{>}}}},
	ustring/.style={draw=#1, postaction={decorate}, decoration={markings,mark=at position .56 with {\arrow[draw=#1]{>}}}},
	oostring/.style={draw=#1, postaction={decorate}, decoration={markings,mark=at position .43 with {\arrow[draw=#1]{>}}}},
	uustring/.style={draw=#1, postaction={decorate}, decoration={markings,mark=at position .59 with {\arrow[draw=#1]{>}}}},
	directed/.style={string=blue!50!black}, 
	odirected/.style={ostring=blue!50!black}, 
	udirected/.style={ustring=blue!50!black}, 
	oodirected/.style={oostring=blue!50!black}, 
	uudirected/.style={uustring=blue!50!black},     
	redirected/.style={costring= blue!50!black},
	redirectedgreen/.style={costring= green!50!black},
	directedgreen/.style={string= green!50!black},
	redirectedred/.style={costring= red!50!black},
	directedred/.style={string= red!50!black}%
}

\tikzset{-dot-/.style={decoration={
			markings,
			mark=at position 0.5 with {\fill circle (2pt);}},postaction={decorate}}}

\tikzset{
	Fdot/.style={circle, draw, fill, inner sep=0pt}, 
	Odot/.style={circle, draw, inner sep=0.1pt, minimum size=0.1cm}
}
\tikzset{partial curve/.style args={%
  from #1 to #2 curve #3 .. controls #4 and #5 .. #6}{insert path={
    #3 coordinate (@1) #4 coordinate (@2)
    #5 coordinate (@3) #6 coordinate (@4)
    \pgfextra{\pgfpathcurvebetweentime{#1}{#2}%
      {\pgfpointanchor{@1}{center}}{\pgfpointanchor{@2}{center}}%
      {\pgfpointanchor{@3}{center}}{\pgfpointanchor{@4}{center}}}
}}}

\makeatletter
\newcommand{\subalign}[1]{%
  \vcenter{%
    \Let@ \restore@math@cr \default@tag
    \baselineskip\fontdimen10 \scriptfont\tw@
    \advance\baselineskip\fontdimen12 \scriptfont\tw@
    \lineskip\thr@@\fontdimen8 \scriptfont\thr@@
    \lineskiplimit\lineskip
    \ialign{\hfil$\m@th##$&$\m@th{}##$\hfil\crcr
      #1\crcr
    }%
  }%
}
\makeatother

\usepackage{thmtools}
\usepackage{mathtools}
\usepackage[colorlinks,citecolor=Myblue,linkcolor=Myblue,urlcolor=Myblue,pdfpagemode=None, linktocpage]{hyperref}
\usepackage{cleveref}

\allowdisplaybreaks
\deffootnote[1em]{1em}{1em}{\textsuperscript{\thefootnotemark}}

\theoremstyle{definition} 
\newtheorem{definition}{Definition}

\newtheorem{lemma}[definition]{Lemma}

\newtheorem{notation}[definition]{Notation}

\numberwithin{equation}{section}
\numberwithin{definition}{section}
\numberwithin{lemma}{section}
\numberwithin{proposition}{section}
\numberwithin{theorem}{section}
\numberwithin{corollary}{section}
\numberwithin{example}{section}
\numberwithin{remark}{section}
\numberwithin{figure}{section}
\numberwithin{table}{section}

\DeclareMathOperator{\pr}{pr}
\DeclareMathOperator{\incl}{incl}

\newcommand*{\longhookrightarrow}{\ensuremath{\lhook\joinrel\relbar\joinrel\rightarrow}}

\newcommand{\C}{\mathds{C}}
\newcommand{\Z}{\mathds{Z}}
\newcommand{\Vect}{\mathrm{Vect}} 
\newcommand{\Grp}{\mathrm{Grp}}
\newcommand{\BG}[1][0]{\ifnum #1<1 {\mathrm{B}G} \else {\mathrm{B}^{#1}G}\fi}
\newcommand{\mg}{m_g}
\newcommand{\Mod}{\mathrm{Mod}}
\DeclareMathOperator{\Rep}{Rep}
\DeclareMathOperator{\U}{U(1)}
\DeclareMathOperator{\SU}{SU}
\newcommand{\Gcbc}{\mathcal{C}^\times_G}

\let\to\undefined
\newcommand{\to}{\longrightarrow}

\newcommand{\ToRoman}[1]{%
  \textup{\uppercase\expandafter{\romannumeral#1}}}%

\usepackage{bbm}

\newcommand{\id}{\mathrm{id}}
\DeclareMathOperator{\Aut}{Aut}
\DeclareMathOperator{\End}{End}
\DeclareMathOperator{\Hom}{Hom}
\newcommand{\ev}{\mathrm{ev}}
\newcommand{\coev}{\mathrm{coev}}

\usepackage[titles]{tocloft}
\title{Examples of Invertible Gauging via Orbifold Data,\,Zesting,\,and Equivariantisation}
\author{Benjamin Haake\\[0.1cm]%
	\normalsize{\texttt{\href{mailto:B.Haake@sms.ed.ac.uk}{B.Haake@sms.ed.ac.uk}}}
	\\[0cm]  %
	{\normalsize\slshape University of Edinburgh, School of Mathematics, James Clerk Maxwell Building,\vspace{-.5em}}\\{ \normalsize\slshape Peter Guthrie Tait Road, Edinburgh EH9\,3FD, United Kingdom}\\
	{\normalsize\slshape Maxwell Institute for Mathematical Sciences, The Bayes Centre,\vspace{-.5em}}\\{\normalsize\slshape 47 Potterrow, Edinburgh EH8 9BT, United Kingdom\vspace{-1em}}}
\date{}

\newcommand{\A}{C}
\newcommand{\TY}{\mathcal{TY}}
\newcommand{\TYAXK}{\TY(\A,\chi,\kappa)}
\newcommand{\TYAQKD}{\TY(\A,q,\kappa,\delta)}

\newcommand{\Ising}{\mathrm{Ising}}
\newcommand{\sgn}{\mathrm{sgn}}

\newcommand{\tsl}[1]{\textsl{#1}}
\begin{document}
\maketitle
\vspace*{-3em}
\begin{abstract}
	We study the gauging of invertible symmetries, particularly in 3 dimensions, using equivariantisation, $G$-crossed braided zesting, and the generalised orbifold construction.
	We discuss how these methods are related and illustrate them in various examples.
	We cover all $\Z_2$-symmetries in Dijkgraaf--Witten $\Z_2$-gauge theory $\mathcal{D}(\Z_2)$, the $\Z_2$-symmetries described by Tambara--Yamagami categories, and obstructions to gauging the central symmetry in Chern--Simons $\SU(2)_k$-gauge theory.
	We introduce zested orbifold data for symmetries related by zesting and show that the two associated orbifold data are Morita-equivalent, i.e.\ they have the same underlying surface defect.
\end{abstract}
\tableofcontents
\newpage

\section{Introduction}
\label{sec:Intro}
Symmetries have historically been of central interest to both physics and mathematics, and the exchange between the two fields has lead to many interesting discoveries on both sides.
As a result, many tools have been developed for their study whose variety makes a comprehensive treatment difficult.
One of these areas are finite group symmetries of (braided) monoidal categories. 
There, the mathematical tools include classification and extension theory \cite{ENO,DGNO,DGPRZ} and related computational results \cite{BN}.
Recent developments \cite{GKSW} in mathematical physics have enhanced the focus on topological defects, resulting in an increased interest in the mathematics of topological quantum field theories and their gauging \cite{CRS1,CH,SW} as well as gauging of potentially non-invertible symmetries in field theories and condensed matter physics \cite{BBCW,BBDR,KZZ}.

This paper has a twofold purpose: 
Initially, we establish the mathematical tools in a language that is common to (mathematical) physics.
We then connect the mathematical and physical developments further by providing simple examples and discussing how the various methods may be applied.

Generally speaking, a (0-form) $G$-symmetry is the action~$\rho$ of a group~$G$ (or higher group) on a vector space (usually the state space), on an object of some category, or even on a (higher) category $\mathcal{C}$.
An example for the latter is the action of a 0-form symmetry on the category $\mathcal{C}$ of line operators.
Given a 0-form symmetry, one can define a codimension 1 topological defect (i.e.\ domain wall) $R(g)$ for each element $g\in G$.
When this defect is deformed, it transforms anything it passes over under the appropriate action, as we illustrate on a local operator $\mathcal{O}$:
\begin{equation}
	    		\begin{tikzpicture}[very thick,scale=0.5,color=black, baseline=0cm]
			\fill [orange!50,opacity=0.545] (-1,-2.5) -- (-1,2.5) -- (3,2.5) -- (3,-2.5);
			\coordinate (hl) at (0,-2.5);
			\coordinate (hlt) at (0,2.5);
			\draw (hl) -- (hlt)
			(hl) node[below] {\scriptsize$R(g)$}; 
			\fill (1,0)  circle (3pt) node[below]{\scriptsize$\mathcal{O}$};
		\end{tikzpicture}\,=\,
    		\begin{tikzpicture}[very thick,scale=0.5,color=black, baseline=0cm]
			\fill [orange!50,opacity=0.545] (-1,-2.5) -- (-1,2.5) -- (3,2.5) -- (3,-2.5);
			\coordinate (hl) at (0,-2.5);
			\coordinate (hlt) at (0,2.5);
			\draw (hl) .. controls +(0,1) and +(0,-2) .. (2,0) .. controls +(0,2) and +(0,-1) .. (hlt)
			(hl) node[below] {\scriptsize$R(g)$}; 
			\fill (1,0)  circle (3pt) node[below]{\scriptsize$\rho_g(\mathcal{O})$};
		\end{tikzpicture}\,.
\end{equation}
Naturally, defects $R(g)$ and $R(h)$ associated to group elements $g,h\in G$ should fuse to produce the defect  $R(gh)$.
This is realised by an invertible junction~$\chi_{g,h}$:
\begin{equation}
        	\begin{tikzpicture}[very thick,scale=0.5,color=black, baseline=0cm]
			\fill [orange!50,opacity=0.545] (-1,-2.5) -- (-1,2.5) -- (3,2.5) -- (3,-2.5);
			\coordinate (hl) at (0,-2.5);
			\coordinate (hr) at (2,-2.5);
			\coordinate (hmt) at (1,2.5);
			\draw (hl) .. controls +(0,1) and +(-1,-1) .. (1,0); 
			\draw (hr) .. controls +(0,1) and +(1,-1) .. (1,0)
			(1,0) -- (hmt)
			(hr) node[below] {\scriptsize$R(h)$}
			(hl) node[below] {\scriptsize$R(g)$}
			(hmt) node[above] {\scriptsize$R(gh)$}; 
			\fill (1,0) circle (4pt) node[right] {\scriptsize$\chi_{g,h}$};
		\end{tikzpicture}\,.
\end{equation}
Overall, this data produces an assignment 
\begin{equation}
R\colon \mathrm{B}\underline{G}\longrightarrow\mathcal{D}
\label{eq:generalSymm}
\end{equation}
into the (higher category of) topological defects $\mathcal{D}$ in a given theory.
This category has domain walls as 1-morphisms, codimension 2 defects interpolate between different domain walls and in general, codimension~$k$ defects are its $k$-morphisms.
Recall that the delooping $\mathrm{B}\underline{G}$ of~$\underline{G}$ is the (higher) category which has only a single object denoted~$\ast$ with morphisms $\ast\longrightarrow\ast$ given by elements of~$G$ and the underline denotes the possible addition of higher identity morphisms (depending on the dimension).
Thus, a functor of the type \eqref{eq:generalSymm} assigns domain walls to group elements and the junction~$\chi_{g,h}$ appears as coherence data for composition (it interpolates between first composing group elements and then taking the associated symmetry defect, and first taking defects and then fusing).
In higher dimensions, additional coherence data is contained in a higher functor to ensure compatibility with fusion in various directions and for defects of all codimensions.

Twisted sectors are additional defects which may lie on the boundary of the domain walls $R(g)$.
Mathematically, these are the objects of additional graded pieces $\mathcal{D}^2_g$ of topological operators of codimension 2, where $\mathcal{D}^2_e:=\End_\mathcal{D}(\id_\mathds{1})$ are codimension 2 (bulk) defects on the trivial surface defect.
These assemble into a $G$-extension $(\mathcal{D}^2)^\times_G:=\bigoplus_{g\in G}\mathcal{D}_g^2$ of the codimension 2 bulk defects.
In 3 dimensions, codimension 2 defects are line defects which form a category using junctions and local operators as morphisms.
Fusion provides a monoidal structure and the lines can braid around each other in 3 dimensions.
It was developed in \cite{ENO} that $G$-crossed braided extensions~$\Gcbc$ of a braided fusion category $\mathcal{C}$ are classified by 3-functors 
\begin{equation}
	\mathrm{B}\underline{G}\longrightarrow \mathrm{B}(\mathrm{Mod}^\times(\mathcal{C}))\,,
	\label{eq:classification}
\end{equation}
where $\mathrm{Mod}^\times(\mathcal{C})$ is the 2-groupoid of invertible module categories, module equivalences and natural isomorphisms.
The module category $\mathcal{C}_g$ assigned to an element $g\in G$ consists of exactly those operators in the $g$-twisted sector, and the module structure determines how they fuse with bulk defects.
It is no coincidence that this structure is so similar to \eqref{eq:generalSymm}: When the 3-category~$\mathcal{D}$ of topological defects is generated by a category of line operators~$\mathcal{C}$ and their ``condensations'' (namely Reshetikhin--Turaev theory), such functors are exactly what describes a symmetry, as we discuss in \Cref{sec:RT}.
This means that symmetries with twisted sectors correspond to $G$-extensions.

Regarding the gauging of (0-form) $G$-symmetries, recall that local operators after gauging are exactly those which are invariant under the symmetry.
In 2 dimensions, this may include invariant (sums of) twisted sector operators which then become bulk local operators in the gauged theory.
Line defects in a gauge theory are similarly given by (sums of) line operators~$L$ (including twisted sectors in 3 dimensions), however, due to the additional structure of junctions (i.e.\ morphisms) between them, the transformed line $\rho_g(L)$ need not be strictly identical but merely connected by an invertible junction (i.e.\ isomorphic) to the original line~$L$.
Specifying a junction $u_g\colon \rho_g(L)\longrightarrow L$ for each $g\in G$ on a line operator~$L$ defines an \textsl{equivariant structure} and objects equipped with this structure form the \textsl{equivariantisation} $(\Gcbc)^G$.
We explain this in detail in \Cref{sec:equivariantisation}.

In topological quantum field theories (TQFT), theories are encoded as functors $\mathrm{Bord}_n\longrightarrow \mathcal{T}$ which traditionally (for $\mathcal{T}=\Vect$) assign state spaces to $(n-1)$-dimensional manifolds and time evolution operators to $n$-dimensional manifolds.
In this context, gauging is described -- among other approaches \cite{GJF} -- by \textsl{orbifolding} \cite{CRS1,carqueville2023orbifoldstopologicalquantumfield}.
This requires an \textsl{orbifold datum} which in our case is an algebra (of line operators) equipped with additional structure.
The algebra itself is condensed to produce the surface defects $R(g)$, while the additional structure corresponds to junctions between these surfaces.
For 0-form $G$-symmetries, this algebra is the group algebra $\C[G]$ which we discuss in detail in \Cref{sec:2groupsymm}.
Heuristically then, the terms ``0-form symmetry with twisted sectors'' and ``$G$-extension'' can be used interchangeably, and similarly ``gauging the 0-form symmetry'', ``equivariantisation'', and ``orbifolding.''

\textsl{Symmetry fractionalisation} and \textsl{defectification} (also known as discrete torsion) describe variations of a given symmetry.
In 3 dimensions, symmetry fractionalisation occurs when the line defect that fuses two surface defects of the $G$-symmetry is enhanced by fusing another topological line defect~$\lambda$ onto it, 
\begin{equation}
\begin{tikzpicture}[thick,scale=2.6,color=blue!50!black, baseline=0.1cm, >=stealth, 
					style={x={(-0.6cm,-0.4cm)},y={(1cm,-0.2cm)},z={(0cm,0.9cm)}}]
					\pgfmathsetmacro{\yy}{0.2}
					\coordinate (T) at (0.5, 0.4, 0);
					\coordinate (L) at (0.5, 0, 0);
					\coordinate (R1) at (0.3, 1, 0);
					\coordinate (R2) at (0.7, 1, 0);
					\coordinate (1T) at (0.5, 0.4, 1);
					\coordinate (1L) at (0.5, 0, 1);
					\coordinate (1R1) at (0.3, 1, );
					\coordinate (1R2) at (0.7, 1, );
					%
					\fill [orange!50,opacity=0.545] (L) -- (T) -- (1T) -- (1L);
					\fill [orange!50,opacity=0.545] (R1) -- (T) -- (1T) -- (1R1);
					\draw [black,opacity=1, very thin] (1T) -- (1R1) -- (R1) -- (T);
					\fill [orange!50,opacity=0.545] (R2) -- (T) -- (1T) -- (1R2);
					\fill[color=black] (0.5,0.17,0.9) node { {\scriptsize$R(gh)$} };
					\fill[color=black] (0.15,0.95,0.835) node[left] { {\scriptsize$R(g)$} };
					\fill[color=black] (0.55,0.95,0.875)  node[left] { {\scriptsize$R(h)$} };
					%
					\draw[string=Mybrown, thick] (T) -- (1T);
					\fill[color=black] (T) node[below] (0up) { {\scriptsize$\chi_{g,h}$} };
					%
					\draw [black,opacity=1, very thin] 
					(1T) -- (1L) -- (L) -- (T)
					(1T) -- (1R2) -- (R2) -- (T);
			\end{tikzpicture}
			\tikz \draw [->, line join=round, decorate, decoration={zigzag, segment length=4, amplitude=.9,post=lineto, post length=2pt}]  (0,0) -- (2.2,0) node[pos=.5,above]{\scriptsize$\text{fractionalisation}$};
\begin{tikzpicture}[thick,scale=2.6,color=blue!50!black, baseline=0.1cm, >=stealth, 
					style={x={(-0.6cm,-0.4cm)},y={(1cm,-0.2cm)},z={(0cm,0.9cm)}}]
					\pgfmathsetmacro{\yy}{0.2}
					\coordinate (T) at (0.5, 0.4, 0);
					\coordinate (L) at (0.5, 0, 0);
					\coordinate (R1) at (0.3, 1, 0);
					\coordinate (R2) at (0.7, 1, 0);
					\coordinate (1T) at (0.5, 0.4, 1);
					\coordinate (1L) at (0.5, 0, 1);
					\coordinate (1R1) at (0.3, 1, );
					\coordinate (1R2) at (0.7, 1, );
					%
					\fill [orange!50,opacity=0.545] (L) -- (T) -- (1T) -- (1L);
					\fill [orange!50,opacity=0.545] (R1) -- (T) -- (1T) -- (1R1);
					\draw [black,opacity=1, very thin] (1T) -- (1R1) -- (R1) -- (T);
					\fill [orange!50,opacity=0.545] (R2) -- (T) -- (1T) -- (1R2);
					\fill[color=black] (0.5,0.17,0.9) node { {\scriptsize$R(gh)$} };
					\fill[color=black] (0.15,0.95,0.835) node[left] { {\scriptsize$R(g)$} };
					\fill[color=black] (0.55,0.95,0.875)  node[left] { {\scriptsize$R(h)$} };
					%
					\draw[string=Mybrown, ultra thick] (T) -- (1T);
					\fill[color=black] (T) node[below] (0up) { {\scriptsize$\chi_{g,h}^{ \lambda}$} };
					%
					\draw [black,opacity=1, very thin] 
					(1T) -- (1L) -- (L) -- (T)
					(1T) -- (1R2) -- (R2) -- (T);
			\end{tikzpicture}\,,
			\label{eq:zestedfusiondiagram1}
\end{equation}
where $\chi^\lambda_{g,h}\cong \chi_{g,h}\otimes \lambda(g,h)$,
and defectification describes the enhancement of the junction between the two ways that three such surfaces can fuse by a local operator~$p$, i.e.\ the associator is modified:
\begin{equation}
			\begin{tikzpicture}[thick,scale=2.321,color=blue!50!black, baseline=0.1cm, >=stealth, 
				style={x={(-0.6cm,-0.4cm)},y={(1cm,-0.2cm)},z={(0cm,0.9cm)}}]
				\pgfmathsetmacro{\yy}{0.2}
				\coordinate (P) at (0.5, \yy, 0);
				\coordinate (R) at (0.375, 0.5 + \yy/2, 0);
				\coordinate (L) at (0.5, 0, 0);
				\coordinate (R1) at (0.25, 1, 0);
				\coordinate (R2) at (0.5, 1, 0);
				\coordinate (R3) at (0.75, 1, 0);
				\coordinate (Pt) at (0.5, \yy, 1);
				\coordinate (Rt) at (0.625, 0.5 + \yy/2, 1);
				\coordinate (Lt) at (0.5, 0, 1);
				\coordinate (R1t) at (0.25, 1, 1);
				\coordinate (R2t) at (0.5, 1, 1);
				\coordinate (R3t) at (0.75, 1, 1);
				\coordinate (alpha) at (0.5, 0.5, 0.5);
				%
				\fill [orange!50,opacity=0.545] (L) -- (P) -- (alpha) -- (Pt) -- (Lt);
				\fill [orange!50,opacity=0.545] (Pt) -- (R1t) -- (R1) -- (R) -- (alpha);
				\draw [black,opacity=1, very thin] (Pt) -- (R1t) -- (R1) -- (R);
				\fill [orange!50,opacity=0.545] (Rt) -- (R2t) -- (R2) -- (R) -- (alpha);
				\draw [black,opacity=1, very thin] (Rt) -- (R2t) -- (R2) -- (R);
				\fill [orange!50,opacity=0.545] (Pt) -- (Rt) -- (alpha);
				\fill [orange!50,opacity=0.545] (P) -- (R) -- (alpha);
				\draw [black,opacity=1, very thin] (P) -- (R);
				\draw[string=Mybrown, thick] (R) -- (alpha);
					\fill[color=black] ($(R)+(0,0,.03)$) circle (0pt) node[below] (0up) { {\scriptsize$\chi_{g,h}$} };
				\fill [orange!50,opacity=0.545] (Rt) -- (R3t) -- (R3) -- (P) -- (alpha);
				\draw[string=Mybrown, thick] (alpha) -- (Rt);
				%
				\draw[string=Mybrown,  thick] (P) -- (alpha);
				\draw[string=Mybrown,  thick] (alpha) -- (Pt);
					\fill[color=Mybrown] (alpha) circle (.8pt);
					\draw [black] (alpha) node[left] (0up) { {\scriptsize$\overline{\alpha}_{g,h,k}$} };
					\fill[color=black] (P) circle (0pt) node[below] (0up) { {\scriptsize$\chi_{gh,k}$} };
					\fill[color=black] ($(Rt)+(0,0.1,-.05)$) circle (0pt) node[above] (0up) { {\scriptsize$\chi_{h,k}$} };
					\fill[color=black] (Pt) circle (0pt) node[above] (0up) { {\scriptsize$\chi_{g,hk}$} };
				%
				\draw [black,opacity=1, very thin] (Pt) -- (Lt) -- (L) -- (P) ;
				\draw [black,opacity=1, very thin] (Pt) -- (Rt);
				\draw [black,opacity=1, very thin] (Rt) -- (R3t) -- (R3) -- (P);
			\end{tikzpicture}\;
			\tikz \draw [->, line join=round, decorate, decoration={zigzag, segment length=4, amplitude=.9,post=lineto, post length=2pt}]  (0,0) -- (2.2,0) node[pos=.5,above]{\scriptsize$\text{defectification}$};\;
			\begin{tikzpicture}[thick,scale=2.321,color=blue!50!black, baseline=0.1cm, >=stealth, 
				style={x={(-0.6cm,-0.4cm)},y={(1cm,-0.2cm)},z={(0cm,0.9cm)}}]
				\pgfmathsetmacro{\yy}{0.2}
				\coordinate (P) at (0.5, \yy, 0);
				\coordinate (R) at (0.375, 0.5 + \yy/2, 0);
				\coordinate (L) at (0.5, 0, 0);
				\coordinate (R1) at (0.25, 1, 0);
				\coordinate (R2) at (0.5, 1, 0);
				\coordinate (R3) at (0.75, 1, 0);
				\coordinate (Pt) at (0.5, \yy, 1);
				\coordinate (Rt) at (0.625, 0.5 + \yy/2, 1);
				\coordinate (Lt) at (0.5, 0, 1);
				\coordinate (R1t) at (0.25, 1, 1);
				\coordinate (R2t) at (0.5, 1, 1);
				\coordinate (R3t) at (0.75, 1, 1);
				\coordinate (alpha) at (0.5, 0.5, 0.5);
				%
				\fill [orange!50,opacity=0.545] (L) -- (P) -- (alpha) -- (Pt) -- (Lt);
				\fill [orange!50,opacity=0.545] (Pt) -- (R1t) -- (R1) -- (R) -- (alpha);
				\draw [black,opacity=1, very thin] (Pt) -- (R1t) -- (R1) -- (R);
				\fill [orange!50,opacity=0.545] (Rt) -- (R2t) -- (R2) -- (R) -- (alpha);
				\draw [black,opacity=1, very thin] (Rt) -- (R2t) -- (R2) -- (R);
				\fill [orange!50,opacity=0.545] (Pt) -- (Rt) -- (alpha);
				\fill [orange!50,opacity=0.545] (P) -- (R) -- (alpha);
				\draw [black,opacity=1, very thin] (P) -- (R);
				\draw[string=Mybrown, thick] (R) -- (alpha);
					\fill[color=black] ($(R)+(0,0,.03)$) circle (0pt) node[below] (0up) { {\scriptsize$\chi_{g,h}$} };
				\fill [orange!50,opacity=0.545] (Rt) -- (R3t) -- (R3) -- (P) -- (alpha);
				\draw[string=Mybrown, thick] (alpha) -- (Rt);
				%
				\draw[string=Mybrown,  thick] (P) -- (alpha);
				\draw[string=Mybrown,  thick] (alpha) -- (Pt);
					\fill[color=Mybrown] (alpha) circle (1.2pt);
					\draw [black] (alpha) node[left] (0up) { {\scriptsize$\overline{\alpha}^p_{g,h,k}$} };
					\fill[color=black] (P) circle (0pt) node[below] (0up) { {\scriptsize$\chi_{gh,k}$} };
					\fill[color=black] ($(Rt)+(0,0.1,-.05)$) circle (0pt) node[above] (0up) { {\scriptsize$\chi_{h,k}$} };
					\fill[color=black] (Pt) circle (0pt) node[above] (0up) { {\scriptsize$\chi_{g,hk}$} };
				%
				\draw [black,opacity=1, very thin] (Pt) -- (Lt) -- (L) -- (P) ;
				\draw [black,opacity=1, very thin] (Pt) -- (Rt);
				\draw [black,opacity=1, very thin] (Rt) -- (R3t) -- (R3) -- (P);
			\end{tikzpicture}\,,
			\label{eq:defectificationGassociator}
\end{equation}
where $\overline{\alpha}^p_{g,h,k}=p(g,h,k)\omega^p_{gh,hk}\cdot\overline{\alpha}_{g,h,k}$ and $\omega^p_{gh,hk}$ is an additional scalar based on~$p$ which we determine in \eqref{eq:simplifiedalpha}.\footnote{%
There is an unfortunate mismatch of conventions between the associators for monoidal categories and the associators used in orbifold data, as they go in opposite directions.
In the diagram \eqref{eq:defectificationGassociator} we use the inverse orbifold datum associator to match monoidal associator conventions.
Note that the actual associator of the category of line operators is modified simply by $p(g,h,k)$, see \eqref{eq:zestedassociator} and \Cref{sec:zestedorbdat}.
}
These configurations of defects are precisely the additional structure that turns an algebra into an orbifold datum.
In the construction \cite[Sect.\,3]{CH} of an orbifold datum from a symmetry \eqref{eq:generalSymm}, this data is directly built from the coherence isomorphisms of the functor~$R$.

In the theory of $G$-extensions, fractionalisation and defectification correspond to so-called \textsl{zesting data}, which can be used to modify a given extension.
\textsl{Zesting} provides an algebraic tool to calculate the new symmetry or extension directly.

In \cite{BBDR} the authors observed that when two 0-form symmetries in 3 dimensions are related by fractionalisation, the algebras of surface defects which gauge the respective symmetries (which we just identified with the orbifold datum) have the same underlying object.
Using the aforementioned identification of \eqref{eq:generalSymm} and \eqref{eq:classification} in Reshetikhin--Turaev theory, we see that codimension 1 defects are identified with module categories.
Furthermore, \cite[Thm.\,3.12]{DGPRZ} states that two $G$-crossed braided extensions are related to each other by zesting (i.e.\ fractionalisation and defectification) if and only if their graded components are given by the same underlying module categories (i.e.\ twisted sectors). 
Therefore, two algebras of surfaces have the same underlying object (i.e.\ are based on the same surface defect) if and only if the symmetries they gauge are related by fractionalisation and defectification.
When the symmetry surfaces are condensation defects, this means that the associated underlying algebras must be Morita-equivalent.
We show this in \Cref{sec:zestedorbdat} and observe the phenomenon explicitly in the examples of \Cref{sec:DZ2}.

A natural generalisation of \eqref{eq:generalSymm} is to replace $\underline{G}$ with another monoidal (higher) category~$\mathcal{A}$.
When this category is of the type $\mathrm{B}\underline{H}$ for some commutative group~$H$, the result $\mathrm{B}^2\underline{H}\longrightarrow \mathcal{D}$ is called a 1-form symmetry.
These are known to arise when a 0-form symmetry is gauged, most notably through the appearance of Wilson lines.
These are of course line defects labelled by representations $\Rep(G)$ of the gauge symmetry group~$G$ and thereby are part of the category of line defects~$\mathcal{C}$ in the gauged theory. 
This emergent (potentially non-invertible) $\Rep(G)$-symmetry can itself be gauged which can be described mathematically by an orbifold datum or equivalently by \textsl{de-equivariantisation}.
The resulting theory is the original theory with its $G$-symmetry, thus the 1-form gauging undoes the 0-form gauging (and vice versa).
\begin{equation}
\begin{tikzcd}[column sep= 80]
    \mathcal{C}\subset \Gcbc
    \arrow[r, bend left, "{\substack{ \text{0-form gauging }\widehat{=}\text{ eq.}\\A_G=\C[G]}}"{above}]
    &
	(\Gcbc)^{G}
    \arrow[l, bend left, "{\substack{B=\C(G)\in\Rep(G)\\\text{1-form gauging }\widehat{=}\text{ de-eq.}}}"{below}]  
\end{tikzcd}
\end{equation}
As for 0-form symmetry, the orbifold construction also treats the gauging of 1-form symmetries using an algebra~$B$ with additional structure (in this case the algebra of functions on~$G$).
More broadly, there is an explicit description of the orbifold datum whenever $\mathcal{A}=\mathcal{G}$ is a (finite) 2-group symmetry \cite{CH}.
Recall that a 2-group can be understood as the combination of a 0-form and a 1-form symmetry, potentially with additional interactions between them.
Therefore, the construction of an orbifold datum for any 2-group symmetry unifies the gauging of 0-form and 1-form symmetries.
We recall the construction in \Cref{sec:2groupsymm} and we compute an explicit orbifold datum for a 2-group symmetry in \Cref{sec:2-groupZ2Z3}.

\medskip

The present paper is structured as follows.
In \Cref{sec:background} we provide a more detailed exposition of the concepts and methods introduced above.
\Cref{sec:equivariantisation,sec:CalculatingEq} define equivariantisations, the intuition behind them and how they can be calculated. The mathematical results are based on \cite{DGNO} and \cite{BN}, respectively.
\Cref{sec:Gcbc} adds twisted sectors and braidings leading to $G$-crossed braided fusion categories whose equivariantisations inherit a braiding.
In Sections \ref{sec:orbifoldDataFromSymmetry} and \ref{sec:2groupsymm} we introduce orbifold data and outline the construction from \cite{CH} that gauges a given 2-group symmetry through such an orbifold datum.
In \Cref{sec:RT} we review this construction in Reshetikhin--Turaev theory which encompasses all our examples (see also \cite{CRS3}).
\Cref{sec:zesting} discusses ($G$-crossed braided) zesting and how it can help us to classify extensions \cite{DGPRZ}.
We relate zesting to the physical concepts of fractionalisation and defectification and illustrate how these are captured by orbifold data.
\Cref{sec:zestedorbdat} combines orbifold data and zesting.
We compute the orbifold datum which gauges the zested $G$-extension, or in other words we compute the effects of fractionalisation and defectification on orbifold data.
We show that the modified orbifold datum is Morita-equivalent to the original one, meaning that they give rise to the same condensation surface defect.
Before turning to the examples, in \Cref{sec:deeq} we discuss de-equivariantisation and its relation to orbifold data \cite{DGNO,CH,HPRW}.
We conclude the section by summarising how gauging 0- and 1-form symmetries are inverse operations to each other.

In \Cref{sec:TY} we discuss Tambara--Yamagami categories as our first class of examples.
These are $\Z_2$-extensions which only contain one (simple) object in the twisted sector making them comparatively easy to describe.
However, due to the fact that this object is non-invertible, these categories still provide some complexity and they exhibit many of the general phenomena that occur during equivariantisation.
Our focus lies on equivariantisation and orbifolding, while leaving zesting to the subsequent examples.
The equivariantisations in this section have been calculated in \cite{GNN,BBCW} and the underlying results for Tambara--Yamagami are based on \cite{Siehler,GLM}.
\Cref{sec:DZ2} treats all 0-form symmetries in $\mathcal{D}(\Z_2)=\mathcal{Z}(\Z_2\text{-}\Vect)$, providing a thorough exposition of zesting and comparing the results of the orbifold construction to those of \cite{BBDR}.
While \cite{BBCW} discussed most of the data of the extensions, we add their equivariantisations in detail, as well as the perspective of zesting, orbifold data, and the aforementioned comparison.
Lastly, \Cref{sec:CS} discusses Chern--Simons theory with gauge group $\SU(2)$ at level~$k$.
We reproduce the obstruction found in \cite{MSzoo} which depends on the level by asking simple questions in the language of category theory.

\medskip 

\noindent
\textbf{Acknowledgements. } 
I am grateful to 
	Mahesh~K.~N.~Balasubramanian,
	Nils Carqueville,
	Hank Chen,
	Tudor Dimofte,
	Subrabalan Murugesan,
	and
	Sean Sanford
for insightful discussions and/or helpful comments on an earlier version of the manuscript. 

\section{$\boldsymbol{G}$-crossed Braided Fusion Categories and Their Gauging}
\label{sec:background}
\subsection{Equivariantisation: Taking Invariants Coherently}
\label{sec:equivariantisation}
In this section we recall the definitions of group actions on monoidal categories and their equivariantisation and motivate their physical interpretations.
A \textsl{$G$-action on a monoidal category~$\mathcal C$} for a group~$G$ is a monoidal functor\footnote{By the superscript ``rev'' we indicate the reversal of the monoidal product. 
This means that we use right $G$-actions in accordance with \cite[Sect.\,5]{CRS3}, \cite{CH,DGPRZ}.}
\begin{equation}
	\rho\colon \underline{G}^\text{rev}\longrightarrow \Aut^{\otimes}(\mathcal{C}) \, . 
\end{equation}
Here~$\underline{G}$ denotes the monoidal category whose objects are group elements $g\in G$, the monoidal product is group multiplication, and~$\underline{G}$ has only identity morphisms. 
The codomain $\Aut^{\otimes}(\mathcal{C})$ is the category of monoidal autoequivalences $\mathcal{C}\xlongrightarrow{\cong}\mathcal{C}$.
For every $g\in G$, we abbreviate $\rho_g := \rho(g)$ and we denote its monoidal structure isomorphisms as
\begin{align}
	(\rho^2_g)_{X,Y}\colon \rho_g(X)\otimes \rho_g(Y)&\longrightarrow \rho_g(X\otimes Y)\,, &\rho^0_g&\colon \mathds{1}\longrightarrow \rho_g(\mathds{1}) \, ,
\end{align}
which interpolate between fusing two objects either before or after acting on them with $\rho_g$, and the unit object and its transformation, respectively.
Since~$\rho$ itself is also monoidal, it comes with structure isomorphisms of its own given by invertible natural transformations 
\begin{equation}
	\rho^2\colon \rho_{(-)} \circ \rho_{(-)} \Longrightarrow\rho_{(-)\cdot(-)}
		\,, \quad 
	\rho^0\colon \id_\mathcal{C}\Longrightarrow \rho_e \,,
\end{equation}
where~$\rho^0$ is monoidal ($e$ is the unit of $G$) and~$\rho^2$ has components which are themselves monoidal natural transformations, 
\begin{align}
	\rho^2_{g,h}\colon \hphantom{\rho_g(}\,\rho_g\circ\rho_h&\Longrightarrow\rho_{hg} \,,\\
	(\rho^2_{g,h})_X\colon \rho_g(\rho_h(X))&\longrightarrow\rho_{hg}(X) \,,
\end{align}
In the language of defects, these components provide the line defects which fuse the surfaces associated to $\rho_g$ and $\rho_h$.
Note that all of (the component morphisms of) these natural transformations can be distinguished by their sub- and superscripts.

Given a fusion category~$\mathcal{C}$ with a monoidal functor $\rho\colon \underline{G}^\text{rev}\longrightarrow \Aut^\otimes(\mathcal{C})$, the equivariantisation is defined as follows:
	\begin{enumerate}
		\item 
		A \textsl{$G$-equivariant object} in~$\mathcal{C}$ is a pair $(X, u:= \lbrace u_g\rbrace_{g\in G})$, where $X \in \mathcal{C}$ and~$u$ is a family of isomorphisms 
		\begin{equation}
			u_g\colon \rho_g(X)\stackrel{\cong}{\longrightarrow} X\,,
		\end{equation}
		such that $u_g\circ\rho_g(u_h)=u_{hg}\circ(\rho^2_{g,h})_X$, i.e.\ the following diagram commutes:
		\begin{equation}
		\label{eq:defining property of equivariant structure}
			\begin{tikzcd}[column sep= 40,row sep= 30, ampersand replacement=\&]
			\rho_g(\rho_h(X))
			\arrow[d, "{(\rho^2_{g,h})_X}"{left}]
			\arrow[r, "{\rho_g(u_h)}"]
			\& \rho_g(X)
			\arrow[d, "{u_g}"]\\
			\rho_{hg}(X)
			\arrow[r, "u_{hg}"{below}]
			\& X			
			\end{tikzcd}\,.
		\end{equation}
		\item 
		The \textsl{equivariantisation} of~$\mathcal{C}$ (with respect to~$\rho$) is the category~$\mathcal{C}^G$ whose objects are $G$-equivariant objects in $\mathcal{C}$ and whose morphisms $(X,u)\longrightarrow (Y,v)$ are morphisms $f\colon X\longrightarrow Y$ in~$\mathcal{C}$ such that $f\circ u_g=v_g\circ\rho_g(f)$, i.e.
		\begin{equation}
		\label{eq:defining property of morphisms in equivariantisation}
		\begin{tikzcd}[column sep= 40,row sep= 30, ampersand replacement=\&]
			\rho_g(X)
			\arrow[d, "{u_g}"{left}]
			\arrow[r, "{\rho_g(f)}"]
			\& \rho_g(Y)
			\arrow[d, "{v_g}"]\\
			X
			\arrow[r, "f"{below}]
			\& Y
		\end{tikzcd} 
		\end{equation}
		commutes for all $g\in G$. 
	\end{enumerate}
		The equivariantisation~$\mathcal{C}^G$ has a monoidal structure $(X,u)\otimes (Y,v):=(X\otimes Y, w)$ where
	\begin{equation}
		w_g := \Big( \!
		\begin{tikzcd}[column sep= 40,row sep= 30, ampersand replacement=\&]
			\rho_g(X\otimes Y)
			\arrow[r, "(\rho_g^2)^{-1}_{X,Y}"]
			\& 
			\rho_g(X)\otimes\rho_g(Y)
			\arrow[r, "u_g\otimes v_g"]
			\&
			X\otimes Y 
		\end{tikzcd} 
		\!\Big) \,.
\label{eq:equivariantStructureOfMonoidalProducts}
	\end{equation}
	The monoidal unit is $(\mathds{1},\lbrace (\rho^0_g)^{-1}\colon \rho_g(\mathds{1})\longrightarrow \mathds{1}\rbrace_{g\in G})$, associators and unitors are given by those of $\mathcal{C}$, i.e.\ they are independent of the choice of equivariant structures.
	
\medskip

To understand the connection between equivariantisation and line operators in a gauge theory, consider a 0-form $G$-symmetry action on a line operator via defects (before gauging):
\begin{equation}
	\begin{tikzpicture}[very thick,scale=0.5,color=black, baseline=0cm]
			\coordinate (hm) at (1,-2.5);
			\coordinate (hmt) at (1,2.5);
			\draw[color=blue!50!black] (hm) -- (hmt);
			\draw[color=blue!50!black] (hm) node[below] {\scriptsize$\rho_g(L)$};
		\end{tikzpicture}
		 =\,
        	\begin{tikzpicture}[very thick,scale=0.5,color=black, baseline=0cm]
        	\fill [orange!50,opacity=0.545] (0,2.5) .. controls +(0,.4) and +(0,.4) .. (2,2.5) .. controls +(0,-.4) and +(0,-.4) .. (0,2.5);
			\coordinate (hm) at (1,-2.5);
			\coordinate (hmt) at (1,2.5);
			\draw[color=blue!50!black] (hm) -- (1,0);
			\draw[color=blue!50!black] ($(hm)+(0,-.1)$) node[below] {\scriptsize$L$};
			\draw[color=blue!50!black] (hmt) -- (1,0);
			\draw[black] (0,0) node[left] {\scriptsize$\rho_g$};
			\draw[densely dotted, thin] (0,0) .. controls +(0,.4) and +(0,.4) .. (2,0);
			\draw[densely dotted, thin] (0,-2.5) .. controls +(0,.4) and +(0,.4) .. (2,-2.5);
			\fill [orange!50,opacity=0.545] (0,2.5) .. controls +(0,-.4) and +(0,-.4) .. (2,2.5) -- (2,-2.5) .. controls +(0,-.4) and +(0,-.4) .. (0,-2.5);
			\draw[densely dotted, thin] (0,0) .. controls +(0,-.4) and +(0,-.4) .. (2,0);
			\draw[very thin] (0,-2.5) .. controls +(0,-.4) and +(0,-.4) .. (2,-2.5);
			\draw[very thin] (0,2.5) .. controls +(0,-.4) and +(0,-.4) .. (2,2.5);
			\draw[very thin] (0,2.5) .. controls +(0,.4) and +(0,.4) .. (2,2.5)
			(0,-2.5) -- (0,2.5)
			(2,-2.5) -- (2,2.5);
		\end{tikzpicture}\, ,
\end{equation}
where the orange surface encodes the 0-form $G$-action, i.e.\ it is a domain wall relating the two sides by the symmetry transformation associated to~$g$. 
Then the gauging of this symmetry is given by condensation of the surface defect in the 3-dimensional bulk (i.e.\ putting it on a ``foam'' filling the entire space, see \Cref{sec:orbifoldDataFromSymmetry}). 
A line operator in the gauged theory has to come equipped with a way to intersect the surface defects associated to the $G$-symmetry.
These points of intersection are the equivariant structure of an equivariant object $(L,u)$:
\begin{equation}
	 	\begin{tikzpicture}[very thick,scale=0.5,color=black, baseline=0cm]
			\coordinate (hm) at (1,-2);
			\coordinate (hmt) at (1,2.5);
			\draw[white, line width=-8pt] (0,0) .. controls +(0,1) and +(0,1) .. (2,0) node [pos=0.5](u){};
        		\draw[color=blue!50!black] (hm) -- (u);
        		\fill [orange!50,opacity=0.545] ($(u)+(-3,1)$) -- ($(u)+(-2,-1)$) -- ($(u)+(3,-1)$) -- ($(u)+(2,1)$);
			\draw[color=blue!50!black] ($(hm)+(0,-.1)$) node[below] {\scriptsize$L$};
			\draw ($(u)+(-2.5,0)$) node[left] {\scriptsize$\rho_g$};
			\draw[very thin] ($(u)+(-3,1)$) -- ($(u)+(-2,-1)$) -- ($(u)+(3,-1)$) -- ($(u)+(2,1)$) -- ($(u)+(-3,1)$);
			\draw[color=blue!50!black] (hmt) -- (u);
			\fill[color=black] (u) circle (2.9pt)
			($(u)+(-.4,.25)$) node {\scriptsize$u_g$};
		\end{tikzpicture}\, \simeq\,
	 	\begin{tikzpicture}[very thick,scale=0.5,color=black, baseline=0cm]
			\coordinate (hm) at (1,-2);
			\coordinate (hmt) at (1,2.5);
			\draw[white, line width=-8pt] (0,0) .. controls +(0,1) and +(0,1) .. (2,0) node [pos=0.5](u){};
        		\draw[color=blue!50!black] (hm) -- (u);
        		\fill [orange!50,opacity=0.545] (0,0) .. controls +(0,1) and +(0,1) .. (2,0) .. controls +(0,-.4) and +(0,-.4) .. (0,0);
			\draw[color=blue!50!black] ($(hm)+(0,-.1)$) node[below] {\scriptsize$L$};
			\draw (0,-1) node[left] {\scriptsize$\rho_g$};
			\draw[densely dotted, thin] (0,0) .. controls +(0,.4) and +(0,.4) .. (2,0);
			\draw[densely dotted, thin] (0,-2) .. controls +(0,.4) and +(0,.4) .. (2,-2);
			\fill [orange!50,opacity=0.545] (0,0) .. controls +(0,-.4) and +(0,-.4) .. (2,0) -- (2,-2) .. controls +(0,-.4) and +(0,-.4) .. (0,-2);
			\draw[densely dotted, thin] (0,0) .. controls +(0,-.4) and +(0,-.4) .. (2,0);
			\draw[very thin] (0,-2) .. controls +(0,-.4) and +(0,-.4) .. (2,-2)
			(0,-2) -- (0,0)
			(2,-2) -- (2,0)
			(0,0) .. controls +(0,1) and +(0,1) .. (2,0);
			\draw[color=blue!50!black] (hmt) -- (u);
			\fill[color=black] (u) circle (2.9pt)
			($(u)+(-.4,.25)$) node {\scriptsize$u_g$};
		\end{tikzpicture}\,.\label{eq:gaugedlines}
\end{equation}
Note that the configuration on the right can be interpreted as a map $\rho_g(L)\longrightarrow L$ read from bottom to top, so inserting~$u_g$ at the junction makes sense.
\eqref{eq:defining property of equivariant structure} then specifies compatibility of these junctions with fusion of the respective symmetry defect surfaces.
As part of (gauging) a symmetry, one also has to include twisted sectors which we introduce in \Cref{sec:Gcbc}.

\subsection{Calculating Equivariantisations}\label{sec:CalculatingEq}
In \cite{BN}, the authors derived algebraic formulas that describe the simple objects of equivariantisations as well as their fusion rules. 
We present a concise summary of these, adapted to right $G$-actions.
This lets us efficiently compute the line operators after gauging a 0-form symmetry.
In the following, let $\mathcal{C}$ be a fusion category with $G$-action.

For a simple object~$Y$ in~$\mathcal{C}$, the \textsl{stabiliser subgroup~$G_Y$} of the $G$-action is given by those group elements which change the line operator only by an isomorphism,
\begin{equation}
	G_Y\equiv\mathrm{Stab}_Y:=\{g\in G\mid \rho_g(Y)\cong Y\}\,.
\end{equation}
For a choice of such isomorphisms $\xi_g\colon \rho_g(Y)\longrightarrow Y$ for $g\in G_Y$, the family of scalars $\beta_Y(g,h)\in\C$ (for $g,h\in G_Y$) defined by
\begin{equation}
	\beta_Y(g,h)^{-1}\id_Y:=\xi_g\circ \rho_g(\xi_h)\circ (\rho^2_{g,h})_Y^{-1}\circ \xi_{hg}^{-1}\colon Y\longrightarrow Y
	\label{eq:projectivitycocycle}
\end{equation}
produces a cocycle $\beta_Y\in \mathrm{H}^2(G_Y,\C^\times )$.
Lastly, we work with $\{Y_1,\ldots,Y_n\}$ a set of representatives of $G$-orbits of equivalence classes of simple objects in~$\mathcal{C}$, i.e.\ none of the $Y_i$, $Y_j$ are related via $\rho_g(Y_i)\cong Y_j$ for any $i\neq j$, and $g\in G$.
\begin{lemma}[label=lem:simplesInEquivar]{\cite[Cor.\,2.13, Lem.\,2.8]{BN}}
	There is a bijection between the set of isomorphism classes of simple objects of~$\mathcal{C}^G$ and the set of pairs $(Y,\pi)$ where~$Y\in\{Y_1,\ldots,Y_n\}$ and~$\pi$ runs over the isomorphism classes of irreducible $\beta_Y$-projective $G_Y$-representations $(V_\pi,\pi)$. 
	The equivariant object $S_{Y,\pi}:=(X,u)$ associated to such a pair $(Y,\pi)$ relies on the choice of $\{\xi_g\}_g$ and is given by the underlying object
	\begin{equation}
		X:= \bigoplus_{t\in \mathcal{R}}\rho_t(V_\pi\otimes Y)\,,
	\end{equation}
	where $\mathcal{R}_Y\equiv\mathcal{R}$ is a set of representatives of $G_Y\backslash G$, the right $G_Y$-cosets in~$G$, and the equivariant structure is
	\begin{equation}
		u_g:=\bigoplus_{t\in \mathcal{R}} \rho_s(\pi_h\otimes \xi_h) \circ (\rho^2_{s,h})^{-1}\circ\rho^2_{g,t} \colon \bigoplus_{t\in \mathcal{R}}\rho_g(\rho_t(V_\pi\otimes Y)) \longrightarrow \bigoplus_{s\in \mathcal{R}}\rho_s(V_\pi\otimes Y)\,,
		\label{eq:equivariantstructurewithprojrep}
	\end{equation}
	for $g\in G$, where $tg=hs$ uniquely defines $h\in G_Y$ and $s\in\mathcal{R}$.
\end{lemma}

Considering the underlying objects, we have
\begin{align}
	S_{Y,\pi}\otimes S_{Z,\gamma}&= \left(\bigoplus_{t\in \mathcal{R}_Y}\rho_t(V_\pi\otimes Y)\right) \otimes \left(\bigoplus_{s\in \mathcal{R}_Z}\rho_s(V_\gamma\otimes Z)\right)\nonumber\\
	&\cong \bigoplus_{(t,s)\in \mathcal{R}_Y\times\mathcal{R}_Z}\left(\rho_t(V_\pi\otimes Y) \otimes \rho_s(V_\gamma\otimes Z)\right)\,.
	\label{eq:eqfusiondecomp}
\end{align}
The equivariant structure is given by \eqref{eq:equivariantStructureOfMonoidalProducts}.
Our goal is to decompose this sum into (equivariant) subobjects in order to determine fusion rules.

There is a right $G$-action on $\mathcal{R}_Y\times\mathcal{R}_Z$ given by right multiplication.
For an orbit~$\mathcal{O}$ of this action, we set
\begin{equation}
	S_\mathcal{O}:= \bigoplus_{(t,s)\in\mathcal{O}}\left(\rho_t(V_\pi\otimes Y) \otimes \rho_s(V_\gamma\otimes Z)\right)\,.
\end{equation}
Since $\mathcal{R}_Y\times\mathcal{R}_Z$ is a disjoint union of such orbits, we have $S_{Y,\pi}\otimes S_{Z,\gamma}\cong \bigoplus_\mathcal{O}S_\mathcal{O}$.

\begin{lemma}[label=lem:FusionRulesOfSimplesInEq]{\cite[Lem.\,3.3]{BN}}
	The object $S_\mathcal{O}$ is an equivariant subobject of $S_{Y,\pi}\otimes S_{Z,\gamma}$ for every orbit $\mathcal{O}\subset\mathcal{R}_Y\times\mathcal{R}_Z$.
\end{lemma}

Note that for $G_Y=G_Z=G$, then there is only a single orbit, and the fusion is given by 
\begin{equation}
	S_{Y,\pi}\otimes S_{Z,\gamma}=(V_\pi\otimes V_\gamma)\otimes (Y\otimes Z)\,,
	\label{eq:eqfusiondecompsimplified}
\end{equation}
and the equivariant structure is once again given by \eqref{eq:equivariantStructureOfMonoidalProducts}. 
If the $G$-action is strict and we chose $\xi_g=\id$ for both~$Y$ and~$Z$, the resulting object can be decomposed into simple objects by simply decomposing the representation $V_\pi\otimes V_\gamma$ into irreducibles and decomposing $Y\otimes Z$ into simple objects.

A general formula for fusion coefficients can be found in \cite[Thm.\,3.9]{BN}, the above is sufficient for our examples.

\subsection[$G$-crossed Braided Extensions: Twisted Sectors]{$\boldsymbol{G}$-crossed Braided Extensions: Twisted Sectors}
\label{sec:Gcbc}
The next step in our exposition is to add twisted sectors, i.e. introduce extensions.
As before, we denote the unit element by $e\in G$.
\begin{definition}[label=def:G-crossed braided category]
	A \textsl{$G$-crossed braided fusion category} is a fusion category $\mathcal{C}^\times_G$ together with
	\begin{enumerate}
		\item 
		a (monoidal) $G$-action $\rho\colon \underline{G}^\textrm{rev}\longrightarrow \Aut^\otimes(\mathcal{C}^\times_G)$,
		\item 
		a decomposition $\mathcal{C}^\times_G=\bigoplus_{g\in G}\mathcal{C}_g$ into twisted sectors, 
		\item 
		a \textsl{$G$-braiding}~$c$ with components (read from bottom to top)
		\begin{equation}
			c_{X,Y}
			\equiv
			\hphantom{\text{{\scriptsize$\rho_{h}($}}}
			\begin{tikzpicture}[very thick,scale=0.7,color=blue!50!black,baseline]
				\draw (-1,-1) node[below] (X) {{\scriptsize$X$}};
				\draw (1,-1) node[below] (Y) {{\scriptsize$Y$}};
				\draw (1,1) node[above] (Xu) {{\scriptsize$\rho_h(X)$}};
				\draw (-1,1) node[above] (Yu) {{\scriptsize$Y\vphantom{\rho_h(X)}$}};
				\draw (Y)  .. controls +(0,1) and +(0,-1) .. (Yu); 
				\draw[color=white, line width=4pt] (X)  .. controls +(0,1) and +(0,-1) .. (Xu); 
				\draw (X)  .. controls +(0,1) and +(0,-1) .. (Xu);
			\end{tikzpicture}
			\colon X\otimes Y\stackrel{\cong}{\longrightarrow} Y\otimes \rho_h(X) 
			\label{eq:Gcbraiding}
		\end{equation}
		for $h\in G$, $X\in\mathcal{C}^\times_G$, and $Y\in \mathcal{C}_h$.
	\end{enumerate}
	These are subject to the compatibility conditions:
	\begin{enumerate}[label=(\alph*)]
		\item
		$\rho_g(\mathcal{C}_h)\subset \mathcal{C}_{g^{-1}hg}$ for all $g,h\in G$,
		\item 
		the isomorphisms $c_{X,Y}$ are natural in~$X$ and~$Y$,
		\item 
		the isomorphisms $c_{X,Y}$ are compatible with the $G$-action in the sense that for all $g\in G$ we have 
		\begin{equation}
			\rho_g(c_{X,Y})=c_{\rho_g(X),\rho_g(Y)} \, ,
		\end{equation}
		suppressing the monoidal structure of~$\rho_g$,
		\item 
		the following identities hold for all $g,h\in G$, $Y\in \mathcal{C}_h$, and $Z\in\mathcal{C}_k$: 
		\begin{align}
		\label{eq:braiding and monoidal product inGcbc}
			\begin{tikzpicture}[very thick,scale=0.7,color=blue!50!black, baseline]
			\draw (-1,-1) node[below] (X) {{\scriptsize$X$}};
			\draw (1,-1) node[below] (Y) {{\scriptsize$Y\otimes Z$}};
			\draw (1,1) node[above] (Xu) {{\scriptsize$\rho_{hk}(X)$}};
			\draw (-1,1) node[above] (Yu) {{\scriptsize$Y\otimes Z\vphantom{\rho_h(X)}$}};
			\draw[ultra thick] (1,-1)  .. controls +(0,1) and +(0,-1) .. (-1,1); 
			\draw[color=white, line width=4pt] (-1,-1)  .. controls +(0,1) and +(0,-1) .. (1,1); 
			\draw (-1,-1)  .. controls +(0,1) and +(0,-1) .. (1,1);
			\end{tikzpicture}
			\, &=\,
			\begin{tikzpicture}[very thick,scale=0.35,color=blue!50!black, baseline=.35cm]
			\draw (-1,-1) node[below] (X) {{\scriptsize$X$}};
			\draw (1,-1) node[below] (Y) {{\scriptsize$Y$}};
			\draw (3,-1) node[below] (Z) {{\scriptsize$Z$}};
			\draw (3,3) node[above] (Xuu) {{\scriptsize$\;\rho_{hk}(X)$}};
			\draw (-1,3) node[above] (Yuu) {{\scriptsize$Y\vphantom{\rho_h(X)}$}};
			\draw (1,3) node[above] (Zuu) {{\scriptsize$Z\vphantom{\rho_h(X)}$}};
			\draw (1,-1)  .. controls +(0,1) and +(0,-1) .. (-1,1); 
			\draw[color=white, line width=4pt] (-1,-1)  .. controls +(0,1) and +(0,-1) .. (1,1); 
			\draw (-1,-1)  .. controls +(0,1) and +(0,-1) .. (1,1)
			(Z)--(3,1)
			(-1,1)--(Yuu);
			\draw (3,1)  .. controls +(0,1) and +(0,-1) .. (1,3); 
			\draw[color=white, line width=4pt] (1,1)  .. controls +(0,1) and +(0,-1) .. (3,3); 
			\draw (1,1)  .. controls +(0,1) and +(0,-1) .. (3,3);
			\end{tikzpicture}\,,&
			\begin{tikzpicture}[very thick,scale=0.7,color=blue!50!black, baseline]
			\draw (-1,-1) node[below] (X) {{\scriptsize$X\otimes Y$}};
			\draw (1,-1) node[below] (Y) {{\scriptsize$Z$}};
			\draw (1,1) node[above] (Xu) {{\scriptsize$\rho_{k}(X\otimes Y)$}};
			\draw (-1,1) node[above] (Yu) {{\scriptsize$Z\vphantom{\rho_h(X)}$}};
			\draw (1,-1)  .. controls +(0,1) and +(0,-1) .. (-1,1); 
			\draw[color=white, line width=4pt] (-1,-1)  .. controls +(0,1) and +(0,-1) .. (1,1); 
			\draw[ultra thick] (-1,-1)  .. controls +(0,1) and +(0,-1) .. (1,1);
			\end{tikzpicture}
			\, &=\,
		\begin{tikzpicture}[very thick,scale=0.35,color=blue!50!black, baseline=.35cm]
			\draw (-1,-1) node[below] (X) {{\scriptsize$X$}};
			\draw (1,-1) node[below] (Y) {{\scriptsize$Y$}};
			\draw (3,-1) node[below] (Z) {{\scriptsize$Z$}};
			\draw (3,3) node[above] (Xuu) {{\scriptsize$\;\;\rho_k(Y)$}};
			\draw (-1,3) node[above] (Yuu) {{\scriptsize$Z\vphantom{\rho_h(X)}$}};
			\draw (1,3) node[above] (Zuu) {{\scriptsize$\rho_k(X)\;$}};
			\draw (3,-1)  .. controls +(0,1) and +(0,-1) .. (1,1); 
			\draw[color=white, line width=4pt] (1,-1)  .. controls +(0,1) and +(0,-1) .. (3,1); 
			\draw (1,-1)  .. controls +(0,1) and +(0,-1) .. (3,1)
			(X)--(-1,1)
			(3,1)--(Xuu);
			\draw (1,1)  .. controls +(0,1) and +(0,-1) .. (-1,3); 
			\draw[color=white, line width=4pt] (-1,1)  .. controls +(0,1) and +(0,-1) .. (1,3); 
			\draw (-1,1)  .. controls +(0,1) and +(0,-1) .. (1,3);
			\end{tikzpicture}\,,
		\end{align}
		where we suppressed the monoidal structure of~$\rho$ and associators.
	\end{enumerate}
	For a given braided fusion category $\mathcal{C}$, a \textsl{$G$-crossed braided extension} is a $G$-crossed braided fusion category $\Gcbc$ such that $\mathcal{C}_e\cong \mathcal{C}$ as braided fusion categories, i.e.\ the line operators~$\mathcal{C}$ are the bulk line operators.
	If $\mathcal{C}$ comes equipped with a $G$-action then the $G$-action of the extension is required to restrict to the given one.
\end{definition}

Physically, the graded components of $\Gcbc$ encode the twisted sectors, i.e.\ line operators on which the symmetry surfaces end.
For $Y\in\mathcal{C}_h$, this is described by the following diagram:
\begin{equation}
	       	\begin{tikzpicture}[very thick,scale=0.5,color=black, baseline=0cm]
			\fill [orange!50,opacity=0.545] (-1,-3.5) -- (-1,1.5) -- (1,2.5) -- (1,-2.5);
			\draw[very thin,black] (1,-2.5) -- (-1,-3.5) -- (-1,1.5) -- (1,2.5);
			\draw[black] (-1,-2.8) node[right] {\scriptsize$\rho_h$};
			\coordinate (hm) at (1,-2.5);
			\coordinate (hmt) at (1,2.5);
			\draw[color=blue!50!black] (hm) -- (hmt)
			(hm) node[below] {\scriptsize$Y$}; 
		\end{tikzpicture}\,.
\end{equation}
The $G$-crossed braiding \eqref{eq:Gcbraiding} then describes a situation where the bottom left line passes through the $h$-surface attached to the right line, thus it is transformed by the symmetry action $\rho_h$:
\begin{equation}
		\begin{tikzpicture}[very thick,scale=0.7,color=blue!50!black, baseline,style={x={(1cm,0cm)},y={(0cm,1cm)},z={(-.6cm,-.2cm)}}]
			\draw (-1,-1,.5) node[below] (X) {{\scriptsize$X$}};
			\draw (1,-1,0) node[below] (Y) {{\scriptsize$Y$}};
			\draw (1,1,.5) node[above] (Xu) {{\scriptsize$\rho_{h}(X)$}};
			\draw (-1,1,0) node[above] (Yu) {{\scriptsize$Y\vphantom{\rho_h(X)}$}};
			\path[line width=-6pt] (-1,1,2) .. controls +(0,-1,0) and +(0,1,0) .. (1,-1,2) node[pos=0.5] (a){};
			\fill [orange!50,opacity=0.545] (1,-1,0) .. controls +(0,1,0) and +(0,-1,0) .. (-1,1,0) -- (-1,1,.5) .. controls +(0,-1,0) and +(0,1,0) .. (1,-1,.5);
			\draw[partial curve={from .4 to 1 curve (1,-1,0)  .. controls (1,0,0) and (-1,0,0) .. (-1,1,0)}];
			\fill [orange!50,opacity=0.545] (-1,-1,.5) .. controls +(0,1) and +(0,-1) .. (1,1,.5) -- (1,1,2.5) .. controls +(0,-1) and +(0,1) .. (-1.,-1,2.5);
			\draw[partial curve={from 0.0 to 0.5 curve (-1,-1,.5)  .. controls (-1,0,.5) and (1,0,.5) .. (1,1,.5)}];
			\fill [orange!50,opacity=0.545] (1,-1,.5) .. controls +(0,1,0) and +(0,-1,0) .. (-1,1,.5) -- (-1,1,2) .. controls +(0,-1,0) and +(0,1,0) .. (1,-1,2);
			\draw[partial curve={from 0 to .4 curve (1,-1,0)  .. controls (1,0,0) and (-1,0,0) .. (-1,1,0)}];
			\draw[partial curve={from 0.5 to 1 curve (-1,-1,.5)  .. controls (-1,0,.5) and (1,0,.5) .. (1,1,.5)}];
			\path[line width=-8pt] (-1,-1,.5)  .. controls (-1,0,.5) and (1,0,.5) .. (1,1,.5) node[pos=0.5] (int){};
			\draw[very thin] (-1,1,0) -- (-1,1,2) .. controls +(0,-1,0) and +(0,1,0) .. (1,-1,2) -- (1,-1,0);
			\draw[very thin] (1,1,.5) -- (1,1,2.5) .. controls +(0,-1) and +(0,1) .. (-1.,-1,2.5) -- (-1,-1,.5);
			\draw[thin,densely dotted] (a) -- (int);
			\fill[color=black] (int) circle (2pt); 
			\draw[black]
			(1,-.8,1.6) node {\scriptsize$\rho_h$}
			(-1,-.8,1.7) node {\scriptsize$\rho_g$};
		\end{tikzpicture}\,,
\end{equation}
where the intersection point of~$X$ with the surface~$\rho_h$ is marked by a black circle and the intersection line of the two surfaces is dotted.
The braiding is the isomorphism which describes this particular junction.

For a $G$-crossed braided fusion category $(\mathcal{C}^\times_G, \rho, c)$, the equivariantisation $(\mathcal{C}^\times_G)^G$ can be equipped with a braiding whose components are
	\begin{align}
		c^G_{(X,u),(Y,v)} 
		\;:= 
		\bigoplus_{h\in G}(\id_{Y_h}\otimes u_h) \circ c_{X,Y} 
		\;= \bigoplus_{h\in G}
		\begin{tikzpicture}[very thick,scale=0.7,color=blue!50!black, baseline]
			\draw (-1,-1) node[below] (X) {{\scriptsize$X$}};
			\draw (1,-1) node[below] (Y) {{\scriptsize$Y_h$}};
			\draw (1,1) node[above] (Xu) {{\scriptsize$X$}};
			\draw (-1,1) node[above] (Yu) {{\scriptsize$Y_h$}};
			\draw (1,-1)  .. controls +(0,1) and +(0,-1) .. (-1,1); 
			\draw[color=white, line width=4pt] (-1,-1)  .. controls +(0,1) and +(0,-1) .. (1,1); 
			\draw (-1,-1)  .. controls +(0,1) and +(0,-1) .. (1,1) node [pos=0.75](u){} ;
			\fill[color=black] (u) circle (2.9pt) node [left] {\scriptsize$u_h$}; 
		\end{tikzpicture}\,,
		\label{eq:eq of Gcbc is braided}
	\end{align}
for $h\in G$, $X\in\mathcal{C}^\times_G$, and $Y=\bigoplus_{h\in G} Y_h$, with $Y_h\in \mathcal{C}_h$. 

\medskip

The term ``gauging'' has been adapted in the mathematical literature to describe the following 2-step process \cite{CGPZ}: 
First, take a $G$-crossed braided extension $\Gcbc$ of the underlying category of line operators $\mathcal{C}$ and then calculate the equivariantisation.
This amounts to first adding twisted sector line operators and then equipping line operators with the relevant junctions with symmetry surface defects (if possible), as required for line operators in the gauged theory (cf. \eqref{eq:gaugedlines}).
Note that in this paper, we prefer to separate the two steps and we sometimes (as above) use the term ``gauging'' to refer to the second step, the equivariantisation.

In our applications we work with categories in which objects~$X$ have duals~$X	^*$ and there is a pivotal structure, i.e.\ isomorphisms $j_X\colon X\longrightarrow X^{**}$.
In this setting, we require the $G$-action to be pivotal as well.
Duals of defects are given by their orientation reversal, and since this should be unique, left and right duals must be identified, which is what the pivotal structure mediates.

\subsection{Orbifold Data: Gauging via Modules over an Algebra}\label{sec:orbifoldDataFromSymmetry}
In this section we review orbifold data and the orbifold construction, providing details only in dimensions 2 and 3.
The resulting algebraic data gives rise to the algebras of lines or surfaces which gauge 1-form and 0-form symmetries, respectively.
We present the algebraic data in 2 dimensions first before briefly explaining how this data is used geometrically in the \tsl{orbifold construction} to produce a new TQFT from a given one (gauging, in particular).
We conclude by presenting 3-dimensional orbifold data.
A thorough treatment of both the algebraic and geometric perspectives in arbitrary dimensions can be found in \cite{CRS1} or the survey \cite{carqueville2023orbifoldstopologicalquantumfield}.

\medskip

To any (collection of) $n$-dimensional defect TQFT~$\mathcal{T}$ there is an associated $n$-category of (topological) defects $\mathcal{D}$.
Its objects are the theories themselves, 1-morphisms are (codimension 1) interfaces between them, 2-morphisms are codimension 2 defects separating these interfaces and so on.
In 3 dimensions, 1-morphisms are thus topological surface defects, 2-morphisms are topological line defects between them, and 3-morphisms are topological junctions between line operators (in particular, local operators are junctions between transparent lines).
In 2 dimensions, 1-morphisms are line defects and 2-morphisms are junctions between them.

A \tsl{$2$-dimensional orbifold datum} is a $\Delta$-separable, symmetric Frobenius algebra.
Recall that a \textsl{Frobenius algebra} is a 1-endomorphism $A\colon \mathcal T\to \mathcal T$ (domain wall in the chosen theory~$\mathcal T$) together with multiplication, comultiplication, unit, and counit, given by 2-morphisms (point defects)
\begin{equation}
	\mu=
	\begin{tikzpicture}[very thick,scale=0.4,color=green!50!black, baseline=0.4cm]
		\coordinate (Yt) at (0,0);
		\coordinate (Xt) at (1,0);
		\draw (0.5, 0.9) -- (0.5,2);
		\draw[-dot-] (Yt) .. controls +(0,1) and +(0,1) .. (Xt);
	\end{tikzpicture}
	\colon A\otimes A \longrightarrow A
	\, ,\quad
	\Delta=
	\begin{tikzpicture}[very thick,scale=0.4,color=green!50!black, baseline=-0.4cm, rotate=180]
		\coordinate (Yt) at (0,0);
		\coordinate (Xt) at (1,0);
		\draw (0.5, 0.9) -- (0.5,2);
		\draw[-dot-] (Yt) .. controls +(0,1) and +(0,1) .. (Xt);
	\end{tikzpicture}
	\colon A \longrightarrow A\otimes A
	\, ,\quad
	\eta=
	\begin{tikzpicture}[very thick,scale=0.4,color=green!50!black, baseline=0.4cm]
		\draw (0.5,0.9) node[Odot] (unit){};
		\draw (unit) -- (0.5,2);
	\end{tikzpicture}
	\colon\id_a \longrightarrow A
	\, ,\quad
	\varepsilon=
	\begin{tikzpicture}[very thick,scale=0.4,color=green!50!black, baseline=-0.55cm, rotate=180]
		\draw (0.5,0.9) node[Odot] (counit){};
		\draw (counit) -- (0.5,2);
	\end{tikzpicture}
	\colon A \longrightarrow \id_a
\end{equation}
subject to the conditions
\begin{equation}
	\label{eq:FrobeniusAlgebra}	
\begin{array}{ccccc}
	\text{associative}\;&\text{unital}\;&\text{coassociative}\;&\text{counital}\;&\text{Frobenius}\;\\
	\tikzzbox{%
		\begin{tikzpicture}[very thick,scale=0.53,color=green!50!black, baseline=0.59cm]
			\draw[-dot-] (3,0) .. controls +(0,1) and +(0,1) .. (2,0);
			\draw[-dot-] (2.5,0.75) .. controls +(0,1) and +(0,1) .. (3.5,0.75);
			\draw (3.5,0.75) -- (3.5,0); 
			\draw (3,1.5) -- (3,2.25); 
		\end{tikzpicture} 
	}%
	=
	\tikzzbox{%
		\begin{tikzpicture}[very thick,scale=0.53,color=green!50!black, baseline=0.59cm]
			\draw[-dot-] (3,0) .. controls +(0,1) and +(0,1) .. (2,0);
			\draw[-dot-] (2.5,0.75) .. controls +(0,1) and +(0,1) .. (1.5,0.75);
			\draw (1.5,0.75) -- (1.5,0); 
			\draw (2,1.5) -- (2,2.25); 
		\end{tikzpicture} 
	}%
	\, , &\;
	\tikzzbox{%
		\begin{tikzpicture}[very thick,scale=0.4,color=green!50!black, baseline]
			\draw (-0.5,-0.5) node[Odot] (unit) {}; 
			\fill (0,0.6) circle (5.0pt) node (meet) {};
			\draw (unit) .. controls +(0,0.5) and +(-0.5,-0.5) .. (0,0.6);
			\draw (0,-1.5) -- (0,1.5); 
		\end{tikzpicture} 
	}%
	=
	\tikzzbox{%
		\begin{tikzpicture}[very thick,scale=0.4,color=green!50!black, baseline]
			\draw (0,-1.5) -- (0,1.5); 
		\end{tikzpicture} 
	}%
	=
	\tikzzbox{%
		\begin{tikzpicture}[very thick,scale=0.4,color=green!50!black, baseline]
			\draw (0.5,-0.5) node[Odot] (unit) {}; 
			\fill (0,0.6) circle (5.0pt) node (meet) {};
			\draw (unit) .. controls +(0,0.5) and +(0.5,-0.5) .. (0,0.6);
			\draw (0,-1.5) -- (0,1.5); 
		\end{tikzpicture} 
	}%
	\, , \;&
	\tikzzbox{%
		\begin{tikzpicture}[very thick,scale=0.53,color=green!50!black, baseline=-0.59cm, rotate=180]
			\draw[-dot-] (3,0) .. controls +(0,1) and +(0,1) .. (2,0);
			\draw[-dot-] (2.5,0.75) .. controls +(0,1) and +(0,1) .. (1.5,0.75);
			\draw (1.5,0.75) -- (1.5,0); 
			\draw (2,1.5) -- (2,2.25); 
		\end{tikzpicture} 
	}%
	=
	\tikzzbox{%
		\begin{tikzpicture}[very thick,scale=0.53,color=green!50!black, baseline=-0.59cm, rotate=180]
			\draw[-dot-] (3,0) .. controls +(0,1) and +(0,1) .. (2,0);
			\draw[-dot-] (2.5,0.75) .. controls +(0,1) and +(0,1) .. (3.5,0.75);
			\draw (3.5,0.75) -- (3.5,0); 
			\draw (3,1.5) -- (3,2.25); 
		\end{tikzpicture} 
	}%
	\, , 
	&\;
	\tikzzbox{%
		\begin{tikzpicture}[very thick,scale=0.4,color=green!50!black, baseline=0, rotate=180]
			\draw (0.5,-0.5) node[Odot] (unit) {}; 
			\fill (0,0.6) circle (5.0pt) node (meet) {};
			\draw (unit) .. controls +(0,0.5) and +(0.5,-0.5) .. (0,0.6);
			\draw (0,-1.5) -- (0,1.5); 
		\end{tikzpicture} 
	}%
	=
	\tikzzbox{%
		\begin{tikzpicture}[very thick,scale=0.4,color=green!50!black, baseline=0, rotate=180]
			\draw (0,-1.5) -- (0,1.5); 
		\end{tikzpicture} 
	}%
	=
	\tikzzbox{%
		\begin{tikzpicture}[very thick,scale=0.4,color=green!50!black, baseline=0cm, rotate=180]
			\draw (-0.5,-0.5) node[Odot] (unit) {}; 
			\fill (0,0.6) circle (5.0pt) node (meet) {};
			\draw (unit) .. controls +(0,0.5) and +(-0.5,-0.5) .. (0,0.6);
			\draw (0,-1.5) -- (0,1.5); 
		\end{tikzpicture} 
	}%
	\, , \;&
	\tikzzbox{%
		\begin{tikzpicture}[very thick,scale=0.4,color=green!50!black, baseline=0cm]
			\draw[-dot-] (0,0) .. controls +(0,-1) and +(0,-1) .. (-1,0);
			\draw[-dot-] (1,0) .. controls +(0,1) and +(0,1) .. (0,0);
			\draw (-1,0) -- (-1,1.5); 
			\draw (1,0) -- (1,-1.5); 
			\draw (0.5,0.8) -- (0.5,1.5); 
			\draw (-0.5,-0.8) -- (-0.5,-1.5); 
		\end{tikzpicture}
	}%
	=
	\tikzzbox{%
		\begin{tikzpicture}[very thick,scale=0.4,color=green!50!black, baseline=0cm]
			\draw[-dot-] (0,1.5) .. controls +(0,-1) and +(0,-1) .. (1,1.5);
			\draw[-dot-] (0,-1.5) .. controls +(0,1) and +(0,1) .. (1,-1.5);
			\draw (0.5,-0.8) -- (0.5,0.8); 
		\end{tikzpicture}
	}%
	=
	\tikzzbox{%
		\begin{tikzpicture}[very thick,scale=0.4,color=green!50!black, baseline=0cm]
			\draw[-dot-] (0,0) .. controls +(0,1) and +(0,1) .. (-1,0);
			\draw[-dot-] (1,0) .. controls +(0,-1) and +(0,-1) .. (0,0);
			\draw (-1,0) -- (-1,-1.5); 
			\draw (1,0) -- (1,1.5); 
			\draw (0.5,-0.8) -- (0.5,-1.5); 
			\draw (-0.5,0.8) -- (-0.5,1.5); 
		\end{tikzpicture}
	}%
	\, .
	\end{array}
\end{equation}
As before, these diagrams are read from bottom to top.

An algebra is \textsl{separable} if there exists a section for the multiplication as a bimodule morphism, and \textsl{$\Delta$-separable} if this section is the comultiplication. 
If there is a braiding, we can speak of commutative algebras, and if the 2-category~$\mathcal{D}$ is pivotal, we have symmetric Frobenius algebras. 
The defining conditions of these properties are as follows: 
\begin{equation}
\begin{array}{ccccc}
	\Delta\text{-separable}&&\text{commutative}&&\text{symmetric }\\
	\label{eq:Symmetry and Commutativity}

	\begin{tikzpicture}[very thick,scale=0.5,color=green!50!black, baseline=-0.2cm]
		\draw[-dot-] (-.5,-.5) .. controls +(0,-1) and +(0,-1) .. (1,-.5);
		\draw[-dot-] (-.5,0) .. controls +(0,1) and +(0,1) .. (1,0);
		\draw (0.25,-1.2) -- (0.25,-2)
		(1,-.5) -- (1,0)
		(-.5,-.5) -- (-.5,0)
		(0.25,0.8) -- (0.25,1.5);
	\end{tikzpicture}
	\, = \, 
	\begin{tikzpicture}[very thick,scale=0.5,color=green!50!black, baseline=-0.2cm]
		\draw (0.5,-2) -- (0.5,1.5); 
	\end{tikzpicture}
	\, ,
	&\quad &
	\begin{tikzpicture}[very thick,scale=0.5,color=green!50!black, baseline=0.2cm]
		\coordinate (X) at (0,-0.5);
		\coordinate (Y) at (1,-0.5);
		\coordinate (Yt) at (0,0.5);
		\coordinate (Xt) at (1,0.5);
		\draw (X)  .. controls +(0,0.5) and +(0,-0.5) .. (Xt); 
		\draw[color=white, line width=4pt] (Y) .. controls +(0,0.5) and +(0,-0.5) .. (Yt); 
		\draw (Y)  .. controls +(0,0.5) and +(0,-0.5) .. (Yt)
		(0.5, 1.5) -- (0.5,1.5);
		\draw[-dot-] (Yt) .. controls +(0,0.7) and +(0,0.7) .. (Xt);
		\draw (X) -- (0,-1.2);
		\draw (Y) -- (1,-1.2);
		\draw (0.5,1) -- (0.5,2.3);
	\end{tikzpicture}
	= 
	\begin{tikzpicture}[very thick,scale=0.5,color=green!50!black, baseline=0.2cm]
		\coordinate (X) at (0,-0.5);
		\coordinate (Y) at (1,-0.5);
		\coordinate (Yt) at (0,0.5);
		\coordinate (Xt) at (1,0.5);
		\draw[-dot-] (Yt) .. controls +(0,0.7) and +(0,0.7) .. (Xt);
		\draw (Xt) -- (1,-1.2);
		\draw (Yt) -- (0,-1.2);
		\draw (0.5,1) -- (0.5,2.3);
	\end{tikzpicture}
	\, ,
	&\quad&
	\begin{tikzpicture}[very thick,scale=0.5,color=green!50!black, baseline=-0.2cm]
		\draw[-dot-] (0,0) .. controls +(0,1) and +(0,1) .. (-1,0);
		\draw[postaction={decorate}, decoration={markings,mark=at position .48 with {\arrow{>}}}] (1,0) .. controls +(0,-1) and +(0,-1) .. (0,0);
		\draw (-1,0) -- (-1,-2); 
		\draw (1,0) -- (1,1.5); 
		\draw (-0.5,1.2) node[Odot] (end) {}; 
		\draw (-0.5,0.8) -- (end); 
	\end{tikzpicture}
	= 
	\begin{tikzpicture}[very thick,scale=0.5,color=green!50!black, baseline=-0.2cm]
		\draw[postaction={decorate}, decoration={markings,mark=at position .48 with {\arrow{>}}}] (-1,0) .. controls +(0,-1) and +(0,-1) .. (0,0);
		\draw[-dot-] (1,0) .. controls +(0,1) and +(0,1) .. (0,0);
		\draw (-1,0) -- (-1,1.5); 
		\draw (1,0) -- (1,-2); 
		\draw (0.5,1.2) node[Odot] (end) {}; 
		\draw (0.5,0.8) -- (end); 
	\end{tikzpicture}
	\, .
\end{array}
\end{equation}

The \textsl{orbifold construction} is a general procedure to produce new TQFTs from a given (oriented) defect TQFT.
This procedure is morally a state sum construction, i.e.\ a process where various ``states'' are summed over to calculate a partition function.
In order to gauge a 0-form $G$-symmetry, we sum over contributions to a partition function that come from different principal $G$-bundles.
Principal bundles can be described by transition functions which we think of as ``gluing together'' patches of the bundle across a manifold.
For finite groups, transition functions correspond to labelling boundaries between these patches by single group elements.
One way to partition a manifold into patches is to triangulate it and take the Poincaré-dual.
By labelling the Poincaré-dual stratification, the orbifold construction produces a network of defects which fills the underlying manifold and can be evaluated by the partition function.
Since we use duals of triangulations, there is a finite set of defects which constitutes an \textsl{orbifold datum}. 
In two dimensions, this includes a domain wall and the two ways a trivalent junction may be oriented:
\begin{equation}
	\begin{tikzpicture}[baseline=0cm]
    \node[name=t,regular polygon, white, regular polygon sides=3, draw,
     inner sep=.5cm] at (6,0) {};
     \draw[postaction={decorate}, decoration={markings,mark=at position .6 with {\arrow[color=black]{<}}},black, thick] (t.corner 2) -- (t.corner 1);
     \draw[postaction={decorate}, decoration={markings,mark=at position .6 with {\arrow[color=black]{<}}},black, thick] (t.corner 1) -- (t.corner 3);
     \draw[postaction={decorate}, decoration={markings,mark=at position .6 with {\arrow[color=black]{<}}},black, thick] (t.corner 2) -- (t.corner 3);
     \draw[postaction={decorate}, decoration={markings,mark=at position .8 with {\arrow[color=green!50!black]{<}}},green!50!black,very thick,shorten >= -0.35cm] (t.center) -- (t.side 1);
     \draw[postaction={decorate}, decoration={markings,mark=at position .8 with {\arrow[color=green!50!black]{>}}},green!50!black,very thick,shorten >= -0.35cm] (t.center) -- (t.side 2);
     \draw[postaction={decorate}, decoration={markings,mark=at position .8 with {\arrow[color=green!50!black]{<}}},green!50!black,very thick,shorten >= -0.35cm] (t.center) -- (t.side 3);
     \fill[green!50!black] (t.center) circle (2.5pt);
     \foreach \anchor/\placement/\lab in
    {corner 1/above/2, corner 2/left/3, corner 3/right/1}
  \draw (t.\anchor) node[\placement] {\scriptsize\lab};
  \draw[green!50!black] ($(t.center)+(.2,-.15)$) node {\scriptsize$\mu$};
\end{tikzpicture}
\qquad\qquad
	\begin{tikzpicture}[baseline=0cm]
    \node[name=t,regular polygon, white, regular polygon sides=3, draw,
     inner sep=.5cm] at (6,0) {};
     \draw[postaction={decorate}, decoration={markings,mark=at position .4 with {\arrow[color=black]{>}}},black, thick] (t.corner 2) -- (t.corner 1);
     \draw[postaction={decorate}, decoration={markings,mark=at position .4 with {\arrow[color=black]{>}}},black, thick] (t.corner 1) -- (t.corner 3);
     \draw[postaction={decorate}, decoration={markings,mark=at position .4 with {\arrow[color=black]{>}}},black, thick] (t.corner 2) -- (t.corner 3);
     \draw[postaction={decorate}, decoration={markings,mark=at position .8 with {\arrow[color=green!50!black]{>}}},green!50!black,very thick,shorten >= -0.35cm] (t.center) -- (t.side 1);
     \draw[postaction={decorate}, decoration={markings,mark=at position .8 with {\arrow[color=green!50!black]{<}}},green!50!black,very thick,shorten >= -0.35cm] (t.center) -- (t.side 2);
     \draw[postaction={decorate}, decoration={markings,mark=at position .8 with {\arrow[color=green!50!black]{>}}},green!50!black,very thick,shorten >= -0.35cm] (t.center) -- (t.side 3);
     \fill[green!50!black] (t.center) circle (2.5pt);
     \foreach \anchor/\placement/\lab in
    {corner 1/above/2, corner 2/left/1, corner 3/right/3}
  \draw (t.\anchor) node[\placement] {\scriptsize\lab};
  \draw[green!50!black] ($(t.center)+(.2,-.15)$) node {\scriptsize$\Delta$};
\end{tikzpicture}\,,
\end{equation}
where we noted the orientation of the triangles by enumerating their corners and the Poincaré-dual stratifications are drawn in green.
For a 0-form symmetry, this data is given by the group algebra and its multiplication and comultiplication.
As an example, consider a theory on a torus.
Given a 2-dimensional orbifold datum, the orbifold construction produces a new theory.
The ``vacuum'' partition function for this theory on the torus would be given by the partition function of the original theory evaluated on the following configuration:
\begin{equation}
		\begin{tikzpicture}[very thick,color=black, baseline=0cm, scale=.8]
			\fill[color=orange!50,opacity=0.545] (4,-1) ellipse (4 and 3);
			\draw[thick] (4,-1) ellipse (4 and 3);
			\coordinate (M1) at (2.5,.5);
			\coordinate (M2) at (5.5,.5);
			\draw[thick] (M1) .. controls +(.6,-1) and +(-.6,-1) .. (M2) node[pos=.3, line width=-7pt](ML){} node[pos=.7, line width=-7pt](MR){} node[pos=.5, line width=-6pt](M3){};
			\fill[white] (ML) .. controls +(.4,.6) and +(-.4,.6) .. (MR) -- (ML);
			\fill[white] ($(ML)+(0,.02)$) .. controls +(.4,-.2) and +(-.4,-.2) .. ($(MR)+(0,.02)$) -- ($(ML)+(0,.02)$);
			\draw[thick] (ML) .. controls +(.4,.6) and +(-.4,.6) .. (MR)
			(M1) .. controls +(.6,-1) and +(-.6,-1) .. (M2);
			\coordinate (T0) at (.5,-1.3);
			\coordinate (T1) at (2,-.8);
			\coordinate (T2) at (2.5,-2.8);
			\coordinate (T3) at (4.5,-1);
			\coordinate (T4) at (4.9,-3.2);
			\coordinate (T5) at (6.5,-2.2);
			\coordinate (T6) at (6,-.2);
			\coordinate (T7) at (1.2,.5);
			\coordinate (T8) at (7.2,-1);
			\coordinate (P0) at ($.3*(T0)+.4*(T1)+.3*(T7)$);
			\coordinate (P1) at ($.3*(T0)+.4*(T1)+.3*(T2)$);
			\coordinate (P2) at ($.3*(T1)+.4*(T3)+.3*(T2)$);
			\coordinate (P3) at ($.3*(T3)+.4*(T4)+.3*(T2)$);
			\coordinate (P4) at ($.3*(T3)+.4*(T4)+.3*(T5)$);
			\coordinate (P5) at ($.3*(T6)+.4*(T3)+.3*(T5)$);
			\coordinate (P6) at ($.3*(T6)+.4*(T8)+.3*(T5)$);
			\foreach \x in {0,1,2,3,4,5,6,7,8} 
			{\fill[color=black] (T\x) circle (.06);}
			\draw[thick,color=black] (T0) -- (T1) -- (T2) -- (T3) -- (T4) -- (T5) -- (T6) -- (T3) -- (T5) -- (T8) -- (T6)
			(T3) -- (T1) -- (T7) -- (T0)
			(T0) -- (T2) -- (T4);
			\draw[thick,densely dashed, color=black] (2,.8) -- (T7) -- (1,.7);
			\draw[thick,densely dashed, color=black] ($.3*(T3)+(0,.2)+.6*(T1)$) -- (T1) -- ($.3*(2.4,.8)+.7*(T1)$)
			($.2*(T1)+(0,.3)+.8*(T3)$) -- (T3) -- ($.33*(T6)+(0,.1)+.6*(T3)$)
			($.2*(T8)+(.3,.3)+.8*(T6)$) -- (T6) -- ($.3*(T3)+(0,.2)+.65*(T6)$)
			(T6) -- ($(T6)+(0,.5)$)
			($.2*(T6)+(.1,.3)+.8*(T8)$) -- (T8) -- ($.3*(T5)+(0.8,-.5)+.65*(T8)$)
			(T8) -- ($(T8)+(.5,0)$)
			($.2*(T4)+(.2,-.3)+.8*(T5)$) -- (T5) -- ($.3*(T8)+(0.5,-.5)+.65*(T5)$)
			($.2*(T5)+(0,-.3)+.8*(T4)$) -- (T4) -- ($.25*(T2)+(0,-.4)+.75*(T4)$)
			($.2*(T0)+(-.3,-.3)+.8*(T2)$) -- (T2) -- ($.25*(T4)+(0,-.4)+.75*(T2)$)
			($.2*(T2)+(-.2,-.3)+.8*(T0)$) -- (T0) -- ($.3*(T7)+(-.2,.3)+.65*(T0)$)
			(T0) -- ($(T0)+(-.3,0)$);
			\foreach \x in {0,1,2,3,4,5,6} 
			{\fill[color=green!50!black] (P\x) circle (.1);}
			\draw[directedgreen] (P1) -- (P0);
			\draw[directedgreen] (P1) -- (P2);
			\draw[directedgreen] (P3) -- (P2);
			\draw[directedgreen] (P3) -- (P4);
			\draw[directedgreen] (P4) -- (P5);
			\draw[redirectedgreen] (P5) -- (P6);
			\draw[color=green!50!black,postaction={decorate}, decoration={markings,mark=at position .35 with {\arrow[color=green!50!black]{<}}}] ($(P0)+(.6,.6)$) -- (P0);
			\draw[color=green!50!black,postaction={decorate}, decoration={markings,mark=at position .35 with {\arrow[color=green!50!black]{<}}}] ($(P0)+(-.8,.4)$) -- (P0);
			\draw[color=green!50!black,postaction={decorate}, decoration={markings,mark=at position .35 with {\arrow[color=green!50!black]{>}}}] ($(P1)+(-.5,-.9)$) -- (P1);
			\draw[redirectedgreen] ($(P2)+(.1,.8)$) -- (P2);
			\draw[green!50!black,postaction={decorate}, decoration={markings,mark=at position .55 with {\arrow[color=green!50!black]{>}}}] ($(P3)+(-.3,-.8)$) -- (P3);
			\draw[redirectedgreen] ($(P4)+(.5,-.7)$) -- (P4);
			\draw[redirectedgreen] ($(P5)+(-.5,.7)$) -- (P5);
			\draw[color=green!50!black,postaction={decorate}, decoration={markings,mark=at position .35 with {\arrow[color=green!50!black]{>}}}] ($(P6)+(.2,.8)$) -- (P6);
			\draw[color=green!50!black,postaction={decorate}, decoration={markings,mark=at position .35 with {\arrow[color=green!50!black]{>}}}] ($(P6)+(.5,-.8)$) -- (P6);
			\end{tikzpicture}
			\label{eq:torustriangulated}
\end{equation}
In order for the result of this construction to be well-defined, it must be invariant under the choice of triangulation.
This results in certain algebraic properties of the orbifold datum, the \textsl{orbifold axioms}. 
In 2 dimensions, these properties are $\Delta$-separable, symmetric, and Frobenius.
For example, the Frobenius property arises from the two ways a four-sided polygon can be triangulated,
\begin{equation}
	\tikzzbox{%
		\begin{tikzpicture}[very thick,scale=0.7,color=green!50!black, baseline=0cm]
			\draw[-dot-] (0,0) .. controls +(0,-1) and +(0,-1) .. (-1,0);
			\draw[-dot-] (1,0) .. controls +(0,1) and +(0,1) .. (0,0);
			\draw (-1,0) -- (-1,1.5)
			(1,0) -- (1,-1.5)
			(0.5,0.8) -- (0.5,1.5)
			(-0.5,-0.8) -- (-0.5,-1.5)
  			(-.5,-.35) node {\scriptsize$\Delta$}
			(.5,.35) node {\scriptsize$\mu$};
		\end{tikzpicture}
	}%
	\;=
	\begin{tikzpicture}[baseline=0cm]
    \node[name=t,regular polygon, white, regular polygon sides=3, draw,
     inner sep=.5cm] at (0,-.35) {};
 	\path (t.center) -- (t.side 3) node[pos=2,name=t2c] {};
    \node[name=t2,regular polygon, white, regular polygon sides=3, draw,
     inner sep=.5cm,rotate=60] at (t2c) {};
     \draw[postaction={decorate}, decoration={markings,mark=at position .4 with {\arrow[color=black]{>}}},black, thick] (t.corner 2) -- (t.corner 1);
     \draw[postaction={decorate}, decoration={markings,mark=at position .4 with {\arrow[color=black]{>}}},black, thick] (t.corner 1) -- (t.corner 3);
     \draw[postaction={decorate}, decoration={markings,mark=at position .4 with {\arrow[color=black]{>}}},black, thick] (t.corner 2) -- (t.corner 3);
     \draw[postaction={decorate}, decoration={markings,mark=at position .8 with {\arrow[color=green!50!black]{>}}},green!50!black,very thick,shorten >= -0.35cm] (t.center) -- (t.side 1);
     \draw[postaction={decorate}, decoration={markings,mark=at position .8 with {\arrow[color=green!50!black]{<}}},green!50!black,very thick,shorten >= -0.35cm] (t.center) -- (t.side 2);
     \draw[postaction={decorate}, decoration={markings,mark=at position .8 with {\arrow[color=green!50!black]{>}}},green!50!black,very thick,shorten >= -0.35cm] (t.center) -- (t.side 3);
     \fill[green!50!black] (t.center) circle (2.5pt);
     \foreach \anchor/\placement/\lab in
    {corner 1/above/2, corner 2/left/1, corner 3/right/3}
  \draw (t.\anchor) node[\placement] {\scriptsize\lab};
  \draw[green!50!black] ($(t.center)+(.2,-.15)$) node {\scriptsize$\Delta$};
     \draw[postaction={decorate}, decoration={markings,mark=at position .4 with {\arrow[color=black]{>}}},black, thick] (t2.corner 1) -- (t2.corner 3);
     \draw[postaction={decorate}, decoration={markings,mark=at position .4 with {\arrow[color=black]{>}}},black, thick] (t2.corner 2) -- (t2.corner 3);
     \draw[postaction={decorate}, decoration={markings,mark=at position .8 with {\arrow[color=green!50!black]{<}}},green!50!black,very thick,shorten >= -0.35cm] (t2.center) -- (t2.side 1);
     \draw[postaction={decorate}, decoration={markings,mark=at position .8 with {\arrow[color=green!50!black]{<}}},green!50!black,very thick,shorten >= -0.35cm] (t2.center) -- (t2.side 2);
     \draw[postaction={decorate}, decoration={markings,mark=at position .8 with {\arrow[color=green!50!black]{>}}},green!50!black,very thick,shorten >= -0.35cm] (t2.center) -- (t2.side 3);
     \fill[green!50!black] (t2.center) circle (2.5pt);
  \draw (t2.corner 3) node[right] {\scriptsize 4};
  \draw[green!50!black] ($(t2.center)+(0,-.3)$) node {\scriptsize$\mu$};
\end{tikzpicture}
\sim
	\begin{tikzpicture}[baseline=0cm]
    \node[name=t,regular polygon, white, regular polygon sides=3, draw,
     inner sep=.5cm] at (0,-.35) {};
 	\path (t.center) -- (t.side 3) node[pos=2,name=t2c] {};
 	\path (t.corner 3) -- (t.corner 1) node[pos=.3,name=mu,line width=-4pt] {} node[pos=.7,name=Del,line width=-4pt] {};
    \node[name=t2,regular polygon, white, regular polygon sides=3, draw,
     inner sep=.5cm,rotate=60] at (t2c) {};
     \draw[postaction={decorate}, decoration={markings,mark=at position .4 with {\arrow[color=black]{>}}},black, thick] (t.corner 2) -- (t.corner 1);
     \draw[postaction={decorate}, decoration={markings,mark=at position .4 with {\arrow[color=black]{>}}},black, thick] (t.corner 2) -- (t2.corner 3);
     \draw[postaction={decorate}, decoration={markings,mark=at position .4 with {\arrow[color=black]{>}}},black, thick] (t.corner 2) -- (t.corner 3);
     \draw[postaction={decorate}, decoration={markings,mark=at position .8 with {\arrow[color=green!50!black]{>}}},green!50!black,very thick,shorten >= -0.35cm] (Del) -- (t.130);
     \draw[postaction={decorate}, decoration={markings,mark=at position .7 with {\arrow[color=green!50!black]{<}}},green!50!black,very thick,shorten >= -0.35cm] (mu) -- (t.290);
     \fill[green!50!black] (Del) circle (2.5pt);
     \foreach \anchor/\placement/\lab in
    {corner 1/above/2, corner 2/left/1, corner 3/right/3}
  \draw (t.\anchor) node[\placement] {\scriptsize\lab};
  \draw[green!50!black] ($(Del)+(.3,0)$) node {\scriptsize$\Delta$};
     \draw[postaction={decorate}, decoration={markings,mark=at position .2 with {\arrow[color=black]{>}}},black, thick] (t2.corner 1) -- (t2.corner 3);
     \draw[postaction={decorate}, decoration={markings,mark=at position .2 with {\arrow[color=black]{>}}},black, thick] (t2.corner 2) -- (t2.corner 3);
     \draw[postaction={decorate}, decoration={markings,mark=at position .8 with {\arrow[color=green!50!black]{>}},mark=at position .3 with {\arrow[color=green!50!black]{>}}},green!50!black,very thick] (mu) -- (Del);
     \draw[postaction={decorate}, decoration={markings,mark=at position .7 with {\arrow[color=green!50!black]{<}}},green!50!black,very thick,shorten >= -0.35cm] (mu) -- (t2.250);
     \draw[postaction={decorate}, decoration={markings,mark=at position .8 with {\arrow[color=green!50!black]{>}}},green!50!black,very thick,shorten >= -0.35cm] (Del) -- (t2.50);
     \fill[green!50!black] (mu) circle (2.5pt);
  \draw (t2.corner 3) node[right] {\scriptsize 4};
  \draw[green!50!black] ($(mu)+(.1,-.25)$) node {\scriptsize$\mu$};
\end{tikzpicture}
=\;
\tikzzbox{%
	\begin{tikzpicture}[very thick,scale=0.7,color=green!50!black, baseline=0cm]
		\draw[-dot-] (0,1.5) .. controls +(0,-1) and +(0,-1) .. (1,1.5);
		\draw[-dot-] (0,-1.5) .. controls +(0,1) and +(0,1) .. (1,-1.5);
		\draw (0.5,-0.8) -- (0.5,0.8); 
  		\draw[green!50!black] 
  		(.5,1.15) node {\scriptsize$\Delta$}
		(.5,-1.15) node {\scriptsize$\mu$};
	\end{tikzpicture}
}\,,
\end{equation}
which should be equal once we evaluate the theory (i.e.\ partition function) on these configurations.
Exchanging the corner labels 2 and 3 (thereby changing the orientation of the triangles on the left) produces the second Frobenius equation \eqref{eq:FrobeniusAlgebra}.

\medskip

A \textsl{condensation defect} is a defect which arises by performing the orbifold construction for a $k$-dimensional orbifold datum on a $k$-dimensional submanifold of an $n$-dimensional theory, where $k<n$.
This means we triangulate and equip only the $k$-dimensional submanifold with defects of the $k$-dimensional orbifold datum.
Most importantly, we can use 2-dimensional orbifold data to construct condensation surface defects from line operators in 3 dimensions.
In \eqref{eq:torustriangulated}, the torus can be seen as a submanifold of a 3-dimensional space on which we have condensed the 2-dimensional orbifold datum.
In this paper, all surface defects are condensation defects.

\medskip

A \tsl{$3$-dimensional orbifold datum~$\mathbb{A}$} consists of a chosen theory $\mathcal{T}\in\mathcal{D}$, and a weakly associative algebra $(A, T, \alpha,\bar{\alpha})$ in the 2-category of surface defects within~$\mathcal{T}$, $\End_\mathcal{D}(\mathcal{T}):=\Hom_\mathcal{D}(\mathcal{T},\mathcal{T})$.
The morphism $A$ specifies the surface defect, $T\colon A\boxtimes A\rightarrow A$ is a line defect fusing two of these surfaces into a third copy, and $\alpha$ and $\bar{\alpha}$ are junctions between the two different ways to fuse three surfaces into one (associators):
\begin{equation}
		\mathbb{A} := \left(\mathcal{T}, \;\;
			\tikzzbox{\begin{tikzpicture}[thick,scale=2.321,color=blue!50!black, baseline=0.0cm, >=stealth, 
					style={x={(-0.6cm,-0.4cm)},y={(1cm,-0.2cm)},z={(0cm,0.9cm)}}]
					\pgfmathsetmacro{\yy}{0.2}
					\coordinate (T) at (0.5, 0.4, 0);
					\coordinate (L) at (0.5, 0, 0);
					\coordinate (R1) at (0.3, 1, 0);
					\coordinate (R2) at (0.7, 1, 0);
					\coordinate (1T) at (0.5, 0.4, 1);
					\coordinate (1L) at (0.5, 0, 1);
					\coordinate (1R1) at (0.3, 1, );
					\coordinate (1R2) at (0.7, 1, );
					\fill [green!50,opacity=0.545] (R1) -- (T) -- (1T) -- (1R1);
					%
					%
					\draw [black,opacity=1, very thin] (R1) -- (T) -- (1T) -- (1R1) -- (R1);
					%
					\fill[color=blue!60!black] (0.5,1.1,0.25) circle (0pt) node[left] (0up) { {\scriptsize$A$} };
					\fill[color=black] (0.45,0.6,-0.05) circle (0pt) node[left] (0up) { {\scriptsize$\mathcal{T}$} };
					\fill[color=black] (0.5,1.1,1.17) circle (0pt) node[left] (0up) { {\scriptsize$\mathcal{T}$} };
			\end{tikzpicture}}
			\, , \;\;
			\tikzzbox{\begin{tikzpicture}[thick,scale=2.321,color=blue!50!black, baseline=0.0cm, >=stealth, 
					style={x={(-0.6cm,-0.4cm)},y={(1cm,-0.2cm)},z={(0cm,0.9cm)}}]
					\pgfmathsetmacro{\yy}{0.2}
					\coordinate (T) at (0.5, 0.4, 0);
					\coordinate (L) at (0.5, 0, 0);
					\coordinate (R1) at (0.3, 1, 0);
					\coordinate (R2) at (0.7, 1, 0);
					\coordinate (1T) at (0.5, 0.4, 1);
					\coordinate (1L) at (0.5, 0, 1);
					\coordinate (1R1) at (0.3, 1, );
					\coordinate (1R2) at (0.7, 1, );
					%
					\fill [green!50,opacity=0.545] (L) -- (T) -- (1T) -- (1L);
					\fill [green!50,opacity=0.545] (R1) -- (T) -- (1T) -- (1R1);
					\draw [black,opacity=1, very thin] (1T) -- (1R1) -- (R1) -- (T);
					\fill [green!50,opacity=0.545] (R2) -- (T) -- (1T) -- (1R2);
					\fill[color=blue!60!black] (0.5,0.25,0.15) circle (0pt) node[left] (0up) { {\scriptsize$A$} };
					\fill[color=blue!60!black] (0.15,0.95,0.065) circle (0pt) node[left] (0up) { {\scriptsize$A$} };
					\fill[color=blue!60!black] (0.55,0.95,0.05) circle (0pt) node[left] (0up) { {\scriptsize$A$} };
					%
					\draw[string=green!30!black, ultra thick] (T) -- (1T);
					\fill[color=green!30!black] (0.5,0.43,0.5) circle (0pt) node[left] (0up) { {\scriptsize$T$} };
					%
					\draw [black,opacity=1, very thin] (1T) -- (1L) -- (L) -- (T);
					\draw [black,opacity=1, very thin] (1T) -- (1R2) -- (R2) -- (T);
			\end{tikzpicture}}
			\, , \;\; 
			\tikzzbox{\begin{tikzpicture}[thick,scale=2.321,color=blue!50!black, baseline=0.0cm, >=stealth, 
					style={x={(-0.6cm,-0.4cm)},y={(1cm,-0.2cm)},z={(0cm,0.9cm)}}]
					\pgfmathsetmacro{\yy}{0.2}
					\coordinate (P) at (0.5, \yy, 0);
					\coordinate (R) at (0.625, 0.5 + \yy/2, 0);
					\coordinate (L) at (0.5, 0, 0);
					\coordinate (R1) at (0.25, 1, 0);
					\coordinate (R2) at (0.5, 1, 0);
					\coordinate (R3) at (0.75, 1, 0);
					\coordinate (Pt) at (0.5, \yy, 1);
					\coordinate (Rt) at (0.375, 0.5 + \yy/2, 1);
					\coordinate (Lt) at (0.5, 0, 1);
					\coordinate (R1t) at (0.25, 1, 1);
					\coordinate (R2t) at (0.5, 1, 1);
					\coordinate (R3t) at (0.75, 1, 1);
					\coordinate (alpha) at (0.5, 0.5, 0.5);
					%
					%
					\fill [green!50,opacity=0.545] (L) -- (P) -- (alpha) -- (Pt) -- (Lt);
					\fill [green!50,opacity=0.545] (Pt) -- (Rt) -- (alpha);
					\fill [green!50,opacity=0.545] (Rt) -- (R1t) -- (R1) -- (P) -- (alpha);
					\draw [black,opacity=1, very thin] (Rt) -- (R1t) -- (R1) -- (P);
					\fill [green!50,opacity=0.545] (Rt) -- (R2t) -- (R2) -- (R) -- (alpha);
					\draw [black,opacity=1, very thin] (Rt) -- (R2t) -- (R2) -- (R);
					\draw[string=green!30!black, ultra thick] (alpha) -- (Rt);
					\fill[color=green!30!black] (0.5,0.77,0.77) circle (0pt) node[left] (0up) { {\scriptsize$T$} };
					\fill [green!50,opacity=0.545] (Pt) -- (R3t) -- (R3) -- (R) -- (alpha);
					\fill [green!50,opacity=0.545] (P) -- (R) -- (alpha);
					%
					\draw[string=green!30!black, ultra thick] (P) -- (alpha);
					\draw[string=green!30!black, ultra thick] (R) -- (alpha);
					\draw[string=green!30!black, ultra thick] (alpha) -- (Pt);
					%
					\fill[color=green!30!black] (alpha) circle (1.2pt) node[left] (0up) { {\scriptsize$\alpha$} };
					\fill[color=green!30!black] (0.5,0.35,0.24) circle (0pt) node[left] (0up) { {\scriptsize$T$} };
					\fill[color=green!30!black] (0.5,0.72,0.21) circle (0pt) node[left] (0up) { {\scriptsize$T$} };
					\fill[color=green!30!black] (0.5,0.4,0.71) circle (0pt) node[left] (0up) { {\scriptsize$T$} };
					%
					\draw [black,opacity=1, very thin] (Pt) -- (Lt) -- (L) -- (P);
					\draw [black,opacity=1, very thin] (Pt) -- (Rt);
					\draw [black,opacity=1, very thin] (Pt) -- (R3t) -- (R3) -- (R);
					\draw [black,opacity=1, very thin] (P) -- (R);
			\end{tikzpicture}}
			\, , \;\;
			\begin{tikzpicture}[thick,scale=2.321,color=blue!50!black, baseline=0.0cm, >=stealth, 
				style={x={(-0.6cm,-0.4cm)},y={(1cm,-0.2cm)},z={(0cm,0.9cm)}}]
				\pgfmathsetmacro{\yy}{0.2}
				\coordinate (P) at (0.5, \yy, 0);
				\coordinate (R) at (0.375, 0.5 + \yy/2, 0);
				\coordinate (L) at (0.5, 0, 0);
				\coordinate (R1) at (0.25, 1, 0);
				\coordinate (R2) at (0.5, 1, 0);
				\coordinate (R3) at (0.75, 1, 0);
				\coordinate (Pt) at (0.5, \yy, 1);
				\coordinate (Rt) at (0.625, 0.5 + \yy/2, 1);
				\coordinate (Lt) at (0.5, 0, 1);
				\coordinate (R1t) at (0.25, 1, 1);
				\coordinate (R2t) at (0.5, 1, 1);
				\coordinate (R3t) at (0.75, 1, 1);
				\coordinate (alpha) at (0.5, 0.5, 0.5);
				%
				\draw[string=green!30!black, ultra thick] (alpha) -- (Rt);
				\fill [green!50,opacity=0.545] (L) -- (P) -- (alpha) -- (Pt) -- (Lt);
				\fill [green!50,opacity=0.545] (Pt) -- (R1t) -- (R1) -- (R) -- (alpha);
				\draw [black,opacity=1, very thin] (Pt) -- (R1t) -- (R1) -- (R);
				\fill [green!50,opacity=0.545] (Rt) -- (R2t) -- (R2) -- (R) -- (alpha);
				\draw [black,opacity=1, very thin] (Rt) -- (R2t) -- (R2) -- (R);
				\fill [green!50,opacity=0.545] (Pt) -- (Rt) -- (alpha);
				\fill [green!50,opacity=0.545] (P) -- (R) -- (alpha);
				\draw [black,opacity=1, very thin] (P) -- (R);
				\fill[color=green!30!black] (0.5,0.82,0.3) circle (0pt) node[left] (0up) { {\scriptsize$T$} };
				\draw[string=green!30!black, ultra thick] (R) -- (alpha);
				\fill [green!50,opacity=0.545] (Rt) -- (R3t) -- (R3) -- (P) -- (alpha);
				\draw[string=green!30!black, ultra thick] (alpha) -- (Rt);
				%
				\draw[string=green!30!black, ultra thick] (P) -- (alpha);
				\draw[string=green!30!black, ultra thick] (alpha) -- (Pt);
				\fill[color=green!30!black] (0.5,0.35,0.24) circle (0pt) node[left] (0up) { {\scriptsize$ T$} };
				\fill[color=green!30!black] (0.5,0.4,0.71) circle (0pt) node[left] (0up) { {\scriptsize$ T$} };
				\fill[color=green!30!black] (0.5,0.73,0.72) circle (0pt) node[left] (0up) { {\scriptsize$ T$} };
				\fill[color=green!30!black] (alpha) circle (1.2pt) node[left] (0up) { {\scriptsize$\overline\alpha$} };
				%
				\draw [black,opacity=1, very thin] (Pt) -- (Lt) -- (L) -- (P) ;
				\draw [black,opacity=1, very thin] (Pt) -- (Rt);
				\draw [black,opacity=1, very thin] (Rt) -- (R3t) -- (R3) -- (P);
			\end{tikzpicture}
			\, , \;\;
			\psi 
			\, , \;\;
			\phi 
		\right) \,.
		\label{eq:3dorbdat}
	\end{equation}
Here, the third dimension in which we fuse with $\boxtimes$ is read front to back.
Additionally, one may specify point insertions $\psi\in\End(\id_A)$ and $\phi\in\End(\id_{\id_\mathcal{T}})$ which decorate the $A$-planes and $\mathcal{T}$-volumes, respectively.
When labelling the dual of a chosen triangulation in the orbifold construction, these carry exponents according to the Euler characteristic of the given region.
Furthermore, a 3-dimensional orbifold datum~$\mathbb{A}$ is subject to the (3-dimensional) orbifold axioms \cite{CRS1,CH}.
These can be separated into four groups, corresponding to a higher associativity for~$\alpha$ (pentagon equation), generalisations of the Frobenius algebra axioms, a generalised symmetry condition, and a generalised separability condition.

As in the 2-dimensional case, this data is precisely suited to label the Poincaré-dual of a triangulation:
The surface~$A$ is dual to a line, the line~$T$ is dual to a triangle (by piercing its center), and the junctions~$\alpha$ and $\bar{\alpha}$ are dual to (variously oriented) tetrahedra.
The process of \textsl{orbifolding} stratifies the 3-dimensional space by the Poincaré-dual of a triangulation, inserts the defects accordingly along these strata, and then evaluates the partition function on the resulting configuration. 
The orbifold axioms are built such that this process is independent of the choice of triangulation (including orientation).

\subsection{2-group Symmetries and Their Orbifold Data}\label{sec:2groupsymm}
In this section, we outline how to construct 3-dimensional orbifold data for 2-group (and thereby 0-form and 1-form) symmetries.
Following our discussion in \Cref{sec:Intro}, a 0-form $G$-symmetry can be neatly summarised in a 3-functor
\begin{equation}
	R\colon 
	\textrm{B}\underline{G} \longrightarrow \mathcal D 
	\label{eq:Gsym1}
\end{equation}
from the delooping of~$\underline{G}$ to~$\mathcal D$.
Note that this includes not only the aforementioned assignment of surface defects to group elements, but also various line and local operators specifying the fusion of those surface defects (where symmetry fractionalisation appears) and junctions between these line operators (associators, i.e.\ defectification).

As discussed previously, the idea behind the construction of an orbifold datum~$\mathbb{A}_R$ from such a functor (i.e.\ a 0-form symmetry) is to use the group algebra $\C[G]$ (or rather its image under~$R$; we identify the two in the following) as the underlying surface defect~$A$ and the line operators and junctions describing fusion (i.e.\ the coherence data of~$R$) to construct~$T$, $\alpha$, and~$\bar{\alpha}$, respectively. 
To build further intuition for this choice, consider the following:
In a theory with only local operators and finite $G$-symmetry, the algebra of local operators forms a $G$-representation, or equivalently a module over the group algebra $\C[G]$.
The local operators in the gauged theory are given by $G$-invariant operators which can be obtained by summing over the orbit of a chosen operator~$\mathcal{O}$ among the local operators:
\begin{equation}
	\bigoplus_{g\in G}\rho_g(\mathcal{O})\,.
\end{equation}
The operators in the orbit span a $\C[G]$-module, which can be identified with $\mathcal{O}\otimes \C[G]$.

For line operators the reasoning is similar.
Note in particular the presence of Wilson lines in the gauged theory, which correspond to $G$-representations and therefore $\C[G]$-modules.
In \Cref{sec:RT} we discuss the fact that operators in the gauged theory are modules of the algebra underlying the orbifold datum.
This explains the use of the group algebra.

\medskip

To extend this procedure to 2-groups (and thereby 1-form symmetries in particular), recall that a 2-group $\mathcal{G}$ is a monoidal category where all objects and morphisms are invertible, so the objects correspond to the 0-form $G$-symmetry and the morphisms correspond to a 1-form $H$-symmetry. 
There are additional interactions between these two:
The 1-form symmetry operators may sit on top of $G$-labelled surface defects (i.e.\ they may not be genuine line operators), and they carry a $G$-action since they are line operators on which the 0-form symmetry may act.
In analogy to \eqref{eq:Gsym1}, a \textsl{$\mathcal{G}$-symmetry in $\mathcal{D}$} is a 3-functor
\begin{equation}
	R\colon 
	\textrm{B}\underline{\mathcal{G}} \longrightarrow \mathcal D\, ,
	\label{eq:Gsym2}
\end{equation}
where~$\underline{\mathcal{G}}$ is~$\mathcal{G}$ viewed as a monoidal 2-category with only identity 2-morphisms. 
Again, to construct an orbifold datum we use the group algebras of the 0- and 1-form symmetries.
However, additional complexity arises from the fact that the two groups interact with each other. 
The group algebra $\C[H]$ of the 1-form symmetry may not be a genuine bulk line operator, instead it may have to sit on top of the $G$-symmetry surface defect $\C[G]$.
Under mild technical assumptions, $\C[H]$ can be equipped with the structure of a $\Delta$-separable symmetric Frobenius algebra, i.e.\ a 2-dimensional orbifold datum.
We can then condense this 1-form group algebra on the surface defect given by the 0-form group algebra $\C[G]$. 
This condensation defect serves as the surface~$A$ in the orbifold datum.
Compared to the 0-form case, we also fuse a copy of $\C[H]$ onto~$T$ and enhance $\alpha$ and $\bar{\alpha}$ similarly.
The details are presented in \cite[Sect.\,3]{CH}, we review the relevant special cases in \Cref{sec:RT,sec:deeq}. 

\subsection{Reshetikhin--Turaev Theory}\label{sec:RT}
In this section, we establish the specific type of TQFT which encompasses our examples, and the details of orbifold data for 0- and 1-form symmetries within them.

Reshetikhin--Turaev theories are 3-dimensional defect TQFTs based on line operators~$\mathcal{C}$ (a modular fusion category).
Its surface defects are given by condensations of 2-dimensional orbifold data (i.e.\ line operators equipped with the structure of a $\Delta$-separable symmetric Frobenius algebra).
Its line defects are line operators in $\mathcal{C}$ equipped with junctions that let the condensed line operators end on them (i.e.\ a bimodule structure).
Consequently, bulk lines are bimodules of the trivial algebra, i.e.\ all lines in~$\mathcal{C}$.
Lastly, local operators on a given line are those point defects (morphisms in $\mathcal{C}$) that commute with the aforementioned junctions.
In summary, the 3-category of topological defects is the delooping of the 2-category of $\Delta$-separable symmetric Frobenius algebras in $\mathcal{C}$, their bimodules and bimodule morphisms,
\begin{equation}
\mathcal{D}_\mathcal{C}:=\mathrm{B}(\mathrm{\Delta ssFrob}(\mathcal{C}))\,.
\end{equation}

An important fact about Reshetikhin--Turaev theories is that the orbifold construction always produces another Reshetikhin--Turaev theory \cite{CMRSS2024}.
The category of line operators~$\mathcal{C}_\mathbb{A}$ underlying the new theory is given by $(A,A)$-bimodules along with junctions~$\tau$ of these modules with~$T$, subject to compatibility conditions. 
See \cite{MR} for details.

\medskip

A $G$-symmetry in a Reshetikhin--Turaev theory is a 3-functor $R\colon \mathrm{B}\underline{G}\longrightarrow \mathcal{D}_\mathcal{C}$ from which the construction of \cite[Sect.\,3]{CH} (outlined in \cref{sec:2groupsymm}) produces an orbifold datum.
Following \cite{ENO}, such a 3-functor also specifies a $G$-crossed braided extension $\Gcbc$ of $\mathcal{C}$ \eqref{eq:classification}.
In \cite[Thm.\,5.1]{CRS3}, the authors constructed a (3-dimensional) orbifold datum for the Reshetikhin--Turaev theory based on~$\mathcal{C}$ for a given $G$-crossed extension $\Gcbc$.
Both of these approaches result in the same orbifold datum \cite[Sect.\,4.2]{CH} which we now describe in detail.

The construction starts by choosing a simple object~$m_g$ in each twisted sector $\mathcal{C}_g$, where $m_e=\mathds{1}$ is the trivial line.
Each object~$m_g$ gives rise to a $\Delta$-separable, symmetric Frobenius algebra~$A_g:=m_g^*\otimes m_g$ whose Frobenius structure is given by 
\begin{align}
\label{eq:AgGextension}
&\mu_g:=\,
\begin{tikzpicture}[very thick,scale=0.9,color=red!50!black, baseline=.9cm]
\draw[line width=0pt] 
(3,0) node[line width=0pt] (D) {{\scriptsize$g$}}
(2,0) node[line width=0pt] (s) {{\scriptsize$g$}}; 
\draw[redirectedred] (D) .. controls +(0,1) and +(0,1) .. (s);
\draw[line width=0pt] 
(3.45,0) node[line width=0pt] (re) {{\scriptsize$g$}}
(1.55,0) node[line width=0pt] (li) {{\scriptsize$g$}}; 
\draw[line width=0pt] 
(2.7,2) node[line width=0pt] (ore) {{\scriptsize$g$}}
(2.3,2) node[line width=0pt] (oli) {{\scriptsize$g$}}; 
\draw[directedred] (2.3,1.25) .. controls +(0,-0.25) and +(0,0.75) .. (li);
\draw (2.3,1.25) -- (oli);
\draw[directedred] (re) .. controls +(0,0.75) and +(0,-0.25) .. (2.7,1.25);
\draw (2.7,1.25) -- (ore);
\end{tikzpicture}
\,, &&\eta_g:=
\begin{tikzpicture}[very thick,scale=0.9,color=red!50!black, baseline=-.4cm,rotate=180]
\draw[line width=0pt] 
(3,0) node[line width=0pt] (D) {{\scriptsize$g$}}
(2,0) node[line width=0pt] (s) {{\scriptsize$g$}}; 
\draw[directedred] (D) .. controls +(0,1) and +(0,1) .. (s);
\end{tikzpicture}
\,,\\
&\Delta_g:=
\begin{tikzpicture}[very thick,scale=0.9,color=red!50!black, baseline=-0.9cm, rotate=180]
\draw[line width=0pt] 
(3,0) node[line width=0pt] (D) {{\scriptsize$g$}}
(2,0) node[line width=0pt] (s) {{\scriptsize$g$}}; 
\draw[directedred] (s) .. controls +(0,1) and +(0,1) .. (D) ;
%
\draw[line width=0pt] 
(3.45,0) node[line width=0pt] (re) {{\scriptsize$g$}}
(1.55,0) node[line width=0pt] (li) {{\scriptsize$g$}}; 
\draw[line width=0pt] 
(2.7,2) node[line width=0pt] (ore) {{\scriptsize$g$}}
(2.3,2) node[line width=0pt] (oli) {{\scriptsize$g$}}; 
\draw[directedred] (2.3,1.25) .. controls +(0,-0.25) and +(0,0.75) .. (li);
\draw (2.3,1.25) -- (oli);
\draw[directedred] (re) .. controls +(0,0.75) and +(0,-0.25) .. (2.7,1.25);
\draw (2.7,1.25) -- (ore);
\end{tikzpicture}\,\cdot \mathrm{dim}(m_g)^{-1}
\,, &&
\varepsilon_g:=
\begin{tikzpicture}[very thick,scale=0.9,color=red!50!black, baseline=.4cm]
\draw[line width=0pt] 
(3,0) node[line width=0pt] (D) {{\scriptsize$g$}}
(2,0) node[line width=0pt] (s) {{\scriptsize$g$}}; 
\draw[redirectedred] (s) .. controls +(0,1) and +(0,1) .. (D);
\end{tikzpicture}\,\cdot \mathrm{dim}(m_g)
\,,
\end{align}
where we labelled lines by~$g$ instead of~$m_g$.
Let us emphasise that the algebra~$A_g$ represents the symmetry surface defect $R(g)= \mathcal{C}_g\cong A_g\text{-}\mathrm{Mod}(\mathcal{C})$ \cite[Thm.\,6.1]{ENO} which is its condensation defect in Reshetikhin--Turaev theory.

For each $g\in G$ we have a left $A_g$-module structure on $m_g^*\in\mathcal{C}_{g^{-1}}$, and a right $A_g$-module structure on $m_g\in\mathcal{C}_g$ given by
\begin{align}
\label{eq:mmodulestructures}
\begin{tikzpicture}[very thick,scale=0.75,color=blue!50!black, baseline]
\draw (0,-1) node[below] (X) {{\scriptsize$m_g^*$}};
\draw[color=green!50!black] (-0.8,-1) node[below] (A1) {{\scriptsize$A_g$}};
\draw (0,1) node[right] (Xu) {};
\draw[color=green!50!black] (A1) .. controls +(0,0.5) and +(-0.5,-0.5) .. (0,0.3);
\draw (0,-1) -- (0,1); 
\fill[color=blue!50!black] (0,0.3) circle (2.9pt) node (meet2) {};
\end{tikzpicture} 
:=
\begin{tikzpicture}[very thick,scale=0.75,color=red!50!black, baseline]
\draw (-1.75,-1) node[below] (mghs) {{\scriptsize$g$}};
\draw (-1,-1) node[below] (mgh) {{\scriptsize$g$}};
\draw (0,-1) node[below] (A1) {{\scriptsize$g$}};
\draw[directedred] (-1,-1) .. controls +(0,0.75) and +(0,0.75) .. (0,-1);
\draw[directedred] (0,1) to[out=-90, in=90] (-1.75,-1);
\end{tikzpicture} 
\, , \qquad 
\begin{tikzpicture}[very thick,scale=0.75,color=blue!50!black, baseline]
\draw (0,-1) node[below] (X) {{\scriptsize$m_g\vphantom{m_g^*}$}};
\draw[color=green!50!black] (0.8,-1) node[below] (A1) {{\scriptsize$A_g$}};
\draw (0,1) node[right] (Xu) {};
\draw[color=green!50!black] (A1) .. controls +(0,0.5) and +(0.5,-0.5) .. (0,0.3);
\draw (0,-1) -- (0,1); 
\fill[color=blue!50!black] (0,0.3) circle (2.9pt) node (meet2) {};
\fill[color=black] (0.2,0.5) circle (0pt) node (meet) {};
\end{tikzpicture} 
:=
\begin{tikzpicture}[very thick,scale=0.75,color=red!50!black, baseline]
\draw (1.75,-1) node[below] (mghs) {{\scriptsize$g$}};
\draw (1,-1) node[below] (mgh) {{\scriptsize$g$}};
\draw (0,-1) node[below] (A1) {{\scriptsize$g$}};
\draw[redirectedred] (1,-1) .. controls +(0,0.75) and +(0,0.75) .. (0,-1);
\draw[directedred] (1.75,-1) to[out=90, in=-90] (0,1);
\end{tikzpicture} 
\,.
\end{align}
\noindent
Using these, we construct $(A_{gh},A_g\otimes A_h)$-bimodules 
\begin{equation}
	\label{eq:chi-def}
	\chi_{g,h} :=m_{gh}^* \otimes( m_g \otimes m_h) \; \in \mathcal C_e 
\end{equation}
with left $A_{gh}$-module structure and right $A_h$-module structure given by \eqref{eq:mmodulestructures}, while the right $A_g$-module structure involves the braiding:
\begin{equation}
	\begin{tikzpicture}[very thick,scale=0.75,color=blue!50!black, baseline]
\draw (0,-1) node[below] (X) {{\scriptsize$\chi_{g,h}$}};
\draw[color=green!50!black] (0.8,-1) node[below] (A1) {{\scriptsize$A_g$}};
\draw (0,1) node[right] (Xu) {};
\draw[color=green!50!black] (A1) .. controls +(0,0.5) and +(0.5,-0.5) .. (0,0.3);
\draw (0,-1) -- (0,1); 
\fill[color=blue!50!black] (0,0.3) circle (2.9pt) node (meet2) {};
\fill[color=black] (0.2,0.5) circle (0pt) node (meet) {{\tiny$1$}};
\end{tikzpicture} 
:=
\begin{tikzpicture}[very thick,scale=0.75,color=red!50!black, baseline]
\draw (1.75,-1) node[below] (mghs) {{\scriptsize$g\vphantom{gh}$}};
\draw (1,-1) node[below] (mgh) {{\scriptsize$g\vphantom{gh}$}};
\draw (-0.75,-1) node[below] (mg) {{\scriptsize$g\vphantom{gh}$}};
\draw (-1.5,-1) node[below] (mh) {{\scriptsize$gh$}};
\draw (0,-1) node[below] (A1) {{\scriptsize$h$}};
\draw[redirectedred] (1,-1) .. controls +(0,0.75) and +(0,0.75) .. (-0.75,-1);
\draw[directedred] (1.75,-1) to[out=90, in=-90] (-0.75,1);
\draw[color=white, line width=4pt] (0,-1) -- (0,1); 
\draw[directedred] (0,-1) -- (0,1);
\draw[redirectedred] (mh) -- (-1.5,1); 
\end{tikzpicture}\, .
\label{eq:chimodulestructure}
\end{equation}
The geometric intuition is that the condensed $A_h$-surface attaches to $\chi_{g,h}$ in front of the $A_g$-surface in the perspective shown here.

\medskip

\noindent
The orbifold datum~$\mathbb{A}_G$ then consists of the following data:
	\begin{align}
		A_G&:=\bigoplus_{g\in G}A_g\,,
		\label{eq:AG}\\
		T_G&:=\bigoplus_{g,h\in G}\chi_{g,h}\,,\\
			\label{eq:string alphaG}
		\alpha_G&=\bigoplus_{g,h,k\in G}
				\begin{tikzpicture}[very thick,scale=0.9,color=red!50!black, baseline=-1.4cm, rotate=180]
					\draw[line width=0pt] 
					(5.5,3) node[line width=0pt] (ghko) {{\scriptsize$ghk\vphantom{ghk}$}}
					(5,3) node[line width=0pt] (go) {{\scriptsize$g\vphantom{ghk}$}}
					(4.5,3) node[line width=0pt] (hkli) {{\scriptsize$hk\vphantom{ghk}$}}
					(3.5,3) node[line width=0pt] (hkre) {{\scriptsize$hk\vphantom{ghk}$}}
					(3,3) node[line width=0pt] (ho) {{\scriptsize$h\vphantom{ghk}$}}
					(2.5,3) node[line width=0pt] (ko) {{\scriptsize$k\vphantom{ghk}$}}
					(5.5,0) node[line width=0pt] (ghku) {{\scriptsize$ghk\vphantom{ghk}$}}
					(5,0) node[line width=0pt] (ghli) {{\scriptsize$gh\vphantom{ghk}$}}
					(4.5,0) node[line width=0pt] (ku) {{\scriptsize$k\vphantom{ghk}$}}
					(3.5,0) node[line width=0pt] (ghre) {{\scriptsize$gh\vphantom{ghk}$}} 
					(3,0) node[line width=0pt] (gu) {{\scriptsize$g\vphantom{ghk}$}} 
					(2.5,0) node[line width=0pt] (hu) {{\scriptsize$h\vphantom{ghk}$}};
					\draw[redirectedred] (ghko) -- (ghku);
					\draw[postaction={decorate}, decoration={markings,mark=at position .65 with {\arrow[color=red!50!black]{<}}}] (ghli) .. controls +(0,1) and +(0,1) .. (ghre);
					\draw[postaction={decorate}, decoration={markings,mark=at position .55 with {\arrow[color=red!50!black]{<}}}] (hkre) .. controls +(0,-1) and +(0,-1) .. (hkli);
					\draw[postaction={decorate}, decoration={markings,mark=at position .65 with {\arrow[color=red!50!black]{<}}}] (gu) .. controls +(0,1.5) and +(0,-1.5) .. (go);
					\draw[redirectedred] (hu) .. controls +(0,1.5) and +(0,-1.5) ..  (ho);
					\draw[color=white, line width=4pt] (ku) .. controls +(0,1.5) and +(0,-1.5) .. (ko);
					\draw[postaction={decorate}, decoration={markings,mark=at position .65 with {\arrow[color=red!50!black]{<}}}] (ku) .. controls +(0,1.5) and +(0,-1.5) .. (ko);
					\end{tikzpicture}\,,\\
		\overline{\alpha}_G&=\bigoplus_{g,h,k\in G}
				\begin{tikzpicture}[very thick,scale=0.9,color=red!50!black, baseline=-1.4cm, rotate=180]
					\draw[line width=0pt] 
					(5.5,0) node[line width=0pt] (ghko) {{\scriptsize$ghk\vphantom{ghk}$}}
					(5,0) node[line width=0pt] (go) {{\scriptsize$g\vphantom{ghk}$}}
					(4.5,0) node[line width=0pt] (hkli) {{\scriptsize$hk\vphantom{ghk}$}}
					(3.5,0) node[line width=0pt] (hkre) {{\scriptsize$hk\vphantom{ghk}$}}
					(3,0) node[line width=0pt] (ho) {{\scriptsize$h\vphantom{ghk}$}}
					(2.5,0) node[line width=0pt] (ko) {{\scriptsize$k\vphantom{ghk}$}}
					(5.5,3) node[line width=0pt] (ghku) {{\scriptsize$ghk\vphantom{ghk}$}}
					(5,3) node[line width=0pt] (ghli) {{\scriptsize$gh\vphantom{ghk}$}}
					(4.5,3) node[line width=0pt] (ku) {{\scriptsize$k\vphantom{ghk}$}}
					(3.5,3) node[line width=0pt] (ghre) {{\scriptsize$gh\vphantom{ghk}$}} 
					(3,3) node[line width=0pt] (gu) {{\scriptsize$g\vphantom{ghk}$}} 
					(2.5,3) node[line width=0pt] (hu) {{\scriptsize$h\vphantom{ghk}$}};
					\draw[directedred] (ghko) -- (ghku);
					\draw[postaction={decorate}, decoration={markings,mark=at position .65 with {\arrow[color=red!50!black]{>}}}] (ghli) .. controls +(0,-1) and +(0,-1) .. (ghre);
					\draw[postaction={decorate}, decoration={markings,mark=at position .55 with {\arrow[color=red!50!black]{>}}}] (hkre) .. controls +(0,1) and +(0,1) .. (hkli);
					\draw[postaction={decorate}, decoration={markings,mark=at position .65 with {\arrow[color=red!50!black]{>}}}] (gu) .. controls +(0,-1.5) and +(0,1.5) .. (go);
					\draw[directedred] (hu) .. controls +(0,-1.5) and +(0,1.5) ..  (ho);
					\draw[color=white, line width=4pt] (ku) .. controls +(0,-1.5) and +(0,1.5) .. (ko);
					\draw[postaction={decorate}, decoration={markings,mark=at position .65 with {\arrow[color=red!50!black]{>}}}] (ku) .. controls +(0,-1.5) and +(0,1.5) .. (ko);
					\end{tikzpicture}\,,\\
		\psi_G&:=\bigoplus_{g\in G} \left(\mathrm{dim}(m_g)\right)^{-\frac{1}{2}}\cdot \id_{A_g}\,, \label{eq:psiInAR}
		\\
		\phi_G&:=\frac{1}{\sqrt{|G|}}\,.
		\label{eq:phiInAR}
	\end{align}
	
Note that, while every $A_g$ has its own algebra structure, the algebra underlying the orbifold datum is $(A_G,T_G,\alpha_G)$, where the surface~$A_G$ decomposes into summands~$A_g$ for each group element $g\in G$ similar to a group algebra, the line defect $\chi_{g,h}$ determines the multiplication of this algebra, fusing~$g$ and~$h$ into~$gh$, and~$\alpha_G$ is the associator.

It was shown in \cite[Thm.\,5.7]{CH} and \cite[Thm.\,5.1]{HPRW} that the category $\mathcal{C}_{\mathbb{A}_G}$ underlying the orbifolded theory is equivalent to the equivariantisation of the extension $\Gcbc$:
\begin{equation}
	 F\colon (\Gcbc)^G\stackrel{\cong}{\longrightarrow}\mathcal{C}_{\mathbb{A}_G}\,,
	\label{eq:orbifoldequivarequivalence}
\end{equation}
and the assignment on objects is given by 
\begin{align}
	\label{eq:FonX}
	 F((X,u))&:=\pr_{\mathcal{C}_e} \Big(\bigoplus_{g,h\in G} m_g^*\otimes X\otimes m_h\Big)
	 	=
	 	\bigoplus_{g,h\in G}m_{gh}^*\otimes X_g\otimes m_h \, , 
\end{align}
where the $A_G$-bimodule structure is induced by the module structures of~$m_{gh}^*$ and~$m_h$.\footnote{The objects in~$\mathcal{C}_{\mathbb{A}_G}$ additionally carry junctions~$\tau$ with $T_G$ which depend on the equivariant structure. We omit this here, see \cite[Sect.\,5]{CH} for details.}
This result is a strong confirmation of the intuition we built in \Cref{sec:equivariantisation} and further supports the mathematical use of the term "gauging" used to describe the combination of extension and equivariantisation.

\subsection{Zesting: Changing extensions through fractionalisation and defectification}
\label{sec:zesting}
In this section, we discuss the classification of $G$-crossed braided extensions through the lens of zesting which lets us relate inequivalent extensions to each other using group cohomological data.

If we compare multiple extensions, a first point of interest is the content of the twisted sectors~$\mathcal{C}_g$ and how twisted sector lines fuse with bulk lines (which comprises a $\mathcal{C}$-module structure on $\mathcal{C}_g$).
In the classification of $G$-crossed braided extensions via 3-functors \cite[Sect.\,7]{ENO} this data is captured on morphism level of the assignment
\begin{equation}
	\mathrm{B}\underline{G}\longrightarrow \mathrm{B}(\mathrm{Mod}^\times(\mathcal{C}))\,.
	\label{eq:classification2}
\end{equation}
It is known (\cite[Sect.\,8]{ENO}, see also \cite[Sect.\,IX]{BBCW}) that the higher coherence data of this classification (i.e.\ how twisted sector operators fuse, their associators, and braiding) can be described by group cohomology for a given symmetry $G$-action~$\rho$ on~$\mathcal{C}_e$.
Assuming all obstructions to vanish, the extensions form a $\mathrm{H}^2_\rho(G,\mathcal{I})\times \mathrm H^3(G,\U)$-torsor where $\mathcal{I}\subset\mathcal{C}_e$ is the group of invertible bulk line operators.
In physical terms, $\lambda\in \mathrm{H}^2_\rho(G,\mathcal{I})$ is the symmetry fractionalisation and $p\in \mathrm H^3(G,\U)$ is the defectification class.
For a given $\lambda$ we additionally require a family of isomorphisms 
\begin{equation}
	\nu_{g,h,k}\colon \rho_{k}(\lambda(g,h))\otimes \lambda (gh,k)\longrightarrow \lambda(h,k)\otimes \lambda(g,hk)\,,
\end{equation}
subject to normalisation when $h=e\in G$ and a pentagon axiom involving the braiding \cite[Def.\,3.1]{DGPRZ}.
Different choices for these families form the aforementioned $\mathrm{H}^3(G,\U)$-torsor and are related by the defectification class~$p\in\mathrm{H}^3(G,\U)$ via $\nu^\prime=p\cdot\nu$.
A pair $(\lambda,\nu)$ satisfying the above conditions is a \textsl{($G$-crossed braided) zesting datum}.
Given a $G$-crossed braided extension~$\Gcbc$ and a zesting datum $(\lambda,\nu)$, there exists another extension $(\Gcbc)^{(\lambda,\nu)}$ which is called its \textsl{($G$-crossed braided) zesting}. 
Its fusion rules for objects~$X_g$ and~$Y_h$ of degrees~$|X_g|=g$ and~$|Y_h|=h$ can be expressed in terms of the original monoidal product according to the following formula \cite[Prop.\,3.3]{DGPRZ}:
\begin{equation}
	X_g\otimes_\lambda Y_h = (X_g\otimes Y_h) \otimes \lambda(g,h)\,.
\end{equation} 
This corresponds to symmetry fractionalisation as illustrated by
\begin{equation}
X_g\otimes Y_h\cong
\begin{tikzpicture}[thick,scale=2.6,color=blue!50!black, baseline=-.1cm, >=stealth, 
					style={x={(-0.6cm,-0.4cm)},y={(1cm,-0.2cm)},z={(0cm,0.9cm)}}]
					\pgfmathsetmacro{\yy}{0.2}
					\coordinate (T) at (0.5, 0.4, 0);
					\coordinate (L) at (0.5, 0, 0);
					\coordinate (R1) at (0.3, 1, 0);
					\coordinate (R2) at (0.7, 1, 0);
					\coordinate (1T) at (0.5, 0.4, 1);
					\coordinate (1L) at (0.5, 0, 1);
					\coordinate (1R1) at (0.3, 1, );
					\coordinate (1R2) at (0.7, 1, );
					%
					\fill [orange!50,opacity=0.545] (L) -- (T) -- (1T) -- (1L);
					\fill [orange!50,opacity=0.545] (R1) -- (T) -- (1T) -- (1R1);
					\draw [black,opacity=1, very thin] (1T) -- (1R1) -- (R1) -- (T);
					\fill [orange!50,opacity=0.545] (R2) -- (T) -- (1T) -- (1R2);
					\fill[color=black] (0.5,0.17,0.9) node { {\scriptsize$R(gh)$} };
					\fill[color=black] (0.15,0.95,0.835) node[left] { {\scriptsize$R(g)$} };
					\fill[color=black] (0.55,0.95,0.875)  node[left] { {\scriptsize$R(h)$} };
					%
					\draw[string=Mybrown, thick] (T) -- (1T);
					\fill[color=black] (T) node[below] (0up) { {\scriptsize$\chi_{g,h}$} };
					%
					\draw [black,opacity=1, very thin] 
					(1T) -- (1L) -- (L) -- (T)
					(1T) -- (1R2) -- (R2) -- (T);
					\draw[ultra thick,costring=blue!50!black]
					(1R1) -- (R1) node[below] {\scriptsize$X_g$};
					\draw[ultra thick,costring=blue!50!black]
					(1R2) -- (R2) node[below] {\scriptsize$Y_h$};
			\end{tikzpicture}
			\tikz \draw [->, line join=round, decorate, decoration={zigzag, segment length=4, amplitude=.9,post=lineto, post length=2pt}]  (0,0) -- (2.2,0) node[pos=.5,above]{\scriptsize$\text{fractionalisation}$};
\begin{tikzpicture}[thick,scale=2.6,color=blue!50!black, baseline=-.1cm, >=stealth, 
					style={x={(-0.6cm,-0.4cm)},y={(1cm,-0.2cm)},z={(0cm,0.9cm)}}]
					\pgfmathsetmacro{\yy}{0.2}
					\coordinate (T) at (0.5, 0.4, 0);
					\coordinate (L) at (0.5, 0, 0);
					\coordinate (R1) at (0.3, 1, 0);
					\coordinate (R2) at (0.7, 1, 0);
					\coordinate (1T) at (0.5, 0.4, 1);
					\coordinate (1L) at (0.5, 0, 1);
					\coordinate (1R1) at (0.3, 1, );
					\coordinate (1R2) at (0.7, 1, );
					%
					\fill [orange!50,opacity=0.545] (L) -- (T) -- (1T) -- (1L);
					\fill [orange!50,opacity=0.545] (R1) -- (T) -- (1T) -- (1R1);
					\draw [black,opacity=1, very thin] (1T) -- (1R1) -- (R1) -- (T);
					\fill [orange!50,opacity=0.545] (R2) -- (T) -- (1T) -- (1R2);
					\fill[color=black] (0.5,0.17,0.9) node { {\scriptsize$R(gh)$} };
					\fill[color=black] (0.15,0.95,0.835) node[left] { {\scriptsize$R(g)$} };
					\fill[color=black] (0.55,0.95,0.875)  node[left] { {\scriptsize$R(h)$} };
					%
					\draw[string=Mybrown, ultra thick] (T) -- (1T);
					\fill[color=black] (T) node[below] (0up) { {\scriptsize$\chi_{g,h}^{ \lambda}$} };
					%
					\draw [black,opacity=1, very thin] 
					(1T) -- (1L) -- (L) -- (T)
					(1T) -- (1R2) -- (R2) -- (T);
					\draw[ultra thick,costring=blue!50!black]
					(1R1) -- (R1) node[below] {\scriptsize$X_g$};
					\draw[ultra thick,costring=blue!50!black]
					(1R2) -- (R2) node[below] {\scriptsize$Y_h$};
			\end{tikzpicture}
			\cong (X_g\otimes Y_h)\otimes \lambda(g,h)\,,
			\label{eq:zestedfusiondiagram}
\end{equation}
where $\chi_{g,h}^{ \lambda}\cong\chi_{g,h}\otimes\lambda(g,h)$, cf. \eqref{eq:zestedfusiondiagram1}.
The choice to put $\lambda(g,h)$ to the right in \eqref{eq:zestedfusiondiagram} is purely conventional \cite[Rem.\,3.5]{DGPRZbraided}, and different orderings are related by the associators and braiding in~$\Gcbc$.

In addition to the fusion rules, $\lambda$ modifies the $G$-action as well as duals (with \textsl{zested (left) duals} denoted by $\overline{X_g}$ instead of the usual $X^*_g$),
\begin{align}
	\rho_g^{(\lambda,\nu)}(Y_h)&:=\rho_g(Y_h)\otimes\lambda(h,g)\otimes \lambda(g,g^{-1}hg)^*\,,
	\label{eq:zestedGaction}\\
	\overline{X_g}&:=X_g^*\otimes\lambda(g,g^{-1})^*\,,
	\label{eq:zesteddual}
\end{align}
 and the associators, $G$-crossed braiding, duality adjunctions and pivotal structure, and monoidal coherence data of the $G$-action have to be adjusted accordingly, cf. \cite[Prop.\,3.3.,Thm.\,3.8]{DGPRZ} and \cite[Sect.\,3.5]{DGPRZ}.
Writing $\overline{g}:=g^{-1}$ and
 \begin{align}
 	\nu^g:=\nu_{g,\overline{g},g}&\colon\rho_g(\lambda(g,\overline{g}))\longrightarrow \lambda(\overline{g},g)\,,
 \intertext{we have }
 	\alpha_{X_{g},Y_{h},Z_{k}}^{(\lambda,\nu)}&:=
	\begin{tikzpicture}[very thick,scale=.8,color=blue!50!black, baseline=0cm,xscale = 1.8]
			\coordinate (X) at (0,-1.5);
			\coordinate (Xt) at (0,1.5);
			\coordinate (Y) at (.75,-1.5);
			\coordinate (Yt) at (.75,1.5);
			\coordinate (Z) at (2.25,-1.5);
			\coordinate (Zt) at (1.5,1.5);
			\coordinate (g12) at (1.5,-1.5);
			\coordinate (g23t) at (2.25,1.5);
			\coordinate (g123) at (3,-1.5);
			\coordinate (g123t) at (3,1.5);
			\draw 
			(X) -- (Xt)
			(X) node[below] {\scriptsize$X_{g}$}
			(Xt) node[above] {\scriptsize$X_{g}$}
			(Y) -- (Yt)
			(Y) node[below] {\scriptsize$Y_{h}$}
			(Yt) node[above] {\scriptsize$Y_{h}$}
			(Z) -- ($(Z)+(0,.5)$) .. controls +(0,.5) and +(0,-.5) .. ($(Zt)+(0,-1.5)$) -- (Zt)
			(Z) node[below] {\scriptsize$Z_{k}$}
			(Zt) node[above] {\scriptsize$Z_{k}$};
			\draw[color=white, line width=4pt]($(g12)+(0,.5)$) .. controls +(0,.5) and +(0,-.5) .. ($(g23t)+(0,-1.5)$);
			\draw[Mybrown] 
			(g12) -- ($(g12)+(0,.5)$) .. controls +(0,.5) and +(0,-.5) .. ($(g23t)+(0,-1.5)$) -- (g23t)
			(g123) -- (g123t);
			\draw[black] 
			(g12) node[below] {\scriptsize$\lambda(g,h)$}
			(g23t) node[above] {\scriptsize$\lambda(h,k)$}
			(g123) node[below] {\scriptsize$\lambda(gh,k)$}
			(g123t) node[above] {\scriptsize$\lambda(g,hk)$};
			\fill[color=white, rounded corners=2pt]($(g23t)+(-.3,-1.5)$)rectangle($(g123t)+(.3,-.5)$);
			\draw[draw=black, name=R, line width=0.5pt, rounded corners=2pt]($(g23t)+(-.3,-1.5)$)rectangle($(g123t)+(.3,-.5)$);
			\draw[black] ($.5*(g23t)+.5*(-.3,-1.5)+.5*(g123t)+.5*(.3,-.5)$) node {$\nu_{g,h,k}$};
		\end{tikzpicture}\,,\label{eq:zestedassociator}\\
		c_{X_{g},Y_{h}}^{(\lambda,\nu)}&:=
		\begin{tikzpicture}[very thick,yscale=.8,xscale=.6,color=blue!50!black,baseline]
			\draw 
			(-1,-1) node[below] (X) {{\scriptsize$X_{g}$}}
			(1,1) node[above] (Xu) {{\scriptsize$\rho_{h}(X_{g})$}}
			(1,-1) node[below] (Y) {{\scriptsize$Y_{h}$}}
			(-1,1) node[above] (Yu) {{\scriptsize$Y_{h}\vphantom{\rho_h(X)}$}};
			\draw[black] 
			(3,-1) node[below] (gh) {{\scriptsize$\lambda(g,h)$}}
			(3,1) node[above] (ghu) {{\scriptsize$\lambda(g,h)$}}
			(5,1) node[above] (gu) {{\scriptsize$\lambda(h,\overline{h}gh)^*$}}
			(7,1) node[above] (g-u) {{\scriptsize$\;\;\lambda(h,\overline{h}gh)$}};
			\draw (Y)  .. controls +(0,1) and +(0,-1) .. (Yu); 
			\draw[color=white, line width=4pt] (X)  .. controls +(0,1) and +(0,-1) .. (Xu); 
			\draw (X)  .. controls +(0,1) and +(0,-1) .. (Xu);
			\draw[string,Mybrown] (gu)  .. controls +(0,-1.5) and +(0,-1.5) .. (g-u);
			\draw[Mybrown] (gh) -- (ghu);
		\end{tikzpicture}\,,\label{eq:zestedbraiding}\\
		\ev_{X_{g}}^{(\lambda,\nu)}&:=
		\begin{tikzpicture}[very thick,scale=.8,color=blue!50!black, baseline=-.7cm,xscale = 1.8]
			\coordinate (Xl) at (0,-1.5);
			\coordinate (Xr) at (1.5,-1.5);
			\draw[directed] (Xr)-- ($(Xr)+(0,.5)$) .. controls +(0,1.5) and +(0,1.5) ..($(Xl)+(0,.5)$) --  (Xl);
			\draw (Xl) node[below] {\scriptsize$X_{g}^*$};
			\draw (Xr) node[below] {\scriptsize$X_{g}$};
			\coordinate (gl) at (.75,-1.5);
			\coordinate (gr) at (2.25,-1.5);
			\draw[color=white, line width=4pt]  (gr) -- ($(gr)+(0,.5)$) .. controls +(0,1.5) and +(0,1.5) ..($(gl)+(0,.5)$) -- (gl);
			\draw[string,Mybrown] (gr) -- ($(gr)+(0,.5)$) .. controls +(0,1.5) and +(0,1.5) ..($(gl)+(0,.5)$) -- (gl);
			 \draw[black]
			 (gl) node[below] {\scriptsize$\lambda(g,\overline{g})^*$}
			(gr) node[below] {\scriptsize$\lambda(\overline{g},g)$};
			\fill[color=white, rounded corners=2pt]($(gr)+(-.3,.8)$)rectangle($(gr)+(.3,.2)$);
			\draw[draw=black, name=R, line width=0.5pt, rounded corners=2pt]($(gr)+(-.3,.8)$)rectangle($(gr)+(.3,.2)$);
			\draw[black] ($(gr)+.5*(-.3,.8)+.5*(.3,.2)$) node {\scriptsize$(\nu^g)^{-1}$};
		\end{tikzpicture}
		\,\colon \overline{X_g}\otimes_\lambda X_g\longrightarrow\mathds{1}\,,\label{eq:zestedev}\\
		\coev_{X_{g}}^{(\lambda,\nu)}&:=
		\begin{tikzpicture}[very thick,scale=.8,color=Mybrown, baseline=1.1cm,xscale = 1.8]
			\coordinate (Xl) at (0,1.5);
			\coordinate (Xr) at (.75,1.5);
			\draw[directed] (Xr)-- ($(Xr)+(0,-.25)$) .. controls +(0,-.75) and +(0,-.75) ..($(Xl)+(0,-.25)$) --  (Xl);
			\draw[blue!50!black] 
			(Xl) node[above] {\scriptsize$X_{g}$}
			(Xr) node[above] {\scriptsize$X_{g}^*$};
			\coordinate (gl) at (1.5,1.5);
			\coordinate (gr) at (2.25,1.5);
			\draw[costring] (gr) -- ($(gr)+(0,-.25)$) .. controls +(0,-.75) and +(0,-.75) ..($(gl)+(0,-.25)$) -- (gl);
			\draw[black] 
			(gl) node[above] {\scriptsize$\lambda(g,\overline{g})^*$}
			(gr) node[above] {\scriptsize$\lambda(g,\overline{g})$};
		\end{tikzpicture}
		\,\colon \mathds{1}\longrightarrow X_g\otimes_\lambda \overline{X_g}\,,\label{eq:zestedcoev}\\
		j_{X_g}^{(\lambda,\nu)}&:=
		\begin{tikzpicture}[very thick,scale=.8,color=Mybrown, baseline=0cm,xscale = 1.8]
			\coordinate (X) at (0,-1.25);
			\coordinate (Xt) at (0,1.25);
			\draw[blue!50!black]
			(X) -- (Xt)
			(X) node[below] {\scriptsize$X_{g}$}
			(Xt) node[above] {\scriptsize$X_{g}^{**}$};
			\coordinate (gtl) at (-.75,1.25);
			\coordinate (gtr) at (.75,1.25);
			\draw[black] 
			(gtl) node[above] {\scriptsize$\lambda(g,\overline{g})$}
			(gtr) node[above] {\scriptsize$\lambda(\overline{g},g)^*$};
			\draw[color=white, line width=4pt]  (gtr) -- ($(gtr)+(0,-.5)$) .. controls +(0,-1.5) and +(0,-1.5) ..($(gtl)+(0,-.5)$) -- (gtl);
			\draw[postaction={decorate}, decoration={markings,mark=at position .48 with {\arrow{>}}}] (gtr) -- ($(gtr)+(0,-.5)$) .. controls +(0,-1.5) and +(0,-1.5) ..($(gtl)+(0,-.5)$) -- (gtl);
			\fill[color=white, rounded corners=2pt]($(gtl)+(-.3,-.8)$)rectangle($(gtl)+(.3,-.2)$);
			\draw[draw=black, name=R, line width=0.5pt, rounded corners=2pt]($(gtl)+(-.3,-.8)$)rectangle($(gtl)+(.3,-.2)$);
			\draw[black] ($(gtl)+.5*(-.3,-.8)+.5*(.3,-.2)$) node {\scriptsize$\nu^{\overline{g}}$};
			\fill[color=white, rounded corners=2pt]($(Xt)+(-.3,-.8)$)rectangle($(Xt)+(.3,-.2)$);
			\draw[draw=black, name=R, line width=0.5pt, rounded corners=2pt]($(Xt)+(-.3,-.8)$)rectangle($(Xt)+(.3,-.2)$);
			\draw[black] ($(Xt)+.5*(-.3,-.8)+.5*(.3,-.2)$) node {\scriptsize$j_{X_g}$};
		\end{tikzpicture}
		\,\colon  X_g\longrightarrow \overline{\overline{X_g}}\,,\label{eq:zestedpivotal}
 \end{align}
 where $j_{X_g}\colon X_g\longrightarrow X_g^{**}$ is the original pivotal structure,
\begin{align}
		(\rho^2_{g})^{(\lambda,\nu)}_{Y_h,Z_k}&:=
		\begin{tikzpicture}[very thick,yscale=.8,color=blue!50!black, baseline=0cm,xscale = 1.7]
			\coordinate (X) at (0,-4.5);
			\coordinate (h) at (.75,-4.5);
			\coordinate (gh) at (1.5,-4.5);
			\coordinate (Y) at (2.25,-4.5);
			\coordinate (k) at (3,-4.5);
			\coordinate (gk) at (3.75,-4.5);
			\coordinate (hk) at (4.5,-4.5);
			\coordinate (ghk) at (5,-3.75);
			\coordinate (Xt) at (0,4.5);
			\coordinate (Yt) at (2.25,4.5);
			\coordinate (ht) at (4.25,4.5);
			\coordinate (hkt) at (5,4.5);
			\coordinate (ghkt) at (5.75,4.5);
			\coordinate (nu1) at ($(hk)+(0,.75)$);
			\coordinate (nu2) at ($(hk)+(0,4.5)$);
			\coordinate (nu3) at ($(ht)+(0,-1.5)$);
			\draw 
			(X) node[below] {\scriptsize$\rho_g(Y_h)$}
			(Xt) node[above] {\scriptsize$\rho_{g}(Y_h)$}
			(Y) node[below] {\scriptsize$\rho_g(Z_k)$}
			(Yt) node[above] {\scriptsize$\rho_{g}(Z_k)$};
			\draw[black]
			(h) node[below] {\scriptsize$\lambda(h,g)$}
			(gh) node[below] {\scriptsize$\lambda(g,\overline{g}hg)^*$}
			(k) node[below] {\scriptsize$\lambda(k,g)$}
			(gk) node[below] {\scriptsize$\lambda(g,\overline{g}kg)^*$}
			(hk) node[below] {\scriptsize$\quad\lambda(\overline{g}hg,\overline{g}kg)$}
			(ht) node[above] {\scriptsize$\rho_g(\lambda(h,k))\;$}
			(hkt) node[above] {\scriptsize$\lambda(hk,g)$}
			(ghkt) node[above] {\scriptsize$\lambda(g,\overline{g}hkg)^*\;$}
			;
			\draw (X) -- (Xt)
			(Y) -- (Yt);
			\draw[Mybrown]
			(k) -- ($(nu3)+(-1.25,-1.5)$)  .. controls +(0,1) and +(0,-1) .. (nu3) -- (ht)
			(ghk) -- (hkt)
			(ghkt) -- ($(ghk)+(.75,0)$)
			(gk) -- ($(gk)+(0,5.5)$)
			($(gh)+(0,1.75)$) -- (gh)
			(h) -- ($(h)+(0,2.25)$)
			(hk) -- ($(hk)+(0,.75)$)
			;
			\draw[Mybrown, costring] ($(gk)+(0,5.5)$) .. controls +(0,.75) and +(0,.75) ..   ($(nu2)+(0,1)$);
			\draw[color=white, line width=4pt] 
			($(gh)+(0,1.75)$) .. controls +(0,1.5) and +(0,1.5) ..   ($(nu1)+(0,1)$)
			($(h)+(0,2.25)$) .. controls +(0,1.5) and +(0,-1.5) .. (nu2)
			;
			\draw[Mybrown] ($(h)+(0,2.25)$) .. controls +(0,1.5) and +(0,-1.5) .. (nu2);
			\draw[Mybrown, postaction={decorate}, decoration={markings,mark=at position .58 with {\arrow{<}}}] ($(gh)+(0,1.75)$) .. controls +(0,1.5) and +(0,1.5) ..   ($(nu1)+(0,1)$);
			\draw[Mybrown, costring] (ghk) .. controls +(0,-.75) and +(0,-.75) ..   ($(ghk)+(.75,0)$);
			\fill[color=white, rounded corners=2pt]($(nu1)+(-.3,0)$)rectangle($(nu1)+(.8,1)$);
			\draw[draw=black, name=R, line width=0.5pt, rounded corners=2pt]($(nu1)+(-.3,0)$)rectangle($(nu1)+(.8,1)$);
			\fill[color=white, rounded corners=2pt]($(nu2)+(-.3,0)$)rectangle($(nu2)+(.8,1)$);
			\draw[draw=black, name=R, line width=0.5pt, rounded corners=2pt]($(nu2)+(-.3,0)$)rectangle($(nu2)+(.8,1)$);
			\fill[color=white, rounded corners=2pt]($(nu3)+(-.3,0)$)rectangle($(nu3)+(1.05,1)$);
			\draw[draw=black, name=R, line width=0.5pt, rounded corners=2pt]($(nu3)+(-.3,0)$)rectangle($(nu3)+(1.05,1)$);
			\draw[black]
			($(nu3)+(.375,.5)$) node {$\nu_{h,k,g}^{-1}$}
			($(nu2)+(.25,.5)$) node {$\nu_{h,g,\overline{g}kg}$}
			($(nu1)+(.25,.5)$) node {$\nu_{g,\overline{g}hg,\overline{g}kg}^{-1}$};
		\end{tikzpicture}\,,\label{eq:tensorator}\\
	(\rho^2_{g,h})^{(\lambda,\nu)}_{Z_k}&:=
		\begin{tikzpicture}[very thick,yscale=.8,color=Mybrown, baseline=-0.7cm,xscale = 2]
			\coordinate (hl) at (-.75,2.75);
			\coordinate (hr) at (0,4.5);
			\coordinate (X) at (.75,5.25);
			\coordinate (k) at (1.5,5.25);
			\coordinate (hg) at (3.5,5.25);
			\coordinate (htl) at (-.75,-4);
			\coordinate (htr) at (0,-5.25);
			\coordinate (Xt) at (0,-6);
			\coordinate (kt) at (.75,-6);
			\coordinate (ht) at (1.5,-6);
			\coordinate (hkht) at (2.25,-6);
			\coordinate (gt) at (3,-6);
			\coordinate (hgt) at (4.5,-4.5);
			\coordinate (nu1) at ($(X)+(0,-2.5)$);
			\coordinate (nu2) at (2.25,-.25);
			\coordinate (nu3) at ($(gt)+(0.75,3)$);
			\draw[color=blue!50!black]
			(X) node[above] {\scriptsize$\rho_{hg}(Z_k)$}
			(Xt) node[below] {\scriptsize$\rho_g(\rho_h(Z_k))$}
			;
			\draw[black]
			(hl) node[right] {\scriptsize$\lambda(h,g)$}
			(k) node[above] {\scriptsize$\lambda(k,hg)$}
			(hg) node[above] {\scriptsize$\lambda(hg,\overline{g}\overline{h}khg)$}
			($(kt)+(0,-.45)$) node[below] {\scriptsize$\rho_g(\lambda(k,h))$}
			(ht) node[below] {\scriptsize$\rho_g(\lambda(h,\overline{h}kh)^*)$}
			($(hkht)+(0,-.45)$) node[below] {\scriptsize$\vphantom{\rho_g}\lambda(\overline{h}kh,g)$}
			(gt) node[below] {\scriptsize$\lambda(g,\overline{g}\overline{h}khg)^*$}
			($(gt)+(.75,4.2)$) node[right] {\scriptsize$\lambda(h,\overline{h}khg)$}
			($(hkht)+(0,7)$) node[right] {\scriptsize$\lambda(kh,g)$}
			 ($(nu1)+(.375,-.5)$) node {$\nu_{k,h,g}$}
			($(nu2)+(.375,-.5)$) node {$\nu_{h,\overline{h}kh,g}^{-1}$}
			($(nu3)+(.375,-.5)$) node {$\nu_{h,g,\overline{g}\overline{h}khg}$}
			;
			\draw[costring] (hl) .. controls +(0,1.5) and +(0,1.5) ..  (nu1);
			\draw (nu1) -- ($(kt)+(0,-.45)$);
			\draw[color=white, line width=4pt] ($(X)+(0,-1)$) .. controls +(0,-.5) and +(0,.5) .. ($(nu1)+(-.75,.25)$);
			\draw[color=blue!50!black] (X) -- ($(X)+(0,-1)$) .. controls +(0,-.5) and +(0,.5) .. ($(nu1)+(-.75,.25)$) -- (Xt);
			\draw (k) -- ($(nu1)+(.75,-1)$) .. controls +(0,-1) and +(0,1) .. ($(nu2)+(.75,0)$) -- ($(nu2)+(.75,-1)$) .. controls +(0,-1) and +(0,1) ..  ($(nu3)+(.75,0)$) -- (hgt);
			\draw[string] (hgt) .. controls +(0,-.75) and +(0,-.75) .. ($(hgt)+(.75,0)$); 
			\draw ($(hgt)+(.75,0)$) --  ($(hgt)+(.75,2)$) .. controls +(0,1.75) and +(0,-1.75) .. ($(hg)+(0,-2)$) -- (hg);
			\draw (ht) -- ($(nu2)+(-.75,0)$);
			\draw[costring] ($(nu2)+(-.75,0)$) .. controls +(0,.75) and +(0,.75) ..  (nu2);
			\draw ($(nu2)+(0,-1)$) -- ($(hkht)+(0,-.45)$);
			\draw (gt) -- ($(nu3)+(-.75,0)$);
			\draw[costring] ($(nu3)+(-.75,0)$) .. controls +(0,.75) and +(0,.75) ..  (nu3);
			\draw[color=white, line width=4pt] ($(nu3)+(0,-1)$) .. controls +(0,-2) and +(0,-2) ..  (htl);
			\draw[postaction={decorate}, decoration={markings,mark=at position .58 with {\arrow{<}}}] ($(nu3)+(0,-1)$) .. controls +(0,-2) and +(0,-2) ..  (htl);
			\draw (htl) -- (hl);
			\fill[color=white, rounded corners=2pt]($(nu1)+(-.3,0)$)rectangle($(nu1)+(1.05,-1)$);
			\draw[draw=black, name=R, line width=0.5pt, rounded corners=2pt]($(nu1)+(-.3,0)$)rectangle($(nu1)+(1.15,-1)$);
			\fill[color=white, rounded corners=2pt]($(nu2)+(-.3,0)$)rectangle($(nu2)+(1.05,-1)$);
			\draw[draw=black, name=R, line width=0.5pt, rounded corners=2pt]($(nu2)+(-.3,0)$)rectangle($(nu2)+(1.15,-1)$);
			\fill[color=white, rounded corners=2pt]($(nu3)+(-.3,0)$)rectangle($(nu3)+(1.05,-1)$);
			\draw[draw=black, name=R, line width=0.5pt, rounded corners=2pt]($(nu3)+(-.3,0)$)rectangle($(nu3)+(1.15,-1)$);
			\draw[black]
			($(nu1)+(.375,-.5)$) node {$\nu_{k,h,g}$}
			($(nu2)+(.375,-.5)$) node {$\nu_{h,\overline{h}kh,g}^{-1}$}
			($(nu3)+(.375,-.5)$) node {$\nu_{h,g,\overline{g}\overline{h}khg}$};
		\end{tikzpicture}\,.\label{eq:compositor}
\end{align}

\begin{notation}[label=nota:conventions]
	Let us briefly remark on how the conventions for coherence isomorphisms relate to each other among the references \cite{BBCW,BBDR,DGPRZ} which are central to our examples.
	For associators, we use the convention of \cite{BBDR,DGPRZ} $\alpha_{X,Y,Z}\colon (X\otimes Y)\otimes Z\longrightarrow X\otimes (Y\otimes Z)$, while the $F$-symbols of \cite{BBCW} are components of $\alpha^{-1}$.
	For the monoidal structure of the $G$-action, \cite{DGPRZ} use $(\mu_g)_{X,Y}:=(\rho^2_g)_{X,Y}^{-1}$ and $(\gamma_{g,h})_X:=(\rho^2_{h,g})_X^{-1}$, while \cite{BBCW} use $U_g(X,Y)=(\rho^2_g)_{X,Y}^{-1}$ and $\eta_X(g,h)=(\rho^2_{g,h})_{X}$.
	We continue to use our notation exclusively.
\end{notation}

If $G=\Z_N$ is the cyclic group of order~$N$, then $\lambda(g,h)\cong\lambda(h,g)$ \cite[Eq.\,(3.18)]{DGPRZ} and since $\lambda(g,h)$ is invertible, $\lambda(h,g)\otimes\lambda(g,h)^*\cong\mathds{1}$.
Thus the $\Z_N$-action on objects is unchanged under $\Z_N$-crossed braided zesting (up to canonical isomorphism), $\rho^{(\lambda,\nu)}_g(Y_h)=\rho_g(Y_h)\otimes\lambda(h,g)\otimes\lambda(g,h)^*\cong \rho_g(Y_h)$.
Similarly, cohomology classes $p\in\mathrm{H}^3(\Z_N,\C^\times)$ can be parametrised to be symmetric in the first two arguments \cite[Eq.\,(3.19)]{DGPRZ}, hence if we only modify the defectification class~$p$ of a chosen zesting datum $(\lambda,\nu)$, the monoidal structure of the $\Z_N$-action satisfies
\begin{align}
	(\rho_g^2)_{Y_h,Z_k}^{(\lambda,p\nu)}&=p(h,k,g)^{-1}(\rho_g^2)_{Y_h,Z_k}^{(\lambda,\nu)}\,,
	\label{eq:zestedtensorator}\\
	(\rho_{g,h}^2)_{Z_k}^{(\lambda,p\nu)}&=p(g,h,k)(\rho_{g,h}^2)_{Z_k}^{(\lambda,\nu)}\,,
	\label{eq:zestedcompositor}
\end{align}
by simplifying \eqref{eq:tensorator} and \eqref{eq:compositor}, cf. \cite[Eqs.\,(3.22),(3.23)]{DGPRZ}.

\medskip

It was shown in \cite[Thm.\,3.12]{DGPRZ} that the underlying module categories are equivalent for two $G$-crossed braided extensions if and only if the latter are related by zesting. 
In other words, once we have fixed the underlying twisted sector line operators and their fusion with the bulk lines (but not fusion between twisted sector operators), zesting allows us to access all possible $G$-extensions.

This result implies in particular that if two extensions are related by zesting, then the surface algebras which gauge the respective $0$-form symmetries have the same underlying objects (cf.\,\cite{BBDR}) which we observe explicitly in \Cref{sec:DZ20}. 
Let us explain this in more detail:
By the classical theorem of Ostrik \cite[Thm.\,1]{Ostrik}, an indecomposable semisimple $\mathcal{C}$-module category is equivalent to a category of $A$-modules for some algebra~$A$ in~$\mathcal{C}$. 
Note that invertible $\mathcal{C}$-module categories are always indecomposable \cite[Cor.\,4.4]{ENO}, hence the theorem applies to twisted sectors of a $G$-extensions of a braided fusion category, since they are always invertible (the inverse of $\mathcal{C}_g$ is $\mathcal{C}_{g^{-1}}$: $\mathcal{C}_g\boxtimes_\mathcal{C}\mathcal{C}_{g^{-1}}\cong \mathcal{C}$).

This fact is used to construct an orbifold datum for a $0$-form $G$-symmetry in Reshetikhin--Turaev theory (cf. \Cref{sec:RT}) and in particular the surface underlying the gauging.
In that construction, the algebra underlying the orbifold datum is given by a direct sum of the algebras~$A_g$ associated to the individual module categories~$\mathcal{C}_g\cong A_g\text{-}\mathrm{Mod}(\mathcal{C}_e)$.
Therefore, if the module categories are unchanged then so is the surface~$A_G$ (i.e.\ $A_G\text{-}\mathrm{Mod}(\mathcal{C}_e)$) underlying the gauging.

\medskip

Alternatively, for a given choice of~$\lambda$ and the associated $G$-crossed braided fusion category $\Gcbc$, one can also view the choice of~$p$ in terms of the fiber product with $G$-$\Vect^p$,\footnote{By using the condensed fiber product, one can construct all zestings this way, not just changes in defectification, see \cite{DGPRZcondensedfiber} for details.} the category of $G$-graded vector spaces twisted by the 3-cocycle~$p$.
This is given by
\begin{equation}
	G\text{-}\Vect^p\stackrel{G}{\boxtimes} \mathcal{C}^\times_G:=\bigoplus_{g\in G} (G\text{-}\Vect^p)_g\boxtimes \mathcal{C}_g\subset G\text{-}\Vect^p\boxtimes \Gcbc\,.
\end{equation}
In physical terms, this amounts to coupling to the bosonic $G$-SPT phase with twist~$p$.
Moreover, we can always generate the trivial $G$-symmetry (i.e.\ the trivial $G$-extension) of~$\mathcal{C}_e$ by gluing in the trivial SPT phase:
\begin{equation}
	\Gcbc= G\text{-}\Vect\boxtimes\mathcal{C}_e\,.\label{eq:trivialExtension}
\end{equation}
This produces a $G$-crossed braided extension where the twisted sectors are given by the bulk line operators enhanced with $G$-charges, i.e.\ $\mathcal{C}_g\cong\mathcal{C}_e$ as categories.
We identify this with the choice of trivial symmetry fractionalisation $\lambda=\mathds{1}$.

Symmetry fractionalisation is encoded in the line operator that fuses surfaces associated to 0-form defects (see e.g. \cite{BBG1}) as illustrated in \eqref{eq:zestedfusiondiagram}.
This is captured in the orbifold datum through the line operator~$T$ \eqref{eq:3dorbdat}. 
For extensions of the type described in \eqref{eq:trivialExtension} and its zestings, we can calculate the symmetry fractionalisation by comparing the line operator \eqref{eq:chi-def} to the one we get from the trivial extension.
We illustrate this in detail in the next section and in the examples of \Cref{sec:DZ20}.

\subsection{Zested Orbifold Data}\label{sec:zestedorbdat}
In this section, we combine the constructions of \Cref{sec:RT} and \Cref{sec:zesting}, i.e.\ we calculate the impact of fractionalisation~$\lambda$ and defectification~$\nu$ on orbifold data.
This section is slightly more technical than previous ones and may be skipped without issue by those who are more interested in the examples and applications.

Starting with a Reshetikhin--Turaev theory with line operators~$\mathcal{C}$, recall that for a given $G$-symmetry described by an extension $\Gcbc$, \Cref{sec:RT} described the orbifold datum that gauges the symmetry through the orbifold construction. 
We can modify the symmetry by a zesting datum $(\lambda,\nu)$ and ask: What is the orbifold datum that gauges the new, zested symmetry expressed through the data of the original extension?
There are three components to this, the Frobenius algebras $A_g^{(\lambda,\nu)}$, the bimodules $\chi_{g,h}^{(\lambda,\nu)}$, and the bimodule morphisms $\alpha^{(\lambda,\nu)}_{g,h,k}$.

As preliminary observations, note that $\nu_{g,e,h}=\id_{\lambda(g,h)}$ by definition and two simple calculations using the pentagon axiom \cite[Eq.\,(3.3)]{DGPRZ} show that\footnote{Specifically, choose $g_1=g_2=e$ and $g_3=g_4=e$ for the first two identities respectively, and $g_1=g_3=g$, $g_2=g_4=\overline{g}$ for the second. In each case we then compose with $\nu^{-1}$ and trace over the right strand.}
\begin{align}
	\nu_{e,g,h}=\nu_{g,h,e}&=\id_{\lambda(g,h)}\,,\\
	\rho_{\overline{g}}(\nu^g)&= \theta_{\lambda(g,\overline{g})}\cdot(\nu^{\overline{g}})^{-1}\circ (\rho_{\overline{g},g}^2)_{\lambda(g,\overline{g})}\,,
	\label{eq:gactsinvertingnu}
\end{align}
where $\theta_{\lambda(g,h)}\in\C$ is the twist of $\lambda(g,h)$.
It was shown in \cite[Eq.\,(3.33)]{DGPRZ} that the twist also contributes to the quantum dimension of objects in the twisted sectors $X_g\in\mathcal{C}_g$:
\begin{equation}
	\mathrm{dim}_{j^{(\lambda,\nu)}}(X_g)= \theta_{\lambda(g,\overline{g})}\cdot \mathrm{dim}_j(X_g)\,.
\end{equation}
This is due to the fact that in a pivotal category, left and right duals are identified via the pivotal structure~$j$,  $^*\!X\cong(^*\!X)^{**}\cong X^*$ where the second isomorphism is canonical, and the adjunction data for one is computed from the other using the pivotal structure, i.e.\ $\widetilde{\ev}_{X}=\ev_{X^*}\circ (j_X\otimes \id_{X^*})$ and $\widetilde{\coev}_X=(\id_{X^*}\otimes j_X^{-1})\circ \coev_{X^*}$ for the right duals.
In the zested extension, the composition of $j_{X_g}^{(\lambda,\nu)}$\eqref{eq:zestedpivotal} and $\ev_{\overline{X_g}}^{(\lambda,\nu)}$ \eqref{eq:zestedev} introduces a factor of $\theta_{\lambda(g,\overline{g})}$ which appears in the definition of the trace and therefore the dimension.

Using these observations, we can evaluate the formulas of \Cref{sec:RT} to compute the orbifold datum using the evaluations, coevaluations, braiding, and associators of \Cref{sec:zesting}.
The Frobenius algebra structures on~$A_g^{(\lambda,\nu)}$ are given by
\begin{align}
	A_g^{(\lambda,\nu)}&:=\overline{\mg}\otimes_\lambda \mg=\left((\mg^*\otimes\lambda(g,\overline{g}))\otimes \mg\right)\otimes \lambda(\overline{g},g)\,,\\
	\label{eq:zestedAgmult}
	\mu_g^{(\lambda,\nu)}&:=
	\begin{tikzpicture}[very thick,scale=1,color=red!50!black, baseline=.9cm]
\draw[line width=0pt] 
(3,0) node[line width=0pt] (D) {{\scriptsize$\overline{\mg}$}}
(2,0) node[line width=0pt] (s) {{\scriptsize$\mg$}}; 
\draw[redirectedred] (D) .. controls +(0,1) and +(0,1) .. (s);
\draw[line width=0pt] 
(3.45,0) node[line width=0pt] (re) {{\scriptsize$\mg$}}
(1.55,0) node[line width=0pt] (li) {{\scriptsize$\overline{\mg}$}}; 
\draw[line width=0pt] 
(2.7,2) node[line width=0pt] (ore) {{\scriptsize$\mg$}}
(2.3,2) node[line width=0pt] (oli) {{\scriptsize$\overline{\mg}$}}; 
\draw[directedred] (2.3,1.25) .. controls +(0,-0.25) and +(0,0.75) .. (li);
\draw (2.3,1.25) -- (oli);
\draw[directedred] (re) .. controls +(0,0.75) and +(0,-0.25) .. (2.7,1.25);
\draw (2.7,1.25) -- (ore);
\end{tikzpicture}
\;=\;
\begin{tikzpicture}[very thick,color=blue!50!black, baseline=0cm,xscale = 1.4,yscale=.8]
			\coordinate (mb1) at (-.75,-2.5);
			\coordinate (lb1) at (0,-2.5);
			\coordinate (m1) at (.75,-2.5);
			\coordinate (l1) at (1.5,-2.5);
			\coordinate (mb2) at (2.75,-2.5);
			\coordinate (lb2) at (3.5,-2.5);
			\coordinate (m2) at (4.25,-2.5);
			\coordinate (l2) at (5,-2.5);
			%
			\coordinate (mbt) at (1,2.5);
			\coordinate (lbt) at (1.75,2.5);
			\coordinate (mt) at (2.5,2.5);
			\coordinate (lt) at (3.25,2.5);
			\coordinate (nu2) at ($(l1)+(.3,1.75)$);
			\draw 
			(mb1) node[below] {\scriptsize$\mg^*$}
			(m1) node[below] {\scriptsize$\mg$}
			(mb2) node[below] {\scriptsize$\mg^*$}
			(m2) node[below] {\scriptsize$\mg$}
			(mbt) node[above] {\scriptsize$\mg^*$}
			(mt) node[above] {\scriptsize$\mg$};
			\draw[black]
			(l1) node[below] {\scriptsize$\lambda(\overline{g},g)$}
			(lb1) node[below] {\scriptsize$\lambda(g,\overline{g})^*$}
			(l2) node[below] {\scriptsize$\lambda(\overline{g},g)$}
			(lb2) node[below] {\scriptsize$\lambda(g,\overline{g})^*$}
			(lbt) node[above] {\scriptsize$\lambda(g,\overline{g})^*$}
			(lt) node[above] {\scriptsize$\lambda(\overline{g},g)$}
			;
			\draw 
			(mb1) .. controls +(0,2) and +(0,-2) .. (mbt)
			(m2) .. controls +(0,2) and +(0,-2) .. (mt)
			;
			\draw[postaction={decorate}, decoration={markings,mark=at position .6 with {\arrow[color=blue!50!black]{>}}}] (m1) .. controls +(0,1.7) and +(0,1.7) .. (mb2);
			\draw[Mybrown]
			(lb1) .. controls +(0,2) and +(0,-2) .. (lbt)
			(l2) .. controls +(0,2) and +(0,-2) .. (lt)
			;
			\draw[color=white, line width=4pt] 
			(l1) .. controls +(0,.5) and +(0,-.5) .. (nu2) 
			;
			\draw[Mybrown]
			(l1) .. controls +(0,.5) and +(0,-.5) .. (nu2) 
			(lb2) .. controls +(0,1) and +(0,-1) .. ($(nu2)+(.75,1)$)
			;
			\draw[Mybrown,string]
			($(nu2)+(0,1)$) .. controls +(0,.75) and +(0,.75) .. ($(nu2)+(.75,1)$)
			;
			\fill[color=white, rounded corners=2pt]($(nu2)+(-.3,0)$)rectangle($(nu2)+(.3,1)$);
			\draw[draw=black, name=R, line width=0.5pt, rounded corners=2pt]($(nu2)+(-.3,0)$)rectangle($(nu2)+(.3,1)$);
			\draw[black]
			($(nu2)+(0,.5)$) node {$\nu^{\overline{g}}$};
\end{tikzpicture}\,,\\
\eta_g^{(\lambda,\nu)}&:=
\begin{tikzpicture}[very thick,scale=0.9,color=red!50!black, baseline=-.4cm,rotate=180]
\draw[line width=0pt] 
(3,0) node[line width=0pt] (D) {{\scriptsize$\overline{\mg}$}}
(2,0) node[line width=0pt] (s) {{\scriptsize$\mg$}}; 
\draw[directedred] (D) .. controls +(0,1) and +(0,1) .. (s);
\end{tikzpicture}
\;=\;
\begin{tikzpicture}[very thick,color=blue!50!black, baseline=1cm,xscale = 1.4,yscale=.8]
			\coordinate (mbt) at (1,2.5);
			\coordinate (lbt) at (1.75,2.5);
			\coordinate (mt) at (2.5,2.5);
			\coordinate (lt) at (3.25,2.5);
			\coordinate (nu1) at ($(mt)+(0,-2)$);
			\draw 
			(mbt) node[above] {\scriptsize$\mg^*$}
			(mt) node[above] {\scriptsize$\mg$};
			\draw[black]
			(lbt) node[above] {\scriptsize$\lambda(g,\overline{g})^*$}
			(lt) node[above] {\scriptsize$\lambda(\overline{g},g)$}
			;
			\draw[string] (mbt) -- ($(mbt)+(0,-2)$) .. controls +(0,-1.9) and +(0,-1.9) .. ($(lt)+(0,-2)$) -- ($(lt)+(0,-1)$) .. controls +(0,.5) and +(0,-.5) .. (mt);
			\draw[color=white, line width=4pt] 
			($(nu1)+(0,1)$) .. controls +(0,.5) and +(0,-.5) .. (lt)
			;
			\draw[Mybrown]
			(lbt) -- ($(lbt)+(0,-2)$)
			($(nu1)+(0,1)$) .. controls +(0,.5) and +(0,-.5) .. (lt)
			;			
			\draw[Mybrown,string]
			($(lbt)+(0,-2)$) .. controls +(0,-.75) and +(0,-.75) .. (nu1);
			\fill[color=white, rounded corners=2pt]($(nu1)+(-.4,0)$)rectangle($(nu1)+(.4,1)$);
			\draw[draw=black, name=R, line width=0.5pt, rounded corners=2pt]($(nu1)+(-.4,0)$)rectangle($(nu1)+(.4,1)$);
			\draw[black]
			($(nu1)+(0,.5)$) node {$(\nu^{\overline{g}})^{-1}$};
\end{tikzpicture}\,,\\
\Delta_g^{(\lambda,\nu)}&:=
\begin{tikzpicture}[very thick,scale=0.9,color=red!50!black, baseline=-0.9cm, rotate=180]
\draw[line width=0pt] 
(3,0) node[line width=0pt] (D) {{\scriptsize$\mg$}}
(2,0) node[line width=0pt] (s) {{\scriptsize$\overline{\mg}$}}; 
\draw[directedred] (s) .. controls +(0,1) and +(0,1) .. (D) ;
\draw[line width=0pt] 
(3.45,0) node[line width=0pt] (re) {{\scriptsize$\overline{\mg}$}}
(1.55,0) node[line width=0pt] (li) {{\scriptsize$\mg$}}; 
\draw[line width=0pt] 
(2.7,2) node[line width=0pt] (ore) {{\scriptsize$\overline{\mg}\;$}}
(2.3,2) node[line width=0pt] (oli) {{\scriptsize$\;\mg$}}; 
\draw[directedred] (2.3,1.25) .. controls +(0,-0.25) and +(0,0.75) .. (li);
\draw (2.3,1.25) -- (oli);
\draw[directedred] (re) .. controls +(0,0.75) and +(0,-0.25) .. (2.7,1.25);
\draw (2.7,1.25) -- (ore);
\end{tikzpicture}\,\cdot \mathrm{dim}_{j^{(\lambda,\nu)}}(m_g)^{-1}\\
\;&=\;
\begin{tikzpicture}[very thick,color=blue!50!black, baseline=0cm,xscale = 1.4,yscale=-.8]
			\coordinate (mb1) at (-.75,-2.5);
			\coordinate (lb1) at (0,-2.5);
			\coordinate (m1) at (.75,-2.5);
			\coordinate (l1) at (1.5,-2.5);
			\coordinate (mb2) at (2.75,-2.5);
			\coordinate (lb2) at (3.5,-2.5);
			\coordinate (m2) at (4.25,-2.5);
			\coordinate (l2) at (5,-2.5);
			\coordinate (mbt) at (1,2.5);
			\coordinate (lbt) at (1.75,2.5);
			\coordinate (mt) at (2.5,2.5);
			\coordinate (lt) at (3.25,2.5);
			\coordinate (nu2) at ($(l1)+(.3,1.75)$);
			\draw 
			(mb1) node[above] {\scriptsize$\mg^*$}
			(m1) node[above] {\scriptsize$\mg$}
			(mb2) node[above] {\scriptsize$\mg^*$}
			(m2) node[above] {\scriptsize$\mg$}
			(mbt) node[below] {\scriptsize$\mg^*$}
			(mt) node[below] {\scriptsize$\mg$};
			\draw[black]
			(l1) node[above] {\scriptsize$\lambda(\overline{g},g)$}
			(lb1) node[above] {\scriptsize$\lambda(g,\overline{g})^*$}
			(l2) node[above] {\scriptsize$\lambda(\overline{g},g)$}
			(lb2) node[above] {\scriptsize$\lambda(g,\overline{g})^*$}
			(lbt) node[below] {\scriptsize$\lambda(g,\overline{g})^*$}
			(lt) node[below] {\scriptsize$\lambda(\overline{g},g)$}
			;
			\draw 
			(mb1) .. controls +(0,2) and +(0,-2) .. (mbt)
			(m2) .. controls +(0,2) and +(0,-2) .. (mt)
			;
			\draw[postaction={decorate}, decoration={markings,mark=at position .6 with {\arrow[color=blue!50!black]{<}}}] (m1) .. controls +(0,1.7) and +(0,1.7) .. (mb2);
			\draw[Mybrown]
			(lb1) .. controls +(0,2) and +(0,-2) .. (lbt)
			(l2) .. controls +(0,2) and +(0,-2) .. (lt)
			(lb2) .. controls +(0,1) and +(0,-1) .. ($(nu2)+(.75,1)$)
			;
			\draw[Mybrown,costring]
			($(nu2)+(0,1)$) .. controls +(0,.75) and +(0,.75) .. ($(nu2)+(.75,1)$)
			;
			\draw[color=white, line width=4pt] 
			(l1) .. controls +(0,.5) and +(0,-.5) .. (nu2) 
			;
			\draw[Mybrown]
			(l1) .. controls +(0,.5) and +(0,-.5) .. (nu2)
			;
			\fill[color=white, rounded corners=2pt]($(nu2)+(-.4,0)$)rectangle($(nu2)+(.4,1)$);
			\draw[draw=black, name=R, line width=0.5pt, rounded corners=2pt]($(nu2)+(-.4,0)$)rectangle($(nu2)+(.4,1)$);
			\draw[black]
			($(nu2)+(0,.5)$) node {$(\nu^{\overline{g}})^{-1}$};
\end{tikzpicture}\,\cdot \mathrm{dim}_j(m_g)^{-1}\,,\\
\varepsilon_g^{(\lambda,\nu)}&:=
\begin{tikzpicture}[very thick,scale=0.9,color=red!50!black, baseline=.4cm]
\draw[line width=0pt] 
(3,0) node[line width=0pt] (D) {{\scriptsize$m_g$}}
(2,0) node[line width=0pt] (s) {{\scriptsize$\overline{m_g}$}}; 
\draw[redirectedred] (s) .. controls +(0,1) and +(0,1) .. (D);
\end{tikzpicture}\,\cdot \mathrm{dim}_{j^{(\lambda,\nu)}}(m_g)
\;=\;
		\begin{tikzpicture}[very thick,yscale=.8,color=blue!50!black, baseline=-.7cm,xscale = 1.4]
			\coordinate (Xl) at (0,-1.5);
			\coordinate (Xr) at (1.5,-1.5);
			\draw[directed] (Xr)-- ($(Xr)+(0,.5)$) .. controls +(0,1.5) and +(0,1.5) ..($(Xl)+(0,.5)$) --  (Xl);
			\draw (Xl) node[below] {\scriptsize$m_{g}^*$};
			\draw (Xr) node[below] {\scriptsize$m_{g}$};
			\coordinate (gl) at (.75,-1.5);
			\coordinate (gr) at (2.25,-1.5);
			\draw[color=white, line width=4pt]  (gr) -- ($(gr)+(0,.5)$) .. controls +(0,1.5) and +(0,1.5) ..($(gl)+(0,.5)$) -- (gl);
			\draw[string,Mybrown] (gr) -- ($(gr)+(0,.5)$) .. controls +(0,1.5) and +(0,1.5) ..($(gl)+(0,.5)$) -- (gl);
			 \draw[black]
			 (gl) node[below] {\scriptsize$\lambda(g,\overline{g})^*$}
			(gr) node[below] {\scriptsize$\lambda(\overline{g},g)$};
			\fill[color=white, rounded corners=2pt]($(gr)+(-.4,1)$)rectangle($(gr)+(.4,.2)$);
			\draw[draw=black, name=R, line width=0.5pt, rounded corners=2pt]($(gr)+(-.4,1)$)rectangle($(gr)+(.4,.2)$);
			\draw[black] ($(gr)+.5*(-.3,1)+.5*(.3,.2)$) node {$(\nu^g)^{-1}$};
		\end{tikzpicture}
\,\cdot \mathrm{dim}(m_g)\theta_{\lambda(g,\overline{g})}\,,
\end{align}
where both the twist coming from the right evaluation in the multiplication $\mu_g^{(\lambda,\nu)}$ and the twist contained in $\mathrm{dim}_{j^{(\lambda,\nu)}}(m_g)^{-1}$ in the coevaluation $\Delta_g^{(\lambda,\nu)}$ are cancelled by a braiding coming from an associator.
As a sanity check, note that composing the comultiplication with the counit $(\varepsilon_g^{(\lambda,\nu)}\otimes\id_{A_g^{(\lambda,\nu)}})\circ \Delta_g^{(\lambda,\nu)}=\id_{A_g^{(\lambda,\nu)}}$ produces the identity since the two instances of $\nu$ cancel via \eqref{eq:gactsinvertingnu} which also handles the remaining factor of $\theta_{\lambda(g,\overline{g})}$ in the counit.

Next, we discuss the line defect~$T_G^{(\lambda,\nu)}$ which fuses the surfaces condensed from $A_G^{(\lambda,\nu)}$.
Recall that this is an $(A_G^{(\lambda,\nu)},A_G^{(\lambda,\nu)}\otimes A_G^{(\lambda,\nu)})$-bimodule which decomposes as follows:
\begin{align}
	T_G^{(\lambda,\nu)}&:=\bigoplus_{g,h\in G} \chi_{g,h}^{(\lambda,\nu)}\,,\\
	\chi_{g,h}^{(\lambda,\nu)}&:=\overline{m_{gh}}\otimes_\lambda (m_g\otimes_\lambda m_h)\nonumber\\
	&=(m_{gh}^*\otimes \lambda(gh,\overline{gh})^*)\otimes (m_g\otimes m_h\otimes \lambda(g,h))\otimes\lambda(\overline{gh},gh)\,,\\
\begin{tikzpicture}[very thick,scale=0.8,color=blue!50!black, baseline]
\draw (0,-2.5) node[below] (X) {{\scriptsize$\chi_{g,h}^{(\lambda,\nu)}$}};
\draw (0,2.5) node[above] (X) {{\scriptsize$\chi_{g,h}^{(\lambda,\nu)}$}};
\draw[color=green!50!black] (-1,-2.5) node[below] (A1) {{\scriptsize$A_{gh}^{(\lambda,\nu)}$}};
\draw (0,2.5) node[right] (Xu) {};
\draw[color=green!50!black] (A1) .. controls +(0,1.25) and +(-.5,-1.25) .. (0,0.75);
\draw (0,-2.5) -- (0,2.5); 
\fill[color=blue!50!black] (0,0.75) circle (2.9pt) node (meet2) {};
\end{tikzpicture} 
&:=
	\begin{tikzpicture}[very thick,color=blue!50!black, baseline=0cm,xscale = 1.4,yscale=.8]
			\coordinate (mb1) at (-.75,-2.5);
			\coordinate (lb1) at (0,-2.5);
			\coordinate (m1) at (.75,-2.5);
			\coordinate (l1) at (1.5,-2.5);
			\coordinate (mb2) at (2.75,-2.5);
			\coordinate (lb2) at (3.5,-2.5);
			\coordinate (m2) at (4.25,-2.5);
			\coordinate (m3) at (5,-2.5);
			\coordinate (l2) at (5.75,-2.5);
			\coordinate (l3) at (6.5,-2.5);
			\coordinate (mbt) at (2.75,2.5);
			\coordinate (lbt) at (3.5,2.5);
			\coordinate (m1t) at (4.25,2.5);
			\coordinate (m2t) at (5,2.5);
			\coordinate (l1t) at (5.75,2.5);
			\coordinate (l2t) at (6.5,2.5);
			\coordinate (nu2) at ($(l1)+(1.25,1.75)$);
			\draw 
			(mb1) node[below] {\scriptsize$m_{gh}^*\vphantom{\overline{gh}}$}
			(m1) node[below] {\scriptsize$m_{gh}\vphantom{\overline{gh}}$}
			(mb2) node[below] {\scriptsize$m_{gh}^*\vphantom{\overline{gh}}$}
			(m2) node[below] {\scriptsize$m_{g}\vphantom{\overline{gh}}$}
			(m3) node[below] {\scriptsize$m_{h}\vphantom{\overline{gh}}$}
			(mbt) node[above] {\scriptsize$m_{gh}^*\vphantom{\overline{gh}}$}
			(m1t) node[above] {\scriptsize$m_g\vphantom{\overline{gh}}$}
			(m2t) node[above] {\scriptsize$m_h\vphantom{\overline{gh}}$};
			\draw[black]
			(lb1) node[below] {\scriptsize$\lambda(gh,\overline{gh})^*$}
			(l1) node[below] {\scriptsize$\lambda(\overline{gh},gh)$}
			(lb2) node[below] {\scriptsize$\lambda(gh,\overline{gh})^*$}
			(l2) node[below] {\scriptsize$\lambda(g,h)\vphantom{\overline{gh}}$}
			(l3) node[below] {\scriptsize$\lambda(\overline{gh},gh)$}
			(lbt) node[above] {\scriptsize$\lambda(gh,\overline{gh})^*\vphantom{\overline{gh}}$}
			(l1t) node[above] {\scriptsize$\lambda(g,h)\vphantom{\overline{gh}}$}
			(l2t) node[above] {\scriptsize$\lambda(\overline{gh},gh)$}
			;
			\draw 
			(mb1) .. controls +(0,2) and +(0,-2) .. (mbt)
			(m2) -- (m1t)
			(m3) -- (m2t)
			;
			\draw[postaction={decorate}, decoration={markings,mark=at position .45 with {\arrow[color=blue!50!black]{>}}}] (m1) .. controls +(0,1.7) and +(0,1.7) .. (mb2);
			\draw[Mybrown]
			(lb1) .. controls +(0,2) and +(0,-2) .. (lbt)
			(l2) -- (l1t)
			(l3) -- (l2t)
			;
			\draw[color=white, line width=4pt] 
			(l1) .. controls +(0,.5) and +(0,-.5) .. (nu2) 
			;
			\draw[Mybrown]
			(l1) .. controls +(0,.5) and +(0,-.5) .. (nu2) 
			(lb2) .. controls +(0,1) and +(0,-1) .. ($(nu2)+(.75,1)$)
			;
			\draw[Mybrown,string]
			($(nu2)+(0,1)$) .. controls +(0,.75) and +(0,.75) .. ($(nu2)+(.75,1)$)
			;
			\fill[color=white, rounded corners=2pt]($(nu2)+(-.3,0)$)rectangle($(nu2)+(.3,1)$);
			\draw[draw=black, name=R, line width=0.5pt, rounded corners=2pt]($(nu2)+(-.3,0)$)rectangle($(nu2)+(.3,1)$);
			\draw[black]
			($(nu2)+(0,.5)$) node {$\nu^{\overline{gh}}$};
\end{tikzpicture}\,,\\
\begin{tikzpicture}[very thick,scale=0.8,color=blue!50!black, baseline]
\draw (0,2.5) node[above] (X) {{\scriptsize$\chi_{g,h}^{(\lambda,\nu)}$}};
\draw (0,-2.5) node[below] (X) {{\scriptsize$\chi_{g,h}^{(\lambda,\nu)}$}};
\draw[color=green!50!black] (1,-2.5) node[below] (A1) {{\scriptsize$A_h^{(\lambda,\nu)}$}};
\draw (0,2.5) node[right] (Xu) {};
\draw[color=green!50!black] (A1) .. controls +(0,1.25) and +(0.5,-1.25) .. (0,0.75);
\draw (0,-2.5) -- (0,2.5); 
\fill[color=blue!50!black] (0,0.75) circle (2.9pt) node (meet2) {};
\fill[color=black] (0.2,1) circle (0pt) node (meet) {{\tiny$2$}};
\end{tikzpicture} 
&:=
	\begin{tikzpicture}[very thick,color=blue!50!black, baseline=0cm,xscale = 1.4,yscale=.8]
			\coordinate (mb1) at (-.75,-2.5);
			\coordinate (lb1) at (0,-2.5);
			\coordinate (m1) at (.75,-2.5);
			\coordinate (m2) at (1.5,-2.5);
			\coordinate (l1) at (2.25,-2.5);
			\coordinate (l2) at (3,-2.5);
			\coordinate (mb2) at (4.25,-2.5);
			\coordinate (lb2) at (5,-2.5);
			\coordinate (m3) at (5.75,-2.5);
			\coordinate (l3) at (6.5,-2.5);
			\coordinate (mbt) at (-.75,2.5);
			\coordinate (lbt) at (0,2.5);
			\coordinate (m1t) at (.75,2.5);
			\coordinate (m2t) at (1.5,2.5);
			\coordinate (l1t) at (2.25,2.5);
			\coordinate (l2t) at (3,2.5);
			\coordinate (nu2) at ($(l3)+(0,.75)$);
			\draw 
			(mb1) node[below] {\scriptsize$m_{gh}^*\vphantom{\overline{gh}}$}
			(m1) node[below] {\scriptsize$m_{g}\vphantom{\overline{gh}}$}
			(m2) node[below] {\scriptsize$m_{h}\vphantom{\overline{gh}}$}
			(mb2) node[below] {\scriptsize$m_{h}^*\vphantom{\overline{gh}}$}
			(m3) node[below] {\scriptsize$m_{h}\vphantom{\overline{gh}}$}
			(mbt) node[above] {\scriptsize$m_{gh}^*\vphantom{\overline{gh}}$}
			(m1t) node[above] {\scriptsize$m_g\vphantom{\overline{gh}}$}
			(m2t) node[above] {\scriptsize$m_h\vphantom{\overline{gh}}$};
			\draw[black]
			(lb1) node[below] {\scriptsize$\lambda(gh,\overline{gh})^*$}
			(l1) node[below] {\scriptsize$\lambda(g,h)\vphantom{\overline{gh}}$}
			(l2) node[below] {\scriptsize$\lambda(\overline{gh},gh)$}
			(lb2) node[below] {\scriptsize$\lambda(h,\overline{h})^*$}
			(l3) node[below] {\scriptsize$\lambda(\overline{h},h)$}
			(lbt) node[above] {\scriptsize$\lambda(gh,\overline{gh})^*\vphantom{\overline{gh}}$}
			(l1t) node[above] {\scriptsize$\lambda(g,h)\vphantom{\overline{gh}}$}
			(l2t) node[above] {\scriptsize$\lambda(\overline{gh},gh)$}
			;
			\draw 
			(mb1) -- (mbt)
			(m1) -- (m1t)
			(m3) .. controls +(0,2) and +(0,-2) .. (m2t)
			;
			\draw[postaction={decorate}, decoration={markings,mark=at position .45 with {\arrow[color=blue!50!black]{>}}}] (m2) .. controls +(0,1.7) and +(0,1.7) .. (mb2);
			\draw[color=white, line width=4pt] 
			(lb2) .. controls +(0,2) and +(0,-.75) .. ($(nu2)+(0,2)$) 
			(l1) -- (l1t)
			(l2) -- (l2t)
			;
			\draw[Mybrown]
			(lb1) -- (lbt)
			(l1) -- (l1t)
			(l2) -- (l2t)
			(lb2)  .. controls +(0,2) and +(0,-.75) .. ($(nu2)+(0,2)$) 
			;
			\draw[color=white, line width=4pt] 
			($(nu2)+(0,1)$) .. controls +(0,.75) and +(0,-.75) .. ($(nu2)+(-.75,2)$)
			;
			\draw[Mybrown]
			(l3) -- (nu2)  
			($(nu2)+(0,1)$) .. controls +(0,.75) and +(0,-.75) .. ($(nu2)+(-.75,2)$)
			;
			\draw[Mybrown,string]
			($(nu2)+(-.75,2)$) .. controls +(0,.75) and +(0,.75) .. ($(nu2)+(0,2)$)
			;
			\fill[color=white, rounded corners=2pt]($(nu2)+(-.4,0)$)rectangle($(nu2)+(.4,1)$);
			\draw[draw=black, name=R, line width=0.5pt, rounded corners=2pt]($(nu2)+(-.4,0)$)rectangle($(nu2)+(.4,1)$);
			\draw[black]
			($(nu2)+(0,.5)$) node {$(\nu^h)^{-1}$};
\end{tikzpicture}
\,,\\
	\begin{tikzpicture}[very thick,scale=0.8,color=blue!50!black, baseline]
\draw (0,-2.5) node[below] (X) {{\scriptsize$\chi_{g,h}^{(\lambda,\nu)}$}};
\draw (0,2.5) node[above] (X) {{\scriptsize$\chi_{g,h}^{(\lambda,\nu)}$}};
\draw[color=green!50!black] (1,-2.5) node[below] (A1) {{\scriptsize$A_g^{(\lambda,\nu)}$}};
\draw (0,2.5) node[right] (Xu) {};
\draw[color=green!50!black] (A1) .. controls +(0,1.25) and +(0.5,-1.25) .. (0,0.75);
\draw (0,-2.5) -- (0,2.5); 
\fill[color=blue!50!black] (0,0.75) circle (2.9pt) node (meet2) {};
\fill[color=black] (0.2,1) circle (0pt) node (meet) {{\tiny$1$}};
\end{tikzpicture} 
&:=
	\begin{tikzpicture}[very thick,color=blue!50!black, baseline=0cm,xscale = 1.4,yscale=.8]
			\coordinate (mb1) at (-.75,-2.5);
			\coordinate (lb1) at (0,-2.5);
			\coordinate (m1) at (.75,-2.5);
			\coordinate (m2) at (1.5,-2.5);
			\coordinate (l1) at (2.25,-2.5);
			\coordinate (l2) at (3,-2.5);
			\coordinate (mb2) at (4.25,-2.5);
			\coordinate (lb2) at (5,-2.5);
			\coordinate (m3) at (5.75,-2.5);
			\coordinate (l3) at (6.5,-2.5);
			\coordinate (mbt) at (-.75,2.5);
			\coordinate (lbt) at (0,2.5);
			\coordinate (m1t) at (.75,2.5);
			\coordinate (m2t) at (1.5,2.5);
			\coordinate (l1t) at (2.25,2.5);
			\coordinate (l2t) at (3,2.5);
			\coordinate (nu2) at ($(l3)+(0,.75)$);
			\draw 
			(mb1) node[below] {\scriptsize$m_{gh}^*\vphantom{\overline{gh}}$}
			(m1) node[below] {\scriptsize$m_{g}\vphantom{\overline{gh}}$}
			(m2) node[below] {\scriptsize$m_{h}\vphantom{\overline{gh}}$}
			(mb2) node[below] {\scriptsize$m_{g}^*\vphantom{\overline{gh}}$}
			(m3) node[below] {\scriptsize$m_{g}\vphantom{\overline{gh}}$}
			(mbt) node[above] {\scriptsize$m_{gh}^*\vphantom{\overline{gh}}$}
			(m1t) node[above] {\scriptsize$m_g\vphantom{\overline{gh}}$}
			(m2t) node[above] {\scriptsize$m_h\vphantom{\overline{gh}}$};
			\draw[black]
			(lb1) node[below] {\scriptsize$\lambda(gh,\overline{gh})^*$}
			(l1) node[below] {\scriptsize$\lambda(g,h)\vphantom{\overline{gh}}$}
			(l2) node[below] {\scriptsize$\lambda(\overline{gh},gh)$}
			(lb2) node[below] {\scriptsize$\lambda(g,\overline{g})^*$}
			(l3) node[below] {\scriptsize$\lambda(\overline{g},g)$}
			(lbt) node[above] {\scriptsize$\lambda(gh,\overline{gh})^*\vphantom{\overline{gh}}$}
			(l1t) node[above] {\scriptsize$\lambda(g,h)\vphantom{\overline{gh}}$}
			(l2t) node[above] {\scriptsize$\lambda(\overline{gh},gh)$}
			;
			\draw 
			(mb1) -- (mbt)
			(m3) .. controls +(0,2) and +(0,-2) .. (m1t)
			;
			\draw[postaction={decorate}, decoration={markings,mark=at position .54 with {\arrow[color=blue!50!black]{>}}}] (m1) .. controls +(0,1.7) and +(0,1.7) .. (mb2);
			\draw[color=white, line width=4pt] 
			(lb2) .. controls +(0,2) and +(0,-.75) .. ($(nu2)+(0,2)$) 
			(l1) -- (l1t)
			(l2) -- (l2t)
			(m2) -- (m2t)
			;
			\draw (m2) -- (m2t);
			\draw[Mybrown]
			(lb1) -- (lbt)
			(l1) -- (l1t)
			(l2) -- (l2t)
			(lb2)  .. controls +(0,2) and +(0,-.75) .. ($(nu2)+(0,2)$) 
			;
			\draw[color=white, line width=4pt] 
			($(nu2)+(0,1)$) .. controls +(0,.75) and +(0,-.75) .. ($(nu2)+(-.75,2)$)
			;
			\draw[Mybrown]
			(l3) -- (nu2)  
			($(nu2)+(0,1)$) .. controls +(0,.75) and +(0,-.75) .. ($(nu2)+(-.75,2)$)
			;
			\draw[Mybrown,string]
			($(nu2)+(-.75,2)$) .. controls +(0,.75) and +(0,.75) .. ($(nu2)+(0,2)$)
			;
			\fill[color=white, rounded corners=2pt]($(nu2)+(-.4,0)$)rectangle($(nu2)+(.4,1)$);
			\draw[draw=black, name=R, line width=0.5pt, rounded corners=2pt]($(nu2)+(-.4,0)$)rectangle($(nu2)+(.4,1)$);
			\draw[black]
			($(nu2)+(0,.5)$) node {$(\nu^g)^{-1}$};
\end{tikzpicture}\,.
\end{align}
By using $\nu^{gh}$ and evaluation, $\chi^{(\lambda,\nu)}_{g,h}\cong m_{gh}\otimes (m_g\otimes m_h\otimes \lambda(g,h))\cong \chi_{g,h}\otimes\lambda(g,h)$. 
This means that, as we claimed before, we can directly detect the fractionalisation in the line defect $\chi$.

The components of the associator $\alpha^{(\lambda,\nu)}_G=\bigoplus_{g,h,k\in G} \alpha_{g,h,k}^{(\lambda,\nu)}$ read
\begin{align}
&\alpha_{g,h,k}^{(\lambda,\nu)}:=
\begin{tikzpicture}[very thick,yscale=0.9,color=red!50!black, baseline=-1.4cm, rotate=180,xscale=1.4]
					\draw[line width=0pt] 
					(5.5,3) node[line width=0pt] (ghko) {{\scriptsize$\overline{m_{ghk}}\vphantom{ghk}$}}
					(5,3) node[line width=0pt] (go) {{\scriptsize$m_g\vphantom{ghk}$}}
					(4.5,3) node[line width=0pt] (hkli) {{\scriptsize$m_{hk}\vphantom{ghk}$}}
					(3.5,3) node[line width=0pt] (hkre) {{\scriptsize$\overline{m_{hk}}\vphantom{ghk}$}}
					(3,3) node[line width=0pt] (ho) {{\scriptsize$m_h\vphantom{ghk}$}}
					(2.5,3) node[line width=0pt] (ko) {{\scriptsize$m_k\vphantom{ghk}$}}
					(5.5,0) node[line width=0pt] (ghku) {{\scriptsize$\overline{m_{ghk}}\vphantom{ghk}$}}
					(5,0) node[line width=0pt] (ghli) {{\scriptsize$m_{gh}\vphantom{ghk}$}}
					(4.5,0) node[line width=0pt] (ku) {{\scriptsize$m_k\vphantom{ghk}$}}
					(3.5,0) node[line width=0pt] (ghre) {{\scriptsize$\overline{m_{gh}}\vphantom{ghk}$}} 
					(3,0) node[line width=0pt] (gu) {{\scriptsize$m_g\vphantom{ghk}$}} 
					(2.5,0) node[line width=0pt] (hu) {{\scriptsize$m_h\vphantom{ghk}$}};
					\draw[redirectedred] (ghko) -- (ghku);
					\draw[postaction={decorate}, decoration={markings,mark=at position .65 with {\arrow[color=red!50!black]{<}}}] (ghli) .. controls +(0,1) and +(0,1) .. (ghre);
					\draw[postaction={decorate}, decoration={markings,mark=at position .55 with {\arrow[color=red!50!black]{<}}}] (hkre) .. controls +(0,-1) and +(0,-1) .. (hkli);
					\draw[postaction={decorate}, decoration={markings,mark=at position .65 with {\arrow[color=red!50!black]{<}}}] (gu) .. controls +(0,1.5) and +(0,-1.5) .. (go);
					\draw[redirectedred] (hu) .. controls +(0,1.5) and +(0,-1.5) ..  (ho);
					\draw[color=white, line width=4pt] (ku) .. controls +(0,1.5) and +(0,-1.5) .. (ko);
					\draw[postaction={decorate}, decoration={markings,mark=at position .65 with {\arrow[color=red!50!black]{<}}}] (ku) .. controls +(0,1.5) and +(0,-1.5) .. (ko);
					\end{tikzpicture}\\
&=\begin{tikzpicture}[very thick,color=blue!50!black, baseline=0cm,xscale = 1.5,yscale=.7]
			\pgfmathsetmacro{\hgt}{11.5}
			\coordinate (mb1) at (-.75,-\hgt);
			\coordinate (lb1) at ($(0,-\hgt)+(0,-.45)$);
			\coordinate (m11) at (.75,-\hgt);
			\coordinate (m12) at (1.5,-\hgt);
			\coordinate (l11) at (2.25,-\hgt);
			\coordinate (l12) at ($(3,-\hgt)+(0,-.45)$);
			\coordinate (mb2) at (4.25,-\hgt);
			\coordinate (lb2) at (5,-\hgt);
			\coordinate (m21) at (5.75,-\hgt);
			\coordinate (m22) at (6.5,-\hgt);
			\coordinate (l21) at (7.25,-\hgt);
			\coordinate (l22) at (8,-\hgt);
			\coordinate (mb1t) at (-.75,\hgt);
			\coordinate (lb1t) at ($(0,\hgt)+(0,.45)$);
			\coordinate (m11t) at (.75,\hgt);
			\coordinate (m12t) at (1.5,\hgt);
			\coordinate (l11t) at (2.25,\hgt);
			\coordinate (l12t) at ($(3,\hgt)+(0,.45)$);
			\coordinate (mb2t) at (4.25,\hgt);
			\coordinate (lb2t) at (5,\hgt);
			\coordinate (m21t) at (5.75,\hgt);
			\coordinate (m22t) at (6.5,\hgt);
			\coordinate (l21t) at (7.25,\hgt);
			\coordinate (l22t) at (8,\hgt);
			\coordinate (nu1) at ($(l11)+(0,.75)$);
			\coordinate (nu2) at ($(l12)+(0,4)$);
			\coordinate (nu3) at ($(l11)+(0,7)$);
			\coordinate (nu4) at ($(l11)+(.75,9)$);
			\coordinate (nu5) at ($(l21)+(0,11.5)$);
			\coordinate (nu6) at ($(m22)+(0,13.25)$);
			\coordinate (nu7) at ($(m21)+(0,17)$);
			\coordinate (nu8) at ($(l11t)+(0,-1.75)$);
			\foreach \coord/\placement/\lab in
			{mb1/below/{ghk}^*,m11/below/g,m12/below/{{hk}},mb2/below/{hk}^*,m21/below/h,m22/below/k,
			mb1t/above/{ghk}^*,m11t/above/{{gh}},m12t/above/k,mb2t/above/{gh}^*,m21t/above/g,m22t/above/h}
			\draw (\coord) node[\placement] {\scriptsize$\vphantom{\overline{h}}m_\lab$};
			\foreach \coord/\placement/\lab in
			{lb1/below/{(ghk,\overline{ghk})^*},l11/below/{(g,hk)},l12/below/{(\overline{ghk},ghk)},lb2/below/{(hk,\overline{hk})^*},l21/below/{(h,k)},l22/below/{(\overline{hk},hk)},
			lb1t/above/{(ghk,\overline{ghk})^*},l11t/above/{(gh,k)},l12t/above/{(\overline{ghk},ghk)},lb2t/above/{(gh,\overline{gh})^*},l21t/above/{(g,h)},l22t/above/{(\overline{gh},gh)}}
			\draw[black] (\coord) node[\placement] {\scriptsize$\vphantom{\overline{h}}\lambda\lab$};
			\draw 
			(mb1) -- (mb1t)
			(m11) -- ($(m11)+(0,2)$) .. controls +(0,3) and +(0,-3) .. ($(nu3)+(-.75,1)$) .. controls +(0,5) and +(0,-5).. (m21t)
			(m21) .. controls +(0,7) and +(0,-7) .. ($(nu7)+(-.75,-.5)$) .. controls +(0,2) and +(0,-2) .. (m22t)
			(m22) .. controls +(0,7) and +(0,-7) .. ($(nu6)+(-.75,.5)$) .. controls +(0,1) and +(0,-1) .. ($(nu6)+(.25,2.25)$)
			;
			\draw[string] (m12) -- ($(m12)+(0,1.25)$) .. controls +(0,2) and +(0,2) .. ($(mb2)+(0,1.25)$) -- (mb2);
			\draw[costring] (m11t) -- ($(m11t)+(0,-8.25)$) .. controls +(0,-1.5) and +(-.5,-2) .. ($(m12t)+(.3,-8.25)$) .. controls +(.5,2) and +(0,-4) .. (mb2t);
			\draw[color=white, line width=4pt] 
			($(nu1)+(.75,1)$) -- (nu2) 
			($(nu1)+(0,1)$) .. controls +(0,4.5) and +(0,-4.5) .. ($(nu4)+(.75,0)$)
			($(nu3)+(0,1)$) .. controls +(0,5) and +(0,-3) .. (nu7)
			($(nu6)+(0,1)$) -- ($(nu7)+(.75,0)$)
			;
			\draw[Mybrown]
			(lb1) -- (lb1t)
			(l12) -- (nu2)
			($(nu2)+(1.25,1)$) .. controls +(0,-2) and +(0,2) .. (lb2)
			(l21)  .. controls +(0,5) and +(0,-5) .. ($(nu6)+(0,1)$) -- ($(nu7)+(.75,0)$) -- ($(nu7)+(.75,1)$) .. controls +(0,2) and +(0,-2).. (l22t)
			($(nu3)+(0,1)$) .. controls +(0,5) and +(0,-3) .. (nu7) -- ($(nu7)+(0,1)$) .. controls +(0,2) and +(0,-2).. (l21t)
			(l22) .. controls +(0,5) and +(0,-5) .. ($(nu5)+(.75,0)$) -- ($(nu5)+(.75,6.5)$) 				($(nu8)+(.75,0)$) -- (l12t)
			($(nu3)+(-1.5,0)$)  .. controls +(0,7) and +(0,-5) .. (lb2t)
			;
			\draw[Mybrown,string] ($(nu2)+(0,1)$) .. controls +(0,1) and +(0,1) .. ($(nu2)+(1.25,1)$);
			\draw[color=white, line width=4pt]
			($(nu4)+(.75,1)$) .. controls +(0,1.5) and +(0,-1.5) .. (nu5)
			($(nu5)+(.75,6.5)$) .. controls +(0,2) and +(0,-2).. ($(nu8)+(.75,0)$)
			(nu3) .. controls +(0,-1.5) and +(0,-1.5) .. ($(nu3)+(-1.5,0)$)
			; 
			\draw[Mybrown]
			(l11) -- ($(nu1)+(0,1)$) .. controls +(0,4.5) and +(0,-4.5) .. ($(nu4)+(.75,0)$) -- ($(nu4)+(.75,1)$) .. controls +(0,1.5) and +(0,-1.5) .. (nu5) -- ($(nu6)+(.75,1)$)
			($(nu5)+(.75,6.5)$) .. controls +(0,2) and +(0,-2).. ($(nu8)+(.75,0)$)
			($(nu3)+(.75,1)$) -- ($(nu4)+(0,1)$) .. controls +(0,4) and +(0,-4) .. (nu8) -- (l11t)
			;
			\draw[costring,Mybrown] (nu3) .. controls +(0,-1.5) and +(0,-1.5) .. ($(nu3)+(-1.5,0)$);
			\draw[color=white, line width=4pt] ($(nu6)+(.25,2.25)$) .. controls +(0,1.5) and +(0,-7) .. (m12t);
			\draw ($(nu6)+(.25,2.25)$) .. controls +(0,1.5) and +(0,-7) .. (m12t);
			\draw[color=white, line width=4pt] ($(nu4)+(0,1)$) .. controls +(0,4) and +(0,-4) .. (nu8);
			\draw[Mybrown] ($(nu4)+(0,1)$) .. controls +(0,4) and +(0,-4) .. (nu8);
			\foreach \coord/\lab in
			{1/{\nu_{\overline{ghk},g,hk}^{-1}},3/{\nu_{gh,\overline{gh},g}^{\vphantom{-1}}},4/{\nu_{\overline{ghk},gh,\overline{h}}^{-1}},5/{\nu_{\overline{k},\overline{h},hk}^{\vphantom{-1}}},6/{\nu_{\overline{h},h,k}^{-1}},7/{\nu_{\overline{gh},g,h}^{\vphantom{-1}}},8/{\nu_{\overline{ghk},gh,k}^{\vphantom{-1}}}}
			{\fill[color=white, rounded corners=2pt]($(nu\coord)+(-.3,0)$)rectangle($(nu\coord)+(1.05,1)$);
			\draw[draw=black, name=R, line width=0.5pt, rounded corners=2pt]($(nu\coord)+(-.3,0)$)rectangle($(nu\coord)+(1.05,1)$);
			\draw[black] ($(nu\coord)+(.375,.5)$) node {$\lab$};}
			\fill[color=white, rounded corners=2pt]($(nu2)+(-.3,0)$)rectangle($(nu2)+(.3,1)$);
			\draw[draw=black, name=R, line width=0.5pt, rounded corners=2pt]($(nu2)+(-.3,0)$)rectangle($(nu2)+(.3,1)$);
			\draw[black] ($(nu2)+(0,.5)$) node {$\nu^{\overline{hk}}$};
			\draw[Mybrown,densely dotted] 
			($(nu3)+(.75,-.4)$) -- ($(nu3)+(.75,0)$)
			($(nu6)+(.75,1)$) -- ($(nu6)+(.75,1.4)$)
			;
			\draw[black] ($(nu3)+(.75,-.7)$) node {\scriptsize$\lambda(e,g)$};
			\draw[black] ($(nu3)+(1.12,1.47)$) node {\scriptsize$\lambda(gh,\overline{h})$};
			\draw[black] ($(nu3)+(1.92,1.47)$) node {\scriptsize$\lambda(\overline{ghk},g)$};
			\draw[black] ($(nu3)+(1.5,3.8)$) node {\scriptsize$\lambda(\overline{k},\overline{h})$};
			\draw[black] ($(nu6)+(1.12,-.37)$) node {\scriptsize$\lambda(\overline{h},hk)$};
			\draw[black] ($(nu6)+(.75,1.57)$) node {\scriptsize$\lambda(e,k)$};
			\draw[black] ($(nu6)+(1.2,3.37)$) node {\scriptsize$\lambda(\overline{k},k)$};
			\draw[black] ($(nu7)+(1.12,-.37)$) node {\scriptsize$\lambda(\overline{h},h)$};
			\draw[black] (4,2.4) node {\scriptsize$\lambda(\overline{gh},g)$};
			\draw[black] ($(nu8)+(-.35,-3.5)$) node {\scriptsize$\lambda(\overline{ghk},gh)$};
\end{tikzpicture}\,,
\end{align}
where $\lambda(e,g)=\mathds{1}=\lambda(k,e)$ were denoted by dotted lines.
In the special case of trivial fractionalisation $\lambda=\mathds{1}$, the cocycle $\nu$ is a scalar and coincides with the defectification~$p\in\mathrm{H}^3(G,\U)$.
Repeated application of the cocycle condition leads to the following identity:
\begin{equation}
	\alpha^{(\mathds{1},p)}_{g,h,k}=p(g,h,k)^{-1}p^{gh}p^{hk}\cdot \alpha_{g,h,k}
	\label{eq:simplifiedalpha}
\end{equation}
Lastly, the point defects $\psi^{(\lambda,\nu)}_G$ and $\phi_G^{(\lambda,\nu)}$ are given by
\begin{align}
	\psi^{(\lambda,\nu)}_G&:=\bigoplus_{g\in G} (\mathrm{dim}_{j^{(\lambda,\nu)}}(m_g))^{-\frac{1}{2}}\cdot \id_{A_g}=\bigoplus_{g\in G} (\mathrm{dim}_{j}(m_g)\theta_{\lambda(g,\overline{g})})^{-\frac{1}{2}}\cdot \id_{A_g}\,,\\
	\phi^{(\lambda,\nu)}_G&:=\phi_G=\frac{1}{\sqrt{G}}\,.
\end{align}

The theorem \cite[Thm.\,3.12]{DGPRZ} states in particular $ \mathcal{C}_g\cong \mathcal{C}_g^{(\lambda,\nu)}$, where $\mathcal{C}_g$ and $\mathcal{C}_g^{(\lambda,\nu)}$ are the line operators in the twisted sector~$g$ for the original and the zested extensions, respectively.
Thus it implies that the algebras $A_g$ and $A_g^{(\lambda,\nu)}$ are Morita-equivalent:
\begin{equation}
	A_g\text{-}\Mod(\mathcal{C})\cong \mathcal{C}_g\cong \mathcal{C}_g^{(\lambda,\nu)}\cong A_g^{(\lambda,\nu)}\text{-}\Mod(\mathcal{C})\,,
	\label{eq:zestedMoritaEquivalence}
\end{equation}
where $\mathcal{C}\equiv\mathcal{C}_e$ is the category of bulk line operators.
This Morita-equivalence can be realised by the $(A_g^{(\lambda,\nu)},A_g)$-bimodule $M:=\overline{m_g}\otimes m_g$ and its inverse $(A_g,A_g^{(\lambda,\nu)})$-bimodule $N:=m_g^*\otimes m_g\otimes \lambda(\overline{g},g)$.
The $A_g$-module structures~$\sigma_N^0$ and~$\sigma_M^1$ are given by \eqref{eq:mmodulestructures} and the $A_g^{(\lambda,\nu)}$-module structures by 
\begin{align}
\label{eq:sigM0}
\sigma_M^0\equiv
\begin{tikzpicture}[very thick,scale=.6,color=blue!50!black, baseline]
\draw (0,-2) node[below] (X) {{\scriptsize$M\vphantom{A_g^{(\lambda,\nu)}}$}};
\draw (0,2) node[above] (X) {{\scriptsize$M$}};
\draw[color=green!50!black] (-1.8,-2) node[below] (A1) {{\scriptsize$A_{g}^{(\lambda,\nu)}$}};
\draw (0,2) node[right] (Xu) {};
\draw[color=green!50!black] (A1) .. controls +(0,1.25) and +(-.5,-1.25) .. (0,0.75);
\draw (0,-2) -- (0,2); 
\fill[color=blue!50!black] (0,0.75) circle (2.9pt) node (meet2) {};
\end{tikzpicture} 
&:=
	\begin{tikzpicture}[very thick,color=blue!50!black, baseline=0cm,xscale = 1.2,yscale=.48]
			\coordinate (mb1) at (-.75,-2.5);
			\coordinate (lb1) at (0,-2.5);
			\coordinate (m1) at (.75,-2.5);
			\coordinate (l1) at (1.5,-2.5);
			\coordinate (mb2) at (2.75,-2.5);
			\coordinate (lb2) at (3.5,-2.5);
			\coordinate (m2) at (4.25,-2.5);
			\coordinate (mbt) at (2.75,2.5);
			\coordinate (lbt) at (3.5,2.5);
			\coordinate (m1t) at (4.25,2.5);
			\coordinate (nu2) at ($(l1)+(1.25,1.5)$);
			\draw 
			(mb1) node[below] {\scriptsize$m_{g}^*\vphantom{A_g^{(\lambda,\nu)}}$}
			(m1) node[below] {\scriptsize$m_{g}\vphantom{A_g^{(\lambda,\nu)}}$}
			(mb2) node[below] {\scriptsize$m_{g}^*\vphantom{A_g^{(\lambda,\nu)}}$}
			(m2) node[below] {\scriptsize$m_{g}\vphantom{A_g^{(\lambda,\nu)}}$}
			(mbt) node[above] {\scriptsize$m_{g}^*\vphantom{\overline{g}}$}
			(m1t) node[above] {\scriptsize$m_g\vphantom{\overline{g}}$};
			\draw[black]
			(lb1) node[below] {\scriptsize$\lambda(g,\overline{g})^*\vphantom{A_g^{(\lambda,\nu)}}$}
			(l1) node[below] {\scriptsize$\lambda(\overline{g},g)\vphantom{A_g^{(\lambda,\nu)}}$}
			(lb2) node[below] {\scriptsize$\lambda(g,\overline{g})^*\vphantom{A_g^{(\lambda,\nu)}}$}
			(lbt) node[above] {\scriptsize$\lambda(g,\overline{g})^*\vphantom{A_g^{(\lambda,\nu)}}$}
			;
			\draw 
			(mb1) .. controls +(0,2) and +(0,-2) .. (mbt)
			(m2) -- (m1t)
			;
			\draw[string] (m1) .. controls +(0,1.7) and +(0,1.7) .. (mb2);
			\draw[Mybrown]
			(lb1) .. controls +(0,2) and +(0,-2) .. (lbt)
			;
			\draw[color=white, line width=4pt] 
			(l1) .. controls +(0,.5) and +(0,-.5) .. (nu2) 
			;
			\draw[Mybrown]
			(l1) .. controls +(0,.5) and +(0,-.5) .. (nu2) 
			(lb2) .. controls +(0,1) and +(0,-1) .. ($(nu2)+(.75,1)$)
			;
			\draw[Mybrown,string]
			($(nu2)+(0,1)$) .. controls +(0,.75) and +(0,.75) .. ($(nu2)+(.75,1)$)
			;
			\fill[color=white, rounded corners=2pt]($(nu2)+(-.3,0)$)rectangle($(nu2)+(.3,1)$);
			\draw[draw=black, name=R, line width=0.5pt, rounded corners=2pt]($(nu2)+(-.3,0)$)rectangle($(nu2)+(.3,1)$);
			\draw[black]
			($(nu2)+(0,.5)$) node {$\nu^{\overline{g}}$};
\end{tikzpicture}\,,\\
\label{eq:sigN1}
\sigma_N^1\equiv
\begin{tikzpicture}[very thick,scale=.6,color=blue!50!black, baseline]
\draw (0,2) node[above] (X) {{\scriptsize$N$}};
\draw (0,-2) node[below] (Xb) {{\scriptsize$N\vphantom{A_g^{(\lambda,\nu)}}$}};
\draw[color=green!50!black] (1.8,-2) node[below] (A1) {{\scriptsize$\;A_g^{(\lambda,\nu)}$}};
\draw[color=green!50!black] (A1) .. controls +(0,1.25) and +(0.5,-1.25) .. (0,0.75);
\draw (0,-2) -- (0,2); 
\fill[color=blue!50!black] (0,0.75) circle (2.9pt) node (meet2) {};
\end{tikzpicture} 
&:=
	\begin{tikzpicture}[very thick,color=blue!50!black, baseline=0cm,xscale = 1.2,yscale=.48]
			\coordinate (mb1) at (0,-2.5);
			\coordinate (m1) at (.75,-2.5);
			\coordinate (l1) at (1.5,-2.5);
			\coordinate (mb2) at (2.75,-2.5);
			\coordinate (lb2) at (3.5,-2.5);
			\coordinate (m2) at (4.25,-2.5);
			\coordinate (l2) at (5,-2.5);
			\coordinate (mbt) at (0,2.5);
			\coordinate (lt) at (1.5,2.5);
			\coordinate (m1t) at (.75,2.5);
			\coordinate (nu2) at ($(m1)+(0.25,1.5)$);
			\draw 
			(mb1) node[below] {\scriptsize$m_{g}^*\vphantom{A_g^{(\lambda,\nu)}}$}
			(m1) node[below] {\scriptsize$m_{g}\vphantom{A_g^{(\lambda,\nu)}}$}
			(mb2) node[below] {\scriptsize$m_{g}^*\vphantom{A_g^{(\lambda,\nu)}}$}
			(m2) node[below] {\scriptsize$m_{g}\vphantom{A_g^{(\lambda,\nu)}}$}
			(mbt) node[above] {\scriptsize$m_{g}^*\vphantom{\overline{g}}$}
			(m1t) node[above] {\scriptsize$m_g\vphantom{\overline{g}}$};
			\draw[black]
			(l1) node[below] {\scriptsize$\lambda(\overline{g},g)\vphantom{A_g^{(\lambda,\nu)}}$}
			(lb2) node[below] {\scriptsize$\lambda(g,\overline{g})^*\vphantom{A_g^{(\lambda,\nu)}}$}
			(l2) node[below] {\scriptsize$\lambda(\overline{g},g)\vphantom{A_g^{(\lambda,\nu)}}$}
			(lt) node[above] {\scriptsize$\lambda(\overline{g},g)\vphantom{A_g^{(\lambda,\nu)}}$}
			;
			\draw 
			(mb1) -- (mbt)
			(m2)  .. controls +(0,2) and +(0,-2) .. (m1t)
			;
			\draw[string] (m1) .. controls +(0,1.7) and +(0,1.7) .. (mb2);
			\draw[Mybrown]
			(l2) .. controls +(0,2) and +(0,-2) .. (lt)
			;
			\draw[color=white, line width=4pt] 
			(l1) .. controls +(0,.5) and +(0,-.5) .. (nu2) 
			;
			\draw[Mybrown]
			(l1) .. controls +(0,.5) and +(0,-.5) .. (nu2) 
			(lb2) .. controls +(0,1) and +(0,-1) .. ($(nu2)+(.75,1)$)
			;
			\draw[Mybrown,string]
			($(nu2)+(0,1)$) .. controls +(0,.75) and +(0,.75) .. ($(nu2)+(.75,1)$)
			;
			\fill[color=white, rounded corners=2pt]($(nu2)+(-.3,0)$)rectangle($(nu2)+(.3,1)$);
			\draw[draw=black, name=R, line width=0.5pt, rounded corners=2pt]($(nu2)+(-.3,0)$)rectangle($(nu2)+(.3,1)$);
			\draw[black]
			($(nu2)+(0,.5)$) node {$\nu^{\overline{g}}$};
\end{tikzpicture}
\,.
\end{align}
	The modules~$M$ and~$N$ defined above witness a Morita equivalence between~$A_g$ and $A_g^{(\lambda,\nu)}$, i.e.\ 
	\begin{align}
		M\otimes_{A_g}N&\cong A_g^{(\lambda,\nu)}\,,\\
		N\otimes_{A_g^{(\lambda,\nu)}}M&\cong A_g\,,
	\end{align}
	as $(A_g^{(\lambda,\nu)},A_g^{(\lambda,\nu)})$- and $(A_g,A_g)$-bimodules, respectively.
	
		For~$A_g^{(\lambda,\nu)}$, the isomorphism $m_g\otimes_{A_g}m_g^*\cong \mathds{1}$ implies $M\otimes_{A_g}N\cong A_g^{(\lambda,\nu)}$ as objects, and the module structures coincide as well, as seen by comparing \eqref{eq:sigM0} and \eqref{eq:sigN1} to \eqref{eq:zestedAgmult}.
		
		The object~$A_g$ has the following retract and section from $N\otimes M$: 
		\begin{equation}
		r:=
		\begin{tikzpicture}[very thick,color=blue!50!black, baseline=0cm,xscale = .8,yscale=.48]
			\coordinate (mb1) at (0,-2.5);
			\coordinate (m1) at (.75,-2.5);
			\coordinate (l1) at (1.5,-2.5);
			\coordinate (mb2) at (2.75,-2.5);
			\coordinate (lb2) at (3.5,-2.5);
			\coordinate (m2) at (4.25,-2.5);
			%
			\coordinate (mbt) at (1.625,2.5);
			\coordinate (mt) at (2.375,2.5);
			\coordinate (nu2) at ($(l1)+(.3,1.75)$);
			\draw 
			(mb1) node[below] {\scriptsize$\mg^*$}
			(m1) node[below] {\scriptsize$\mg\vphantom{\mg^*}$}
			(mb2) node[below] {\scriptsize$\mg^*$}
			(m2) node[below] {\scriptsize$\mg\vphantom{\mg^*}$}
			(mbt) node[above] {\scriptsize$\mg^*$}
			(mt) node[above] {\scriptsize$\mg\vphantom{\mg^*}$};
			\draw[black]
			(l1) node[below] {\scriptsize$\lambda_{\overline{g}}$}
			(lb2) node[below] {\scriptsize$\lambda_g^*$}
			;
			\draw 
			(mb1) .. controls +(0,2) and +(0,-2) .. (mbt)
			(m2) .. controls +(0,2) and +(0,-2) .. (mt)
			;
			\draw[postaction={decorate}, decoration={markings,mark=at position .7 with {\arrow[color=blue!50!black]{>}}}] (m1) .. controls +(0,1.7) and +(0,1.7) .. (mb2);
			\draw[color=white, line width=4pt] 
			(l1) .. controls +(0,.5) and +(0,-.5) .. (nu2) 
			;
			\draw[Mybrown]
			(l1) .. controls +(0,.5) and +(0,-.5) .. (nu2) 
			(lb2) .. controls +(0,1) and +(0,-1) .. ($(nu2)+(.75,1)$)
			;
			\draw[Mybrown,string]
			($(nu2)+(0,1)$) .. controls +(0,.75) and +(0,.75) .. ($(nu2)+(.75,1)$)
			;
			\fill[color=white, rounded corners=2pt]($(nu2)+(-.3,0)$)rectangle($(nu2)+(.3,1)$);
			\draw[draw=black, name=R, line width=0.5pt, rounded corners=2pt]($(nu2)+(-.3,0)$)rectangle($(nu2)+(.3,1)$);
			\draw[black]
			($(nu2)+(0,.5)$) node {$\nu^{\overline{g}}$};
		\end{tikzpicture}\,,\qquad 
		s:=\frac{1}{d_g}\cdot
		\begin{tikzpicture}[very thick,color=blue!50!black, baseline=0cm,xscale = .8,yscale=-.48]
			\coordinate (mb1) at (0,-2.5);
			\coordinate (m1) at (.75,-2.5);
			\coordinate (l1) at (1.5,-2.5);
			\coordinate (mb2) at (2.75,-2.5);
			\coordinate (lb2) at (3.5,-2.5);
			\coordinate (m2) at (4.25,-2.5);
			%
			\coordinate (mbt) at (1.625,2.5);
			\coordinate (mt) at (2.375,2.5);
			\coordinate (nu2) at ($(l1)+(.3,1.75)$);
			\draw 
			(mb1) node[above] {\scriptsize$\mg^*$}
			(m1) node[above] {\scriptsize$\mg\vphantom{\mg^*}$}
			(mb2) node[above] {\scriptsize$\mg^*$}
			(m2) node[above] {\scriptsize$\mg\vphantom{\mg^*}$}
			(mbt) node[below] {\scriptsize$\mg^*$}
			(mt) node[below] {\scriptsize$\mg\vphantom{\mg^*}$};
			\draw[black]
			(l1) node[above]  {\scriptsize$\lambda_{\overline{g}}$}
			(lb2) node[above] {\scriptsize$\lambda_g^*$}
			;
			\draw 
			(mb1) .. controls +(0,2) and +(0,-2) .. (mbt)
			(m2) .. controls +(0,2) and +(0,-2) .. (mt)
			;
			\draw[postaction={decorate}, decoration={markings,mark=at position .7 with {\arrow[color=blue!50!black]{<}}}] (m1) .. controls +(0,1.7) and +(0,1.7) .. (mb2);
			\draw[color=white, line width=4pt] 
			(l1) .. controls +(0,.5) and +(0,-.5) .. (nu2) 
			;
			\draw[Mybrown]
			(l1) .. controls +(0,.5) and +(0,-.5) .. (nu2) 
			(lb2) .. controls +(0,1) and +(0,-1) .. ($(nu2)+(.75,1)$)
			;
			\draw[Mybrown,costring]
			($(nu2)+(0,1)$) .. controls +(0,.75) and +(0,.75) .. ($(nu2)+(.75,1)$)
			;
			\fill[color=white, rounded corners=2pt]($(nu2)+(-.65,0)$)rectangle($(nu2)+(.65,1)$);
			\draw[draw=black, name=R, line width=0.5pt, rounded corners=2pt]($(nu2)+(-.65,0)$)rectangle($(nu2)+(.65,1)$);
			\draw[black]
			($(nu2)+(0,.5)$) node {$(\nu^{\overline{g}})^{-1}$};
		\end{tikzpicture}\,,
		\end{equation}
		 where we abbreviated $\lambda_g:=\lambda(g,\overline{g})$, $\lambda_{\overline{g}}:= \lambda(\overline{g},g)$, and $d_g:=\dim(m_g)$.
		 These are $(A_g,A_g)$-bimodule morphisms satisfying $r\circ s=\id_{A_g}$, and they witness~$A_g$ as the relative tensor product $N\otimes_{A_g^{(\lambda,\nu)}}M$.

\subsection{De-equivariantisation: Undoing the 0-form Gauging}
\label{sec:deeq}
In this section we introduce de-equivariantisation which corresponds to gauging a 1-form symmetry.
This undoes the gauging of the 0-form symmetry which we calculated in \Cref{sec:equivariantisation} as an equivariantisation. 

In any gauge theory, there are line operators labelled by representations~$(V,\pi)$ of the gauge group, namely Wilson lines $W_{(V,\pi)}$.
In the equivariantisation~$(\Gcbc)^G$, these are given by $\mathrm{dim}(V)$ copies of the trivial line $\mathds{1}$ equipped with an equivariant structure where $u_g=\pi_g$.
In the notation of \Cref{sec:CalculatingEq}, these are the objects $S_{\mathds{1},\pi}$ for an irreducible representation~$(V,\pi)$.
In this way, the category of $G$-representations is always a full subcategory of any equivariantisation, $\Rep(G)\subset (\Gcbc)^G$ (cf. \cite[Sect.\,4.2.2,~Sect.\,4.4.4]{DGNO}).
Gauging this (potentially non-invertible) $\Rep(G)$-symmetry reverses the gauging of the 0-form $G$-symmetry, which we now describe.

The algebra underlying the orbifold datum is the regular representation~$B=\C(G)$ (the algebra of functions on~$G$) which is a commutative $\Delta$-separable symmetric Frobenius algebra in $\Rep(G)$ and hence in $(\Gcbc)^G$.
Any algebra~$A$ with these properties produces an orbifold datum in $\mathcal{D}_{(\Gcbc)^G}$ \cite[Prop.\,3.15]{CRS3}, given by 
\begin{equation}
	\label{eq:orbdat from cssFrob}
		\mathbb{A}_A 
			:=
			\big(\mathcal{C}, \, A, \, T=A, \, \alpha=\overline{\alpha}=\Delta\circ\mu, \,  \psi=\id, \, \phi=1\big)\,,
	\end{equation}
	where the $(A,A\otimes A)$-bimodule structure of $T$ is given by the multiplication~$\mu$.
	For~$B$, this orbifold datum recovers the original category of line operators \cite[Prop.\,4.8]{CH}:
		\begin{equation}
	\label{eq:non-commutative deequivariantisation via orbifold}
		\left((\mathcal{C}^\times_G)^G\right)_{\mathbb{A}_B}
		\; \cong \; 
		\mathcal{C} \, ,
	\end{equation}
	thus ``undoing the gauging''.
	Note that for this particular type of orbifold datum, the orbifold is described by so-called \tsl{local} modules of the algebra~$B$ \cite[Thm.\,4.1]{Mulevicius2022}, in this case
	\begin{equation}
		\label{eq:orbifold category of a condensable algebra}
		\big((\mathcal{C}^\times_G)^G\big)_{\mathbb{A}_B}
		\;\cong\; 
		B\text{-}\Mod^\textrm{loc}\big((\mathcal{C}^\times_G)^G\big)\,.
	\end{equation}
	If we also want to recover the twisted sectors, we need to take all $B$-modules \cite[Prop.\,4.19]{DGNO} which is the definition of \textsl{de-equivariantisation}:
\begin{align}
	((\Gcbc)^G)_G\; :=\; B\text{-}\Mod\big((\Gcbc)^G\big)\;\cong\;\Gcbc \label{eq:eq-deeq equivalence}\, .
\end{align}
This system of gauging and ungauging is summarised in the following diagram:
\begin{equation}
\begin{tikzcd}[column sep= 20, ampersand replacement=\&]
	(\Gcbc)^{G}
    \arrow[rrrrr, "{\substack{\text{1-form gauging }\widehat{=}\text{ de-eq.}\\ B=\C(G)}}"{above}]
    \arrow[d,phantom,"\cong"{sloped},"{\scriptsize\eqref{eq:orbifoldequivarequivalence}\;}"{left} ]
    \&\&\&\&\&
    ((\Gcbc)^{G})_{\mathbb{A}_B}
    \arrow[d,phantom,"\cong"{sloped},shift right=3,"{\;\scriptsize\eqref{eq:non-commutative deequivariantisation via orbifold}}"{right}]
    \arrow[rr,phantom, "\subset"]
    \&\&
     ((\Gcbc)^{G})_{G}
    \arrow[d,phantom,"\cong"{sloped},shift right=3,"{\;\scriptsize\eqref{eq:eq-deeq equivalence}}"{right}]
    \\
    	\mathcal{C}_{\mathbb{A}_G}
     \&\&\&\&\&
     \hphantom{(}\mathcal{C}\hphantom{)_{\mathbb{A}_B}}
    \arrow[lllll, "{\substack{\mathbb{A}_G\\\text{0-form gauging }\widehat{=}\text{ eq.}}}"{below}]
    \arrow[rr,phantom, "\subset"]
    \&\&
     \Gcbc\hphantom{)^{G})_{G}}
\end{tikzcd}\, .
\label{eq:gaugingUngauging}
\end{equation}

\medskip

In the case where~$G$ is commutative, the orbifold datum $\mathbb{A}_B$ arises from a 1-form $\widehat{G}$-symmetry, where $\widehat{G}:=\Hom_\Grp(G,\U)$ is the Pontryagin-dual group.
The 1-form symmetry is then given by the composition 
\begin{equation}
	\widehat{R}\colon \underline{\widehat{G}}\longhookrightarrow \widehat{G}\text{-}\Vect\cong\Rep(G)\subset (\Gcbc)^G\,,
	\label{eq:1formFunctor}
\end{equation}
which we can deloop twice to get a 3-functor, and its associated orbifold datum is $\mathbb{A}_{\widehat{R}}=\mathbb{A}_B$ by \cite[Thm.\,4.12]{CH}.
In particular, the group algebra underlying the orbifold datum is equivalent to~$B$, $\C[\widehat{G}]\cong\C(G)=B$, as algebras in $\Rep(G)$.

\section{Tambara--Yamagami Categories}\label{sec:TY}
	In this section we introduce Tambara--Yamagami categories as our first example.
	We discuss their $\Z_2$-crossed braided structures and for $\A$ of odd order and $\A=\Z_2$, we spell out their orbifold data and equivariantisations.
	
	Throughout this section, $\A$ is a finite commutative group. 
	We denote $\Z_2=\{1,-1\}$ multiplicatively and write~$e$ for the unit of~$\A$.
	
\subsection{Definitions}
\label{sec:TYdef}
In this section, we review Tambara--Yamagami categories with $\Z_2$-crossed braided structures.

Tambara--Yamagami categories are minimal $\Z_2$-extensions of $\Vect_{\A}$ (where $\A$ is a finite commutative group), adding only a single, non-invertible simple object in degree $-1$, called~$\sigma$.
The fusion rules of simple objects are
	\begin{align}
		a\otimes b&:=ab\,,& a\otimes \sigma&:=\sigma=:\sigma\otimes b\,,& \sigma\otimes \sigma&:=\bigoplus_{x\in \A}x\,.
	\end{align}
Given these, the monoidal unit is $\mathds{1}=e$, and unitors are trivial, but the associators are not unique. 
To specify them, we have to choose a (non-degenerate, symmetric) \textsl{bicharacter on~$\A$} which is a map $\chi\colon \A\times \A\longrightarrow \C^\times$ that is multiplicative in each argument, i.e.\ for all $a,b,c\in \A$
	\begin{align}
		\chi(ab,c)&=\chi(a,c)\chi(b,c)\,,\\
		\chi(a,bc)&=\chi(a,b)\chi(a,c)\,.
	\end{align}
	It is \textsl{symmetric} if $\chi(a,b)=\chi(b,a)$ for all $a,b\in \A$, and \textsl{non-degenerate} if its \textsl{radical}
	\begin{equation}
		\mathrm{Rad}(\chi):=\lbrace b\in \A\mid \chi(a,b)=1\;\; \forall a\in \A\rbrace 
	\end{equation}
	is trivial: $\mathrm{Rad}(\chi)=\{e\}$.
	For such a non-degenerate symmetric bicharacter~$\chi$ on~$\A$ and~$\kappa$ a square root of $1/|\A|$, the \textsl{Tambara--Yamagami category} $\TYAXK$ is the pivotal $\Z_2$-graded fusion category with simple objects given by elements of~$\A$ in degree~$1$ and a single simple object in degree $-1$, denoted~$\sigma$.
	In addition to the fusion rules above, the non-trivial associators are given by
	\begin{align}
		&&a_{a,\sigma,b}&:=\hphantom{\bigoplus_{x,y\in \A}}\,\chi(a,b)\id_\sigma\colon&\hphantom{\bigoplus_{x\in \A}} \sigma&\longrightarrow \sigma,&&\\
		&&a_{\sigma,a,\sigma}&:=\bigoplus_{x,y\in \A}\chi(a,x)\delta_{x,y}\id_x\colon& \bigoplus_{x\in \A}x\;&\longrightarrow\bigoplus_{y\in \A}y,&&\\
		&&a_{\sigma,\sigma,\sigma}&:=\bigoplus_{x,y\in \A}\kappa\chi(x,y)^{-1}\id_\sigma \colon &\bigoplus_{x\in \A}\sigma&\longrightarrow\bigoplus_{y\in \A}\sigma\,.&&
	\end{align}
	The (left) duals are
	\begin{align}
		a^*&:= a^{-1}\,,&\ev_a&:=\id_e\colon a^*\otimes a\longrightarrow e\,,& \coev_a&:=\id_e\colon e\longrightarrow a\otimes a^*\,,\\
		\sigma^*&:= \sigma\,,&\ev_\sigma&:=\kappa^{-1}\pr\colon \sigma\otimes \sigma\longrightarrow e\,,& \coev_\sigma&:=\incl\colon e\longrightarrow \sigma\otimes \sigma\,,
	\end{align}
	where $\pr\colon \bigoplus_{x\in \A}x\longrightarrow e$ and $\incl\colon e\longrightarrow\bigoplus_{x\in \A}x$ are the projection and inclusion, respectively. 
The pivotal structure $j\colon  \id_{\TYAXK}\longrightarrow(-)^{**}$ has components
	\begin{align}
		j_a&:= \id_a\,,\\
		j_\sigma&:=\mathrm{sgn}(\kappa)\id_\sigma\,.
	\end{align}
We note here that $\sgn(\kappa)$ corresponds to the defectification class~$p\in\mathrm{H}^3(\Z_2,\U)=\{\pm 1\}$ of the extension, and the adjustments to the associator, evaluation, coevaluation, and pivotal structure match the formulas given by zesting (see \cite[Eqs.\,(3.4), (3.30) - (3.32)]{DGPRZ}, or \eqref{eq:zestedassociator}, \eqref{eq:zestedev}, \eqref{eq:zestedcoev}, and \eqref{eq:zestedpivotal}).

The dimensions of objects in $\TYAXK$ are as follows:
\begin{equation}
	\label{eq:dimension in TY}
	\dim_{\TY}(a)
	=1\,,
	\qquad 
	\dim_{\TY}(\sigma)
	=\sqrt{|\A|}\,,
\end{equation}
confirming the invertibility of~$\sigma$.
Therefore, the (global) dimension is $\dim\TYAXK=2|\A|$.

While a braiding exists on a Tambara--Yamagami category if and only if every element in~$\A$ has order 2 \cite{Siehler}, a $\Z_2$-action and a $\Z_2$-crossed braiding can always be defined \cite{GLM}:
\begin{lemma}{\cite[Thm.\,3.2\,part\,1]{GLM}}
	Assuming  
	\begin{align}
		\rho_{-1}(a)&=a^{-1}\,,
		\label{eq:TYaction}
	\end{align}
	and strictness of the action on $\A$-$\Vect$ (i.e.\ all the coherence data consists of identities), there are two $\Z_2$-actions on $\TYAXK$ (up to equivalence). 
	For these two actions, all additional components of the coherence data are also identities, with the exception of
	\begin{equation}
		\left(\rho_{-1,-1}^2\right)_\sigma=\pm\id_\sigma\colon \quad\rho_{-1}(\rho_{-1}(\sigma))\longrightarrow \sigma=\rho_{(-1)^2}(\sigma)\,.
	\end{equation}
\end{lemma}
	We refer to the choice of $\id_\sigma$ as the strict action and to the choice of $-\id_\sigma$ as the non-strict action.
	In order to specify the full $\Z_2$-crossed braided structure, we need to choose a square root of $\kappa\sum_{a\in\A}q(a)^{-1}$ called~$\delta$ and a \textsl{quadratic form on~$\A$} associated to the chosen bicharacter.
	This is a map $q\colon \A\longrightarrow\C^\times$ such that $q(a)=q(a^{-1})$ and the assignment $(a,b)\longmapsto\frac{q(ab)}{q(a)q(b)}$ is the bicharacter~$\chi$. 
\begin{lemma}{\cite[Thm.\,3.2\,parts\,2,3]{GLM}}
	The Tambara--Yamagami category $\TYAXK$ can be equipped with a $\Z_2$-crossed braided structure if and only if it carries the strict $\Z_2$-action. $\Z_2$-braidings are classified by pairs $(q,\delta)$ of a quadratic form $q\colon\A\longrightarrow \C^\times$ such that $\chi(a,b)=\frac{q(ab)}{q(a)q(b)}$ for all $a,b\in\A$, and $\delta^2=\kappa\sum_{a\in\A}q(a)^{-1}$. The components of the $\Z_2$-crossed braiding are then defined as
	\begin{align}
		c_{a,b}&:=\chi(a,b)\id_{ab}\,,\nonumber\\
		c_{a,\sigma}&:=q(a)^{-1}\id_\sigma\,,\nonumber\\
		c_{\sigma,a}&:=q(a)^{-1}\id_\sigma\,,\nonumber\\
		c_{\sigma,\sigma}&:=\bigoplus_{x\in\A}\delta q(x)\id_x\,.
	\end{align}
	We denote the resulting pivotal $\Z_2$-crossed braided fusion category by $\TYAQKD$.
\end{lemma}

Note that there are additional $\Z_2$-actions if we allow the restriction of the monoidal structure of~$\rho$ to $\A$-$\Vect$ to be non-strict. One of these choices was discussed in \cite[Sect.\,4.2,~Sect.\,5.1]{GNN}:
	\begin{equation}
		(\rho^2_{-1,-1})_Y:=
		\begin{cases}
			\chi(a,a^{-1})\id_a&\text{for }Y=a\,,\\
			\left(\kappa \sum_{a\in\A}q(a)\right)\id_\sigma&\text{for }Y=\sigma\,,
		\end{cases}
	\end{equation}
	To adjust the option described here, one has to modify the braiding (which no longer satisfies $c_{a,\sigma}= c_{\sigma,a}$) and the equivariantisation features genuine projective representations.
\subsection{Orbifolding and Equivariantisation}
In this section, we compute the equivariantisations for Tambara--Yamagami categories for odd order $|\A|$ and for $\A=\Z_2$, and provide the associated orbifold datum.

In order to apply the methods of \Cref{sec:CalculatingEq}, we have to determine the stabiliser subgroup $(\Z_2)_Y$ for each simple object~$Y$ and calculate the associated cocycle~$\beta_Y$ according to \eqref{eq:projectivitycocycle}.
The latter is always trivial since we use a strict $\Z_2$-action, thus we can work exclusively with linear representations.
The strictness of the action also implies that we only need to spell out $u_{-1}$ for an equivariant object $(X,u)$, since $u_1=\id$ for any simple equivariant object. 

\subsubsection{Groups of Odd Order}
\label{sec:oddOrderGroups} 
In this section, we discuss the equivariantisation and orbifold data for the Tambara--Yamagami categories of (commutative) groups~$\A$ of odd order $n:=|\A|$.

We have $G$-orbits $[e]=\{e\}$, $[a]=\{a,a^{-1}\}$, and $[\sigma]=\{\sigma\}$, where $a\in\A\setminus\{e\}$. 
Their stabiliser subgroups are $(\Z_2)_e=\Z_2=(\Z_2)_\sigma$ and $(\Z_2)_a=\{1\}$.
We can equip~$e$ with the two 1-dimensional irreducible representations of~$\Z_2$, the trivial one and the sign representation, leading to two equivariant structures given by $u_{-1}^e=\pm\id_e$. 
The $\Z_2$-orbit~$[a]$ has trivial stabiliser, so it pairs with the trivial representation of the trivial group.
The object underlying the resulting simple in the equivariantisation is $a\oplus a^{-1}$ and $-1\in\Z_2$ acts by exchanging the two summands with equivariant structure $u_{-1}^{[a]}=\id_{[a]}$. 
Lastly, we can equip~$\sigma$ with the two 1-dimensional irreducible $\Z_2$-representations as well. 
In summary, this leads to the following: 
Following \cite[Sect.\,5.1]{GNN}), the equivariantisation $(\TYAQKD)^{\Z_2}$ contains the following simple objects: 
	\begin{itemize}
		\item 2 invertible objects $X_\pm=(e,\pm\id_e)$,
		\item $\frac{n-1}{2}$ 2-dimensional objects $Y_a=(a\oplus a^{-1},\id)$ for $a\in\A\setminus \{e\}$ (with $Y_{a^{-1}}=Y_a$),
		\item 2 $\sqrt{n}$-dimensional objects $Z_\pm=(\sigma,\pm\id_\sigma)$.
	\end{itemize}
	The monoidal structure is given by \Cref{tab:TYeqfusion}, and $X_+$ is the monoidal unit with strict unitors. 
	Associators are given by those in $\TYAQKD$.
	\begin{table}[h]
	{\renewcommand{\arraystretch}{1.5}\setlength{\tabcolsep}{12pt}
	\begin{center}
		\begin{tabular}{c!{\vrule width 1pt} c|c|c|c|c}
			&$X_+$&$X_-$&$Y_b$&$Z_+$&$Z_-$\\\noalign{\hrule height 1pt}
			$X_+$&$X_+$&$X_-$&$Y_b$&$Z_+$&$Z_-$\\\hline
			$X_-$&$X_-$&$X_+$&$Y_b$&$Z_-$&$Z_+$\\\hline
			$Y_a$&$Y_a$&$Y_a$&$Y_{ab}\oplus Y_{ab^{-1}}$&$Z_+\oplus Z_-$&$Z_+\oplus Z_-$\\\hline
			$Z_+$&$Z_+$&$Z_-$&$Z_+ \oplus Z_-$&$X_+\oplus\left(\bigoplus Y_x\right)$&$X_-\oplus \left(\bigoplus Y_x\right)$\\\hline
			$Z_-$&$Z_-$&$Z_+$&$Z_+ \oplus Z_-$&$X_-\oplus\left(\bigoplus Y_x\right)$&$X_+\oplus\left(\bigoplus Y_x\right)$
		\end{tabular}
	\end{center}}
	\caption{Fusion rules for the equivariantisation $(\TYAQKD)^{\Z_2}$ under the strict $\Z_2$-action. 
	An entry in the table is to be understood as ``row label $\otimes$ column label.''
	For the case $a=b$ we have set $Y_{ab^{-1}}=Y_e:=X_+\oplus X_-$ and the sums of~$Y_x$ are over orbits of the $\Z_2$-action on $\A\setminus\{e\}$ such that every $Y_x$ appears exactly once.}
	\label{tab:TYeqfusion}
	\end{table}
	
The fusion rules in \Cref{tab:TYeqfusion} between~$X_\pm$ and~$Z_\pm$ can be calculated easily from \eqref{eq:eqfusiondecompsimplified}.
Specifically, when at most one factor of~$Z$ is involved, we simply have $X\otimes Z=Z\otimes X=Z$, $X\otimes X=X$, and the representation attached to the product is determined by the product of the signs in the subscripts.
When both factors are~$Z_\pm$, the underlying object is the sum $\bigoplus_{x\in\A}x$ which splits into the sum over~$Y_x$ and one summand~$e$ whose associated representation is once again determined by the product of the subscripts.
For those involving~$Y_a$ we use \eqref{eq:equivariantStructureOfMonoidalProducts} noting again that~$\rho^2_g$ is trivial. 
More specifically, an isomorphism $X_-\otimes Y_b\cong Y_b$ is given by $f:=(-\id_b)\oplus\id_{b^{-1}}$, then $\rho_{-1}(f)=\id_b\oplus(-\id_{b^{-1}})$ (where~$\rho_{-1}$ exchanges the identity morphisms on~$b$ and~$b^{-1}$ thereby carrying the sign over) and thus \eqref{eq:defining property of morphisms in equivariantisation} commutes, since $f\circ u_{-1}^{X_-\otimes Y_b}=f\circ -\id_{b\oplus b^{-1}}=\id_b\oplus(-\id_{b^{-1}})=u_{-1}^{Y_b}\circ \rho_{-1}(f)$.
The fusion $Y_a\otimes Y_b=(a\oplus a^{-1})\otimes (b\oplus b^{-1})=(ab\oplus (ab)^{-1})\oplus (ab^{-1}\oplus a^{-1}b)=Y_{ab}\oplus Y_{ab^{-1}}$ is straightforward for $a\neq b$.
For $a=b$, the second term is $(ab^{-1}\oplus a^{-1}b)={e\oplus e}$ and the two copies of~$e=\C$ are swapped under $\rho_{-1}$, i.e.\ the equivariant structure is 
\begin{equation}
	u_{-1}^{e\oplus e}=\begin{pmatrix}
	0&1\\
	1&0
	\end{pmatrix}\colon \C^2\longrightarrow\C^2\,.
\end{equation}
We choose $f\colon (e\oplus e,u^{e\oplus e}) \stackrel{\cong}{\longrightarrow} X_+\oplus X_-$ to be given by the basis change matrix
\begin{equation}
\frac{1}{\sqrt{2}}
\begin{pmatrix}
	1&1\\
	1&-1
\end{pmatrix}\,,
\end{equation}
and note $\rho_{-1}(f)=f$, then a simple multiplication shows $f\circ u_{-1}^{e\oplus e}=u_{-1}^{X_+\oplus X_-}\circ \rho_{-1}(f)$.
A similar calculation also proves $Y_a\otimes Z_\pm\cong Z_+\oplus Z_-$.

Lastly, the braiding on the equivariantisation is given by that of the original Tambara--Yamagami category determined by the underlying objects unless the second factor is $Z_\pm$.
In the latter case the braiding is enhanced by the equivariant structure of the first factor.
Since only $X_-$ and $Z_-$ have non-trivial equivariant structure, the modifications appear exclusively for $c^{\Z_2}_{X_-,Z_\pm}$ and $c^{\Z_2}_{Z_-,Z_\pm}$.
The components are given by
\begin{align}
	c^{\Z_2}_{X_\pm,X_\pm}&=\id_e\,,\\
	c^{\Z_2}_{X_\pm,Y_a}=c^{\Z_2}_{Y_a,X_\pm}&=\id_{a\oplus a^{-1}}\,,\\
	c^{\Z_2}_{Y_a,Y_b}
	&=
	\begin{cases}
	\chi(a,b)\id_{ab\oplus (ab)^{-1}}\oplus \chi(a,b^{-1})\id_{ab^{-1}\oplus a^{-1}b}\qquad &a\neq b\,,\\
	\chi(a,a)\id_{a^2\oplus a^{-2}}\oplus \chi(a,a^{-1})\id_{e\oplus e}\qquad &a=b\,,
	\end{cases}\\
	c^{\Z_2}_{Z_\pm,X_\pm}&=\id_\sigma\,,\\
	c^{\Z_2}_{Z_\pm,Y_a}=c^{\Z_2}_{Y_a,Z_\pm}&=q(a)^{-1}\id_{\sigma\oplus\sigma}\,,\\
	c^{\Z_2}_{X_+,Z_\pm}&=\hphantom{-}\id_\sigma\,,\\
	c^{\Z_2}_{X_-,Z_\pm}&=-\id_\sigma\,,\\
	c^{\Z_2}_{Z_+,Z_\pm}&=\hphantom{-}\id_{\bigoplus_{x\in \A}x}\,,\\
	c^{\Z_2}_{Z_-,Z_\pm}&=-\id_{\bigoplus_{x\in \A}x}\,.
\end{align}

\medskip

We turn to describe the orbifold datum and the associated orbifold category. 
Recall (\Cref{sec:RT}) that we need to choose a simple object~$m_g$ in each degree $g\in\Z_2$ and $m_1=e$.
Thus, we are forced to pick $m_{-1}=\sigma$ and the orbifold datum is given by
\begin{align}
	A_{\Z_2}&= (e\otimes e)\oplus (\sigma\otimes\sigma)=e\oplus \bigoplus_{a\in\A}a=:A_1\oplus A_{-1}\,,\\
	T_{\Z_2}&= (e\otimes e\otimes e) \oplus (\sigma\otimes e\otimes \sigma)\oplus (\sigma\otimes \sigma\otimes e)\oplus (e\otimes \sigma\otimes \sigma)\\
	&= \hphantom{(e\otimes\vphantom{e}}e\hphantom{\vphantom{e}\otimes e)}\oplus \hphantom{(\sigma}\,\;\bigoplus_{x\in\A}x\hphantom{\sigma)}\oplus\hphantom{(\sigma}\;\,\bigoplus_{y\in\A}y\hphantom{\sigma)}\oplus\hphantom{(\sigma}\;\,\bigoplus_{z\in\A}z\,,\\
		\alpha_{\Z_2}&=\bigoplus_{g,h,k\in \Z_2}
				\begin{tikzpicture}[very thick,scale=0.9,color=red!50!black, baseline=-1.4cm, rotate=180]
					\draw[line width=0pt] 
					(5.5,3) node[line width=0pt] (ghko) {{\scriptsize$ghk\vphantom{ghk}$}}
					(5,3) node[line width=0pt] (go) {{\scriptsize$g\vphantom{ghk}$}}
					(4.5,3) node[line width=0pt] (hkli) {{\scriptsize$hk\vphantom{ghk}$}}
					(3.5,3) node[line width=0pt] (hkre) {{\scriptsize$hk\vphantom{ghk}$}}
					(3,3) node[line width=0pt] (ho) {{\scriptsize$h\vphantom{ghk}$}}
					(2.5,3) node[line width=0pt] (ko) {{\scriptsize$k\vphantom{ghk}$}}
					(5.5,0) node[line width=0pt] (ghku) {{\scriptsize$ghk\vphantom{ghk}$}}
					(5,0) node[line width=0pt] (ghli) {{\scriptsize$gh\vphantom{ghk}$}}
					(4.5,0) node[line width=0pt] (ku) {{\scriptsize$k\vphantom{ghk}$}}
					(3.5,0) node[line width=0pt] (ghre) {{\scriptsize$gh\vphantom{ghk}$}} 
					(3,0) node[line width=0pt] (gu) {{\scriptsize$g\vphantom{ghk}$}} 
					(2.5,0) node[line width=0pt] (hu) {{\scriptsize$h\vphantom{ghk}$}};
					\draw[redirectedred] (ghko) -- (ghku);
					\draw[postaction={decorate}, decoration={markings,mark=at position .65 with {\arrow[color=red!50!black]{<}}}] (ghli) .. controls +(0,1) and +(0,1) .. (ghre);
					\draw[postaction={decorate}, decoration={markings,mark=at position .55 with {\arrow[color=red!50!black]{<}}}] (hkre) .. controls +(0,-1) and +(0,-1) .. (hkli);
					\draw[postaction={decorate}, decoration={markings,mark=at position .65 with {\arrow[color=red!50!black]{<}}}] (gu) .. controls +(0,1.5) and +(0,-1.5) .. (go);
					\draw[redirectedred] (hu) .. controls +(0,1.5) and +(0,-1.5) ..  (ho);
					\draw[color=white, line width=4pt] (ku) .. controls +(0,1.5) and +(0,-1.5) .. (ko);
					\draw[postaction={decorate}, decoration={markings,mark=at position .65 with {\arrow[color=red!50!black]{<}}}] (ku) .. controls +(0,1.5) and +(0,-1.5) .. (ko);
					\end{tikzpicture}\,,\label{eq:alphaTY}\\
		\overline{\alpha}_{\Z_2}&=\bigoplus_{g,h,k\in \Z_2}
				\begin{tikzpicture}[very thick,scale=0.9,color=red!50!black, baseline=-1.4cm, rotate=180]
					\draw[line width=0pt] 
					(5.5,0) node[line width=0pt] (ghko) {{\scriptsize$ghk\vphantom{ghk}$}}
					(5,0) node[line width=0pt] (go) {{\scriptsize$g\vphantom{ghk}$}}
					(4.5,0) node[line width=0pt] (hkli) {{\scriptsize$hk\vphantom{ghk}$}}
					(3.5,0) node[line width=0pt] (hkre) {{\scriptsize$hk\vphantom{ghk}$}}
					(3,0) node[line width=0pt] (ho) {{\scriptsize$h\vphantom{ghk}$}}
					(2.5,0) node[line width=0pt] (ko) {{\scriptsize$k\vphantom{ghk}$}}
					(5.5,3) node[line width=0pt] (ghku) {{\scriptsize$ghk\vphantom{ghk}$}}
					(5,3) node[line width=0pt] (ghli) {{\scriptsize$gh\vphantom{ghk}$}}
					(4.5,3) node[line width=0pt] (ku) {{\scriptsize$k\vphantom{ghk}$}}
					(3.5,3) node[line width=0pt] (ghre) {{\scriptsize$gh\vphantom{ghk}$}} 
					(3,3) node[line width=0pt] (gu) {{\scriptsize$g\vphantom{ghk}$}} 
					(2.5,3) node[line width=0pt] (hu) {{\scriptsize$h\vphantom{ghk}$}};
					\draw[directedred] (ghko) -- (ghku);
					\draw[postaction={decorate}, decoration={markings,mark=at position .65 with {\arrow[color=red!50!black]{>}}}] (ghli) .. controls +(0,-1) and +(0,-1) .. (ghre);
					\draw[postaction={decorate}, decoration={markings,mark=at position .55 with {\arrow[color=red!50!black]{>}}}] (hkre) .. controls +(0,1) and +(0,1) .. (hkli);
					\draw[postaction={decorate}, decoration={markings,mark=at position .65 with {\arrow[color=red!50!black]{>}}}] (gu) .. controls +(0,-1.5) and +(0,1.5) .. (go);
					\draw[directedred] (hu) .. controls +(0,-1.5) and +(0,1.5) ..  (ho);
					\draw[color=white, line width=4pt] (ku) .. controls +(0,-1.5) and +(0,1.5) .. (ko);
					\draw[postaction={decorate}, decoration={markings,mark=at position .65 with {\arrow[color=red!50!black]{>}}}] (ku) .. controls +(0,-1.5) and +(0,1.5) .. (ko);
					\end{tikzpicture}\,,\label{eq:alphabarTY}\\
	\psi_{\Z_2}&=\id_{A_1}\oplus|\kappa|\id_{A_{-1}}\,,\\
	\phi_{\Z_2}&=|\kappa|\,.
\end{align}
Here and in the following, we denote~$A_1=e$ which has trivial Frobenius algebra structure and~$A_{-1}=\bigoplus_{a\in\A}a=\C[\A]\in\A\text{-}\Vect$ with Frobenius algebra structure given by \eqref{eq:AgGextension}:
\begin{align}
	&&\mu_{-1}&:=\sgn(\kappa)\bigoplus_{\substack{a,b,c\in\A\\ab=c}}\id_{ab} \colon& \bigoplus_{a\in\A}a\otimes \bigoplus_{b\in\A}b&\longrightarrow \bigoplus_{c\in\A}c\,,\label{eq:TYZ2algstructure}\\
	&&\eta_{-1}&:=\sgn(\kappa)\,\mathrm{incl}\colon& e&\longrightarrow \bigoplus_{a\in\A}a\,,\\
	&&\Delta_{-1}&:=\sgn(\kappa)\kappa^2\bigoplus_{\substack{a,b,c\in\A\\ab=c}}\id_{ab} \colon& \bigoplus_{c\in\A}c &\longrightarrow \bigoplus_{a\in\A}a\otimes \bigoplus_{b\in\A}b\,,\\
	&&\varepsilon_{-1}&:=\sgn(\kappa)\kappa^{-2}\mathrm{proj}\colon& \bigoplus_{a\in\A}a&\longrightarrow e\,.
\end{align}
Note that the algebra structure above coincides with the group algebra structure of $\C[\A]$ up to $\sgn(\kappa)$.
Once again, viewing the latter as defectification, these factors match the formulas of \Cref{sec:zestedorbdat}.

The $A_{\Z_2}$-module structure on~$T_{\Z_2}$ is given by \eqref{eq:mmodulestructures} and \eqref{eq:chimodulestructure} and the factors of $\sgn(\kappa)$ can be deduced from \Cref{sec:zestedorbdat} here as well.
The parts of the module structure with respect to $A_1$ are trivial, those with respect to $A_{-1}$ are given by
\begin{align}
	\begin{tikzpicture}[very thick,scale=0.75,color=red!50!black, baseline]
\draw (-1.75,-1) node[below] (mghs) {{\scriptsize$\sigma$}};
\draw (-1,-1) node[below] (mgh) {{\scriptsize$\sigma$}};
\draw (0,-1) node[below] (A1) {{\scriptsize$\sigma$}};
\draw (0.75,-1) node[below] (mg) {{\scriptsize$\sigma$}};
\draw (1.5,-1) node[below] (mh) {{\scriptsize$e$}};
\draw[directedred] (-1,-1) .. controls +(0,0.75) and +(0,0.75) .. (0,-1);
\draw[directedred] (0,1) to[out=-90, in=90] (-1.75,-1);
\draw[directedred] (mg) -- (.75,1); 
\draw[directedred, dashed] (mh) -- (1.5,1); 
\end{tikzpicture} 
=
\begin{tikzpicture}[very thick,scale=0.75,color=red!50!black, baseline]
\draw (-1.75,-1) node[below] (mghs) {{\scriptsize$\sigma$}};
\draw (-1,-1) node[below] (mgh) {{\scriptsize$\sigma$}};
\draw (0,-1) node[below] (A1) {{\scriptsize$\sigma$}};
\draw (1.5,-1) node[below] (mg) {{\scriptsize$\sigma$}};
\draw (.75,-1) node[below] (mh) {{\scriptsize$e$}};
\draw[directedred] (-1,-1) .. controls +(0,0.75) and +(0,0.75) .. (0,-1);
\draw[directedred] (0,1) to[out=-90, in=90] (-1.75,-1);
\draw[directedred] (mg) -- (1.5,1); 
\draw[directedred, dashed] (mh) -- (.75,1); 
\end{tikzpicture} 
=
\begin{tikzpicture}[very thick,scale=0.75,color=red!50!black, baseline]
\draw (1.75,-1) node[below] (mghs) {{\scriptsize$\sigma$}};
\draw (1,-1) node[below] (mgh) {{\scriptsize$\sigma$}};
\draw (0,-1) node[below] (A1) {{\scriptsize$\sigma$}};
\draw (-0.75,-1) node[below] (mg) {{\scriptsize$\sigma$}};
\draw (-1.5,-1) node[below] (mh) {{\scriptsize$e$}};
\draw[redirectedred] (1,-1) .. controls +(0,0.75) and +(0,0.75) .. (0,-1);
\draw[directedred] (1.75,-1) to[out=90, in=-90] (0,1);
\draw[directedred] (mg) -- (-.75,1); 
\draw[redirectedred, dashed] (mh) -- (-1.5,1); 
\end{tikzpicture} 
=
\begin{tikzpicture}[very thick,scale=0.75,color=red!50!black, baseline]
\draw (1.75,-1) node[below] (mghs) {{\scriptsize$\sigma$}};
\draw (1,-1) node[below] (mgh) {{\scriptsize$\sigma$}};
\draw (0,-1) node[below] (A1) {{\scriptsize$\sigma$}};
\draw (-0.75,-1) node[below] (mg) {{\scriptsize$e$}};
\draw (-1.5,-1) node[below] (mh) {{\scriptsize$\sigma$}};
\draw[redirectedred] (1,-1) .. controls +(0,0.75) and +(0,0.75) .. (0,-1);
\draw[directedred] (1.75,-1) to[out=90, in=-90] (0,1);
\draw[directedred,dashed] (mg) -- (-.75,1); 
\draw[redirectedred] (mh) -- (-1.5,1); 
\end{tikzpicture}\nonumber
\end{align}
\begin{align}
=\begin{tikzpicture}[very thick,scale=0.75,color=red!50!black, baseline]
\draw (1.75,-1) node[below] (mghs) {{\scriptsize$\sigma$}};
\draw (1,-1) node[below] (mgh) {{\scriptsize$\sigma$}};
\draw (-0.75,-1) node[below] (mg) {{\scriptsize$\sigma$}};
\draw (-1.5,-1) node[below] (mh) {{\scriptsize$\sigma$}};
\draw (0,-1) node[below] (A1) {{\scriptsize$e$}};
\draw[redirectedred] (1,-1) .. controls +(0,0.75) and +(0,0.75) .. (-0.75,-1);
\draw[directedred] (1.75,-1) to[out=90, in=-90] (-0.75,1);
\draw[color=white, line width=4pt] (0,-1) -- (0,1); 
\draw[directedred,dashed] (0,-1) -- (0,1);
\draw[redirectedred] (mh) -- (-1.5,1); 
\end{tikzpicture}
&=
\sgn(\kappa)\bigoplus_{\substack{a,b,c\in\A\\ab=c}}\id_{ab} \colon& \bigoplus_{a\in\A}a\otimes \bigoplus_{b\in\A}b&\longrightarrow \bigoplus_{c\in\A}c\,,\\
\begin{tikzpicture}[very thick,scale=0.75,color=red!50!black, baseline]
\draw (1.75,-1) node[below] (mghs) {{\scriptsize$\sigma$}};
\draw (1,-1) node[below] (mgh) {{\scriptsize$\sigma$}};
\draw (-0.75,-1) node[below] (mg) {{\scriptsize$\sigma$}};
\draw (-1.5,-1) node[below] (mh) {{\scriptsize$e$}};
\draw (0,-1) node[below] (A1) {{\scriptsize$\sigma$}};
\draw[redirectedred] (1,-1) .. controls +(0,0.75) and +(0,0.75) .. (-0.75,-1);
\draw[directedred] (1.75,-1) to[out=90, in=-90] (-0.75,1);
\draw[color=white, line width=4pt] (0,-1) -- (0,1); 
\draw[directedred] (0,-1) -- (0,1);
\draw[dashed,redirectedred] (mh) -- (-1.5,1); 
\end{tikzpicture}
&=
\sgn(\kappa)\bigoplus_{\substack{z,b,z^\prime\in\A\\ zb=z^\prime}}\frac{q(z^\prime)}{q(z)}\delta_{zb,z^\prime}\id_{zb} \colon &\bigoplus_{z\in\A}z\otimes \bigoplus_{b\in\A}b&\longrightarrow \bigoplus_{z^\prime\in\A}z^\prime\,,
\label{eq:modulestructure1-1}
\end{align}
where in the last line the sum over~$z$ comes from $e\otimes(\sigma\otimes\sigma)$ in~$T_{\Z_2}$ and the sum over~$b$ from~$A_{-1}$.

The components of $\alpha$ \eqref{eq:alphaTY} with $k=1$ are given by 
\begin{align}
					\begin{tikzpicture}[very thick,scale=0.9,color=red!50!black, baseline=-1.4cm, rotate=180]
					\draw[line width=0pt] 
					(5.5,3) node[line width=0pt] (ghko) {{\scriptsize$e\vphantom{ghk}$}}
					(5,3) node[line width=0pt] (go) {{\scriptsize$e\vphantom{ghk}$}}
					(4.5,3) node[line width=0pt] (hkli) {{\scriptsize$e\vphantom{ghk}$}}
					(3.5,3) node[line width=0pt] (hkre) {{\scriptsize$e\vphantom{ghk}$}}
					(3,3) node[line width=0pt] (ho) {{\scriptsize$e\vphantom{ghk}$}}
					(2.5,3) node[line width=0pt] (ko) {{\scriptsize$e\vphantom{ghk}$}}
					(5.5,0) node[line width=0pt] (ghku) {{\scriptsize$e\vphantom{ghk}$}}
					(5,0) node[line width=0pt] (ghli) {{\scriptsize$e\vphantom{ghk}$}}
					(4.5,0) node[line width=0pt] (ku) {{\scriptsize$e\vphantom{ghk}$}}
					(3.5,0) node[line width=0pt] (ghre) {{\scriptsize$e\vphantom{ghk}$}} 
					(3,0) node[line width=0pt] (gu) {{\scriptsize$e\vphantom{ghk}$}} 
					(2.5,0) node[line width=0pt] (hu) {{\scriptsize$e\vphantom{ghk}$}};
					\draw[dashed,redirectedred] (ghko) -- (ghku);
					\draw[dashed,postaction={decorate}, decoration={markings,mark=at position .65 with {\arrow[color=red!50!black]{<}}}] (ghli) .. controls +(0,1) and +(0,1) .. (ghre);
					\draw[dashed,postaction={decorate}, decoration={markings,mark=at position .55 with {\arrow[color=red!50!black]{<}}}] (hkre) .. controls +(0,-1) and +(0,-1) .. (hkli);
					\draw[dashed,postaction={decorate}, decoration={markings,mark=at position .65 with {\arrow[color=red!50!black]{<}}}] (gu) .. controls +(0,1.5) and +(0,-1.5) .. (go);
					\draw[dashed,redirectedred] (hu) .. controls +(0,1.5) and +(0,-1.5) ..  (ho);
					\draw[color=white, line width=4pt] (ku) .. controls +(0,1.5) and +(0,-1.5) .. (ko);
					\draw[dashed,postaction={decorate}, decoration={markings,mark=at position .65 with {\arrow[color=red!50!black]{<}}}] (ku) .. controls +(0,1.5) and +(0,-1.5) .. (ko);
					\end{tikzpicture}
					&= \id_e\colon & e&\longrightarrow e\,,\\
				\begin{tikzpicture}[very thick,scale=0.9,color=red!50!black, baseline=-1.4cm, rotate=180]
					\draw[line width=0pt] 
					(5.5,3) node[line width=0pt] (ghko) {{\scriptsize$e\vphantom{ghk}$}}
					(5,3) node[line width=0pt] (go) {{\scriptsize$\sigma\vphantom{ghk}$}}
					(4.5,3) node[line width=0pt] (hkli) {{\scriptsize$\sigma\vphantom{ghk}$}}
					(3.5,3) node[line width=0pt] (hkre) {{\scriptsize$\sigma\vphantom{ghk}$}}
					(3,3) node[line width=0pt] (ho) {{\scriptsize$\sigma\vphantom{ghk}$}}
					(2.5,3) node[line width=0pt] (ko) {{\scriptsize$e\vphantom{ghk}$}}
					(5.5,0) node[line width=0pt] (ghku) {{\scriptsize$e\vphantom{ghk}$}}
					(5,0) node[line width=0pt] (ghli) {{\scriptsize$e\vphantom{ghk}$}}
					(4.5,0) node[line width=0pt] (ku) {{\scriptsize$e\vphantom{ghk}$}}
					(3.5,0) node[line width=0pt] (ghre) {{\scriptsize$e\vphantom{ghk}$}} 
					(3,0) node[line width=0pt] (gu) {{\scriptsize$\sigma\vphantom{ghk}$}} 
					(2.5,0) node[line width=0pt] (hu) {{\scriptsize$\sigma\vphantom{ghk}$}};
					\draw[redirectedred,dashed] (ghko) -- (ghku);
					\draw[dashed, postaction={decorate}, decoration={markings,mark=at position .65 with {\arrow[color=red!50!black]{<}}}] (ghli) .. controls +(0,1) and +(0,1) .. (ghre);
					\draw[postaction={decorate}, decoration={markings,mark=at position .55 with {\arrow[color=red!50!black]{<}}}] (hkre) .. controls +(0,-1) and +(0,-1) .. (hkli);
					\draw[postaction={decorate}, decoration={markings,mark=at position .65 with {\arrow[color=red!50!black]{<}}}] (gu) .. controls +(0,1.5) and +(0,-1.5) .. (go);
					\draw[redirectedred] (hu) .. controls +(0,1.5) and +(0,-1.5) ..  (ho);
					\draw[color=white, line width=4pt] (ku) .. controls +(0,1.5) and +(0,-1.5) .. (ko);
					\draw[dashed,postaction={decorate}, decoration={markings,mark=at position .65 with {\arrow[color=red!50!black]{<}}}] (ku) .. controls +(0,1.5) and +(0,-1.5) .. (ko);
					\end{tikzpicture}
					&=\sgn(\kappa)\bigoplus_{\substack{y,z\in\A\\yz=z^\prime}}\id_{yz} \colon& \bigoplus_{z\in\A}z\otimes \bigoplus_{y\in\A}y&\longrightarrow \bigoplus_{z^\prime\in\A}z^\prime\,,\\
					\begin{tikzpicture}[very thick,scale=0.9,color=red!50!black, baseline=-1.4cm, rotate=180]
					\draw[line width=0pt] 
					(5.5,3) node[line width=0pt] (ghko) {{\scriptsize$\sigma\vphantom{ghk}$}}
					(5,3) node[line width=0pt] (go) {{\scriptsize$\sigma\vphantom{ghk}$}}
					(4.5,3) node[line width=0pt] (hkli) {{\scriptsize$e\vphantom{ghk}$}}
					(3.5,3) node[line width=0pt] (hkre) {{\scriptsize$e\vphantom{ghk}$}}
					(3,3) node[line width=0pt] (ho) {{\scriptsize$e\vphantom{ghk}$}}
					(2.5,3) node[line width=0pt] (ko) {{\scriptsize$e\vphantom{ghk}$}}
					(5.5,0) node[line width=0pt] (ghku) {{\scriptsize$\sigma\vphantom{ghk}$}}
					(5,0) node[line width=0pt] (ghli) {{\scriptsize$\sigma\vphantom{ghk}$}}
					(4.5,0) node[line width=0pt] (ku) {{\scriptsize$e\vphantom{ghk}$}}
					(3.5,0) node[line width=0pt] (ghre) {{\scriptsize$\sigma\vphantom{ghk}$}} 
					(3,0) node[line width=0pt] (gu) {{\scriptsize$\sigma\vphantom{ghk}$}} 
					(2.5,0) node[line width=0pt] (hu) {{\scriptsize$e\vphantom{ghk}$}};
					\draw[redirectedred] (ghko) -- (ghku);
					\draw[postaction={decorate}, decoration={markings,mark=at position .65 with {\arrow[color=red!50!black]{<}}}] (ghli) .. controls +(0,1) and +(0,1) .. (ghre);
					\draw[dashed,postaction={decorate}, decoration={markings,mark=at position .55 with {\arrow[color=red!50!black]{<}}}] (hkre) .. controls +(0,-1) and +(0,-1) .. (hkli);
					\draw[postaction={decorate}, decoration={markings,mark=at position .65 with {\arrow[color=red!50!black]{<}}}] (gu) .. controls +(0,1.5) and +(0,-1.5) .. (go);
					\draw[dashed,redirectedred] (hu) .. controls +(0,1.5) and +(0,-1.5) ..  (ho);
					\draw[color=white, line width=4pt] (ku) .. controls +(0,1.5) and +(0,-1.5) .. (ko);
					\draw[dashed,postaction={decorate}, decoration={markings,mark=at position .65 with {\arrow[color=red!50!black]{<}}}] (ku) .. controls +(0,1.5) and +(0,-1.5) .. (ko);
					\end{tikzpicture}
					&=\kappa\bigoplus_{\substack{y,y^\prime,y^{\prime\prime}\in\A\\y^\prime y^{\prime\prime}=y}}\id_{y} \colon& \bigoplus_{y\in\A}y &\longrightarrow \bigoplus_{y^\prime \in\A}y^\prime \otimes \bigoplus_{y^{\prime\prime}\in\A}y^{\prime\prime}\,,\\
				\begin{tikzpicture}[very thick,scale=0.9,color=red!50!black, baseline=-1.4cm, rotate=180]
					\draw[line width=0pt] 
					(5.5,3) node[line width=0pt] (ghko) {{\scriptsize$\sigma\vphantom{ghk}$}}
					(5,3) node[line width=0pt] (go) {{\scriptsize$e\vphantom{ghk}$}}
					(4.5,3) node[line width=0pt] (hkli) {{\scriptsize$\sigma\vphantom{ghk}$}}
					(3.5,3) node[line width=0pt] (hkre) {{\scriptsize$\sigma\vphantom{ghk}$}}
					(3,3) node[line width=0pt] (ho) {{\scriptsize$\sigma\vphantom{ghk}$}}
					(2.5,3) node[line width=0pt] (ko) {{\scriptsize$e\vphantom{ghk}$}}
					(5.5,0) node[line width=0pt] (ghku) {{\scriptsize$\sigma\vphantom{ghk}$}}
					(5,0) node[line width=0pt] (ghli) {{\scriptsize$\sigma\vphantom{ghk}$}}
					(4.5,0) node[line width=0pt] (ku) {{\scriptsize$e\vphantom{ghk}$}}
					(3.5,0) node[line width=0pt] (ghre) {{\scriptsize$\sigma\vphantom{ghk}$}} 
					(3,0) node[line width=0pt] (gu) {{\scriptsize$e\vphantom{ghk}$}} 
					(2.5,0) node[line width=0pt] (hu) {{\scriptsize$\sigma\vphantom{ghk}$}};
					\draw[redirectedred] (ghko) -- (ghku);
					\draw[postaction={decorate}, decoration={markings,mark=at position .65 with {\arrow[color=red!50!black]{<}}}] (ghli) .. controls +(0,1) and +(0,1) .. (ghre);
					\draw[postaction={decorate}, decoration={markings,mark=at position .55 with {\arrow[color=red!50!black]{<}}}] (hkre) .. controls +(0,-1) and +(0,-1) .. (hkli);
					\draw[dashed,postaction={decorate}, decoration={markings,mark=at position .65 with {\arrow[color=red!50!black]{<}}}] (gu) .. controls +(0,1.5) and +(0,-1.5) .. (go);
					\draw[redirectedred] (hu) .. controls +(0,1.5) and +(0,-1.5) ..  (ho);
					\draw[color=white, line width=4pt] (ku) .. controls +(0,1.5) and +(0,-1.5) .. (ko);
					\draw[dashed,postaction={decorate}, decoration={markings,mark=at position .65 with {\arrow[color=red!50!black]{<}}}] (ku) .. controls +(0,1.5) and +(0,-1.5) .. (ko);
					\end{tikzpicture}
					&=|\kappa|\bigoplus_{\substack{x,y,x^\prime,y^\prime\in\A\\xy=y^\prime x^\prime}}\id_{xy} \colon& \bigoplus_{x\in\A}x\otimes \bigoplus_{y\in\A}y&\longrightarrow \bigoplus_{y^\prime\in\A}y^\prime\otimes \bigoplus_{x^\prime\in\A}x^\prime\,.
\end{align}
The components with $k=-1$ are given by 
\begin{align}
				\begin{tikzpicture}[very thick,scale=0.9,color=red!50!black, baseline=-1.4cm, rotate=180]
					\draw[line width=0pt] 
					(5.5,3) node[line width=0pt] (ghko) {{\scriptsize$e\vphantom{ghk}$}}
					(5,3) node[line width=0pt] (go) {{\scriptsize$e\vphantom{ghk}$}}
					(4.5,3) node[line width=0pt] (hkli) {{\scriptsize$e\vphantom{ghk}$}}
					(3.5,3) node[line width=0pt] (hkre) {{\scriptsize$e\vphantom{ghk}$}}
					(3,3) node[line width=0pt] (ho) {{\scriptsize$\sigma\vphantom{ghk}$}}
					(2.5,3) node[line width=0pt] (ko) {{\scriptsize$\sigma\vphantom{ghk}$}}
					(5.5,0) node[line width=0pt] (ghku) {{\scriptsize$e\vphantom{ghk}$}}
					(5,0) node[line width=0pt] (ghli) {{\scriptsize$\sigma\vphantom{ghk}$}}
					(4.5,0) node[line width=0pt] (ku) {{\scriptsize$\sigma\vphantom{ghk}$}}
					(3.5,0) node[line width=0pt] (ghre) {{\scriptsize$\sigma\vphantom{ghk}$}} 
					(3,0) node[line width=0pt] (gu) {{\scriptsize$e\vphantom{ghk}$}} 
					(2.5,0) node[line width=0pt] (hu) {{\scriptsize$\sigma\vphantom{ghk}$}};
					\draw[dashed,redirectedred] (ghko) -- (ghku);
					\draw[postaction={decorate}, decoration={markings,mark=at position .65 with {\arrow[color=red!50!black]{<}}}] (ghli) .. controls +(0,1) and +(0,1) .. (ghre);
					\draw[dashed,postaction={decorate}, decoration={markings,mark=at position .55 with {\arrow[color=red!50!black]{<}}}] (hkre) .. controls +(0,-1) and +(0,-1) .. (hkli);
					\draw[dashed,postaction={decorate}, decoration={markings,mark=at position .65 with {\arrow[color=red!50!black]{<}}}] (gu) .. controls +(0,1.5) and +(0,-1.5) .. (go);
					\draw[redirectedred] (hu) .. controls +(0,1.5) and +(0,-1.5) ..  (ho);
					\draw[color=white, line width=4pt] (ku) .. controls +(0,1.5) and +(0,-1.5) .. (ko);
					\draw[postaction={decorate}, decoration={markings,mark=at position .65 with {\arrow[color=red!50!black]{<}}}] (ku) .. controls +(0,1.5) and +(0,-1.5) .. (ko);
					\end{tikzpicture}
					&=\kappa\bigoplus_{\substack{z,x^\prime,z^\prime\in\A\\z^\prime x^\prime=z}}\frac{q(z^\prime)}{q(z)}\id_{z^\prime x^\prime} \colon& \bigoplus_{z\in\A}z&\longrightarrow \bigoplus_{z^\prime\in\A}z^\prime\otimes \bigoplus_{x^\prime\in\A}x^\prime\,,\\
				\begin{tikzpicture}[very thick,scale=0.9,color=red!50!black, baseline=-1.4cm, rotate=180]
					\draw[line width=0pt] 
					(5.5,3) node[line width=0pt] (ghko) {{\scriptsize$\sigma\vphantom{ghk}$}}
					(5,3) node[line width=0pt] (go) {{\scriptsize$e\vphantom{ghk}$}}
					(4.5,3) node[line width=0pt] (hkli) {{\scriptsize$\sigma\vphantom{ghk}$}}
					(3.5,3) node[line width=0pt] (hkre) {{\scriptsize$\sigma\vphantom{ghk}$}}
					(3,3) node[line width=0pt] (ho) {{\scriptsize$e\vphantom{ghk}$}}
					(2.5,3) node[line width=0pt] (ko) {{\scriptsize$\sigma\vphantom{ghk}$}}
					(5.5,0) node[line width=0pt] (ghku) {{\scriptsize$\sigma\vphantom{ghk}$}}
					(5,0) node[line width=0pt] (ghli) {{\scriptsize$e\vphantom{ghk}$}}
					(4.5,0) node[line width=0pt] (ku) {{\scriptsize$\sigma\vphantom{ghk}$}}
					(3.5,0) node[line width=0pt] (ghre) {{\scriptsize$e\vphantom{ghk}$}} 
					(3,0) node[line width=0pt] (gu) {{\scriptsize$e\vphantom{ghk}$}} 
					(2.5,0) node[line width=0pt] (hu) {{\scriptsize$e\vphantom{ghk}$}};
					\draw[redirectedred] (ghko) -- (ghku);
					\draw[dashed,postaction={decorate}, decoration={markings,mark=at position .65 with {\arrow[color=red!50!black]{<}}}] (ghli) .. controls +(0,1) and +(0,1) .. (ghre);
					\draw[postaction={decorate}, decoration={markings,mark=at position .55 with {\arrow[color=red!50!black]{<}}}] (hkre) .. controls +(0,-1) and +(0,-1) .. (hkli);
					\draw[dashed,postaction={decorate}, decoration={markings,mark=at position .65 with {\arrow[color=red!50!black]{<}}}] (gu) .. controls +(0,1.5) and +(0,-1.5) .. (go);
					\draw[dashed,redirectedred] (hu) .. controls +(0,1.5) and +(0,-1.5) ..  (ho);
					\draw[color=white, line width=4pt] (ku) .. controls +(0,1.5) and +(0,-1.5) .. (ko);
					\draw[postaction={decorate}, decoration={markings,mark=at position .65 with {\arrow[color=red!50!black]{<}}}] (ku) .. controls +(0,1.5) and +(0,-1.5) .. (ko);
					\end{tikzpicture}
					&=\sgn(\kappa)\bigoplus_{\substack{x,x^\prime\in\A\\ xx^\prime=x^{\prime\prime}}}\id_{xx^\prime} \colon& \bigoplus_{x\in\A}x\otimes \bigoplus_{x^\prime\in\A}x^\prime&\longrightarrow \bigoplus_{x^{\prime\prime}\in\A}x^{\prime\prime}\,,\\
				\begin{tikzpicture}[very thick,scale=0.9,color=red!50!black, baseline=-1.4cm, rotate=180]
					\draw[line width=0pt] 
					(5.5,3) node[line width=0pt] (ghko) {{\scriptsize$e\vphantom{ghk}$}}
					(5,3) node[line width=0pt] (go) {{\scriptsize$\sigma\vphantom{ghk}$}}
					(4.5,3) node[line width=0pt] (hkli) {{\scriptsize$\sigma\vphantom{ghk}$}}
					(3.5,3) node[line width=0pt] (hkre) {{\scriptsize$\sigma\vphantom{ghk}$}}
					(3,3) node[line width=0pt] (ho) {{\scriptsize$e\vphantom{ghk}$}}
					(2.5,3) node[line width=0pt] (ko) {{\scriptsize$\sigma\vphantom{ghk}$}}
					(5.5,0) node[line width=0pt] (ghku) {{\scriptsize$e\vphantom{ghk}$}}
					(5,0) node[line width=0pt] (ghli) {{\scriptsize$\sigma\vphantom{ghk}$}}
					(4.5,0) node[line width=0pt] (ku) {{\scriptsize$\sigma\vphantom{ghk}$}}
					(3.5,0) node[line width=0pt] (ghre) {{\scriptsize$\sigma\vphantom{ghk}$}} 
					(3,0) node[line width=0pt] (gu) {{\scriptsize$\sigma\vphantom{ghk}$}} 
					(2.5,0) node[line width=0pt] (hu) {{\scriptsize$e\vphantom{ghk}$}};
					\draw[dashed,redirectedred] (ghko) -- (ghku);
					\draw[postaction={decorate}, decoration={markings,mark=at position .65 with {\arrow[color=red!50!black]{<}}}] (ghli) .. controls +(0,1) and +(0,1) .. (ghre);
					\draw[postaction={decorate}, decoration={markings,mark=at position .55 with {\arrow[color=red!50!black]{<}}}] (hkre) .. controls +(0,-1) and +(0,-1) .. (hkli);
					\draw[postaction={decorate}, decoration={markings,mark=at position .65 with {\arrow[color=red!50!black]{<}}}] (gu) .. controls +(0,1.5) and +(0,-1.5) .. (go);
					\draw[dashed,redirectedred] (hu) .. controls +(0,1.5) and +(0,-1.5) ..  (ho);
					\draw[color=white, line width=4pt] (ku) .. controls +(0,1.5) and +(0,-1.5) .. (ko);
					\draw[postaction={decorate}, decoration={markings,mark=at position .65 with {\arrow[color=red!50!black]{<}}}] (ku) .. controls +(0,1.5) and +(0,-1.5) .. (ko);
					\end{tikzpicture}
					&=|\kappa|\bigoplus_{\substack{x,z,y^\prime,z^\prime\in\A\\zx=z^\prime y^\prime}}\frac{q(z^\prime)}{q(zx)}\id_{zx}\colon& \bigoplus_{z\in\A}z\otimes \bigoplus_{x\in\A}x&\longrightarrow \bigoplus_{z^\prime\in\A}z^\prime\otimes \bigoplus_{y^\prime\in\A}y^\prime\,,\\
				\begin{tikzpicture}[very thick,scale=0.9,color=red!50!black, baseline=-1.4cm, rotate=180]
					\draw[line width=0pt] 
					(5.5,3) node[line width=0pt] (ghko) {{\scriptsize$\sigma\vphantom{ghk}$}}
					(5,3) node[line width=0pt] (go) {{\scriptsize$\sigma\vphantom{ghk}$}}
					(4.5,3) node[line width=0pt] (hkli) {{\scriptsize$e\vphantom{ghk}$}}
					(3.5,3) node[line width=0pt] (hkre) {{\scriptsize$e\vphantom{ghk}$}}
					(3,3) node[line width=0pt] (ho) {{\scriptsize$\sigma\vphantom{ghk}$}}
					(2.5,3) node[line width=0pt] (ko) {{\scriptsize$\sigma\vphantom{ghk}$}}
					(5.5,0) node[line width=0pt] (ghku) {{\scriptsize$\sigma\vphantom{ghk}$}}
					(5,0) node[line width=0pt] (ghli) {{\scriptsize$e\vphantom{ghk}$}}
					(4.5,0) node[line width=0pt] (ku) {{\scriptsize$\sigma\vphantom{ghk}$}}
					(3.5,0) node[line width=0pt] (ghre) {{\scriptsize$e\vphantom{ghk}$}} 
					(3,0) node[line width=0pt] (gu) {{\scriptsize$\sigma\vphantom{ghk}$}} 
					(2.5,0) node[line width=0pt] (hu) {{\scriptsize$\sigma\vphantom{ghk}$}};
					\draw[redirectedred] (ghko) -- (ghku);
					\draw[dashed,postaction={decorate}, decoration={markings,mark=at position .65 with {\arrow[color=red!50!black]{<}}}] (ghli) .. controls +(0,1) and +(0,1) .. (ghre);
					\draw[dashed,postaction={decorate}, decoration={markings,mark=at position .55 with {\arrow[color=red!50!black]{<}}}] (hkre) .. controls +(0,-1) and +(0,-1) .. (hkli);
					\draw[postaction={decorate}, decoration={markings,mark=at position .65 with {\arrow[color=red!50!black]{<}}}] (gu) .. controls +(0,1.5) and +(0,-1.5) .. (go);
					\draw[redirectedred] (hu) .. controls +(0,1.5) and +(0,-1.5) ..  (ho);
					\draw[color=white, line width=4pt] (ku) .. controls +(0,1.5) and +(0,-1.5) .. (ko);
					\draw[postaction={decorate}, decoration={markings,mark=at position .65 with {\arrow[color=red!50!black]{<}}}] (ku) .. controls +(0,1.5) and +(0,-1.5) .. (ko);
					\end{tikzpicture}
					&=\kappa \bigoplus_{\substack{y,z,x^\prime,z^\prime\in\A\\yz=x^\prime z^\prime}}\gamma(y,z,z^\prime)\id_{yz}
					\colon& \bigoplus_{y\in\A}y\otimes \bigoplus_{z\in\A}z&\longrightarrow \bigoplus_{x^\prime\in\A}x^\prime\otimes \bigoplus_{z^\prime\in\A}z^\prime\,,
\end{align}
where $\gamma(y,z,z^\prime)=\frac{q(y)q(z^\prime yz)q(z^\prime)}{q(yz)q(yz^\prime)}$. 
The components of~$\overline{\alpha}$ are the inverses of those of~$\alpha$ up to insertions of~$\psi$, such that $\alpha\circ(\id\otimes \psi_0^2)\circ \overline{\alpha}=\id$ where $\psi_0^2$ is the insertion of $\psi^2$ on the $hk$-line in \eqref{eq:alphabarTY}.
Viewing $\sgn(\kappa)$ as the defectification~$p$ again, the above signs match \eqref{eq:simplifiedalpha}.

For completeness, we briefly mention that the $\mathbb{A}_{\Z_2}$-orbifold $(\A$-$\Vect)_{\mathbb{A}_{\Z_2}}$ is the fusion category whose simple objects are 
\begin{itemize}
	\item 2 invertible objects $X_\pm^R=(A_{\Z_2},\tau^{e,\pm})$,
	\item $\frac{n-1}{2}$ 2-dimensional objects $Y_a^R=((a\oplus A_{-1})\oplus (a^{-1}\oplus A_{-1}),\tau^a)$,
	\item 2 $\sqrt{n}$-dimensional objects $Z_\pm^R=((e\otimes A_{-1})\oplus (A_{-1}\otimes e),\tau^{\sigma,\pm})$,
\end{itemize} 
according to \eqref{eq:FonX} with the same fusion rules and braiding as above.

Lastly, the de-equivariantising (1-form symmetry) algebra is $B=X_+\oplus X_-$, with the obvious algebra structure given by a sum of projections.
\begin{equation}
	(X_+\oplus X_-)\otimes (X_+\oplus X_-)=(X_+\oplus X_-)\oplus (X_-\oplus X_+)\longrightarrow X_+\oplus X_-\,.
\end{equation}
The comultiplication is given by $1/2$ times the diagonal map, and unit and counit are given by the inclusion and twice the projection, respectively.
Note that calculating the de-equivariantising algebra is very simple when we describe the gauging of the 0-form symmetry via equivariantisation, since we know immediately which equivariant objects are based on the unit~$\mathds{1}$.

\subsubsection{Ising Categories}
\label{sec:Ising}
For $\A=\Z_2$, the $\Z_2$-action \eqref{eq:TYaction} is trivial and thus the $\Z_2$-crossed braiding is a genuine braiding. 
Additionally, there is only a single non-degenerate bicharacter and it allows 2 options for the quadratic form, determined by $q(-1)=\pm i$.
In combination with the signs of $\kappa$ and $\delta$, both of the Ising categories $\TY(\Z_2,\chi,\pm\frac{1}{\sqrt{2}})$ therefore admit 4 braided structures each.
	Let us denote 
\begin{align}
	\varepsilon_q&=-iq(-1)\,, &&\text{s.t. }q(-1)=\varepsilon_q \cdot i\,,\\
	\varepsilon_\kappa&=\sgn(\kappa)\,,&&\\
	\varepsilon_\delta&\in\{\pm1\}\,,&
	&\text{s.t. }\delta=\exp\left( \frac{2\pi i}{16}(-\varepsilon_q+2(1-\varepsilon_\kappa)+4(1-\varepsilon_\delta)\right) \,.
\end{align}
	We write $\TY(\Z_2,\varepsilon_q,\varepsilon_\kappa,\varepsilon_\delta)$ for these modular fusion categories.
	
	Ising type modular fusion categories can also be classified by an odd integer $\eta$ mod 16 corresponding to the twist $\theta_\sigma=\exp(\frac{2\pi i}{16} \eta)$.
This description is used in \cite{BBCW}, and it relates to our parametrisation via $\eta=6+\varepsilon_q-2\varepsilon_\kappa-4\varepsilon_\delta$ mod 16.
	
Since the $\Z_2$-action is strict and -- in this case -- also trivial, all simple objects pair with linear $\Z_2$-representations and the equivariantisation is simply given by
	\begin{equation}
		\TY(\Z_2,\varepsilon_q,\varepsilon_\kappa,\varepsilon_\delta)^{\Z_2}\cong \Rep(\Z_2)\boxtimes\TY(\Z_2,\varepsilon_q,\varepsilon_\kappa,\varepsilon_\delta)\,.
	\end{equation}
The braiding is $c_{(X,\pi),(Y,\pi^\prime)}^{\Z_2}:=\pi(|Y|)c_{X,Y}$ , so it differs from the Tambara--Yamagami braiding by a sign if~$X$ comes with the sign representation and $Y=\sigma$.

The 0-form orbifold datum is described by the formulas we gave in \Cref{sec:oddOrderGroups}.
The de-equivariantising (1-form symmetry) algebra is $B=X_+\oplus X_-$ with the previous Frobenius algebra structure.

\section{Symmetries in $\boldsymbol{\mathcal{D}(\Z_2)}$}
\label{sec:DZ2}
In this section we gauge symmetries in $\mathcal{D}(\Z_2)$ which is 3-dimensional Dijkgraaf--Witten $\Z_2$-gauge theory.
The fusion rules of the line defects are identical to the anions in the toric code and we adopt the notation $\mathds{1}$, $e$, $m$, and $f=e\otimes m$.
Our treatment includes gauging a 1-form symmetry to obtain the trivial theory, gauging all possible 0-form symmetries, and combining these two techniques in a 2-group symmetry. 
Among these examples are some of those of \cite{BBDR} to which we add the perspective of orbifolds and extend them by adding defectification, see also \cite[Sect.\,XI.I]{BBCW} for parts of the 0-form symmetries. 

In \cite{BBDR}, the authors take the following approach: 
\begin{enumerate}
\item They start in a ``larger'' theory with topological defects (i.e.\ symmetry) described by a fusion category $\mathcal{B}_1$. 
\item They then gauge a 1-form symmetry by finding a corresponding algebra~$A_L$ (of \textsl{L}ine operators) in~$\mathcal{B}_1$ which is condensed into a surface $S_{A_L}$ which in turn is used to gauge. $A_L$ is precisely the algebra which we use for the orbifold datum, and when this algebra comes from an invertible symmetry (described by the group $H=\widehat{G}$), it is the group algebra we use for de-equivariantisation.
\item Upon gauging, the new topological defects are described by $\mathcal{B}_2$.
\item They then construct an algebra corresponding to the dual 0-form symmetry, given by a \textsl{S}urface~$\hat{\mathcal{A}}_S\in \Mod(\mathcal{B}_2)$.
\end{enumerate}
While this is the way they present their examples, to compute them, they instead start with module categories of $\mathcal{D}(\Z_2)$ (i.e.\ surface defects) and try to find algebra structures on them.
As before, the resulting surfaces $\hat{\mathcal{A}}_S$ are condensation defects and thus represented by algebras, and these are the algebras which underlie the corresponding orbifold datum (cf. \Cref{sec:RT}).

Among their examples are non-invertible symmetries which -- by virtue of being non-invertible -- cannot be covered by the methods laid out in this paper, such as equivariantisation and de-equivariantisation. 
However, the algebras they use still constitute orbifold data and so they can also be gauged through the orbifold construction. 
In fact, any gauging of a condensable algebra (1-form symmetry) corresponds to an orbifold datum and in Reshetikhin--Turaev theory this can always be undone by another orbifold datum, the ``condensation inversion'' described in \cite{Mulevicius2022}. 
In order to keep this section in line with the others, we restrict our attention to the examples given by invertible symmetries. 

When the algebra describes an invertible symmetry, the gauged theory is once again given by the equivariantisation of the $G$-crossed braided fusion category $(\mathcal{B}_2)^\times_G$ (including twisted sectors) \cite{CH,HPRW}.
We collect the correspondence between the two approaches in \Cref{tab:terminology}.
\begin{table}[h]
\begin{center}
\begin{tabular}{l|c|c}
	& \cite{BBDR} & This paper\\\hline
	Topological defects & $\mathcal{B}_1$& category after eq.\\\hline
	1-form $\widehat{G}$-symmetry algebra & $A_L$& $B:=\C[\widehat{G}]\in\Rep(G)\subset\mathcal{B}_1$\\
	1-form symmetry gauging&$\mathcal{B}_2:=(\mathcal{B}_1)^\text{loc}_{A_L}$&$(\mathcal{B}_1)_{\mathbb{A}_B}$ orbifold\\\hline
	Including twisted sectors&$(\mathcal{B}_1)_{A_L}$& $(\mathcal{B}_2)^\times_G\cong(\mathcal{B}_1)_G$ de-eq.\\\hline
	0-form $G$-symmetry algebra& $\hat{\mathcal{A}}_S\cong(\mathcal{B}_2)_{A_{\scaleto{G}{4pt}}}$& $A_G$\\
	0-form symmetry gauging & \eqref{eq:0-formGaugingViaMoritaDual}& $\mathcal{B}_1\cong ((\mathcal{B}_2)^\times_G)^G$ eq.
\end{tabular}
\caption{Comparison of terminology. Categories of (local) $A$-modules in~$\mathcal{C}$ are denoted~$\mathcal{C}_A^{\text{(loc)}}$.}\label{tab:terminology}
\end{center}
\end{table}
Note that in the third line, the orbifold datum ${\mathbb{A}_B}$ is obtained from the algebra~$B$ via \eqref{eq:orbdat from cssFrob} \cite[Prop.\,3.15]{CRS3}.
We shall point out that the first four rows of this table are equalities (or equivalences), i.e.\ both approaches describe the same data.
Instead of using explicit algebras, in \cite{BBDR}, 0-form gauging is treated via Morita-duality of module categories which gives the relation 
\begin{equation} 
	\Mod(\mathcal{B}_2)^*_{\Mod(\hat{\mathcal{A}}_S)}\simeq\Mod(\mathcal{B}_1)\label{eq:0-formGaugingViaMoritaDual}
\end{equation} 
in this setting, which is due to \cite[Ex.\,5.3.8]{DecoppetFus2}.

Let us unpack this briefly, a more detailed treatment can be found in \cite[Sect.\,3.6, App.\,A]{BBDR}. 
Just as the category of $A$-modules for an algebra~$A$ in a fusion category $\mathcal{C}$ is a $\mathcal{C}$-module category by the fusion product in $\mathcal{C}$, so is the 2-category $\Mod(\hat{\mathcal{A}}_S)$ of $\mathcal{A}_S$-module categories a $\Mod(\mathcal{B}_2)$-module 2-category. 
Physically, $\Mod(\mathcal{B}_2)$ corresponds to (condensed) surface defects in the original theory whereas $\Mod(\hat{\mathcal{A}}_S)$ corresponds to interfaces between the gauged theory and the original theory. 
Naturally, then, surfaces in $\Mod(\mathcal{B}_2)$ can be inserted and fused with the interface, giving rise to the $\Mod(\mathcal{B}_2)$-module structure.

The Morita-dual is given by $\Mod(\mathcal{B}_2)$-module endofunctors of $\Mod(\hat{\mathcal{A}}_S)$.
These tranform one interface into another in a way that is compatible with fusing with surfaces from $\Mod(\mathcal{B}_2)$.
Any such transformation can be identified with a surface defect ``on the other side'', i.e.\ in the gauged theory, hence the transformations coincide with $\Mod(\mathcal{B}_1)$.
Conversely, each surface in $\Mod(\mathcal{B}_1)$ can be fused with any interface in $\Mod(\hat{\mathcal{A}}_S)$, and this fusion produces a functor $\Mod(\hat{\mathcal{A}}_S)\longrightarrow \Mod(\hat{\mathcal{A}}_S)$ for each surface.

\medskip

In the following sections, we present the results of gauging through the following diagram (where we set $H\cong G\cong \Z_2$ compared to \Cref{tab:terminology}):
\begin{equation}
\begin{tikzcd}[column sep= 20, ampersand replacement=\&]
	\mathcal{B}_1
    \arrow[rrrrr, "{\substack{\text{1-form gauging }\widehat{=}\text{ de-eq.}\\A_L=B=\C(\Z_2)}}"{above}]
    \arrow[d,phantom,sloped, "\cong"]
    \&\&\&\&\&
    (\mathcal{B}_1)^\text{loc}_{A_L}\stackrel{\eqref{eq:orbifold category of a condensable algebra}}{\cong} (\mathcal{B}_1)_{\mathbb{A}_B}
    \arrow[d,phantom,sloped, "\cong",shift right=14]
    \arrow[rr,phantom, "\subset"]
    \&\&
     (\mathcal{B}_1)_{\Z_2}\equiv \,(\mathcal{B}_1)_{A_L}
    \arrow[d,phantom,sloped, "\cong",shift right=11]
    \\
     \hspace{-20pt}((\mathcal{B}_2)^\times_{\Z_2})^{\Z_2}
     \&\&\&\&\&
     \hphantom{(}\mathcal{B}_2\hphantom{)_{A_L}\stackrel{\eqref{eq:orbifold category of a condensable algebra}}{\cong} (\mathcal{B}_1)_{\mathbb{A}_B}}
    \arrow[lllll, "{\substack{A_{\Z_2}\\ \hat{\mathcal{A}}_S\cong(\mathcal{B}_2)_{A_{\scaleto{\Z_2}{4pt}}}\\\text{0-form gauging }\widehat{=}\text{ eq.}}}"{below}]
    \arrow[rr,phantom, "\subset"]
    \&\&
     (\mathcal{B}_2)^\times_{\Z_2}\hphantom{\equiv \,(\mathcal{B}_1)_{A_L}}
\end{tikzcd}
\label{eq:gaugingUngaugingBBDR}
\end{equation}

In these examples the initial category is the double $\mathcal{B}_2=\mathcal{D}(\Z_2):=\mathcal{Z}(\Z_2\text{-}\Vect)$, i.e.\ 3-dimensional Dijkgraaf--Witten $\Z_2$-gauge theory. 
The only exception is \Cref{sec:Z2-Vect} where we start from the trivial theory $\mathcal{B}_2=\Vect$ to obtain $\mathcal{B}_1=\mathcal{D}(\Z_2)$ by gauging a (degenerate) 0-form symmetry.
We note that $\mathcal{D}(\Z_2)=\mathcal{Z}(\Z_2\text{-}\Vect)\cong (\Z_2\oplus\hat{\Z}_2)\text{-}\Vect\cong \Z_2\text{-}\Vect\boxtimes \Rep(\Z_2)$ where elements of the Pontryagin-dual $\hat{\Z}_2:=\Hom_{\Grp}(\Z_2,\U)$ correspond to the choices of half-braidings in the Drinfeld center.

The simple objects are $\mathds{1}=(\C,\C_+)$, $m:=(\tilde\C,\C_+)$, $e:=(\C,\C_-)$, and $f:=(\tilde\C,\C_-)$, where the simple objects of $\Z_2$-$\Vect$ are denoted by $\C$ (degree 1) and~$\tilde\C$ (degree -1), and the simple objects of $\Rep(\Z_2)$ by $\C_+$ (trivial representation) and $\C_-$ (sign representation).
The monoidal structure is given by the canonical choices for $\Z_2$-$\Vect$ and $\Rep(\Z_2)$, i.e.\ associators and unitors are trivial.
The braiding is given by 
\begin{equation}
	c^{\mathcal{D}(\Z_2)}_{X,Y}=
	\begin{cases}
	-\id\qquad &X\in\{e,f\}, Y\in\{m,f\}\,,\\
	\hphantom{-}\id\qquad&\text{else}.
	\end{cases}
	\label{eq:braidingInDZ2}
\end{equation} 
All objects are their own inverses and therefore self-dual, with identity evaluation and coevaluation morphisms and strict pivotal structure.
We write $S_\mathds{1}$ for the module category over the trivial algebra $\mathds{1}$, and $S_\gamma$ for the module category over the algebra $\mathds{1}\oplus \gamma$ for $\gamma\in \lbrace m,e,f\rbrace$.
The latter are different incarnations of the group algebra~$\C[\Z_2]$.
 
The following sections now proceed as follows: We present a diagram similar to \eqref{eq:gaugingUngaugingBBDR} for each example and show how these results are obtained using orbifold data. 
For the 0-form symmetries, we focus on the underlying algebra to compare to the surfaces $\hat{\mathcal{A}}_S$ and the line fusing two of these surfaces which detects symmetry fractionalisation.
For the 1-form symmetry, we provide the underlying algebra, which gives rise to an orbifold datum as discussed in \Cref{sec:deeq}.
Note that in contrast to \cite{BBDR}, we start with the theory $(\mathcal{B}_2)^\times_{\Z_2}$ (including twisted sectors) and then recover~$\mathcal{B}_1$ via equivariantisation. 
The section titles read $\mathcal{B}_2\longrightarrow\mathcal{B}_1$, i.e.\ we gauge a 0-form symmetry going from left to right.

\subsection[$\Vect\rightarrow\mathcal D(\Z_2)$]{$\boldsymbol{\Vect\rightarrow\mathcal{D}^{(p)}(\Z_2)}$}
\label{sec:Z2-Vect}
In this section, we gauge a 0-form $\Z_2$-symmetry in $\Vect$, given by the $\Z_2$-extension of~$\Vect$, $\Z_2\text{-}\Vect$, i.e.\ there is one simple object in the twisted sector, the $(-1)$-graded vector space~$\tilde\C\widehat{=}m$.
The results are as follows: 

\begin{equation}
\begin{tikzcd}[column sep= 20, ampersand replacement=\&]
	\mathcal{D}^{(p)}(\Z_2)
    \arrow[rrrrr, "{\substack{\text{1-form gauging }\widehat{=}\text{ de-eq.}\\A_L=B=\C(\Z_2)=\mathds{1}\oplus e}}"{above}]
    \arrow[d,phantom,sloped, "\cong"]
    \&\&\&\&\&
    (\mathcal{D}^{(p)}(\Z_2))^\text{loc}_{A_L}
    \arrow[d,phantom,sloped, "\cong",shift right=3]
    \arrow[r,phantom, "\subset"]
    \&
     (\mathcal{D}^{(p)}(\Z_2))_{\Z_2}\equiv \,(\mathcal{D}^{(p)}(\Z_2))_{A_L}
    \arrow[d,phantom,sloped, "\cong",shift right=14]
    \\
     \hspace{-20pt}(\Z_2\text{-}\Vect^{(p)})^{\Z_2}
     \&\&\&\&\&
     \hphantom{(}\Vect\hphantom{)_{A_L}}
    \arrow[lllll, "{\substack{A_{\Z_2}=\,\C\oplus\C\\ \hat{\mathcal{A}}_S\cong(\Vect)_{A_{\scaleto{\Z_2}{4pt}}}=\, S_\mathds{1}\boxplus S_\mathds{1}\\\text{0-form gauging }\widehat{=}\text{ eq.}}}"{below}]
    \arrow[r,phantom, "\subset"]
    \&
     (\Vect)^\times_{\Z_2}\hphantom{)}\equiv \,\Z_2\text{-}\Vect^{(p)}
\end{tikzcd}
\label{tab:Z2-Vect}
\end{equation}

\medskip

\noindent\textbf{Trivial defectification} $p=1$: This means the associator of $\Z_2\text{-}\Vect$ is trivial.
The category has a $\Z_2$-crossed braided structure with the trivial strict $\Z_2$-action and the trivial braiding.

The 0-form symmetry is gauged via the algebra \eqref{eq:AG}
\begin{equation}
	A_{\Z_2}=(\C\otimes\C)\oplus(\tilde\C\otimes \tilde\C)=\C\oplus\C\,.
\end{equation}
This is a direct sum of two copies of the trivial algebra, so the associated category of modules corresponds indeed to $S_\mathds{1}\boxplus S_\mathds{1}$. 
We can then compute the equivariantisation of $\Z_2\text{-}\Vect$: Since the action is trivial, all simple objects have~$\Z_2$ as their stabilizers and because it is strict, we can choose $\xi=\id$ in \eqref{eq:projectivitycocycle}.
Thus, the simple objects of the equivariantisation $\left(\Z_2\text{-}\Vect\right)^{\Z_2}$ are pairs $(X,V)$ where $X\in\Z_2\text{-}\Vect$ is simple and $V\in\Rep(\Z_2)$ is an irreducible representation (cf. \Cref{lem:simplesInEquivar}), so we recover $\mathcal{D}(\Z_2)$. 
The fact that this statement holds for the fusion rules can be calculated as $(X_1,V_1)\otimes (X_2,V_2)=(X_1\otimes X_2,V_1\otimes V_2)$ (following \Cref{lem:FusionRulesOfSimplesInEq}). 
For the braiding, recall that the braidings of $\Z_2$-$\Vect$ and $\Rep(\Z_2)$ are trivial, so the equivariant structures are the only source of non-trivial braiding components in the equivariantisation.
More precisely, the braiding $c^{\Z_2}_{(X_1,V_1),(X_2,V_2)}$ of \eqref{eq:eq of Gcbc is braided} between simple objects contains a factor of $u_{|X_2|}$ where $u$ is the $\Z_2$-action on $V_1$.
This is non-trivial if and only if $V_1=\C_-$ is the sign representation and $X_2=\tilde\C$ is of non-trivial degree, in which case the braiding obtains a factor of $u_{|X_2|}=\sgn(-1)=-1$, precisely matching \eqref{eq:braidingInDZ2}.

Let us briefly turn to the emergent 1-form $\widehat{\Z}_2$- or equivalently $\Z_2$-symmetry.
The inclusion $\Rep(\Z_2)\subset \mathcal{D}(\Z_2)$ is given by pairs of the type $(\C,\C_\pm)$, i.e.\ $\mathds{1}$ and $e$. 
Thus, $\C(\Z_2)\cong \mathds{1}\oplus e$ is the algebra associated to the 1-form symmetry.
As a sanity check, we can also calculate its gauging.
As expected, $\C(\Z_2)$-$\Mod\cong \Z_2$-$\Vect$ recovers the extension of $\Vect$ we started with.

\medskip

\noindent\textbf{Non-trivial defectification} $p=-1$: We obtain the twisted extension $\Z_2$-$\Vect^p$, where $p$ represents the non-trivial element of $\mathrm{H}^3(\Z_2,\U)=\Z_2$, which has $p(-1,-1,-1)=-1$ as its only non-trivial component.
This cocycle modifies the usual associator $\alpha_{X,Y,Z}$ by the factor $p(|X|,|Y|,|Z|)$.
Understanding this as a $G$-crossed braided zesting of $\Z_2$-$\Vect$ (with trivial $\lambda$), we can read off from \eqref{eq:zestedtensorator} and \eqref{eq:zestedcompositor} that the monoidal structure of the $G$-action is also modified, leading to non-trivial cocycles for the projective representations in the equivariantisation (cf.\,\eqref{eq:projectivitycocycle}).
In this case, we can make the trivial choice $\xi_{-1}=\xi_1=\id_{\tilde\C}$ to obtain 
\begin{equation}
	\beta_{Y}(g,h)=p(g,h,|Y|)\,,
\end{equation}
resulting in genuinely projective representations for $Y=\tilde\C$. 
Therefore, simple objects in the gauged theory are given by $\{\C_\pm,\tilde\C_{\pm i}\}$.
The conventional labels for these objects are $\mathds{1}:=\C_+$, $s:=\tilde\C_{+i}$, $\overline{s}:=\tilde\C_{-i}$, $s\overline{s}:=\C_-$ where the letter ``s'' is chosen to indicate that, physically, these are \textsl{semions}. 
This means they braid with $\pm i$ with themselves -- halfway between bosonic $+1$ and fermionic $-1$ -- as we show below. 
Note that in the equivariant structure \eqref{eq:equivariantStructureOfMonoidalProducts}, the non-trivial cocycle~$p$ appears once again, such that the equivariant structure on $s\otimes \overline{s}$ is not simply given by the product of its irreps $u^s_{-1}\otimes u^{\overline{s}}_{-1}=i\cdot(-i)=1$, but obtains an additional factor of $(\rho^2_{-1})^{-1}_{s,\overline{s}}=p(-1,|s|,|\overline{s}|)=-1$.
Thus, we indeed have $s\otimes \overline{s}=s\overline{s}$ and all simple objects are self-inverse (for the same reason).
Associators are again solely dependent on the underlying objects, so the only non-trivial components are $\alpha_{X,Y,Z}=-\id$ whenever $X,Y,Z\in\{s,\overline{s}\}$.
The resulting category $\mathcal{D}^p(\Z_2)$ is known as the \textsl{double semion} model.\footnote{There is another convention for this model, where only the associator $\alpha_{s,s,s}$ is non-trivial, while the mixed associators and $\alpha_{\overline{s},\overline{s},\overline{s}}$ are trivial.}
Note that the fusion ring (\Cref{tab:DoubleSemionFusion}) is the same as that of $\mathcal{D}(\Z_2)$, hence the notation.
\begin{table}[h]
	{\renewcommand{\arraystretch}{1.5}\setlength{\tabcolsep}{12pt}
	\begin{center}
		\begin{tabular}{c!{\vrule width 1pt} c|c|c|c}
			&$\mathds{1}$&$s$&$\overline{s}$&$s\overline{s}$\\\noalign{\hrule height 1pt}
			$\mathds{1}$&$\mathds{1}$&$s$&$\overline{s}$&$s\overline{s}$\\\hline
			$s$&$s$&$\mathds{1}$&$s\overline{s}$&$\overline{s}$\\\hline
			$\overline{s}$&$\overline{s}$&$s\overline{s}$&$\mathds{1}$&$s$\\\hline
			$s\overline{s}$&$s\overline{s}$&$\overline{s}$&$s$&$\mathds{1}$
		\end{tabular}
	\end{center}}
	\caption{Fusion rules for the double semion model $\mathcal{D}^p(\Z_2)$.}
	\label{tab:DoubleSemionFusion}
	\end{table}\noindent
	
Lastly, the braiding is given by
\begin{equation}
	c^{\mathcal{D}^p(\Z_2)}_{X,Y}=
	\begin{cases}
		\hphantom{\pm i}\,\id\qquad& Y\in\{\mathds{1}, s\overline{s}\}\,,\\
		\hphantom{\pm i}\,\id \qquad & X=\mathds{1},  Y\in\{s,\overline{s}\}\,,\\
		-\hphantom{i}\,\id \qquad & X=s\overline{s}, Y\in\{s,\overline{s}\}\,,\\
		\hphantom{\pm}i\,\id \qquad & X=s, Y\in\{s,\overline{s}\}\,,\\
		-i\,\id \qquad &  X=\overline{s}, Y\in\{s,\overline{s}\}\,,
	\end{cases}
	\label{eq:semionBraiding}
\end{equation}
justifying the name ``semion'' for this object, as it satisfies $c^{\Z_2}_{s,s}=i\,\id_\mathds{1}$.

From the point of view of orbifolds, the 0-form algebra is $A_{\Z_2}^p=\C\oplus\C$ where the first term carries the trivial Frobenius algebra structure as is the case in $\Z_2$-$\Vect$, but the second term obtains an additional sign, such that $\mu_{-1}=-\mu_1$ and similarly $\eta_{-1}=-\eta_1$, $\Delta_{-1}=-\Delta_1$, and $\varepsilon_{-1}=-\varepsilon_1$ (cf. the sign appearing in \eqref{eq:TYZ2algstructure} which corresponds exactly to~$p$). 
However, this algebra is Morita-equivalent to the trivial one, i.e.\ the categories of modules in $\Vect$ of $\mathbb{A}_{\Z_2}$ and~$\mathbb{A}_{\Z_2}^p$ are equivalent, explicitly confirming our discussion in \Cref{sec:zesting} and \eqref{eq:zestedMoritaEquivalence}.
The associator of $\Vect_{\mathbb{A}_{\Z_2}^p}\cong \mathcal{D}^p(\Z_2)$ is then inherited from $\Z_2\text{-}\Vect^p$ through the algebra structure, thereby distinguishing the two gaugings (as does the braided fusion structure). 

Conversely, the 1-form algebra is $B=\mathds{1}\oplus s\overline{s}$ and the non-trivial associator of its module category $\Z_2\text{-}\Vect^p$ (i.e.\ de-equivariantisation) spanned by~$B$ and $s\oplus\overline{s}$ is inherited from that in $\mathcal{D}^p(\Z_2)$.

\subsection[0-form Symmetries in $\mathcal D(\Z_2)$]{0-form Symmetries in $\boldsymbol{\mathcal{D}(\Z_2)}$}
\label{sec:DZ20}
In this section, we gauge all 0-form $\Z_2$-symmetries of $\mathcal{D}(\Z_2)$. 
We start with the trivial $\Z_2$-action on objects and consider possible extensions before turning to electric-magnetic duality which exchanges~$e$ and~$m$. 
Since the braided autoequivalences of $\mathcal{D}(\Z_2)$ are $\Aut^\text{br}(\mathcal{D}(\Z_2))=\Z_2$ \cite{BBCW}, there are no other $\Z_2$-actions.

For the trivial $\Z_2$-action, there are eight inequivalent $\Z_2$-crossed braided extensions.
These are determined by $\lambda(-1,-1)\in\lbrace \mathds{1},m,e,f\rbrace$ and $p\in\lbrace \pm 1\rbrace$.
For electric-magnetic duality, we have a unique $\lambda(-1,-1)= \mathds{1}$ and $p\in\lbrace \pm1\rbrace$.

While the $\Z_2$-action is trivial on objects and morphisms (even in the twisted sectors), the monoidal structure of the action depends on the choice of fractionalisation. 
More precisely, the morphism $(\rho^2_{g,h})_{X_k}$ carries a factor determined by the S-matrix, such that (cf. \cite[Eq.\,(604)]{BBCW})
\begin{equation}
	(\rho_{g,h}^2)_{X_k}=2S_{X,\lambda(h,g)}\cdot p(g,h,k)\,,
	\label{eq:compositorconventionBBCW}
\end{equation}
where for $X_{-1}=\tilde{X}$ in the twisted sector, the factor $S_{X,\lambda(g,h)}$ is determined by the underlying object~$X$ in $\mathcal{D}(\Z_2)$.
This can also be calculated directly from \eqref{eq:compositor}, where the factor arises from the strand labelled $\lambda(h,g)$ which wraps around~$X$.
The S-matrix for $\mathcal{D}(\Z_2)$ (with entries multiplied by $2$) is given by \Cref{tab:SmatrixDZ2}.
Note that this only comes into play once we consider non-trivial fractionalisation.
\begin{table}[h]
\begin{center}
\begin{tabular}[t]{|c|c|c|c|c|} 
		\hline
		&$\mathds{1}$&$m$&$e$&$f$\\
		\hline
        $\mathds{1}$&$1$&$1$&$1$&$1$ \\\hline 
        $m$&$1$&$1$&$-1$&$-1$ \\\hline
        $e$&$1$&$-1$&$1$&$-1$\\\hline 
        $f$&$1$&$-1$&$-1$&$1$\\\hline 
\end{tabular}
\caption{$S$-matrix of $\mathcal{D}(\Z_2)$. Each value has been multiplied by $2$ to avoid fractions.}
\label{tab:SmatrixDZ2}
\end{center}
\end{table}
Lastly, recall from \Cref{nota:conventions} how our notation for the monoidal structure of the $G$-action compares to that of \cite{BBCW,DGPRZ}.

\subsubsection{Trivial Symmetry Fractionalisation: $\boldsymbol{\mathcal{D}(\Z_2)\rightarrow \mathcal{D}(\Z_2\times\Z_2)}$ and $\boldsymbol{\mathcal{D}(\Z_2)\rightarrow \mathcal{D}^p(\Z_2)\boxtimes\mathcal{D}(\Z_2)}$}
\label{sec:D4}
In this section we gauge the ``simplest'' 0-form $\Z_2$-symmetry in $\mathcal{D}(\Z_2)$. 
Starting from a trivial $\Z_2$-action and trivial extension (trivial fractionalisation with potentially non-trivial defectification~$p$), we obtain a theory with symmetry $\mathcal{D}^{(p)}(\Z_2)\boxtimes\mathcal{D}(\Z_2)$ upon gauging. 
We denote objects of $(\Z_2\times\Z_2)\text{-}\Vect$ as $\mathds{1}$, $m_1$, $m_2$, and $m_1m_2$, objects of $\Rep(\Z_2\times\Z_2)$ as $\mathds{1}$, $e_1$, $e_2$, and $e_1e_2$, and objects of $\mathcal{D}(\Z_2)\boxtimes\mathcal{D}(\Z_2)\cong\mathcal{D}(\Z_2\times\Z_2)\cong (\Z_2\times\Z_2)$-$\Vect\boxtimes \Rep(\Z_2\times\Z_2)$ accordingly. 
For $\mathcal{D}^p(\Z_2)\boxtimes\mathcal{D}(\Z_2)$ we use the previous notation for $\mathcal{D}(\Z_2)$ without subscripts and the notation of \Cref{sec:Z2-Vect} for $\mathcal{D}^p(\Z_2)$.
Note that in this case all objects are self-dual. 
We obtain the following results:
\begin{equation}
\begin{tikzcd}[column sep= 20, ampersand replacement=\&]
	\mathcal{D}(\Z_2^{(p)}\times\Z_2)
    \arrow[rrrr, "{\substack{\text{1-form gauging }\widehat{=}\text{ de-eq.}\\A_L=B=\C(\Z_2)=\mathds{1}\oplus e_1}}"{above}]
    \arrow[d,phantom,sloped, "\cong"]
    \&\&\&\&
    (\mathcal{D}(\Z_2^{(p)}\times\Z_2))^\text{loc}_{A_L}
    \arrow[d,phantom,sloped, "\cong",shift right=3]
    \arrow[r,phantom, "\subset"]
    \&
     (\mathcal{D}(\Z_2^{(p)}\times\Z_2))_{A_L}
    \arrow[d,phantom,sloped, "\cong",shift right=11]
    \\
     \hspace{-12pt}(\Z_2^{(p)}\boxtimes\mathcal{D}(\Z_2))^{\Z_2}
     \&\&\&\&
     \hphantom{(}\mathcal{D}(\Z_2)\hphantom{)_{A_L}}
    \arrow[llll, "{\substack{A_{\Z_2}=\,\mathds{1}\oplus\mathds{1}\\ \hat{\mathcal{A}}_S\cong(\mathcal{D}(\Z_2))_{A_{\scaleto{\Z_2}{4pt}}}=\, S_\mathds{1}\boxplus S_\mathds{1}\\\text{0-form gauging }\widehat{=}\text{ eq.}}}"{below}]
    \arrow[r,phantom, "\subset"]
    \&
     (\mathcal{D}(\Z_2))^\times_{\Z_2}\hphantom{)}\equiv \,\Z_2^{(p)}\boxtimes\mathcal{D}(\Z_2)
\end{tikzcd}
\label{tab:D4}
\end{equation}
where we used the shorthand $\mathcal{D}(\Z_2^{(p)}\times\Z_2):=\mathcal{D}^{(p)}(\Z_2)\boxtimes\mathcal{D}(\Z_2)$ and $\Z_2^{(p)}\boxtimes\mathcal{D}(\Z_2):=\Z_2\text{-}\Vect^{(p)}\boxtimes\mathcal{D}(\Z_2)$.

\medskip

\noindent\textbf{Trivial defectification} $p=1$:
The $\Z_2$-extension of $\mathcal{D}(\Z_2)\cong \Z_2\text{-}\Vect\boxtimes \Rep(\Z_2)$ we start with is given by 
\begin{equation}
	\Z_2\text{-}\Vect\boxtimes \mathcal{D}(\Z_2)\cong (\Z_2\times\Z_2)\text{-}\Vect\boxtimes \Rep(\Z_2).
\end{equation}
where the first factor of $\Z_2$ denotes the extension.
The inclusion of $\mathcal{D}(\Z_2)\subset (\Z_2\times\Z_2)\text{-}\Vect\boxtimes \Rep(\Z_2)$ is thus given by the subcategory $\Z_2\text{-}\Vect\subset (\Z_2\times\Z_2)\text{-}\Vect$ with simple objects $\mathds{1}$ and $m_2$, and the objects which involve~$m_1$ are the twisted sector line operators.
The $\Z_2$-action is trivial on all of $(\Z_2\times\Z_2)\text{-}\Vect\boxtimes \Rep(\Z_2)$.

For concreteness, let us choose~$m_1$ as the simple object of degree $-1$, required for the construction in \Cref{sec:RT}, and we have $\mathds{1}$ as the simple object of trivial degree $1$.
The algebra underlying the orbifold datum is given by $A_{\Z_2}=\mathds{1}\oplus \mathds{1}$ with trivial Frobenius algebra structure on both summands, leading to the surface $S_\mathds{1}\oplus S_\mathds{1}$.

Consider two surfaces which correspond to the symmetry action of $-1\in\Z_2$.
The line operator which fuses them is given by $\chi_{-1,-1}:=\mathds{1}\otimes m_{1}\otimes m_{1}\cong \mathds{1}$ (eq. \eqref{eq:chi-def}). 
This coincides with our choice of trivial symmetry fractionalisation, which furthermore corresponds to a trivial extension $\Z_2\times\Z_2$ of the group~$\Z_2$ by~$\Z_2$:
Just as the gauging of the 0-form symmetry associated to the $\Z_2$-extension $\Z_2$-$\Vect$ of $\Vect$ in \Cref{sec:Z2-Vect} produced $\mathcal{D}(\Z_2)$, here we obtain $\mathcal{D}(\Z_2\times\Z_2)\cong \mathcal{D}(\Z_2)\boxtimes \mathcal{D}(\Z_2)$ from the extension $\Z_2$-$\Vect\boxtimes \mathcal{D}(\Z_2)$.

Regarding the emergent 1-form symmetry, the inclusion $\Rep(\Z_2)\subset \mathcal{D}(\Z_2\times\Z_2)$ maps the irreducible representations $\C_\pm\in\Rep(\Z_2)$ onto $\mathds{1}$ and $e_1$, aligned with the fact that the first factor of $\Z_2\times\Z_2$ corresponds to the extension.
This reproduces the algebra $B=\mathds{1}\oplus e_1$.

\medskip 

\noindent \textbf{Non-trivial defectification} $p=-1$:
For non-trivial  defectification, the fusion rules of the extension $\Z_2$-$\Vect^p\boxtimes \mathcal{D}(\Z_2)$ are the same, but as in \Cref{sec:Z2-Vect}, associators and the monoidal structure of the group action \eqref{eq:compositorconventionBBCW} are modified, and all twisted sector objects now carry the non-trivial 2-cocycle, $\beta(-1,-1)=p(-1,-1,-1)=-1$.
Therefore, the simple objects in the equivariantisation come in two types, either characterised by simple objects in $\mathcal{D}(\Z_2)$ with a choice of sign (corresponding to the trivial and sign $\Z_2$-representations) or by simple objects in (the twisted sector copy of) $\mathcal{D}(\Z_2)$ with projective $\Z_2$-representations defined by a choice of $\sqrt{-1}$, i.e.\ $\pm i$. 
The fusion rules of the equivariantisation are those of $\mathcal{D}^p(\Z_2)\boxtimes\mathcal{D}(\Z_2)$:
Denoting the twisted sector objects as $\tilde{\mathds{1}}$, $\tilde{e}$, $\tilde{m}$, and $\tilde{f}$, and the simple objects in $\mathcal{D}^p(\Z_2)\boxtimes\mathcal{D}(\Z_2)$ as in \Cref{sec:Z2-Vect}, the identification is given in \Cref{tab:simpleobjectsZ2p1} (omitting $\boxtimes$ between objects for brevity).
As before, the naive fusion product of representations has to be modified by the monoidal structure of the $\Z_2$-action via \eqref{eq:zestedtensorator}.
\begin{table}[h]
\centering
\begin{tabular}{cc}
$\mathcal{D}^p(\Z_2)\boxtimes\mathcal{D}(\Z_2)$&$(\Z_2\text{-}\Vect^p\boxtimes \mathcal{D}(\Z_2))^{\Z_2}$\\
\begin{tabular}[t]{@{}|c|c|c|c|@{}} 
		\hline
        $\mathds{1}$&$e$&$m$&$f$ \\\hline 
        $s$&$s e$&$s m$&$s f$ \\\hline
        $\overline{s}$&$\overline{s} e$&$\overline{s}m$&$\overline{s} f$\\\hline 
        $s\overline{s}$&$s\overline{s}e$&$s\overline{s} m$&$s\overline{s} f$\\\hline
      \end{tabular}&
      \begin{tabular}[t]{@{}|c|c|c|c|@{}} 
		\hline
        $\mathds{1}_+$&$e_+$&$m_+$&$f_+$ \\\hline 
        $\tilde{\mathds{1}}_{+i}$&$\tilde{e}_{+i}$&$\tilde{m}_{+i}$&$\tilde{f}_{+i}$\\\hline
        $\tilde{\mathds{1}}_{-i}$&$\tilde{e}_{-i}$&$\tilde{m}_{-i}$&$\tilde{f}_{-i}$\\\hline
        $\mathds{1}_-$&$e_-$&$m_-$&$f_-$\\\hline 
      \end{tabular} 
\end{tabular}
\caption{Identification of simple objects in the fusion rings of $\mathcal{D}^p(\Z_2)\boxtimes\mathcal{D}(\Z_2)$ and $(\Z_2\text{-}\Vect^p\boxtimes \mathcal{D}(\Z_2))^{\Z_2}$ for non-trivial~$p=-1$.}\label{tab:simpleobjectsZ2p1}
\end{table}

Associator and braiding are both non-trivial.
In particular, the associator descends directly from that of $\Z_2\text{-}\Vect^{p}\boxtimes \mathcal{D}(\Z_2)$ (i.e.\ it depends only on the underlying object and not on the choice of representation), thereby corresponding to that in $\mathcal{D}^p(\Z_2)\boxtimes\mathcal{D}(\Z_2)$.
The braiding of simple objects is given by
\begin{align}
	c^{p=-1,\Z_2}_{(X,u),(Y,v)}=
	\begin{cases}
		\hphantom{\pm} c_{X,Y}\qquad & |Y|=\hphantom{-}1\,,\\
		\pm c_{X,Y} \qquad & |Y|=-1, (X,u)\in\{\mathds{1}_\pm,e_\pm,m_\pm,f_\pm\}\,,\\
		\pm ic_{X,Y}\qquad & |Y|=-1, (X,u)\in\{\tilde{\mathds{1}}_{\pm i},\tilde{e}_{\pm i},\tilde{m}_{\pm i},\tilde{f}_{\pm i}\}\,,
	\end{cases}
\end{align}
which also corresponds to the braiding \eqref{eq:semionBraiding} of $\mathcal{D}^p(\Z_2)$ for $X,Y\in \{\mathds{1}_\pm,\tilde{\mathds{1}}_{\pm i}\}$.

As in \Cref{sec:Z2-Vect}, the algebra underlying the orbifold datum is now given by $A_{\Z_2}=\mathds{1}\oplus\mathds{1}$ where the second term carries the algebra structure with additional sign.
Once again, this leads to the same modules, just the associator is twisted by the cocycle~$p$.

Regarding the emergent 1-form symmetry, the inclusion $\Rep(\Z_2)\subset \mathcal{D}^p(\Z_2)\times\mathcal{D}(\Z_2)$ maps the irreducible representations $\C_\pm\in\Rep(\Z_2)$ onto $\mathds{1}_\pm\in \mathcal{D}^p(\Z_2)$.
The resulting algebra is $B=\mathds{1}\oplus s\overline{s}$ in the notation of $\mathcal{D}^p(\Z_2)$.

\subsubsection{Non-trivial Symmetry Fractionalisation: $\boldsymbol{\lambda=m}$, $\boldsymbol{\mathcal{D}(\Z_2)\rightarrow \mathcal{D}^{(\tilde{p})}(\Z_4)}$}
\label{sec:Z4}
In this section we introduce non-trivial symmetry fractionalisation to the 0-form $\Z_2$-symmetry in $\mathcal{D}(\Z_2)$. 
We obtain a theory with symmetry $\mathcal{D}^{(\tilde{p})}(\Z_4)$. 
Adopting the notation of \cite{BBDR}, the objects of $\Z_4$-$\Vect$ are denoted as $\mathds{1}$, $m$, $m^2$, and $m^3$, objects of $\Rep(\Z_4)$ as $\mathds{1}$, $e$, $e^2$, and $e^3$, and objects of $\mathcal{D}(\Z_4)\cong\Z_4$-$\Vect\boxtimes \Rep(\Z_4)$ accordingly. 
Note that $m^*=m^3$ and $e^*=e^3$, and vice versa. 
More precisely, we set $e^l$ to be the 1-dimensional representation $(\C,\varphi^l)$ where $\varphi^l(k)=\exp(\frac{2\pi i}{4}kl)=i^{kl}$ for $k\in\Z_4$.
We also set $m^0:=\mathds{1}\in\Z_4$-$\Vect$ and $e^0:=\mathds{1}\in\Rep(\Z_4)$ and all superscripts are understood as mod 4.

Similar to $\mathcal{D}(\Z_2)$, the braided structure of $\mathcal{D}(\Z_4)$ is given by 
\begin{equation}
	c^{\Z_4}_{(m^{k_1},e^{l_1}),(m^{k_2},e^{l_2})}:= \varphi^{l_1}(k_2)\id_{(m^{k_1+k_2},e^{l_1+l_2})}\,,
	\label{eq:braidingDZ4}
\end{equation}
i.e.\ the coefficient is determined by evaluating the representation of the left factor at the degree of the right factor.
As for $\mathcal{D}^{(p)}(\Z_2)$, the category $\mathcal{D}^{(\tilde{p})}(\Z_4)$ arises as an equivariantisation $(\Z_4\text{-}\Vect^{(\tilde{p})})^{\Z_4}$ of the trivial $\Z_4$-extension (of $\Vect$).

\begin{equation}
\begin{tikzcd}[column sep= 20, ampersand replacement=\&]
	\mathcal{D}^{(\tilde{p})}(\Z_4)
    \arrow[rrrr, "{\substack{\text{1-form gauging }\widehat{=}\text{ de-eq.}\\A_L=B=\C(\Z_2)=\mathds{1}\oplus e^2}}"{above}]
    \arrow[d,phantom,sloped, "\cong"]
    \&\&\&\&
    (\mathcal{D}^{(\tilde{p})}(\Z_4))^\text{loc}_{A_L}
    \arrow[d,phantom,sloped, "\cong",shift right=3]
    \arrow[r,phantom, "\subset"]
    \&
     (\mathcal{D}^{(\tilde{p})}(\Z_4))_{A_L}
    \arrow[d,phantom,sloped, "\cong",shift right=11]
    \\
     \hspace{-12pt}((\Z_2\boxtimes\mathcal{D}(\Z_2))^{(\lambda=m,p)})^{\Z_2}
     \&\&\&\&
     \hphantom{(}\mathcal{D}(\Z_2)\hphantom{)_{A_L}}
    \arrow[llll, "{\substack{A_{\Z_2}=\,\mathds{1}\oplus\mathds{1}\\ \hat{\mathcal{A}}_S\cong(\mathcal{D}(\Z_2))_{A_{\scaleto{\Z_2}{4pt}}}=\, S_\mathds{1}\boxplus S_\mathds{1}\\\text{0-form gauging }\widehat{=}\text{ eq.}}}"{below}]
    \arrow[r,phantom, "\subset"]
    \&
     (\Z_2\boxtimes\mathcal{D}(\Z_2))^{(\lambda=m,p)}
\end{tikzcd}
\label{tab:Z4}
\end{equation}

The $\Z_2$-extension of $\mathcal{D}(\Z_2)\cong \Z_2\text{-}\Vect\boxtimes \Rep(\Z_2)$ we start with is given by $\Z_4\text{-}\Vect\boxtimes \Rep(\Z_2)$.\footnote{In \cite{BBDR}, this category was denoted $\Rep(\Z_2\times \Z_4)$. We prefer this notation since it makes the inclusion more obvious, however, while the resulting equivariantisation is $ \mathcal{D}(\Z_4)$, the extension does \textbf{not} inherit the braiding from $ \mathcal{D}(\Z_4)=\Z_4\text{-}\Vect\boxtimes \Rep(\Z_4)$.}
The inclusion of $\mathcal{D}(\Z_2)\subset \Z_4\text{-}\Vect\boxtimes \Rep(\Z_2)$ is given by the subcategory 
\begin{align}
	\Z_2\text{-}\Vect&\hookrightarrow \Z_4\text{-}\Vect\,,\nonumber\\
	\mathds{1}&\longmapsto \mathds{1}\,,\nonumber\\
	m&\longmapsto m^2\,.\label{eq:Z2Z4inclusion}
\end{align}
This encodes the zested fusion rules with $\lambda(-1,-1)=m$ under the identification given in \Cref{tab:simpleobjectsZ4zesting} (omitting $\boxtimes$ for brevity). 
In the notation of $\Z_4\text{-}\Vect\boxtimes\Rep(\Z_2)$, an object~$X$ lies in the twisted sector ($|X|=-1$) if and only if~$X$ contains an odd power of~$m$, i.e.\ $|m^ke^l|=(-1)^k$.
We use the notation of the left-hand side of \Cref{tab:simpleobjectsZ4zesting} to describe the remaining structure of the extension.
\begin{table}[h]
\centering
\begin{tabular}{cc}
$\Z_4$-$\Vect\boxtimes\Rep(\Z_2)$&\makecell{$\lambda(-1,-1)=m$, \\$p=\pm 1$}\\
\begin{tabular}[t]{@{}|c|c|@{}} 
		\hline
        $\mathds{1}$&$e$ \\\hline 
        $m$&$m e$ \\\hline
        $m^2$&$m^2e$\\\hline
        $m^3$&$m^3e$\\\hline
      \end{tabular}&
      \begin{tabular}[t]{@{}|c|c|@{}} 
		\hline
        $\mathds{1}$&$e$ \\\hline 
        $\tilde{\mathds{1}}$&$\tilde{e}$\\\hline
        $m$&$f$\\\hline
        $\tilde{m}$&$\tilde{f}$\\\hline
      \end{tabular} 
\end{tabular}
\caption{Identification of simple objects in the fusion rings of $\Z_4$-$\Vect\boxtimes\Rep(\Z_2)$ and zestings of $\Z_2$-$\Vect\boxtimes\mathcal{D}(\Z_2)$. Objects in the twisted sector (the non-trivial component of $\Z_2$-$\Vect$) carry a tilde.}\label{tab:simpleobjectsZ4zesting}
\end{table}

The associator of this extension is given by (cf. \eqref{eq:zestedassociator} or \cite[Eq.\,(601)]{BBCW})\footnote{Since in our case, all objects are invertible and hence the involved $\Hom$-spaces are 1-dimensional, we treat associators as a number $z$, such that $\alpha_{X,Y,Z}=z\cdot \id$.}
\begin{equation}
	\alpha^{(\lambda,p)}_{X,Y,Z}= p(|X|,|Y|),|Z|)\cdot c_{\lambda(|X|,|Y|), Z}=p(|X|,|Y|),|Z|)\,,
	\label{eq:assocDZ2lambda}
\end{equation}
since even if $|X|=|Y|=-1$, all objects~$Z$ braid trivially $c_{m, Z}=\id$ with~$\lambda(-1,-1)=m$.

The $\Z_2$-crossed braiding of the extension is given by
\begin{align}
	c_{X,Y}^{(\lambda=m,p)}&=
	\begin{cases}
		-\id\qquad & X\in\{e,\tilde{e},f,\tilde{f}\},Y\in\{m,\tilde{m},f,\tilde{f}\}\,,\\
		\hphantom{-}\id\qquad & \text{otherwise\,,}
	\end{cases}\\
	&=(-1)^{l_1\lfloor \frac{k_2}{2} \rfloor}\id\qquad \text{ for }X=m^{k_1}e^{l_1},Y=m^{k_2}e^{l_2}\,,
	\label{eq:mzestedbraiding}
\end{align}
where we used the floor function to round down $k_2/2$.
Note that $\lfloor k_2/2 \rfloor$ detects whether~$Y$ contains a factor of $m\in\mathcal{D}(\Z_2)$, and~$l_1$ detects whether~$X$ contains a factor of $e\in\mathcal{D}(\Z_2)$, and then the braiding is that of $\mathcal{D}(\Z_2)$ \eqref{eq:braidingInDZ2}.

All simple objects in the $(-1)$-graded part of $\Z_4\text{-}\Vect\boxtimes \Rep(\Z_2)$ satisfy $X^*\otimes X\cong \mathds{1}$ ($(m^k,e^l)^*=(m^{4-k},e^{2-l})$, see \eqref{eq:zesteddual}), so we obtain the algebra $A_{\Z_2}=\mathds{1}\oplus \mathds{1}$. 
For concreteness, let us choose~$m$ as the simple object of homogeneous degree $-1$ required for the orbifold construction.
The fusion of two surfaces realising the symmetry action of $-1\in\Z_2$ is then mediated via the line operator given by the module $\chi_{-1,-1}=\mathds{1}\otimes m\otimes m=m^2$ which corresponds to the chosen symmetry fractionalisation $m\in\mathcal{D}(\Z_2)$ via \eqref{eq:Z2Z4inclusion}.
This is associated to a non-trivial $\Z_2$-extension ($\Z_4$) of the group~$\Z_2$.
Note that the module $\chi_{-1,-1}$ is independent of the simple object we chose from the twisted sector.
The dependence on the defectification~$p$ appears only in the Frobenius algebra structure of $A_{-1}$, with $\mu_{-1}=p\mu_1$ and similarly for the remaining components (with coefficients $p^{\pm 1}$, see \Cref{sec:zestedorbdat}). 
As before, these are Morita-equivalent, so the surface $\hat{\mathcal{A}}_S$ is unchanged and coincides with the surface underlying the gauging of the 0-form symmetry for trivial fractionalisation, explicitly confirming our discussion in \Cref{sec:zesting} and \eqref{eq:zestedMoritaEquivalence}.

To calculate the equivariantisation, note that the $\Z_2$-action is trivial on objects, but its monoidal structure is not. 
From \eqref{eq:compositorconventionBBCW}, we can calculate the cocycles associated to the projective representations for each object:
\begin{equation}
\beta_Y(-1,-1)=
\begin{cases}
	\hphantom{-}1\qquad & Y\in\lbrace \mathds{1},m^2\rbrace\,,\\
	-1\qquad & Y\in\lbrace e,m^2 e\rbrace\,,\\
	\hphantom{-}p\qquad & Y\in\lbrace m,m^3\rbrace\,,\\
	-p\qquad & Y\in\lbrace m e,m^3 e\rbrace\,.
\end{cases}
\end{equation}
For $p=1$, the simple objects of the equivariantisation are thus $m^k_\pm$ and $m^k e_{\pm i}$, for $0\leq k\leq 3$. 
For $p=-1$, we have $\mathds{1}_\pm$, $m^2_\pm$, $m^k_{\pm i}$, $e_{\pm i}$, $m^2 e_{\pm i}$, and $m^k e_{\pm}$, for $k= 1,3$. 
Their fusion rules are those of $\mathcal{D}(\Z_4)$, given by the identification in \Cref{tab:simpleobjectsZ4} (noting that $(\rho_g^2)_{X,Y}=p(g,|X|,|Y|)^{-1}$ enters the fusion rules as before).
\begin{table}[h]
\begin{tabular}{ccc}
$\mathcal{D}(\Z_4)$&$\lambda(-1,-1)=m,p=1$&$\lambda(-1,-1)=m,p=-1$\\
\begin{scriptsize}
\begin{tabular}[t]{@{}|c|c|c|c|@{}} 
		\hline
        $\mathds{1}$&$e$&$e^2$&$e^3$ \\\hline 
        $m$&$m e$&$m e^2$&$m e^3$ \\\hline
        $m^2$&$m^2 e$&$m^2 e^2$&$m^2 e^3$\\\hline 
        $m^3$&$m^3 e$&$m^3 e^2$&$m^3 e^3$\\\hline
      \end{tabular}
\end{scriptsize} &
\begin{scriptsize}
      \begin{tabular}[t]{@{}|c|c|c|c|@{}} 
		\hline
        $\mathds{1}_+$&$e_{+i}$&$\mathds{1}_-$&$e_{-i}$ \\\hline 
        $m_+$&$m e_{+i}$&$m_-$&$m e_{-i}$ \\\hline
        $m^2_+$&$m^2 e_{+i}$&$m^2_-$&$m^2 e_{-i}$\\\hline 
        $m^3_+$&$m^3 e_{+i}$&$m^3_-$&$m^3 e_{-i}$\\\hline
      \end{tabular} 
\end{scriptsize}&
\begin{scriptsize}
      \begin{tabular}[t]{@{}|c|c|c|c|@{}} 
		\hline
        $\mathds{1}_+$&$e_{+i}$&$\mathds{1}_-$&$e_{-i}$ \\\hline 
        $m_{+i}$&$m e_{-}$&$m_{-i}$&$m e_{+}$ \\\hline
        $m^2_{+}$&$m^2 e_{+i}$&$m^2_-$&$m^2 e_{-i}$\\\hline 
        $m^3_{+i}$&$m^3 e_{-}$&$m^3_{-i}$&$m^3 e_{+}$\\\hline
      \end{tabular} 
\end{scriptsize}
\end{tabular}
\caption{Identification of the fusion rings of $((\Z_2\text{-}\Vect\boxtimes\mathcal{D}(\Z_2))^{(\lambda=m,p)})^{\Z_2}$ and $\mathcal{D}(\Z_4)$.}\label{tab:simpleobjectsZ4}
\end{table}

Regardless of the choice of defectification, due to the identification of $\mathds{1}_-$ with $e^2$, the resulting 1-form symmetry algebra is $\C(\Z_2)=\mathds{1}\oplus e^2$, matching the result of \cite{BBDR}.

The associators of the equivariantisations are (again) trivial for $p=1$ (thus coinciding with $\mathcal{D}(\Z_4)$), for $p=-1$ the $G$-crossed extension and the braided fusion category resulting from equivariantisation are discussed below, and they are inequivalent to those of $p=1$.

\medskip

\noindent\textbf{Trivial defectification} $p=1$: 
The braiding is given by \eqref{eq:eq of Gcbc is braided}.
We can calculate it by simply mutliplying \eqref{eq:mzestedbraiding} with the appropriate factor from the equivariant structure.
Recall that an object is in the twisted sector when it has an odd power of~$m$, and its equivariant structure is given by the subscript, or alternatively by the exponent of~$e$ in the notation of $\mathcal{D}(\Z_4)$ from \Cref{tab:simpleobjectsZ4}. 
The braiding is then given by
\begin{align}
	c_{m^{k_1}e^{l_1},m^{k_2}e^{l_2}}^{(\lambda=m,p=1),\Z_2}
	&=(-1)^{l_1\lfloor \frac{k_2}{2} \rfloor}\cdot i^{l_1\pi(k_2)}\cdot \id_{m^{k_1+k_2}e^{l_1+l_2}}\,,
	\label{eq:braidingEqZ4}
\end{align}
where $\pi(k_2)=(1-(-1)^n)/2\in\{0,1\}$ is the parity.
Note that $2\lfloor k_2/2 \rfloor+\pi(k_2)=k_2$, hence this matches \eqref{eq:braidingDZ4} $c_{(m^{k_1},e^{l_1}),(m^{k_2},e^{l_2})}^{\Z_4}:= \varphi^{l_1}(k_2)\cdot \id=i^{l_1k_2}\id$, the braiding of $\mathcal{D}(\Z_4)$.
Therefore 
\begin{equation}
	((\Z_2\text{-}\Vect\boxtimes\mathcal{D}(\Z_2))^{(\lambda=m,p=1)})^{\Z_2}\cong \mathcal{D}(\Z_4)
\end{equation}
as braided fusion categories.

\medskip

\noindent\textbf{Non-trivial defectification} $p=-1$:
The associators $\alpha_{X,Y,Z}$ are non-trivial (equal to $-1$) whenever $X$, $Y$, and $Z$ are in the non-trivially graded component (i.e.\ contain an odd power of $m$).
The braiding is given by two factors coming from \eqref{eq:braidingInDZ2} and the equivariant structure \eqref{eq:eq of Gcbc is braided}:
\begin{align}
	c_{m^{k_1}e^{l_1},m^{k_2}e^{l_2}}^{(\lambda=m,p=-1),\Z_2}
	&= (-1)^{l_1\lfloor \frac{k_2}{2} \rfloor}\cdot i^{(l_1+\pi(k_1))\pi(k_2)} \id=i^{l_1k_2+\pi(k_1)\pi(k_2)}\id\,,
	\label{eq:braidingEqZ4-}
\end{align}

This braided fusion category is equivalent to $\mathcal{D}^{\tilde{p}}(\Z_4)=(\Z_4\text{-}\Vect^{\tilde{p}})^{\Z_4}$, where the cocycle $\tilde{p}\in\mathrm{H}^3(\Z_4,\U)$ is
\begin{equation}
	{\tilde{p}}(k,l,m)=(-1)^{klm}\,,
\end{equation}
and the resulting associator in $\Z_4\text{-}\Vect^{\tilde{p}}$ is
\begin{equation}
\alpha^{\Z_4\text{-}\Vect^{\tilde{p}}}_{m^{k_1},m^{k_2},m^{k_3}}=\tilde{p}(k_1,k_2,k_3)\,,
\end{equation}
which is~$-1$ when all involved factors have an odd power of~$m$, and~$1$ otherwise.
We then have 
\begin{equation}
	\beta_Y(k,l)=\tilde{p}(k,l,|Y|)=
	\begin{cases}
		\hphantom{(-}1\qquad &Y\in\{\mathds{1}, m^2\}\,,\\
		(-1)^{kl}\qquad & Y\in\{m,m^3\}\,.
	\end{cases}
\end{equation}
In the equivariantisation, $\mathds{1}$ and $m^2$ therefore pair with linear representations of $\Z_4$, while~$m$ and~$m^3$ carry projective representations. 
Similar to the linear ones, there is an irreducible $\beta_m$-representation $\tilde{\varphi}^l$ for each $l\in\Z_4$ and these are determined by $\tilde{\varphi}^l(1)$ being a fourth root of unity.
We choose to parametrise them as $\tilde{\varphi}^l(1)=i^{l+1}$, then $\tilde{\varphi}^l(2)=(-1)^{l}$ and $\tilde{\varphi}^l(3)=i^{3l+1}$, which can be summarised as $\tilde{\varphi}^l(k)=i^{lk+\pi(k)}$.
If we denote $\tilde{\varphi}^l$ by $e^l$, we have $me^{l_1}\otimes me^{l_2}=m^2e^{l_1+l_2}$, since that product carries the linear representation $\tilde{\varphi}^{l_1}(1)\cdot\tilde{\varphi}^{l_2}(1)\cdot (\rho^2_1)_{me^{l_1},me^{l_2}}=i^{l_1+1}i^{l_2+1}(-1)=i^{l_1+l_2}=\varphi^{l_1+l_2}(1)$, and the values on $2$ and $3\in\Z_4$ are determined by this.
One can similarly check that fusions between different powers of~$m$ and~$e$ amount to adding the exponents, leading to the expected $\Z_4\times \widehat{\Z}_4$ fusion rules.
The braiding is then calculated as before and the result is 
\begin{equation}
	c^{\mathcal{D}^{\tilde{p}}(\Z_4)}_{(m^{k_1},e^{l_1}),(m^{k_2},e^{l_2})}=
	\begin{cases}	
		\tilde{\varphi}^{l_1}(k_2)\id_{(m^{k_1+k_2},e^{l_1+l_2})}\qquad& k_1 \text{ odd}\,,\\
		\varphi^{l_1}(k_2)\id_{(m^{k_1+k_2},e^{l_1+l_2})}\qquad& k_1 \text{ even}\,.
	\end{cases}
\end{equation}

Given this description of $\mathcal{D}^{\tilde{p}}(\Z_4)$, \Cref{tab:simpleobjectsZ4} induces a (strict) braided monoidal equivalence 
\begin{equation} 
	(\Z_2\text{-}\Vect\boxtimes\mathcal{D}(\Z_2)^{(\lambda=m,p=-1)})^{\Z_2}\cong\mathcal{D}^{\tilde{p}}(\Z_4)\,.
\end{equation}

\subsubsection{Non-trivial Symmetry Fractionalisation: $\boldsymbol{\lambda=e}$}
In this section we apply zesting for $\lambda(-1,-1)=e$ and calculate equivariantisation and the algebras underlying the orbifold data.
The extension is $\mathcal{D}(\Z_2)\subset \Z_2\text{-}\Vect\boxtimes\Rep(\Z_4)$, with the canonical inclusion $\Rep(\Z_2)\longhookrightarrow \Rep(\Z_4)$, and the full identification is given in \Cref{tab:simpleobjectsZ4ezesting}.

\begin{table}[h]
\centering
\begin{tabular}{cc}
$\Z_2$-$\Vect\boxtimes\Rep(\Z_4)$&\makecell{$\lambda(-1,-1)=e$, \\$p=\pm 1$}\\
\begin{tabular}[t]{@{}|c|c|c|c|@{}} 
		\hline
        $\mathds{1}$&$e$&$e^2$&$e^3$ \\\hline 
        $m$&$m e$&$me^2$&$me^3$ \\\hline
      \end{tabular}&
      \begin{tabular}[t]{@{}|c|c|c|c|@{}} 
		\hline
        $\mathds{1}$&$\tilde{\mathds{1}}$&$e$&$\tilde{e}$ \\\hline 
        $m$&$\tilde{m}$&$f$&$\tilde{f}$\\\hline
      \end{tabular} 
\end{tabular}
\caption{Identification of simple objects in the fusion rings of $\Z_4$-$\Vect\boxtimes\Rep(\Z_2)$ and zestings of $\Z_2$-$\Vect\boxtimes\mathcal{D}(\Z_2)$. Objects in the twisted sector (the non-trivial component of $\Z_2$-$\Vect$) carry a tilde.}\label{tab:simpleobjectsZ4ezesting}
\end{table}

However, the extension is equipped with the following non-trivial associator (as in \eqref{eq:assocDZ2lambda}) and braiding
\begin{align}
	\alpha^{(\lambda=e,p)}_{X_1,X_2,X_3}=(-1)^{l_1l_2k_3}p(l_1,l_2,l_3)\,,
	\label{eq:Z4eassociator}\\
	c^{(\lambda=e,p)}_{X_1,X_2}=(-1)^{\lfloor \frac{l_1}{2} \rfloor k_2}\id_{(m^{k_1+k_2},e^{l_1+l_2})}\,,
\end{align}
where $X_i=m^{k_i}e^{l_i}$, $k_i\in\Z_2$, $l_i\in\Z_4$ and we used the floor function again to round down $l_1/2$.
This means the braiding is $-1$ whenever $X\in\{e,\tilde{e},f,\tilde{f}\}$ and $Y\in\{m,\tilde{m},f,\tilde{f}\}$ (cf. \eqref{eq:braidingInDZ2}).

The orbifold datum for the 0-form symmetry once again has the underlying algebra $A_{\Z_2}=\mathds{1}\oplus\mathds{1}$ with algebra structure of the second term depending on the defectification. 
As before, the resulting surface is $\hat{\mathcal{A}}_S=S_\mathds{1}\oplus S_\mathds{1}$, but we can detect the fractionalisation in $\chi_{-1,-1}=\mathds{1}\otimes (e\otimes e)=e^2$ which corresponds to~$e\in\mathcal{D}(\Z_2)$ under the inclusion $\Rep(\Z_2)\longhookrightarrow\Rep(\Z_4)$.

\medskip

The cocycles for the projective representations in the equivariantisation are
\begin{equation}
\beta_Y(-1,-1)=
\begin{cases}
	\hphantom{-}1\qquad & Y\in\lbrace \mathds{1},e^2\rbrace\,,\\
	-1\qquad & Y\in\lbrace m,m e^2\rbrace\,,\\
	\hphantom{-}p\qquad & Y\in\lbrace e,e^3\rbrace\,,\\
	-p\qquad & Y\in\lbrace m e,m e^3\rbrace\,.\\
\end{cases}
\end{equation}
Denoting the projective and linear representations by their value on $-1\in\Z_2$ as a subscript (as above), we obtain the identification with a $\Z_4\times\Z_4$ fusion ring given in \Cref{tab:simpleobjectsZ4e}.
\begin{table}[h]
\begin{tabular}{ccc}
$\mathcal{D}(\Z_4)$&$\lambda(-1,-1)=e,p=1$&$\lambda(-1,-1)=e,p=-1$\\
\begin{scriptsize}
\begin{tabular}[t]{@{}|c|c|c|c|@{}} 
		\hline
        $\mathds{1}$&$e$&$e^2$&$e^3$ \\\hline 
        $m$&$m e$&$m e^2$&$m e^3$ \\\hline
        $m^2$&$m^2 e$&$m^2 e^2$&$m^2 e^3$\\\hline 
        $m^3$&$m^3 e$&$m^3 e^2$&$m^3 e^3$\\\hline
      \end{tabular}
\end{scriptsize} &
\begin{scriptsize}
      \begin{tabular}[t]{@{}|c|c|c|c|@{}} 
		\hline
        $\mathds{1}_+$&$e_+$&$e^2_+$&$e^3_+$ \\\hline 
        $m_{+i}$&$m e_{+i}$&$me^2_{+i}$&$m e^3_{+i}$ \\\hline
        $\mathds{1}_-$&$e_-$&$e^2_-$&$e^3_-$\\\hline 
        $m_{-i}$&$me_{-i}$&$me^2_{-i}$&$me^3_{-i}$\\\hline
      \end{tabular} 
\end{scriptsize}&
\begin{scriptsize}
      \begin{tabular}[t]{@{}|c|c|c|c|@{}} 
		\hline
        $\mathds{1}_+$&$e_{+i}$&$e^2_+$&$e^3_{+i}$ \\\hline 
        $m_{+i}$&$m e_-$&$me^2_{+i}$&$m e^3_-$ \\\hline
        $\mathds{1}_-$&$e_{-i}$&$e^2_-$&$e^3_{-i}$\\\hline 
        $m_{-i}$&$me_+$&$me^2_{-i}$&$me^3_+$\\\hline
      \end{tabular} 
\end{scriptsize}
\end{tabular}
\caption{Identification of the fusion rings of $((\Z_2\text{-}\Vect\boxtimes\mathcal{D}(\Z_2))^{(\lambda=e,p)})^{\Z_2}$ and $\mathcal{D}(\Z_4)$.}\label{tab:simpleobjectsZ4e}
\end{table}

Note that these are essentially the transpositions of \Cref{tab:simpleobjectsZ4}, i.e.\ we have exchanged the roles of~$e$ and~$m$.
These two categories are not equivalent, however, due to the different associators and braidings.
In particular, the first factor $(-1)^{l_1l_2k_3}$ in \eqref{eq:Z4eassociator} cannot be obtained by equivariantisation of a category of the type $\Z_4\text{-}\Vect^{\tilde{p}}$, since it depends on the powers of both~$m$ and~$e$.

\medskip 

\noindent\textbf{Trivial defectification} $p=1$:
Using the notation of $\mathcal{D}(\Z_4)$ for simple objects, the braiding on $((\Z_2\text{-}\Vect\boxtimes\mathcal{D}(\Z_2))^{(\lambda=e,p=1)})^{\Z_2}$ is given by
\begin{equation}
	c^{(\lambda=e,p=1),\Z_2}_{m^{k_1}e^{l_1},m^{k_2}e^{l_2}}=(-1)^{\lfloor \frac{l_1}{2} \rfloor k_2}\cdot i^{\pi(l_2)k_1}\id
\end{equation}
Here, the first factor comes from the braiding $c^{(\lambda=e,p)}$ and the second factor arises from the equivariant structure as $u^{(k_1,l_1)}_{|e^{l_2}|}$.

\medskip

\noindent\textbf{Non-trivial defectification} $p=-1$:
Using the notation of $\mathcal{D}(\Z_4)$ for simple objects, the braiding on $((\Z_2\text{-}\Vect\boxtimes\mathcal{D}(\Z_2))^{(\lambda=e,p=-1)})^{\Z_2}$ is given by
\begin{equation}
	c^{(\lambda=e,p=-1),\Z_2}_{m^{k_1}e^{l_1},m^{k_2}e^{l_2}}=(-1)^{\lfloor \frac{l_1}{2} \rfloor k_2}\cdot i^{\pi(l_2)(k_1+\pi(l_1))}\id
\end{equation}
Again, the first factor comes from the braiding $c^{(\lambda=e,p)}_{e,m}$ and the remaining factors arise from the equivariant structure as $u^{(k_1,l_1)}_{|e^{l_2}|}$.

\subsubsection{Non-trivial Symmetry Fractionalisation: $\boldsymbol{\lambda=f}$}

For the choice of fractionalisation $\lambda(-1,-1)=f$, there is no canonical choice of defectification, but the two possible choices still form an $\mathrm{H}^3(\Z_2,\U)$-torsor.
These are the defectifications $\nu\equiv\nu_{-1,-1,-1}=\pm {i}$ \cite{BBCW}, where the sign arises from $\mathrm{H}^3(\Z_2,\U)\cong\{\pm 1\}$ as before.

We can identify the fusion ring of the zesting $(\lambda=f,\nu=\pm {i})$ of $\Z_2\text{-}\Vect\boxtimes\mathcal{D}(\Z_2)$ once again with $\Z_2\times\Z_4$ fusion rules, see \Cref{tab:simpleobjectsZ4fzesting}.
\begin{table}[h]
\centering
\begin{tabular}{cc}
$\Z_2$-$\Vect\boxtimes\Z_4\text{-}\Vect$&\makecell{$\lambda(-1,-1)=f$, \\$\nu=\pm {i}$}\\
\begin{tabular}[t]{@{}|c|c|c|c|@{}} 
		\hline
        $\mathds{1}$&$f$&$f^2$&$f^3$ \\\hline 
        $m$&$m f$&$mf^2$&$mf^3$ \\\hline
      \end{tabular}&
      \begin{tabular}[t]{@{}|c|c|c|c|@{}} 
		\hline
        $\mathds{1}$&$\tilde{\mathds{1}}$&$f$&$\tilde{f}$ \\\hline 
        $m$&$\tilde{m}$&$e$&$\tilde{e}$\\\hline
      \end{tabular} 
\end{tabular}
\caption{Identification of simple objects in the fusion rings of $\Z_4$-$\Vect\boxtimes\Rep(\Z_2)$ and zestings of $\Z_2$-$\Vect\boxtimes\mathcal{D}(\Z_2)$. Objects in the twisted sector (the non-trivial component of $\Z_2$-$\Vect$) carry a tilde.}\label{tab:simpleobjectsZ4fzesting}
\end{table}

The extension is equipped with the following associator and braiding
\begin{align}
	a^{(\lambda=f,\nu)}_{X_1,X_2,X_3}&=(-1)^{j_1j_2(k_3+\lfloor\frac{j_3}{2}\rfloor )}\nu(j_1,j_2,j_3)\id\,,
	\label{eq:Z4fassociator}\\
	c^{(\lambda=f,\nu)}_{X_1,X_2}&=(-1)^{\lfloor \frac{j_1}{2} \rfloor (k_2+\lfloor\frac{j_2}{2}\rfloor)}\id\,,
\end{align}
where $X_i=m^{k_i}f^{j_i}$, $k_i\in\Z_2$, $j_i\in\Z_4$ and the $j_i$ are understood mod $2$ when we evaluate $p$.
The braiding is~$-1$ whenever $X\in\{e,\tilde{e},f,\tilde{f}\}$ and $Y\in\{m,\tilde{m},f,\tilde{f}\}$ (cf. \eqref{eq:braidingInDZ2}).

The orbifold datum for the 0-form symmetry once again has the underlying algebra $A_{\Z_2}=\mathds{1}\oplus\mathds{1}$ with algebra structure of the second term depending on the defectification.
In this case, we have $\mu_{-1}=\pm {i}\,\id$, $\eta_{-1}=\mp {i}\,\id$,  $\Delta_{-1}=\mp {i}\,\id$,  $\varepsilon_{-1}=\pm {i}\,\id$ for $\nu=\pm {i}$.
As before, the resulting surface is $\hat{\mathcal{A}}_S=S_\mathds{1}\oplus S_\mathds{1}$, but we can detect the fractionalisation in $\chi_{-1,-1}=\mathds{1}\otimes (f\otimes f)=f^2$ which corresponds to~$f\in\mathcal{D}(\Z_2)$ under the identification of \Cref{tab:simpleobjectsZ4fzesting}.

The factor $\nu=\pm {i}$ also appears in $(\rho^2_g)_{X,Y}=\nu(g,|X|,|Y|)^{-1}$ and $(\rho^2_{g,h})_X=2S_{X,f}\cdot \nu(g,h,|X|)$.
We can get rid of it in the cocycles $\beta_Y$ for the projective representations (in favour of $\pm 1$) by choosing $\xi_{-1}=\zeta\id$ in \eqref{eq:projectivitycocycle} on objects in the twisted sector, where $\zeta^2=-{i}$.
We then obtain cocycles $\beta_Y(-1,-1)= (\xi_{-1}^Y)^{-2}\cdot 2S_{Y,f}\cdot \nu(-1,-1,|Y|)$ as follows:
\begin{equation}
\beta_Y(-1,-1)=
\begin{cases}
	\hphantom{-}1\qquad & Y\in\lbrace \mathds{1},f\rbrace\,,\\
	-1\qquad & Y\in\lbrace m, e\rbrace\,,\\
	\hphantom{-}{i}\nu\qquad & Y\in\lbrace \tilde{\mathds{1}},\tilde{f}\rbrace\,,\\
	-{i}\nu\qquad & Y\in\lbrace \tilde{m},\tilde{e}\rbrace\,.\\
\end{cases}
\end{equation}
The non-trivial $\xi$ also appears in the equivariantisation's fusion rules through \eqref{eq:equivariantstructurewithprojrep}.
In the monoidal product, we then have 
\begin{equation}
	(\xi^X_g\otimes\xi^Y_g)\circ (\rho^2_g)^{-1}_{X,Y}=
	\begin{cases}
		-1 \qquad & |X|=|Y|=g=-1\,,\\
		\hphantom{-}\zeta\qquad & \text{either $|X|=-1$ or $|Y|=-1$, and $g=-1$,}\\
		\hphantom{-}1\qquad &\text{otherwise}\,.
	\end{cases}
\end{equation}
In the second case, the resulting object lies in the twisted sector again $|X\otimes Y|=-1$ and we identify the factor of $\zeta$ with $\xi_{-1}^{X\otimes Y}$, which reproduces the patterns of fusion rules from the previous two cases $\lambda\in\{m,e\}$, and the choice of~$\zeta$ only appears in the braiding.
The resulting fusion rules match $\mathcal{D}(\Z_4)$ as before, and we choose the identification in \Cref{tab:simpleobjectsZ4f}.

\begin{table}[h]
\begin{tabular}{ccc}
$\mathcal{D}(\Z_4)$&$\lambda(-1,-1)=f,\nu=-{i}$&$\lambda(-1,-1)=f,\nu={i}$\\
\begin{scriptsize}
\begin{tabular}[t]{@{}|c|c|c|c|@{}} 
		\hline
        $\mathds{1}$&$f$&$f^2$&$f^3$ \\\hline 
        $m$&$m f$&$m f^2$&$m f^3$ \\\hline
        $m^2$&$m^2 f$&$m^2 f^2$&$m^2 f^3$\\\hline 
        $m^3$&$m^3 f$&$m^3 f^2$&$m^3 f^3$\\\hline
      \end{tabular}
\end{scriptsize} &
\begin{scriptsize}
      \begin{tabular}[t]{@{}|c|c|c|c|@{}} 
		\hline
        $\mathds{1}_+$&$f_+$&$f^2_+$&$f^3_+$ \\\hline 
        $m_{+{i}}$&$m f_{+{i}}$&$mf^2_{+{i}}$&$mf^3_{+{i}}$ \\\hline
        $\mathds{1}_-$&$f_-$&$f^2_-$&$f^3_-$\\\hline 
        $m_{-{i}}$&$mf_{-{i}}$&$mf^2_{-{i}}$&$mf^3_{-{i}}$\\\hline
      \end{tabular} 
\end{scriptsize}&
\begin{scriptsize}
      \begin{tabular}[t]{@{}|c|c|c|c|@{}} 
		\hline
        $\mathds{1}_+$&$f_{+{i}}$&$f^2_+$&$f^3_{+{i}}$ \\\hline 
        $m_{+{i}}$&$m f_-$&$mf^2_{+{i}}$&$m f^3_-$ \\\hline
        $\mathds{1}_-$&$f_{-{i}}$&$f^2_-$&$f^3_{-{i}}$\\\hline 
        $m_{-{i}}$&$mf_+$&$mf^2_{-{i}}$&$mf^3_+$\\\hline
      \end{tabular} 
\end{scriptsize}
\end{tabular}
\caption{Identification of the fusion rings of $((\Z_2\text{-}\Vect\boxtimes\mathcal{D}(\Z_2))^{(\lambda=f,\nu)})^{\Z_2}$ and $\mathcal{D}(\Z_4)$.}\label{tab:simpleobjectsZ4f}
\end{table}

\noindent
The braidings for the two choices of defectification are given by
\begin{align}
	c^{(\lambda=f,\nu=-{i}),\Z_2}_{m^{k_1}f^{j_1},m^{k_2}f^{j_2}}&=(-1)^{\lfloor \frac{j_1}{2} \rfloor (k_2+\lfloor\frac{j_2}{2}\rfloor)}\cdot \zeta^{\pi(j_2)}{i}^{k_1\pi(j_2)}\,,\\
	\intertext{and}
	c^{(\lambda=f,\nu=i),\Z_2}_{m^{k_1}f^{j_1},m^{k_2}f^{j_2}}&=(-1)^{\lfloor \frac{j_1}{2} \rfloor (k_2+\lfloor\frac{j_2}{2}\rfloor)}\cdot \zeta^{\pi(j_2)}{i}^{(k_1+\pi(j_1))\pi(j_2)}\,.
\end{align}

\subsubsection{Electric-Magnetic Duality: $\boldsymbol{\mathcal{D}(\Z_2)\rightarrow \mathcal{Z}(\Ising)}$}
In this section, we gauge the electric-magnetic duality, and obtain the gauged theory $\mathcal{Z}(\Ising)\cong\Ising\boxtimes\overline{\Ising}$.
Denoting the simple objects of $\Ising$ as~$\mathds{1}$,~$\psi$, and~$\sigma$ with fusion rules as in \Cref{sec:Ising}, we have the following results:
\begin{equation}
\begin{tikzcd}[column sep= 20, ampersand replacement=\&]
	\mathcal{Z}(\Ising)
    \arrow[rrrr, "{\substack{\text{1-form gauging }\widehat{=}\text{ de-eq.}\\A_L=B=\C(\Z_2)=\mathds{1}\boxtimes\overline{\mathds{1}}\,\oplus\, \psi\boxtimes\overline{\psi}}}"{above}]
    \arrow[d,phantom,sloped, "\cong"]
    \&\&\&\&
    (\mathcal{Z}(\Ising))^\text{loc}_{A_L}
    \arrow[d,phantom,sloped, "\cong",shift right=3]
    \arrow[r,phantom, "\subset"]
    \&
     (\mathcal{Z}(\Ising))_{A_L}
    \arrow[d,phantom,sloped, "\cong",shift right=11]
    \\
     \hspace{-12pt}((\mathcal{D}(\Z_2)\oplus \Vect^{\oplus 2})^{(p)})^{\Z_2}
     \&\&\&\&
     \hphantom{(}\mathcal{D}(\Z_2)\hphantom{)_{A_L}}
    \arrow[llll, "{\substack{A_{\Z_2}=\,\mathds{1}\oplus (\mathds{1}\oplus f)\\ \hat{\mathcal{A}}_S\cong(\mathcal{D}(\Z_2))_{A_{\scaleto{\Z_2}{4pt}}}=\, S_\mathds{1}\boxplus S_f\\\text{0-form gauging }\widehat{=}\text{ eq.}}}"{below}]
    \arrow[r,phantom, "\subset"]
    \&
     (\mathcal{D}(\Z_2)\oplus \Vect^{\oplus 2})^{(p)}
\end{tikzcd}
\label{tab:Ising}
\end{equation}

Here, $(\mathcal{D}(\Z_2))^\times_{\Z_2}$ is the $\Z_2$-crossed braided extension of $\mathcal{D}(\Z_2)$ with twisted sector given by a category with two (self-dual) simple objects, $c_1$ and $c_2$. 
They satisfy the following fusion rules (and module structure with respect to $\mathcal{D}(\Z_2)$):
\begin{align}
	c_1\otimes c_1&=c_2\otimes c_2\cong \mathds{1}\oplus f\,,\\
	c_1\otimes c_2&=c_2\otimes c_1\cong e\oplus m\,,\\
	c_{1/2}\otimes e\cong e\otimes c_{1/2}&\cong m\otimes c_{1/2}\cong c_{1/2}\otimes m\cong c_{2/1}\,.
\end{align}
The $\Z_2$-action acts as electric-magnetic duality, so it exchanges~$e$ and~$m$, but~$\mathds{1}$ and~$f$ (and~$c_{1/2}$) are invariant under it.

The $G$-crossed braided structure of this category (for which there are 2 inequivalent options) and its equivariantisation have been discussed in detail in \cite[Sect.\,XI.I]{BBCW}, so we do not repeat them here.
The two options correspond to the zestings $p=\pm 1\in \mathrm{H}^3(\Z_2,\U)$ as before,%
\footnote{While the calculations of \cite{BBCW} use a convention for the two options $\kappa_\sigma=\pm 1$ which differs from zesting (notably with trivial $(\rho_{g,h}^2)_Y$ for all simples in both cases), using their data for $\kappa_\sigma=1$ and then zesting produces equivalent results. 
In that case, the simple objects $c_{1/2}$ carry a non-trivial cocycle, and we get projective $\Z_2$-representations with $\varphi(-1)=\pm i$ as before.
 We identify $c_1=\sigma_+$ and $c_2=\sigma_-$ in their notation for the extension, and then $(c_2)_{-i}$ corresponds to $(c_2)_+$ in \Cref{tab:simpleobjectsEMGauged} and $(c_1)_{+i}$ to $(c_1)_+$.
 } 
 since $\lambda\in\mathrm{H}^2_{\rho^{\text{e-m}}}(\Z_2,\mathcal{D}(\Z_2))=0$ is necessarily trivial for electric-magnetic duality.
We obtain the following simple objects in the equivariantisation:
There is $e\oplus m$ (with trivial equivariant structure) and four pairs of simple objects $X_\pm$ where $X\in\lbrace \mathds{1}, f, c_1,c_2\rbrace$ and $\pm$ indicates the trivial ($+$) and sign ($-$) representations as before. 
As was worked out in \cite[Sect.\,XI.I]{BBCW}, the resulting braided category is equivalent to $\mathcal{Z}(\Ising)$ under the identification of simple objects given in \Cref{tab:simpleobjectsEMGauged}.
\begin{table}[h]
\centering
\begin{tabular}{cc}
$(\mathcal{D}(\Z_2)^\times_{\Z_2})^{\Z_2}$&$\mathcal{Z}(\Ising)$\\
\begin{tabular}[t]{@{}|c|c|c|@{}} 
		\hline
        $\mathds{1}_+$&$f_+$&$(c_1)_+$ \\\hline 
        $f_-$&$\mathds{1}_-$&$(c_1)_-$\\\hline 
        $(c_2)_+$&$(c_2)_-$&$e\oplus m$\\\hline
      \end{tabular}&
      \begin{tabular}[t]{@{}|c|c|c|@{}} 
		\hline
        $\mathds{1}\boxtimes\overline{\mathds{1}}$&$\mathds{1}\boxtimes\overline{\psi}$&$\mathds{1}\boxtimes\overline{\sigma}$ \\\hline
        $\psi\boxtimes\overline{\mathds{1}}$&$\psi\boxtimes\overline{\psi}$&$\psi\boxtimes\overline{\sigma}$ \\\hline
        $\sigma\boxtimes\overline{\mathds{1}}$&$\sigma\boxtimes\overline{\psi}$&$\sigma\boxtimes\overline{\sigma}$ \\\hline
      \end{tabular} 
\end{tabular}
\caption{Identification of simple objects in $(\mathcal{D}(\Z_2)^\times_{\Z_2})^{\Z_2}$ and $\mathcal{Z}(\Ising)$.}\label{tab:simpleobjectsEMGauged}
\end{table}

In the notation of \Cref{sec:Ising} ($\Ising=\TY(\Z_2,\varepsilon_q,\varepsilon_\kappa,\varepsilon_\delta)$), the choice $p=1$ corresponds to $\mathcal{Z}(\TY(\Z_2,i,1,1))\cong\TY(\Z_2,i,1,1)\boxtimes \TY(\Z_2,-i,1,1)$ and $p=-1$ corresponds to $\mathcal{Z}(\TY(\Z_2,i,-1,-1))\cong \TY(\Z_2,i,-1,-1)\boxtimes\TY(\Z_2,-i,-1,1)$.
As before, the left factor in these products is generated by the leftmost column and the right factor is generated by the first line in \Cref{tab:simpleobjectsEMGauged}.

\medskip

The orbifold datum gauging the 0-form symmetry is given by the algebra
\begin{equation}
	A_{\Z_2}=\mathds{1}\oplus (c_1\otimes c_1)=\mathds{1}\oplus (\mathds{1}\oplus f)\,,
\end{equation}
which corresponds to the surface $S_\mathds{1}\oplus S_f$.
As before, the Frobenius algebra structure of $A_{-1}=\mathds{1}\oplus f$ depends on the defectification (it is that of a $\Z_2$-group algebra up to the sign~$p$), but the two algebras are Morita-equivalent by \eqref{eq:zestedMoritaEquivalence}.

The induced inclusion of $\Rep(\Z_2)$ (the 1-form symmetry) is then given by $\mathds{1}\boxtimes\overline{\mathds{1}}$ and $\psi\boxtimes\overline{\psi}$, therefore $A_L=\C(\Z_2)\cong\mathds{1}\boxtimes\overline{\mathds{1}}\oplus \psi\boxtimes\overline{\psi}$ is the algebra associated to the 1-form symmetry.

As was calculated in \cite[Sect.\,6.2]{BBDR}, the equivalence between $\mathcal{Z}(\Ising)_{A_L}$ and $\mathcal{D}(\Z_2)^\times_{\Z_2}$ is given by the identification of simple objects in \Cref{tab:simpleobjectsEMExtension}.
\begin{table}[h]
\centering
\begin{tabular}{cc}
$\mathcal{D}(\Z_2)^\times_{\Z_2}$&$\mathcal{Z}(\Ising)_{A_L}$\\
\begin{tabular}[t]{@{}|c|c|@{}} 
		\hline
        $\mathds{1}$&$e$\\\hline 
        $m$&$f$\\\hline 
        $c_1$&$c_2$\\\hline
      \end{tabular}&
      \begin{tabular}[t]{@{}|c|c|@{}} 
		\hline
        $A_L$&$\sigma\boxtimes\overline{\sigma}$\\\hline
        $ (\sigma\boxtimes\overline{\sigma})\otimes (\psi\boxtimes\overline{\psi})$&$\psi\boxtimes\overline{\mathds{1}}\oplus \mathds{1}\boxtimes\overline{\psi}$ \\\hline 
        $\mathds{1}\boxtimes\overline{\sigma}\oplus \psi\boxtimes\overline{\sigma}$&$\sigma\boxtimes\overline{\mathds{1}}\oplus \sigma\boxtimes \overline{\psi}$\\\hline
      \end{tabular} 
\end{tabular}
\caption{Identification of simple objects in $\mathcal{D}(\Z_2)^\times_{\Z_2}$ and $\mathcal{Z}(\Ising)_{A_L}$.}\label{tab:simpleobjectsEMExtension}
\end{table}

\subsection[2-group Symmetry: $\mathcal D(\Z_2)\longleftrightarrow \mathcal D(\Z_3)$]{2-group Symmetry: $\boldsymbol{\mathcal{D}(\Z_2)\longleftrightarrow \mathcal{D}(\Z_3)}$}
\label{sec:2-groupZ2Z3}
In this section, we gauge a split 2-group symmetry in $\mathcal{D}(\Z_2)$. 
The fact that the 2-group is split means it is given by independent 0-form and 1-form symmetries, so the 0-form symmetry acts trivially on the line defects that constitute the 1-form symmetry and the line defects are genuine bulk lines.
This allows us to gauge its two components in succession rather than gauging the entire 2-group at once.
In fact, the 1-form part is the symmetry we gauged in \Cref{sec:Z2-Vect}.
The discussion there applies to any cyclic group~$\Z_n$: By gauging the 0-form $\Z_n$-symmetry associated to the extension $\Z_n$-$\Vect$ we obtain $\mathcal{D}(\Z_n)$.
Here we combine the 1-form $\Z_2$-symmetry in $\mathcal{D}(\Z_2)$ with the 0-form $\Z_3$-symmetry associated to the extension $\Z_3\text{-}\Vect$. 
Overall, this produces a 2-group symmetry with topological defects including twisted sectors given by $\Z_3\text{-}\Vect\boxtimes \mathcal{D}(\Z_2)$. 
\begin{equation}
\begin{tikzcd}[column sep= 94, ampersand replacement=\&, row sep=30]
\&
\hspace{-1pt}
\&
\\\\
	\mathcal{D}(\Z_2)
    \arrow[r, "{\substack{\text{1-form gauging }\widehat{=}\text{ de-eq.}\\A_L^{\Z_2}=B=\C(\Z_2)=\mathds{1}\oplus e}}"{above}]
    \arrow[d,phantom,sloped, "\cong"]
    \&
    (\mathcal{D}(\Z_2))^\text{loc}_{A_L^{\Z_2}}
    \arrow[d,phantom,sloped, "\cong",shift right=3]
    \arrow[uu, phantom, ""{coordinate, name=Z}]
    \&
    \\
     (\Z_2\text{-}\Vect)^{\Z_2}
     \&
     \hphantom{(}\Vect\hphantom{)_{A_L}}
    \arrow[l, "{\substack{A_{\Z_2}=\,\C\oplus\C\\ \hat{\mathcal{A}}_S\cong(\Vect)_{A_{\scaleto{\Z_2}{4pt}}}=\, S_\mathds{1}\boxplus S_\mathds{1}\\\text{0-form gauging }\widehat{=}\text{ eq.}}}"{below}]
    \arrow[r, "{\substack{\text{0-form gauging }\widehat{=}\text{ eq.}\\ \hat{\mathcal{A}}_S\cong(\Vect)_{A_{\scaleto{\Z_3}{4pt}}}=\, S_\mathds{1}\boxplus S_\mathds{1}\boxplus S_\mathds{1}\\ A_{\Z_3}=\,\C\oplus\C\oplus\C}}"{above}]
    \arrow[d,phantom,sloped, "\cong",shift right=3]
    \&
    (\Z_3\text{-}\Vect)^{\Z_3}
    \arrow[d,phantom,sloped, "\cong"]
    \arrow[llu,leftarrow,"\substack{\text{2-group gauging}\\A_{\Z_3\times\Z_2}=(\mathds{1}\oplus \omega\oplus\omega^2)^{\oplus 2}\\\hat{\mathcal{A}}_S=\,S_\omega\oplus S_\omega}"{above,pos=1}, rounded corners, to path={ -- ([yshift=2ex]\tikztostart.north) |- (Z) [near end]\tikztonodes -| ([yshift=2ex]\tikztotarget.north) -- (\tikztotarget)}]
    \\
     \&
    (\mathcal{D}(\Z_3))^\text{loc}_{A_L^{\Z_3}}
    \arrow[dd, phantom, ""{coordinate, name=Y}]
     \&
     \mathcal{D}(\Z_3)
    \arrow[l, "{\substack{A_L^{\Z_3}=B=\C(\Z_3)=\mathds{1}\oplus \omega\oplus\omega^2\\\text{1-form gauging }\widehat{=}\text{ de-eq.}}}"{below}]
        \arrow[llu,"\substack{ A_{\Z_2\times\Z_3}=(\mathds{1}\oplus e)^{\oplus 3}\\ \hat{\mathcal{A}}_S=\,S_e\oplus S_e\oplus S_e\\ \text{2-group gauging}}"{below,pos=1}, rounded corners,from=5-3, to=4-1, to path={ -- ([yshift=-2ex]\tikztostart.south) |- (Y) [near end]\tikztonodes -| ([yshift=-2ex]\tikztotarget.south) -- (\tikztotarget)}]
\\
\\
\&
\hspace{-1pt}
\end{tikzcd}
\label{tab:Z3}
\end{equation}

The orbifold datum which gauges this 2-group symmetry can be calculated by carrying out the procedure described in \Cref{sec:2groupsymm}.
Recall that the surface defects in this theory are all given by condensation, where the 1-form $\Z_2$-symmetry is generated by the algebra $A_L=\mathds{1}\oplus e$ (cf. \eqref{tab:Z2-Vect}) producing the surface defect $S_e$, and the 0-form $\Z_3$-symmetry is given by three copies of $S_\mathds{1}$ (generalising \Cref{sec:Z2-Vect}), coming from the trivial algebra $\mathds{1}$.
To determine the surface defect that underlies the orbifold datum for this 2-group gauging, we need to fuse these two defects.
From the 3-categorical perspective $\mathcal{D}_{\mathcal{D}(\Z_2)}=\mathrm{B}(\Delta\mathrm{ssFrob}(\mathcal{D}(\Z_2)))$, this fusion is the composition of 1-morphisms, i.e.\ the monoidal structure in $\mathcal{D}(\Z_2)$.
This leads us to the surface $S_e\oplus S_e\oplus S_e$ which is the condensation of $A_L\otimes (\mathds{1}\oplus\mathds{1}\oplus\mathds{1})$.

Naturally, undoing this gauging is possible through the same method with $\Z_2$ and $\Z_3=\{1,\omega,\omega^2\}$ exchanged, hence we arrive at the full algebra being $(\mathds{1}\oplus\omega\oplus\omega^2)\otimes (\mathds{1}\oplus\mathds{1})$ giving rise to a surface $S_\omega\oplus S_\omega$.

\medskip

The general question of how to ``compose'' consecutive orbifoldings is not fully understood.
The problem amounts to providing an equivalence $(\mathcal{D}_\text{orb})_\text{orb}\cong \mathcal{D}_\text{orb}$, where $\mathcal{D}_\text{orb}$ is the ``orbifold completion'' of the category of defects $\mathcal{D}$, i.e.\ the (higher) category of all orbifolds (a.k.a.\ gaugings) of the original theory and defects within and between them.
In \cite[Sect.\,4.2.2]{CMcompletion} the authors give the precise answer in 2 dimensions and a conjectural answer in any dimension.

\section{Chern--Simons Theory}
\label{sec:CS}
In this section we briefly look at Chern--Simons theory for the group $\SU(2)$ at level~$k\in\Z_{\geq 0}$.
This theory can be modelled by Reshetikhin--Turaev theory using the category of quantum group representations $\Rep_q(\SU(2))$ at a $2(k+2)$th root of unity.
In this context, the level~$k$ is an obstruction to gauging the central 1-form $\Z_2$-symmetry, first observed in \cite{MSzoo}.
As discussed in e.g. \cite{PS}, the existence of an obstruction depends on the 3-manifold the theory is evaluated on.
However, a TQFT like Reshetikhin--Turaev theory is specified by evaluating it on every (oriented) 3-manifold, so the obstruction in our case ensures gaugeability in any scenario.
The condition for $\SU(2)_k$ was determined to be $k\equiv 0\mod 4$ \cite{MSzoo}.
The following simple calculation recovers this fact in our formalism.

\medskip

We only recall the details of the category $\Rep_q(\SU(2))$ as required, and we refer the interested reader to \cite{Kassel1995} for an introduction to quantum groups.
We always work with $q=\exp(\frac{\pi i}{k+2})$ for simplicity.
Like $\Rep(\SU(2))$, the simple objects of $\Rep_q(\SU(2))$ are associated to integers (but restricted to $0\leq a\leq k$), and their underlying vector spaces are $(a+1)$-dimensional.
However, as objects in $\Rep_q(\SU(2))$, their (quantum) dimensions are given by $[a+1]_q$, defined as
\begin{equation}
	[n]_q:= \frac{q^n-q^{-n}}{q-q^{-1}}\,.
\end{equation}
Fusion rules are given by (cf. \cite[Eq.\,(107)]{AS})
\begin{equation}
	a\otimes b= \bigoplus_{c=|a-b|}^{\mathrm{min}(a+b,2k-a-b)}c\,,
\end{equation}
and the braiding by 
\begin{equation}
	c_{a,b}=\bigoplus_{c=|a-b|}^{\mathrm{min}(a+b,2k-a-b)} (-1)^\frac{a+b-c}{2}q^{\frac{1}{4}(c(c+2)-a(a+2)-b(b+2))}\,\id_c\,.
\end{equation}

Our goal is to identify $\Rep(\Z_2)$ as a full braided fusion subcategory of $\Rep_q(\SU(2))$ which would give us the generator for the 1-form symmetry and the 1-form symmetry functor \eqref{eq:1formFunctor}.
Note that $a\otimes k=k\otimes a=k-a$ for any $0\leq a\leq k$ and in particular, $k\otimes k=0$ is the trivial 1-dimensional representation.
This is the only invertible object aside from $0$, since all other objects have quantum dimension $\neq 1$.
The braiding $c_{k,k}=(-1)^{k}\exp(-2\pi i\frac{k}{4})$ immediately provides the desired constraint:
For $\Rep(\Z_2)$ to be a full braided monoidal subcategory, this braiding has to be trivial $c_{k,k}=\id$ which is true if and only if $k\equiv 0\mod 4$. 
One can similarly calculate the associator from \cite[Eq.\,(110)]{AS} and all components involving only~$0$ and~$k$ are trivial.

\printbibliography

@article{DGNO,
archivePrefix = {arXiv},
arxivId = {0906.0620},
author = {Drinfeld, Vladimir and Gelaki, Shlomo and Nikshych, Dmitri and Ostrik, Victor},
doi = {10.1007/S00029-010-0017-Z},
eprint = {0906.0620}, 
   archivePrefix={arXiv},
   primaryClass={math.QA},
issn = {10221824},
journal = {Selecta Mathematica, New Series},
month = {jun},
number = {1},
pages = {1--119},
title = {{On braided fusion categories I}},
url = {https://arxiv.org/abs/0906.0620v3 http://arxiv.org/abs/0906.0620},
volume = {16},
year = {2010}
}

@article{BN,
author = {Burciu, Sebastian and Natale, Sonia},
doi = {10.1063/1.4774293},
eprint = {1206.6625},
archivePrefix={arXiv},
primaryClass={math.QA},
journal = {Journal of Mathematical Physics},
month = {jun},
number = {1},
title = {{Fusion rules of equivariantizations of fusion categories}},
url = {http://arxiv.org/abs/1206.6625 http://dx.doi.org/10.1063/1.4774293},
volume = {54},
year = {2012}
}

@article{MR,
author = {Mulevi{\v{c}}ius, Vincentas and Runkel, Ingo},
doi = {10.4171/qt/170},
eprint = {2002.00663},
   archivePrefix={arXiv},
   primaryClass={math.QA}, 
issn = {1664073X},
journal = {Quantum Topology},
month = {feb},
number = {3},
pages = {459--523},
publisher = {European Mathematical Society Publishing House},
title = {{Constructing modular categories from orbifold data}},
url = {https://arxiv.org/abs/2002.00663v1},
volume = {13},
year = {2022}
}

@article{Mulevicius2022,
eprint = {2206.02611},
   archivePrefix={arXiv},
   primaryClass={math.QA}, 
author = {Mulevi{\v{c}}ius, Vincentas},
journal = {Theory and Applications of Categories},
pages = {1203--1292},
publisher = {European Mathematical Society Publishing House},
volume = {41},
title = {{Condensation inversion and Witt equivalence via generalised orbifolds}},
url = {https://arxiv.org/abs/2206.02611v1},
year = {2024}
}

@article{CRS3,
eprint = {1809.01483}, 
   archivePrefix={arXiv},
   primaryClass={math.QA}, 
author = {Carqueville, Nils and Runkel, Ingo and Schaumann, Gregor},
issn = {1201561X},
journal = {Theory and Applications of Categories},
month = {sep},
number = {20},
pages = {513--562},
publisher = {Mount Allison University},
title = {{Orbifolds of Reshetikhin--Turaev TQFTs}},
volume = {35},
year = {2020}
}

@article{CRS1,
eprint = {1705.06085},
archivePrefix={arXiv},
   primaryClass={math.QA}, 
author = {Carqueville, Nils and Runkel, Ingo and Schaumann, Gregor},
doi = {10.2140/gt.2019.23.781},
journal = {Geometry and Topology},
month = {may},
number = {2},
pages = {781--864},
publisher = {Mathematical Sciences Publishers},
title = {{Orbifolds of $n$-dimensional defect TQFTs}},
url = {http://arxiv.org/abs/1705.06085 http://dx.doi.org/10.2140/gt.2019.23.781},
volume = {23},
year = {2017}
}

@article{CMcompletion,
eprint={2307.06485},
   archivePrefix={arXiv},
   primaryClass={math.QA}, 
author = {Carqueville, Nils and M{\"{u}}ller, Lukas},
month = {jul},
title = {{Orbifold completion of 3-categories}},
url = {https://arxiv.org/abs/2307.06485v2},
year = {2023}
}

@article{GKSW,
   title={Generalized global symmetries},
   eprint = {1412.5148}, 
   archivePrefix={arXiv},
   primaryClass={hep-th},
   volume={2015},
   ISSN={1029-8479},
   url={http://dx.doi.org/10.1007/JHEP02(2015)172},
   DOI={10.1007/jhep02(2015)172},
   number={2},
   journal={Journal of High Energy Physics},
   publisher={Springer Science and Business Media LLC},
   author={Gaiotto, Davide and Kapustin, Anton and Seiberg, Nathan and Willett, Brian},
   year={2015},
   month=feb 
}

@article{carqueville2023orbifoldstopologicalquantumfield,
      title={Orbifolds of topological quantum field theories}, 
      author={Nils Carqueville},
      eprint={2307.16674},
      archivePrefix={arXiv},
      primaryClass={math-ph},
      url={https://arxiv.org/abs/2307.16674}, 
      DOI={10.1016/B978-0-323-95703-8.00008-2},
   number={3},
   journal={Encyclopedia of Mathematical Physics (Second Edition)},
   year={2025},
   pages={618--634}
}

@article{ENO,
      title={Fusion categories and homotopy theory}, 
      author={Pavel Etingof and Dmitri Nikshych and Victor Ostrik and with an appendix by Ehud Meir},
      DOI={10.4171/QT/6},
   number={3},
   volume = {1},
   journal={Quantum Topology},
   year={2010},
   pages={209--273}, 
      eprint={0909.3140},
      archivePrefix={arXiv},
      primaryClass={math.QA},
      shorthand = {ENO}
}

@article{Ostrik,
    author = "Ostrik, Viktor",
    title = "{Module categories, weak Hopf algebras and modular invariants}",
    eprint = "math/0111139",
    archivePrefix = "arXiv",
    doi = "10.1007/s00031-003-0515-6",
    journal = "Transform. Groups",
    volume = "8",
    number = "2",
    pages = "177--206",
    year = "2003"
}

@article{HPRW,
      title={Generalised Orbifolds and G-equivariantisation}, 
      author={Sebastian Heinrich and Julia Plavnik and Ingo Runkel and Abigail Watkins},
      year={2025},
      eprint={2506.08154},
      archivePrefix={arXiv},
      primaryClass={math.QA},
      url={https://arxiv.org/abs/2506.08154}, 
}

@article{CH,
 title={2-Group Symmetries of 3-dimensional Defect TQFTs and Their Gauging}, 
      author={Nils Carqueville and Benjamin Haake},
      year={2025},
      eprint={2506.08178},
      archivePrefix={arXiv},
      primaryClass={math.QA},
      url={https://arxiv.org/abs/2506.08178}, 
}

@article{BBDR,
      title={Gauging Non-Invertible Symmetries in (2+1)d Topological Orders}, 
      author={Mahesh K. N. Balasubramanian and Matthew Buican and Clement Delcamp and Rajath Radhakrishnan},
      year={2025},
      eprint={2507.01142},
      archivePrefix={arXiv},
      primaryClass={hep-th},
      url={https://arxiv.org/abs/2507.01142}, 
}

@article{BBCW,
   title={Symmetry fractionalization, defects, and gauging of topological phases},
   volume={100},
   ISSN={2469-9969},
   url={http://dx.doi.org/10.1103/PhysRevB.100.115147},
   DOI={10.1103/physrevb.100.115147},
   number={11},
   journal={Physical Review B},
   publisher={American Physical Society (APS)},
   author={Barkeshli, Maissam and Bonderson, Parsa and Cheng, Meng and Wang, Zhenghan},
   year={2019},
   month=sep }

@article{MSzoo,
    author = "Moore, Gregory W. and Seiberg, Nathan",
    title = "{Taming the Conformal Zoo}",
    reportNumber = "IASSNS-HEP-89/6",
    doi = "10.1016/0370-2693(89)90897-6",
    journal = "Phys. Lett. B",
    volume = "220",
    pages = "422--430",
    year = "1989"
}

@article{GJF,
      title={Condensations in higher categories}, 
      author={Davide Gaiotto and Theo Johnson-Freyd},
      year={2025},
      eprint={1905.09566},
      archivePrefix={arXiv},
      primaryClass={math.CT},
      url={https://arxiv.org/abs/1905.09566}, 
}

@misc{BBG1,
      title={Higher representations for extended operators}, 
      author={Thomas Bartsch and Mathew Bullimore and Andrea Grigoletto},
      year={2023},
      eprint={2304.03789},
      archivePrefix={arXiv},
      primaryClass={hep-th},
      url={https://arxiv.org/abs/2304.03789}, 
}

@article{CMRSS2024,
   title={Reshetikhin–Turaev TQFTs Close Under Generalised Orbifolds},
   volume={405},
   ISSN={1432-0916},
   url={http://dx.doi.org/10.1007/s00220-024-05068-6},
   DOI={10.1007/s00220-024-05068-6},
   number={10},
   journal={Communications in Mathematical Physics},
   publisher={Springer Science and Business Media LLC},
   author={Carqueville, Nils and Mulevičius, Vincentas and Runkel, Ingo and Schaumann, Gregor and Scherl, Daniel},
   year={2024},
   month=sep }

@article{DGPRZ,
   title={G‐crossed braided zesting},
   volume={109},
   ISSN={1469-7750},
   url={http://dx.doi.org/10.1112/jlms.12816},
   DOI={10.1112/jlms.12816},
   number={1},
   journal={Journal of the London Mathematical Society},
   publisher={Wiley},
   author={Delaney, Colleen and Galindo, César and Plavnik, Julia and Rowell, Eric C. and Zhang, Qing},
   year={2023},
   month=oct }

@misc{DGPRZcondensedfiber,
      title={The Condensed Fiber Product and Zesting}, 
      author={Colleen Delaney and César Galindo and Julia Plavnik and Eric C. Rowell and Qing Zhang},
      year={2024},
      eprint={2410.09025},
      archivePrefix={arXiv},
      primaryClass={math.QA},
      url={https://arxiv.org/abs/2410.09025}, 
}

@article{DGPRZbraided,
   title={Braided Zesting and Its Applications},
   volume={386},
   ISSN={1432-0916},
   url={http://dx.doi.org/10.1007/s00220-021-04002-4},
   DOI={10.1007/s00220-021-04002-4},
   number={1},
   journal={Communications in Mathematical Physics},
   publisher={Springer Science and Business Media LLC},
   author={Delaney, Colleen and Galindo, César and Plavnik, Julia and Rowell, Eric C. and Zhang, Qing},
   year={2021},
   month=July, pages={1–55} }

@article{GNN,
      title={Centers of graded fusion categories}, 
      author={Shlomo Gelaki and Deepak Naidu and Dmitri Nikshych},
      year={2009},
      eprint={0905.3117},
      archivePrefix={arXiv},
      primaryClass={math.QA},
      url={https://arxiv.org/abs/0905.3117}, 
      journal={Algebra \& Number Theory},
      number={3},
      pages={959--990},
      DOI={10.2140/ant.2009.3.959},
      url={https://doi.org/10.2140/ant.2009.3.959},
}

@article{Siehler,
   title={Near-group categories},
   volume={3},
   ISSN={1472-2747},
   url={http://dx.doi.org/10.2140/agt.2003.3.719},
   DOI={10.2140/agt.2003.3.719},
   number={2},
   journal={Algebraic \& Geometric Topology},
   publisher={Mathematical Sciences Publishers},
   author={Siehler, Jacob},
   year={2003},
   month=aug, pages={719–775} }

@article{GLM,
      title={Computing $G$-Crossed Extensions and Orbifolds of Vertex Operator Algebras}, 
      author={César Galindo and Simon Lentner and Sven Möller},
      year={2024},
      eprint={2409.16357},
      archivePrefix={arXiv},
      primaryClass={math.QA},
      url={https://arxiv.org/abs/2409.16357}, 
}

@article{CGPZ,
   title={On Gauging Symmetry of Modular Categories},
   volume={348},
   ISSN={1432-0916},
   url={http://dx.doi.org/10.1007/s00220-016-2633-8},
   DOI={10.1007/s00220-016-2633-8},
   number={3},
   journal={Communications in Mathematical Physics},
   publisher={Springer Science and Business Media LLC},
   author={Cui, Shawn X. and Galindo, César and Plavnik, Julia Yael and Wang, Zhenghan},
   year={2016},
   month=may, pages={1043–1064} }

@article{AS,
   title={Clebsch–Gordan and 6j-coefficients for rank 2 quantum groups},
   volume={43},
   ISSN={1751-8121},
   url={http://dx.doi.org/10.1088/1751-8113/43/39/395205},
   DOI={10.1088/1751-8113/43/39/395205},
   number={39},
   journal={Journal of Physics A: Mathematical and Theoretical},
   publisher={IOP Publishing},
   author={Ardonne, Eddy and Slingerland, Joost},
   year={2010},
   month=aug, pages={395205} }

@article{PS,
   title={Decomposition in Chern–Simons theories in three dimensions},
   volume={37},
   ISSN={1793-656X},
   url={http://dx.doi.org/10.1142/S0217751X2250227X},
   DOI={10.1142/s0217751x2250227x},
   number={36},
   journal={International Journal of Modern Physics A},
   publisher={World Scientific Pub Co Pte Ltd},
   author={Pantev, Tony and Sharpe, Eric},
   year={2022},
   month=dec }

@misc{SW,
      title={Extended Homotopy Quantum Field Theories and their Orbifoldization}, 
      author={Christoph Schweigert and Lukas Woike},
      year={2019},
      eprint={1802.08512},
      archivePrefix={arXiv},
      primaryClass={math.QA},
      url={https://arxiv.org/abs/1802.08512}, 
}

@book{Kassel1995,
   author = {Christian Kassel},
   city = {New York, NY},
   doi = {10.1007/978-1-4612-0783-2},
   isbn = {978-1-4612-6900-7},
   publisher = {Springer New York},
   title = {Quantum Groups},
   year = {1995}
}

@article{KZZ,
      title={Higher condensation theory}, 
      author={Liang Kong and Zhi-Hao Zhang and Jiaheng Zhao and Hao Zheng},
      year={2025},
      eprint={2403.07813},
      archivePrefix={arXiv},
      primaryClass={cond-mat.str-el},
      url={https://arxiv.org/abs/2403.07813}, 
}

@article{DecoppetFus2,
   title={The Morita Theory of Fusion 2-Categories},
   volume={7},
   url={http://dx.doi.org/10.21136/HS.2023.07},
   DOI={10.21136/hs.2023.07},
   number={1},
   journal={Higher Structures},
   publisher={Institute of Mathematics, Czech Academy of Sciences},
   author={Décoppet, Thibault D.},
   year={2023},
   month=May, pages={234–292} }
\end{document}